\documentclass[floatfix,twocolumn,showpacs,preprintnumbers,amsmath,amssymb,pra,superscriptaddress,longbibliography]{revtex4-1}
\usepackage{color}
\usepackage[usenames,dvipsnames,svgnames,table]{xcolor}
\usepackage[colorlinks=true,linkcolor=blue,urlcolor=blue,citecolor=blue]{hyperref}
\usepackage{siunitx}
\usepackage{mathtools}
\usepackage{graphicx}
\usepackage{dcolumn}
\usepackage{array}
\usepackage{lipsum}
\usepackage{bm}
\usepackage{subfigure}
\usepackage{amssymb}
\usepackage{multirow}
\usepackage{tabularx}
\usepackage{amsmath}
\usepackage{braket}
\usepackage{ulem}
\usepackage{verbatim}
\usepackage{epstopdf}
\graphicspath{{plots/}}
 \usepackage{lipsum}
\usepackage{mathrsfs}

\DeclareSIUnit{\angstrom}{\textup{\AA}}
\DeclareSIUnit\bar{bar}
\DeclareMathOperator*{\SumInt}{%
\mathchoice%
  {\ooalign{$\displaystyle\sum$\cr\hidewidth$\displaystyle\int$\hidewidth\cr}}
  {\ooalign{\raisebox{.14\height}{\scalebox{.7}{$\textstyle\sum$}}\cr\hidewidth$\textstyle\int$\hidewidth\cr}}
  {\ooalign{\raisebox{.2\height}{\scalebox{.6}{$\scriptstyle\sum$}}\cr$\scriptstyle\int$\cr}}
  {\ooalign{\raisebox{.2\height}{\scalebox{.6}{$\scriptstyle\sum$}}\cr$\scriptstyle\int$\cr}}
}
\newcommand{\beq}{\begin{equation}}
\newcommand{\eeq}{\end{equation}}
\newcommand{\bea}{\begin{eqnarray}}
\newcommand{\eea}{\end{eqnarray}}

\newcommand{\qv}{{\bf q}}

\renewcommand{\vec}[1]{\mathbf{#1}}

\begin{document}
%\underline{Draft version:}
\title{
%\underline{Work in progress:} \textit{Ab initio} 
First principles simulations of dense hydrogen
}

\author{Michael Bonitz}
\email{bonitz@theo-physik.uni-kiel.de}
\affiliation{Institut f\"ur Theoretische Physik und Astrophysik, Christian-Albrechts-Universit\"at zu Kiel, D-24098 Kiel, Germany}
\affiliation{Kiel Nano, Surface and Interface Science KiNSIS, Christian-Albrechts Universität Kiel, D-24098 Kiel, Germany}

\author{Jan Vorberger}
\affiliation{Helmholtz-Zentrum Dresden-Rossendorf (HZDR), D-01328 Dresden, Germany}

\author{Mandy Bethkenhagen}
\affiliation{Ecole Polytechnique, Laboratoire Pour l'Utilisation des Lasers Intenses (LULI), France}

\author{Maximilian~P.~B\"ohme}
\affiliation{Center for Advanced Systems Understanding (CASUS), D-02826 G\"orlitz, Germany}
\affiliation{Helmholtz-Zentrum Dresden-Rossendorf (HZDR), D-01328 Dresden, Germany}
\affiliation{Technische  Universit\"at  Dresden,  D-01062  Dresden,  Germany}

\author{David M. Ceperley}
\affiliation{Department of Physics, University of Illinois at Urbana-Champaign, Urbana, Illinois 61801, USA}

\author{Alexey Filinov}
\affiliation{Institut f\"ur Theoretische Physik und Astrophysik, Christian-Albrechts-Universit\"at zu Kiel, D-24098 Kiel, Germany}

\author{Thomas Gawne}
\affiliation{Center for Advanced Systems Understanding (CASUS), D-02826 G\"orlitz, Germany}
\affiliation{Helmholtz-Zentrum Dresden-Rossendorf (HZDR), D-01328 Dresden, Germany}

\author{Frank Graziani}
\affiliation{Lawrence Livermore National Laboratory, Livermore, California, USA}

\author{Gianluca Gregori}
\affiliation{Department of Physics, University of Oxford, Oxford OX1 3PU, UK}

\author{Paul Hamann}
\affiliation{Institut f\"ur Theoretische Physik und Astrophysik, Christian-Albrechts-Universit\"at zu Kiel, D-24098 Kiel, Germany}

\author{Stephanie B. Hansen}
\affiliation{Sandia National Laboratories, Albuquerque, NM 87185, USA}

\author{Markus Holzmann}
\affiliation{Univ.~Grenoble Alpes, CNRS, LPMMC, 38000 Grenoble, France}

\author{S.X. Hu}
\affiliation{Laboratory for Laser Energetics, University of Rochester, Rochester, New York 14623-1299, USA}
\affiliation{Department of Physics and Astronomy, University of Rochester, Rochester, New York 14627, USA}
\affiliation{Department of Mechanical Engineering, University of Rochester, Rochester, New York 14627, USA}

\author{Hanno Kählert}
\affiliation{Institut f\"ur Theoretische Physik und Astrophysik, Christian-Albrechts-Universit\"at zu Kiel, D-24098 Kiel, Germany}
\affiliation{Kiel Nano, Surface and Interface Science KiNSIS, Christian-Albrechts Universität Kiel, D-24098 Kiel, Germany}

\author{Valentin Karasiev}
\affiliation{Laboratory for Laser Energetics, University of Rochester, Rochester, New York 14623-1299, USA}

\author{Uwe Kleinschmidt}
\affiliation{Institut f\"ur Physik, Universit\"at Rostock, D-18051 Rostock, Germany}

\author{Linda Kordts}
\affiliation{Institut f\"ur Theoretische Physik und Astrophysik, Christian-Albrechts-Universit\"at zu Kiel, D-24098 Kiel, Germany}

\author{Christopher Makait}
\affiliation{Institut f\"ur Theoretische Physik und Astrophysik, Christian-Albrechts-Universit\"at zu Kiel, D-24098 Kiel, Germany}

\author{Burkhard Militzer}
\affiliation{Department of Earth and Planetary Science, Department of Astronomy, University of California, Berkeley, California 94720, USA}

\author{Zhandos A.~Moldabekov}
\affiliation{Center for Advanced Systems Understanding (CASUS), D-02826 G\"orlitz, Germany}
\affiliation{Helmholtz-Zentrum Dresden-Rossendorf (HZDR), D-01328 Dresden, Germany}

\author{Carlo Pierleoni}
\affiliation{Department of Physical and Chemical Sciences, University of L’Aquila, Via Vetoio 10, I-67010 L’Aquila, Italy}

\author{Martin Preising}
\affiliation{Institut f\"ur Physik, Universit\"at Rostock, D-18051 Rostock, Germany}

\author{Kushal~Ramakrishna}
\affiliation{Center for Advanced Systems Understanding (CASUS), D-02826 G\"orlitz, Germany}
\affiliation{Helmholtz-Zentrum Dresden-Rossendorf (HZDR), D-01328 Dresden, Germany}

\author{Ronald Redmer}
\affiliation{Institut f\"ur Physik, Universit\"at Rostock, D-18051 Rostock, Germany}

\author{Sebastian Schwalbe}
\affiliation{Center for Advanced Systems Understanding (CASUS), D-02826 G\"orlitz, Germany}
\affiliation{Helmholtz-Zentrum Dresden-Rossendorf (HZDR), D-01328 Dresden, Germany}

\author{Pontus Svensson}
\affiliation{Department of Physics, University of Oxford, Oxford OX1 3PU, UK}

\author{Tobias Dornheim}
\email{t.dornheim@hzdr.de}
\affiliation{Center for Advanced Systems Understanding (CASUS), D-02826 G\"orlitz, Germany}
\affiliation{Helmholtz-Zentrum Dresden-Rossendorf (HZDR), D-01328 Dresden, Germany}

\begin{abstract}
Accurate knowledge of the properties of hydrogen at high compression is crucial for astrophysics (e.g. planetary and stellar interiors, brown dwarfs, atmosphere of compact stars) and laboratory experiments, including inertial confinement fusion. 
There exists 
experimental data for the equation of state, conductivity, and Thomson scattering spectra. However, the analysis of the measurements at extreme pressures and temperatures typically involves additional model assumptions, which makes it difficult to assess the accuracy of the experimental data.  rigorously. 
On the other hand, theory and modeling have produced extensive collections of data. They originate from a very large variety of models and simulations including path integral Monte Carlo (PIMC) simulations, density functional theory (DFT), chemical models, machine-learned models, and combinations thereof. At the same time, each of these methods has fundamental limitations (fermion sign problem in PIMC, approximate exchange-correlation functionals of DFT, inconsistent interaction energy contributions in chemical models, etc.), so for some parameter ranges accurate predictions are difficult. Recently, a number of breakthroughs in first principle PIMC and DFT simulations were achieved which are discussed in this review.
Here we use these results to benchmark different simulation methods.
We present an update of the hydrogen phase diagram at high pressures, the expected phase transitions, and thermodynamic properties including the equation of state and momentum distribution. Furthermore, we discuss available dynamic results for warm dense hydrogen, including the conductivity, dynamic structure factor, plasmon dispersion, imaginary-time structure, and density response functions. We conclude by outlining strategies to combine different simulations to achieve accurate theoretical predictions.
\end{abstract}

\maketitle

\tableofcontents

\section{Introduction\label{sec:introduction}}

Hydrogen was the first element that formed after the Big Bang and it has remained the most abundant species in the Universe. Its properties therefore determine the structure and evolution of astrophysical objects including stars, planets, and interstellar gas clouds. These objects exist in a huge range of temperatures -- spanning from a few Kelvin to billions of Kelvin -- and pressures that extend from zero to trillions of atmospheres in neutron stars. Astrophysical observations have revealed extensive though indirect information about the properties of hydrogen. For example, observations of the oscillations of the sun's surface are employed to constrain its interior properties (``helioseismology''), e.g.~\cite{gough_sp_85,christensen-calsgaard_nat_85,gough_sc_96}. More recently, a major source of knowledge has come from laboratory experiments that reach ever higher pressures. Examples include static compression using diamond anvil cells (DAC), e.g.~\cite{hemley_prl_88}, dynamic compression using shock waves that are generated by impactors \cite{nellis_prl_92, weir_prl_96}, explosives, e.g. \cite{grigoriev_jeptl_72,fortov_prl_07} or high-intensity lasers \cite{dasilva_prl_97}. Extracting data from these measurements and translating them into static, dynamic, or optical properties of hydrogen can be challenging. The accuracy of the  diagnostics is usually severely limited due to the short observation time and relaxation processes, unknown intrinsic properties of the equipment, and so forth. These gaps in knowledge are often bridged with simple models for the equation of state or hydrodynamics the validity of which may be questionable. All of this renders the accuracy of many measurements unclear. Moreover, high-pressure experiments are challenging and costly, and available only at a limited number of laboratories.
Therefore, there is a very high demand for theoretical analysis and computer experiments that would allow for an improved interpretation of experimental data but also independent and reliable predictions.% of novel experiments.

Aside from experiments that focus on basic science aspects of the properties of hydrogen at high pressure, there is a rapidly increasing interest in technological applications of hydrogen. This includes inertial confinement fusion (ICF) \cite{Nature_ICF_1972, Lindl_PoP_1994, Craxton_PoP_2015} and the use of hydrogen in green energy applications. 
In particular, the ignition campaign at the NIF has recently reported major breakthroughs. Remarkably, the fusion gain exceeded unity (not accounting for wasted energy)~\cite{icf-collab_prl_22,icf-collab_prl_24,hurricane_prl_24,pak-icf_pre_24}.  
Technological advances, in particular, ICF, rely heavily on theoretical modeling that allows one to understand the relevant physical processes, make reliable predictions, and enable the optimization of experiments.
Moreover, the direct measurement even of basic parameters such as the temperature or density is often not possible; instead, they have to be inferred indirectly from other observations, which, in turn, requires theoretical results for different observables~\cite{Dornheim2023}.

The theory and modeling of dense hydrogen has a long history. The hydrogen atom and molecule were the focus of the early developments of quantum mechanics, and hydrogen was the first many-body system for which the consequences of the Fermi statistics of the electrons and the Pauli principle were explored by R.H.~Fowler~\cite{fowler_mnras_26}. Pressure ionization of atoms was already predicted by F.~Hund~\cite{hund_36}. At about the same time, E.~Wigner and H.B.~Huntington predicted that atomic hydrogen under high pressure would be metallic \cite{wigner-huntington_35}. Since then the questions about the liquid and solid phases of hydrogen as well as of transitions between insulating, conducting, and, possibly, a superconducting phase predicted by N.W.~Ashcroft \cite{ashcroft_prl_68}, have been a driver of both experimental and theoretical progress. There have been many reports of the experimental observation of hydrogen metallization at room temperature, see Refs.~\cite{hemley_science_89,hemley_science_90,loubeyre_nat_20} for example but there is no consensus that metallization has been reproducibly observed.

On the other hand, it is well accepted that the reverberating shock experiments by S.T.~Weir \textit{et al.}~\cite{We96} achieved the metallization of hydrogen at {\it high} temperature and pressure. The conditions are broadly consistent with those in the interior of Jupiter. There is indeed strong evidence that a thick layer of metallic hydrogen exists in the planet's interior because Jupiter has a strong magnetic field, even though the precise location of the dynamo active layer is still being debated~\cite{tsang2020characterising}. 

\begin{figure}[h]\centering
\includegraphics[width=0.51\textwidth]{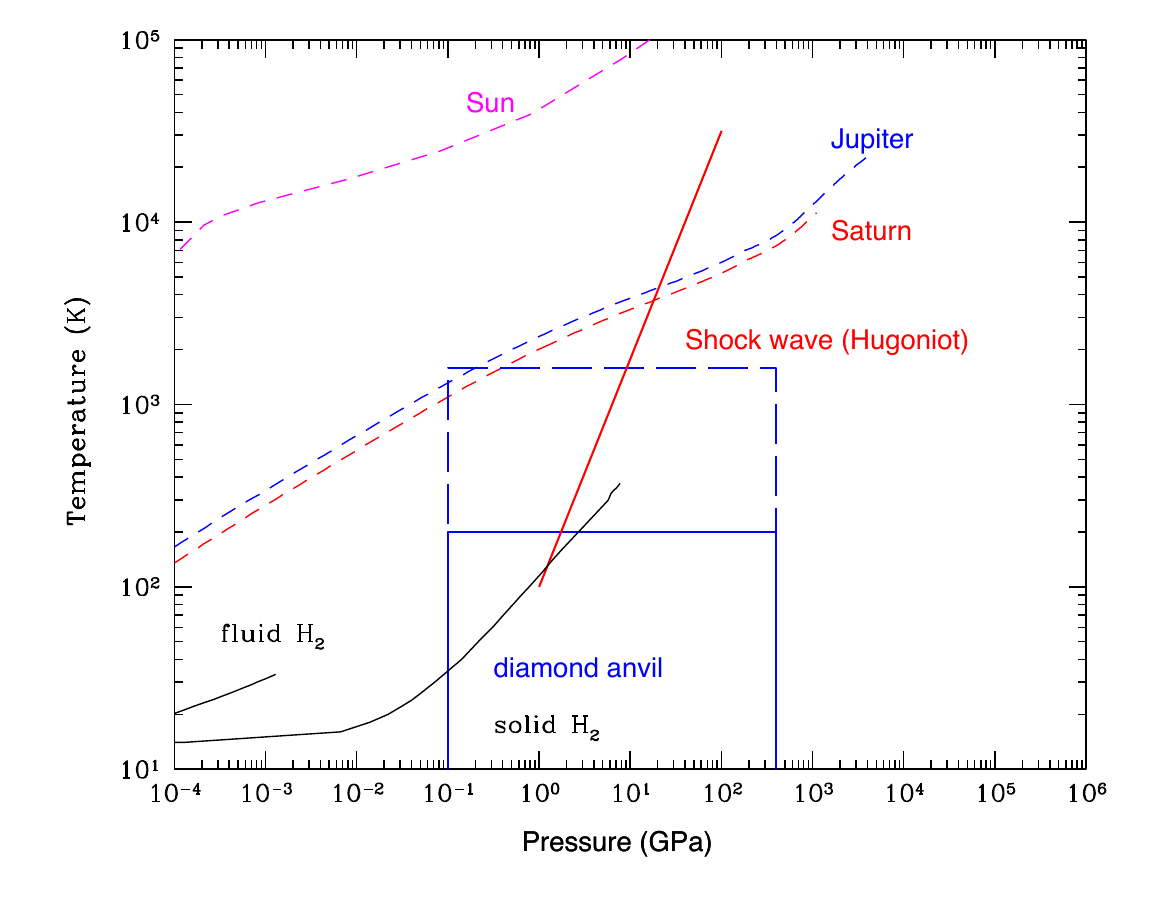}
\caption{ Regions of temperatures and pressures accessed by experiments. Solid blue lines:  static Diamond anvil cell (DAC), dashed blue lines: dynamic DAC. The solid red line is the Hugoniot starting from cold solid hydrogen at ambient pressure. Dashed lines are the conditions at the interiors of Jupiter, Saturn, and the Sun. }

\label{fig:h-phase-diag-exp} 
\end{figure}

The equation of state of hydrogen (and helium) is of crucial importance~\cite{MilitzerJGR2016,Helled2020b} when measurements of Jupiter's and Saturn's gravity fields by orbiting spacecrafts like {\it Juno} and {\it Cassini} are interpreted~\cite{Durante2020,Iess2019}. Interior models typically start from the isentropic pressure-density relationship for hydrogen (see Fig.~\ref{fig:h-phase-diag-exp}) before helium and heavier elements are introduced. The gravity measurements have shown that Jupiter's interior cannot be perfectly homogeneous and almost all models have introduced a step in composition into the planet's interior, which requires a physical justification. Earlier models thus invoked a first-order plasma phase transition (PPT) in hydrogen~\cite{saumon-apj-04} until first-principles computer simulations could not confirm its existence and instead predicted a liquid-liquid phase transition (LLPT)~\cite{scandolo_pnas_03,Morales2010,Lorenzen2010}. However, the latter is believed to occur at too low temperatures to matter for Jupiter's interior. Modern Jupiter models now invoke the phase separation of hydrogen-helium mixtures~\cite{Morales2009,Lorenzen2009,Schoettler2018} or a dilute core~\cite{Wahl2017a,DiluteCore} to justify an inhomogeneous interior structure. Still, a major challenge in understanding Jupiter's interior is remaining. Measurements of the planet's atmosphere have revealed a heavy element abundance of $\sim$$4\%$. If one assumes these measurements to represent the bulk of the planet and if one invokes common EOSs for hydrogen and helium~\cite{saumon_aps_95,MH13,Chabrier2019}, then gravity measurements of the {\it Juno} spacecraft cannot be reproduced~\cite{HubbardMilitzer2016,Helled_2022,Militzer_2023}. Various hypotheses have been proposed to reconcile this discrepancy, which include lowering the density of hydrogen~\cite{Howard2023}, increasing the planet's interior temperature~\cite{Miguel2022}, or reducing the deep abundance of the heavy elements~\cite{Debras2021}. 
Besides the equation of state, further thermodynamic (e.g.\ compressibility, Gr\"uneisen parameter) and transport properties (e.g.\ electrical and thermal conductivity, viscosity) are required to model the thermal evolution (cooling) and dynamo processes (magnetic field generation) in giant planets like Jupiter~and Saturn; for recent work, see~\cite{french_ApJS_2012,preising_ApJS_2023}.

The relevance to planetary science makes it important to characterize how hydrogen transitions from a molecular, insulating state to an atomic and conducting state. As noted above, an early conjecture was the plasma phase transition (PPT), predicted by K.H.~Schramm~\cite{schramm_zphys_61} and later by G.~Norman and A.~Starostin \cite{norman-starostin_68}. Even though no longer favored today, this issue has led to considerable development in theory and simulations. The first models were based on the thermodynamics of chemically reacting gases applied to many-particle systems of electrons, protons, atoms, and molecules (``chemical models''), e.g., Ref.~\cite{norman-starostin_68}. They were systematically extended to take into account interactions between all constituents (``nonideality effects'') by W.~Ebeling and co-workers \cite{ebeling_ap_73,ebeling-richert_85}, D.~Mihalas \textit{et al.} \cite{hummer-mihalas_88,mihalas2_aj_88,daeppen_aj_88},  G.~Chabrier and co-workers \cite{saumon1,saumon-chabrier_92,saumon_aps_95}, and the Rostock school around D.~Kremp, W.D.~Kraeft, M.~Schlanges, T.~Bornath,  M.~Bonitz \cite{schlanges-etal.95cpp,bezkrovny_pre_4}, G.~Röpke, and R.~Redmer and co-workers \cite{reinholz_PRE_95,juranek_jcp_00,juranek_jcp_02}. This has led to important advances in the many-body theory of Coulomb systems in the framework of kinetic theory and the theory of fluctuations~\cite{klimontovich_1982}, Statistical physics \cite{klimontovich_stat-phys, IIT,fortov-book}, nonequilibrium Green functions \cite{red-book,green-book}, density operator theory \cite{bonitz_qkt} and linear response theory \cite{zubarev-book1,zubarev-book2}.

Chemical models rely on effective interaction potentials between the considered ``species''. Their density and temperature dependence are usually treated within simple approximations such as static screening and perturbation theory. 
With the emergence of modern computers, many-body simulations such as quantum Monte Carlo~\cite{Foulkes_RMP_2001} (QMC) and density functional theory~\cite{Jones_RMP_2015} (DFT) became feasible. These methods are based on the fundamental properties of electrons and nuclei and work in the ``physical picture''. They avoid any artificial classification into different chemical species. QMC simulations of hydrogen and helium, based on Feynman's imaginary-time path integral representation of quantum mechanics (PIMC) were pioneered by L.D.~Fosdick and H.F.~Jordan \cite{Fosdick_PR_1966,Jordan_PR_1968} and
V.~Filinov \textit{et al.} \cite{filinov-norman_75,zamalin_77}. However, when applied to fermions, these simulations, even though being free of systematic errors, are severely hampered by the fermion sign problem~\cite{troyer,Dornheim2019a,Loh_PRB_1990} (FSP) confining them to regions of low % moderate 
electron degeneracy. To address the FSP, D.M.~Ceperley introduced the restricted PIMC (RPIMC) method~\cite{Ceperley1991}, which is completely sign-problem free, as long as diagonal density matrix elements are evaluated, which is sufficient to compute the internal energy, pressure, pair correlation functions, and structure factor. However, some negative contributions need to be included when the momentum distribution is computed~\cite{Militzer2019}. While being formally exact, RPIMC requires information about the nodal structure of the thermal density matrix, which has to be approximated in practice.
This has allowed D.M.~Ceperley, B.~Militzer, and co-workers to perform simulations for temperatures where hydrogen is dominated by atoms and molecules~\cite{Pierleoni1994,magro_prl_96,PhysRevB.55.R11907,militzer_pre_00,militzer1}.

The second approach within the ``physical picture'' is Kohn-Sham DFT coupled to molecular dynamics for the ions (DFT-MD or Born-Oppenheimer MD, BOMD, also known as \textit{ab initio} MD or AIMD) which have dramatically extended the scope of problems that can be studied while yielding a higher accuracy compared to simpler models that require the introduction of assumptions specific to hydrogen. While DFT-MD and RPIMC methods rely on some approximations, these do not depend on a particular material. DFT-MD simulations of hot, dense hydrogen were performed by T.~Lenosky~\cite{PhysRevB.55.R11907}, S.~Scandolo \cite{scandolo_pnas_03}, M.~Desjarlais \cite{desjarlais_prb_03} and S.A.~Bonev \textit{et al.}~\cite{Bonev2004}.  

\begin{figure*}[ht]
    \centering
    \begin{minipage}[t]{0.45\textwidth}
     \centering
     \includegraphics[width=\textwidth]{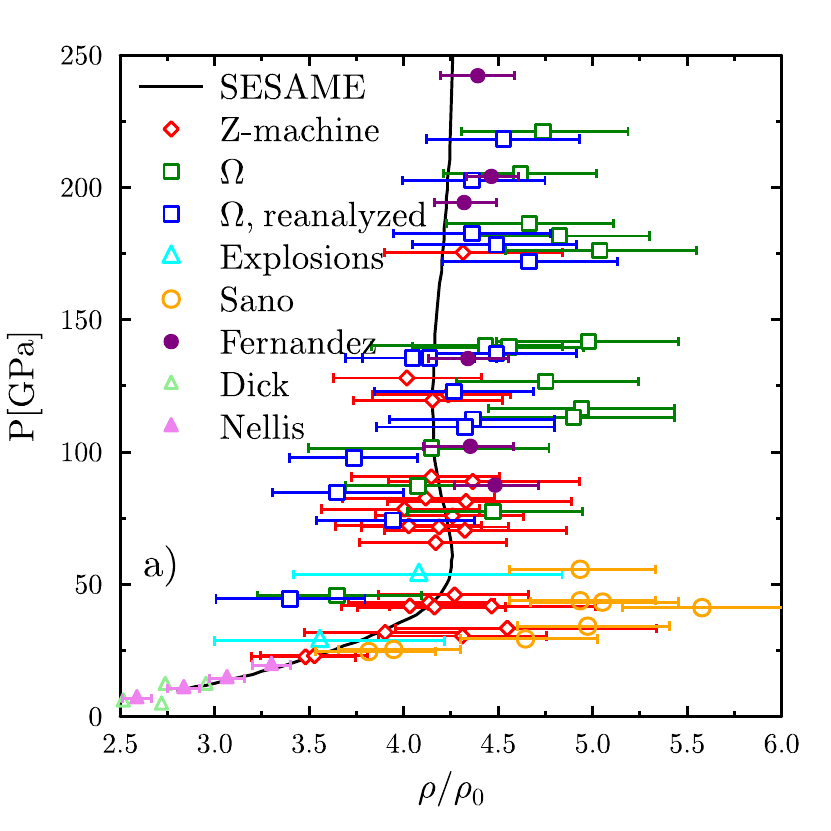}
    \end{minipage}
    \begin{minipage}[t]{0.45\textwidth}
     \centering
     \includegraphics[width=\textwidth]{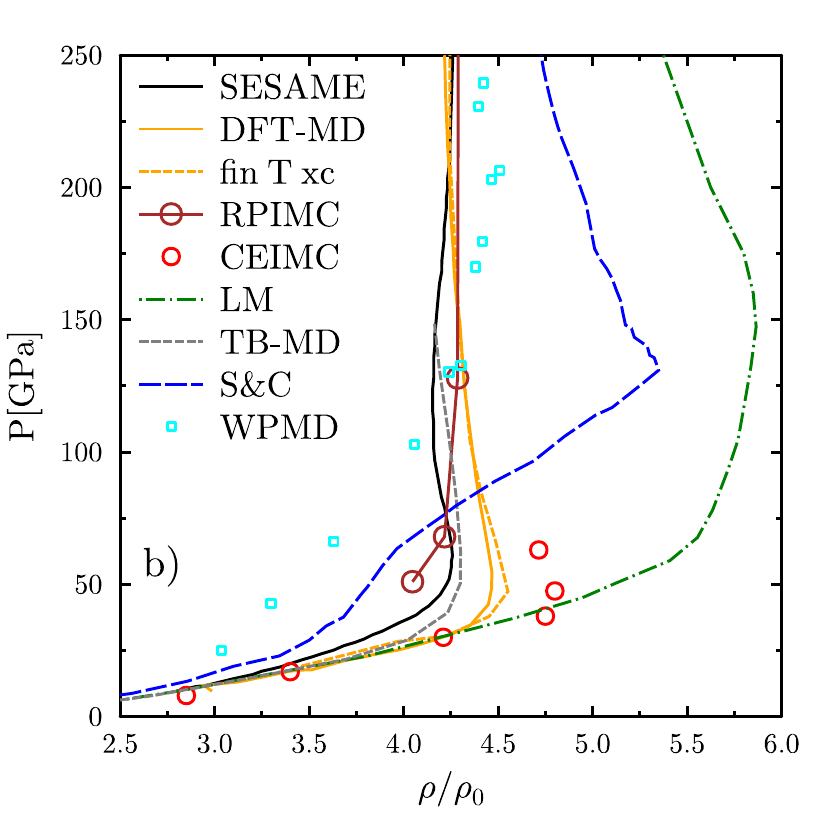}
    \end{minipage}
    \medskip
    \begin{minipage}[t]{\textwidth}
    \begin{minipage}[t]{0.45\textwidth}
     \centering
     \includegraphics[width=\textwidth]{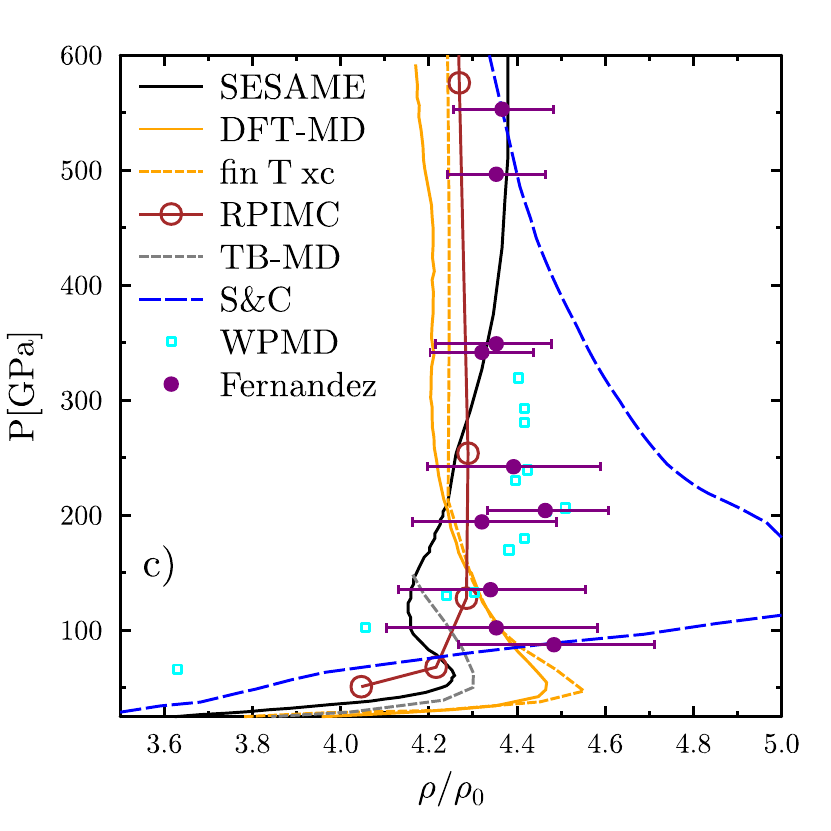}
    \end{minipage}
    \begin{minipage}[b]{.45\textwidth}
    \caption{Pressure versus compression along the deuterium Hugoniot. 
%    \jan{
    \textbf{Panel a)}: Experimental results up to \SI{250}{\giga\Pa}:   
%    NOVA laser (grey filled circles)~\cite{Collins_1998};
    Z-machine (red diamonds)~\cite{Knudson_2004}; $\Omega$ laser shocks (green and blue squares)~\cite{Hicks_2009,Knudson_2009}; explosions (cyan triangles)~\cite{Boriskov_2005}; Sano laser shock (orange circles)~\cite{Sano_2011}; Fernandez laser shock~\cite{fernandez_prl_19}; Nellis gas gun~\cite{Nellis_1983}; Dick gas gun~\cite{Dick_1980}. \textbf{Panel b)}: Theory results up to \SI{250}{\giga\Pa}: SESAME (solid black line)~\cite{sesame}; DFT-MD (solid orange)~\cite{Holst_PRB_2008,Caillabet_2011,Wang_2013}; DFT-MD including finite temperature xc functionals (dashed orange line)~\cite{Karasiev_hug_2019}; RPIMC (brown circles, solid line)~\cite{militzer_prl_00}; CEIMC (red circles)~\cite{Tubman_2015,Ruggeri2020}; tight binding MD (grey dashed line)~\cite{lenosky_prb_97}; linear mixing model (green dash-dotted line)~\cite{ross_prb_98}, Saumon \textit{et al.} (S\&C, blue dashed)~\cite{saumon_aps_95}; WPMD~(blue dots)~\cite{Lavrinenko_2021}. 
    \textbf{Panel c)}: high pressure regime up to \SI{600}{\giga\Pa}: curves as introduced in panels a) and b) (note the different scales).
    Notice that experiments do not measure temperature directly which gives rise to additional uncertainties when comparing to simulations.}
    \label{fig:hugoniot}
    \end{minipage}
    \end{minipage}
\end{figure*}

The ``first principles'' character (formally exact quantum mechanical approach within the physical picture) of RPIMC and DFT simulations~\cite{BO_note} rests on two fundamental approximations: the choice of the nodal surface of the N-particle density matrix and the exchange-correlation functional, respectively. The exact versions of these quantities are, in general, not known and the accuracy of a particular choice is difficult to assess from within the methods, and it will depend on the phase of hydrogen.
Of course, the ultimate accuracy arbitrator is given by experimental measurements, however, in the case of highly compressed matter such comparisons are difficult and afflicted by large uncertainties and error bars. An instructive example provides the laser-driven shock wave measurements of liquid deuterium of L.B.~Da Silva\textit{ et al.}~\cite{dasilva_prl_97,collins_science_98}. They used the Nova laser and reported a very high compression ratio, $\rho/\rho_0$ of up to $5.9$, significantly larger than predicted by the SESAME model~\cite{sesame}. This discrepancy sparked an intense discussion and led to a series of additional experiments and many theoretical investigations. Already the first two articles that reported results from first-principles simulations with RPIMC~\cite{militzer_prl_00} and DFT-MD~\cite{PhysRevB.55.R11907}, consistently predicted a much smaller compression ratio of $\sim$4.3 at \SI{200}{\giga\Pa} pressure (right panel of Fig.~\ref{fig:hugoniot}). It was argued that the uncertainty in the simulation results was smaller than an energy increase of \SI{\sim 3}{\eV}/atom that would be required to bring the simulation results in agreement with the Nova measurements. That conclusion was supported by many following articles that reported simulation results. Using chemical models, on the other hand, it is possible to find agreement with the Nova measurements. Examples are the linear mixing model of M. Ross~\cite{ross_prb_98} and the equation of state of D.~Saumon and G.~Chabrier \cite{saumon_aps_95}, as Fig.~\ref{fig:hugoniot} shows. Note that both models predated the experiments and were not fit to the results. 

This discrepancy and the resulting uncertainty in the EOS led to a number of new measurements such as the Z-pinch experiments by M.~Knudson \textit{et al.} \cite{knudson_prl_01} and shock experiments at the Omega laser \cite{Hicks_2009,Knudson_2009}, see Fig.~\ref{fig:hugoniot}a. 
The difference in maximum compression ($4\dots 4.5$-fold \cite{knudson_prl_01} vs. $5\dots 6$-fold \cite{dasilva_prl_97, collins_science_98, Sano_2011}) is striking. Interestingly, the theoretical results may also be sorted into two groups: high compression, close to a maximum of 6 has been predicted by chemical models, such as W.~Ebeling's Pad\'e formulae \cite{ebeling-richert_85}, in addition to the two models mentioned above. On the other hand, there are the first-principle RPIMC and DFT-MD simulations and also the SESAME tables \cite{sesame}, that predicted significantly lower compressibilities. It is one aim of the present article to resolve these discrepancies. Testing the accuracy of the different simulations for dense partially ionized hydrogen allows us to confirm and quantify the reliability of RPIMC and DFT-MD.

Still, more work will be needed to resolve the remaining disagreements between various experimental and theoretical predictions for the Hugoniot state in Fig.~\ref{fig:hugoniot}. For example, measurements at the Z-machine indicate a lower compression ratio, at \SI{50}{\giga\Pa}, than is predicted by CEIMC simulations which, on the other hand, are in good agreement with the laser measurements by T.~Sano \textit{et al}.~\cite{Sano_2011}. 
Compared to the Nova results, the more recent laser experiments at the $\Omega$ facility have yielded much lower compression ratios, which brought them into better agreement with predictions from the $Z$-facility and from first principles simulations. 

Finally, we note that a comparatively small discrepancy has remained between first principles simulations and the recent laser shock experiments by A.~Fernandez-Panella \textit{et al.} \cite{fernandez_prl_19} on cryogenic liquid deuterium at very high pressures up to \SI{550}{\giga\pascal}, see Fig.~\ref{fig:hugoniot}c. R.~Rygg \textit{et al.} \cite{rygg_prl_23} suggested that collective modes (plasmons) would be missing from existing first principles calculations and might be the reason for the discrepancy. They performed a Debye-type calculation of the specific heat of electron plasma waves~\cite{rygg_prl_23} and simply added their contributions to the RPIMC results of S.~Hu \textit{et al.}~\cite{hu_ICF}. This brought the theoretical results within the one-sigma error bars of the measurements above \SI{300}{\giga\Pa} but could not explain the deviations at lower pressures. 
The central question is whether such plasmon waves are excited under such conditions and whether they would introduce a sufficiently large correction to the computed PIMC energies. 
Note that PIMC simulations take into account the entire Hamiltonian, so that plasmons are automatically accounted for, as long as the simulation cell is sufficiently large. By comparing results featuring 32 and 64 atoms, B.~Militzer \textit{et al.}~\cite{militzer_prl_00} determined that the compression ratio shifts by less than 0.01 at \SI{\sim 580}{\giga\Pa} and renders it unlikely that finite size effects or plasmons can explain the deviation of $\sim$0.1 between the measurements by A.~Fernandez-Panella \textit{et al.} and the RPIMC predictions. In addition, in this paper, we present novel independent tests of the RPIMC data of Ref.~\cite{militzer_prl_00}. A comparison with the first-principle fermionic PIMC (FPIMC) results of A.~Filinov and M.~Bonitz \cite{filinov_pre_23} shows that the deviations in energy and pressure between RPIMC and FPIMC, for the parameters of interest, are less than $1\%$, cf. Sec.~\ref{sss:rpimc-vs-fpimc}, which rules out the plasmon hypothesis. Note that the FPIMC data have undergone a finite size extrapolation, an issue that will be discussed in Sec.~\ref{ss:simulations-intro}.

As an alternative to laboratory measurements, new insights into the quality of theoretical models might be obtained with novel simulation methods that involve fewer or no uncontrolled approximations, such as exact diagonalization for model systems. Another tool for benchmarks that was already mentioned is fermionic PIMC simulations (featuring the FSP) under conditions for which they are feasible. Indeed, progress has been made recently with first-principles PIMC simulations for jellium (the uniform electron gas model~\cite{loos,dornheim_physrep_18}, UEG) by M.~Bonitz, T.~Dornheim and co-workers who were able to avoid the FSP by a combination of two complementary approaches -- configuration PIMC (PIMC simulations in occupation number representation, CPIMC) \cite{schoof_cpp11,cpimc_springer_14,schoof_cpp15,schoof_prl15} and permutation blocking PIMC (an advanced coordinate space approach, PB-PIMC) \cite{Dornheim2015a,dornheim_jcp15,dornheim_cpp_19}. Simulations turned out to be feasible for all densities and temperatures exceeding half of the Fermi energy \cite{groth_prb16,dornheim_prb16,dornheim_prl16,groth_prl17}. Both were employed to benchmark RPIMC simulations \cite{schoof_prl15,dornheim_pop17} and popular chemical models \cite{groth_cpp17}, including the Padé formulas of W.~Ebeling \cite{ebeling-richert_85}, the STLS parametrization of S.~Ichimaru \textit{et al.} \cite{stls,tanaka_pra_85,stls2}, and the models of Vashishta and Singwi \cite{ichimaru_rmp_93} and F.~Perrot and M.W.C.~Dharma-wardana \cite{perrot}. Another result of these simulations is the highly accurate parametrization of the exchange--correlation free energy -- the functionals of S.~Groth \textit{et al.}~(GDSMFB) \cite{groth_prl17,dornheim_physrep_18} and of V.~Karasiev \textit{et al.}~(KSDT) \cite{ksdt} -- which constitute explicitly thermal exchange--correlation functionals for DFT simulations on the level of the local density approximation~\cite{karasiev_importance_2016,kushal}. This demonstrates the high potential of parameter-free computer experiments for theory and model development in the field of warm dense matter.

Similar benefits can be expected from recent first-principles fermionic PIMC simulations -- further improvements of the PB-PIMC approach -- for partially ionized dense hydrogen by A.~Filinov and M.~Bonitz \cite{filinov_pre_23,filinov_cpp_21,bonitz_cpp_23} that cover temperatures above \SI{15000}{\kelvin} and densities corresponding to values of the Wigner-Seitz (Brueckner) coupling parameter $r_s\gtrsim3$, see Sec.~\ref{sec:wd-hydrogen}. Reference \cite{filinov_pre_23} presented tables of benchmark data and comparisons with several alternative simulations. Another %important 
step forward is fermionic PIMC simulations for the static density response of hydrogen for fixed ionic configurations by M.~B\"ohme \textit{et al.} \cite{Bohme_PRL_2022, Bohme_PRE_2023}, which extend previous first principle results for jellium~\cite{groth_jcp17,dornheim_pre17,dornheim_ML,Dornheim_PRL_2020_ESA,Dornheim_PRB_ESA_2021}.
Finally, we mention very recent works by T.~Dornheim \textit{et al.}~\cite{Dornheim_LFC_2024,Dornheim_JCP_2024} who have presented the first dependable PIMC results for a number of structural, imaginary-time, and density-response properties for hydrogen at the electronic Fermi temperature considering the cases of $r_s=1$, $r_s=2$, and $r_s=3.23$.

When considering hydrogen at temperatures well below the electronic degeneracy temperature, $T_F=E_F/k_B$, fermionic PIMC (FPIMC) becomes highly inefficient because of the FSP. RPIMC simulations have been applied to temperatures as low as $0.1\,T_F$, but nodal restriction makes sampling new paths and moving the nuclei inefficient below such temperatures (see below). On the other hand, electronic temperature effects become less relevant here, and one can resort to a finite-temperature treatment of the nuclei, combined with a ground state description of the electrons. This is the realm of coupled electron ion Monte Carlo (CEIMC). Usually, the electronic problem is solved by DFT methods (BOMD), but for hydrogen, even more controlled electronic ground state QMC methods have been applied: the CEIMC method uses electronic energies from ground state QMC to sample nuclear degrees of freedom using Monte Carlo ~\cite{Pierleoni2004,Pierleoni2006}, while a similar method using QMC-derived forces in a Langevin dynamics has also been proposed~\cite{Mazzola2018}. Despite being heavy in terms of computer resources, those methods are more controlled than DFT since their accuracy can be judged based on the variational principle.
\begin{figure}[ht]\centering
\includegraphics[width=0.5\textwidth]{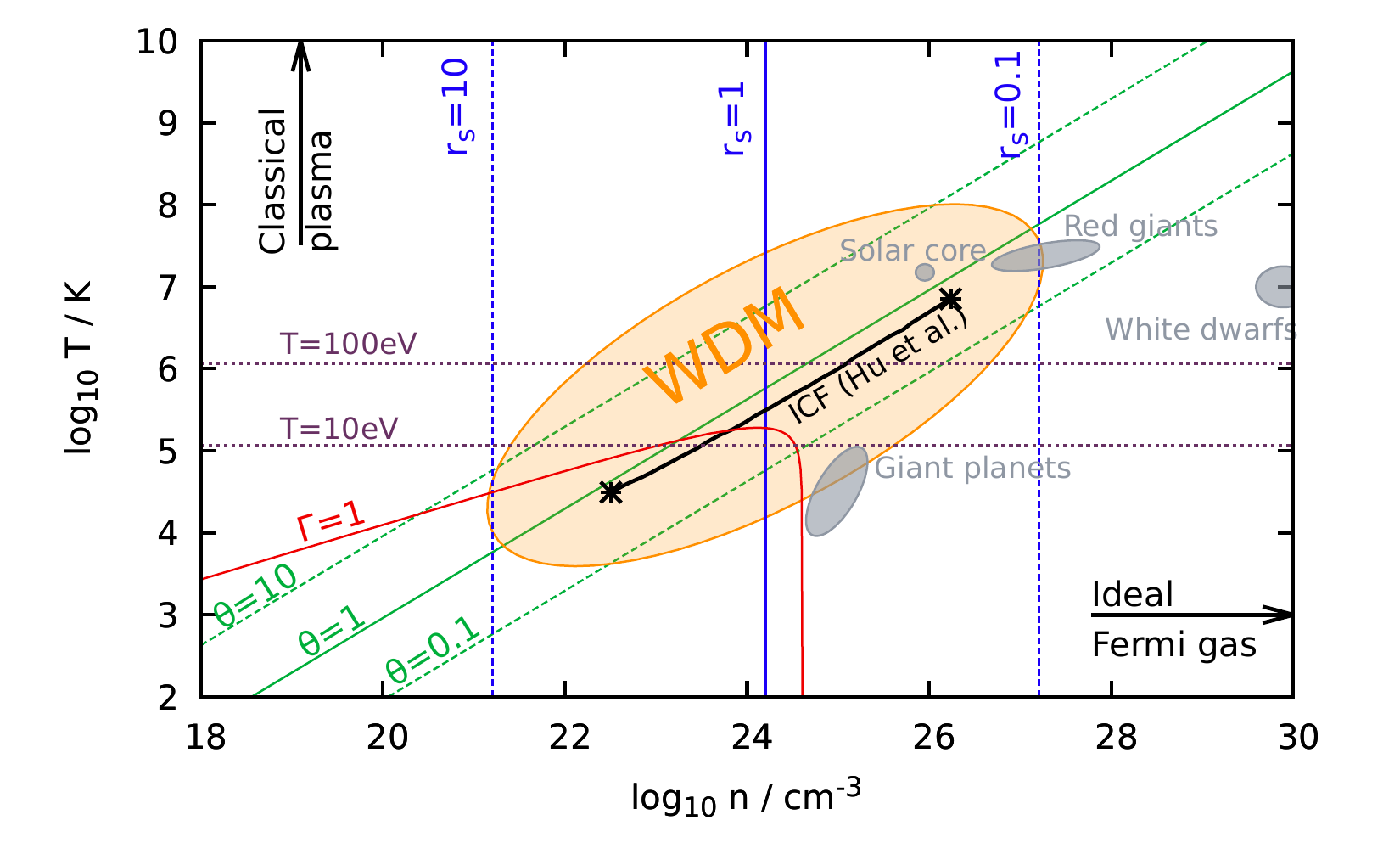}
\caption{\label{fig:overview} Qualitative phase diagram of warm dense matter (WDM), introducing a number of relevant parameters that are defined in Sec.~\ref{sec:wd-hydrogen}. In addition, we include examples of warm dense hydrogen in selected astrophysical objects such as giant planet interiors and ICF compression path from Ref.~\cite{Hu_PRL_2010, hu_ICF, Hu_PoP_2015}.
Adapted from Refs.~\cite{dornheim_physrep_18,Dornheim_HEDP_2022}.
}
\end{figure} 

The first goal of the present paper is to summarize the current knowledge of the phase diagram of hydrogen, in particular at low temperatures. Here, many new results have been obtained recently but many details still remain open. We update the review of Ref.~\cite{mc-mahon_RevModPhys.84.1607}. The analysis is also extended to higher temperatures of partially ionized hydrogen over a wide range of densities.

The second goal of this paper is to give an overview of the diverse arsenal of simulation methods that can be applied to dense hydrogen and range from first-principles methods, such as QMC and DFT, to simpler models, such as chemical models and hydrodynamic equations. We discuss which quantities are accessible by the different methods and compare them with respect to their accuracy and applicability range. To this end, we perform comparisons for the equation of state, pair distributions, and degree of ionization between recent first-principles QMC results and DFT simulations, semiclassical MD with quantum potentials, average atom codes, and chemical models.

The third goal of this article is to present an overview of recent developments in simulations of dense hydrogen. These include the influence of finite-temperature XC-functionals in DFT simulations for hydrogen \cite{ksdt,Karasiev_hug_2019, kushal,moldabekov2023bound} and the development of wave packet molecular dynamics simulations. 
%RR We have calculated XRTS spectra solely using DFT-MD for WDM for a decade. I have added some papers here:
Today, ab-initio approaches can be used to analyze X-ray Thomson scattering (XRTS) signals from warm dense matter experiments~\cite{siegfried_review}. It is for instance possible to infer the experimental plasma parameters by comparing DFT-MD data for the dynamic structure factor to the XRTS scattering signal~\cite{Witte_PRL_2017}. A novel first principle approach to accurately analyze XRTS experiments was developed by T.~Dornheim \textit{et al.}~\cite{Dornheim_T_2022,Dornheim_T_follow_up,dornheim2023xray,Dornheim2023,dornheim2024unraveling,Dornheim_LFC_2024,Dornheim_MRE_2023}. Thus, it has become possible to compare not just integrated quantities like the EOS but also dynamic (frequency dependent) quantities from first principles simulations to experimental measurements. PIMC results exist for jellium for the dynamic structure factor~\cite{groth_prb_19,Dornheim2018b}, the dynamic local field correction \cite{hamann_prb_20}, the plasmon spectrum \cite{hamann_cpp_20}, the dielectric function and the conductivity \cite{hamann_prb_20}. These can be extended in a semi-quantitative way to hydrogen and predicted a similar roton-type feature in hydrogen as in strongly correlated jellium that might be observable~\cite{Dornheim2018b,hamann_prr_23}. 

Given the high interest in ICF, we also reevaluate the open issues for the accurate theoretical modeling of the compression path of the deuterium-tritium fuel.  
%\fg{
The density temperature trajectory of the fuel and ablator materials traverses the warm dense matter into the high dense matter regimes as the ICF capsule implodes and then undergoes thermonuclear ingition, burns its fuel, and ultimately decompresses, cf. Fig.~\ref{fig:overview}. Pressures range from around a Megabar, at a density of $\SI{e24}{\centi\meter^{-3}}$ and temperatures of $\SI{1}{\eV}$, to tens of Gigabars of pressure, during ignition and burn, where densities are on the order of $\SI{e26}{\centi\meter^{-3}}$  and temperatures are many keV. 
It is expected that the novel results obtained from first-principles simulations presented in this article will be valuable also for ICF modeling.
\\

The structure of this article is as follows:  In Sec.~\ref{sec:phase-dia} we summarize the current knowledge as well as open questions on the phase diagram of hydrogen, including a discussion on metallization, the plasma phase transition, and the liquid-liquid phase transition(s). In Sec.~\ref{sec:theory} we present an overview of important simulation approaches for dense partially ionized hydrogen including the regimes of low and high temperatures. In Sec.~\ref{sec:results} we present our numerical results and comparisons of methods, including thermodynamic properties of hydrogen, density response functions, and transport properties. We also make suggestions for the future developments of the different methods and conclude with a summary and outlook in Sec.~\ref{sec:summary}.

\section{Phase diagram of hydrogen at high pressure}\label{sec:phase-dia}
\subsection{Parameters of dense hydrogen}\label{sec:wd-hydrogen}
Despite its chemical simplicity, warm dense hydrogen exists in a broad variety of phases, ranging from solid to liquid, gas, and plasma, for an overview see Figs.~\ref{fig:overview} and \ref{fig:h-phase-diag}. For systematics, it is useful to introduce dimensionless parameters. In the following, we will use CGS units, i.e. set $4\pi\epsilon_0=1$.

The properties of the electrons and protons in thermal equilibrium are characterized by a combination of standard jellium parameters and Coulomb one-component plasma (OCP) parameters~\cite{ott_epjd18}:
\begin{itemize}
    \item the  Wigner-Seitz  (Brueckner) coupling parameter of electrons, 
    \begin{align}
    r_s=\frac{d}{a_B}\,,
    \label{eq:rs}
    \end{align}
     the ratio of the mean interparticle distance (or Wigner-Seitz radius), $d$, given by 
     \begin{align}
     \frac{4\pi}{3}d^3=\frac{1}{n}    \,,
     \label{eq:ws-radius}
     \end{align}
      to the Bohr radius, Eq.~(\ref{eq:bohr-radius}), where $n$ denotes the number density;
     \item the degeneracy parameter
     \begin{align}
         \Theta = \frac{k_B T}{E_F} = \frac{T}{T_F}\,,
         \label{eq:theta}
     \end{align}
     the ratio of thermal energy to the Fermi energy, $E_F=(\hbar q_F)^2/2m$, where the Fermi wave number is $q_F = \left( 3\pi^2n \right)^{1/3}$. The parameter $\Theta$ is closely related to the degeneracy parameter $\chi_a$ given in terms of the thermal de Broglie wavelength $\Lambda$ of electrons and protons,
     \begin{align}\label{eq:chi}
         \chi_a &= n_a \Lambda_a^3\,,\\
         \Lambda_a &=\frac{h}{(2\pi m_a k_BT)^{1/2}}\,,\quad a=e, p\,,
         \label{eq:lambda-debroglie}
     \end{align}
     This expression for $\Lambda$ holds in thermal equilibrium for ideal particles, its modification due to interaction effects is discussed in Sec.~\ref{ssubsec:pair-pot-md}.
     \item the generalized coupling parameter
     \begin{align}\label{eq:gamma}
         \Gamma = \frac{e^2}{dk_BT\frac{2\sigma+1}{n\Lambda^3}I_{3/2}(\beta\mu)}\,,
     \end{align}
     which takes into account the quantum kinetic energy via the Fermi-Integral $I_{3/2}$ and the spin degeneracy factor $2\sigma+1$~\cite{blue-book}. It is shown as the red line in Fig.~\ref{fig:overview}. For non-degenerate particles in the gas (plasma) phase, i.e. for $\chi_a \ll 1$, this simplifies to the well-known classical Coulomb coupling parameter,
     $\Gamma = e^2/(d k_BT)$. For high degeneracy, $\Gamma$ becomes constant when the zero-temperature quantum coupling parameter, $r_s$, is constant.
     \item screening and dynamic properties of the free electrons are characterized by the Debye length, $r_{\rm D}$ and the plasma frequency,
     \begin{align}\label{eq:debye-length}
         r_{\rm D} &= \left(\frac{k_BT}{4\pi n e^2}\right)^{1/2}\,,\\
         \omega_p &= \left(4\pi\frac{n e^2}{ m}\right)^{1/2}\,,
         \label{eq:om-plasma}\\
         r_{\rm TF} & =\frac{\hbar}{2m^{1/2}}\left( \frac{\pi}{3n} \right)^{1/6}\,.\label{eq:tf-length}
     \end{align}
\end{itemize}
    For strong degeneracy, the Debye length is replaced by the Thomas-Fermi screening length, $r_{\rm TF}$.

%\textbf{Partially ionized plasma:}
\begin{itemize}
    \item  In the case of partial ionization, it is convenient to characterize the gas and plasma phases by finite fractions of free particles (degree of ionization), atoms and molecules, $\alpha, x_A$ and $x_M$ which relate the densities of free electrons and  electrons bound in atoms and molecules to each other,  
    \begin{align}
    n_e & =n_e^* + n_e^{\rm bound}\,,
    \label{eq:nestar}\\
    n_e^{\rm bound} & =n_A+2n_M\,,
    \label{eq:nebound}\\
    \label{eq:alpha-def}
       \alpha &= \frac{n_e^*}{n_e^*+n_e^{\rm bound}} \,,\\
       x_A &= \frac{n_A}{n_e^*+n_e^{\rm bound}} \,,\\
       x_M &= \frac{2 n_M}{n_e^*+n_e^{\rm bound}} \,, \label{eq:xm-equation}
    \end{align}
    and $\alpha+x_A+2x_M = 1$.
    Such a subdivision into ``free'' and ``bound'' electrons is called ``chemical picture'', cf. Sec.~\ref{subsec:chem-model}, and is somewhat artificial. The results may depend on the procedure; different criteria lead to different results. This issue will be discussed below in Sec.~\ref{sssec:alpha}.
    \item Often, instead of the particle density, the mass density is used which we give for hydrogen ($\rho$) and deuterium ($\rho_D$):
    \begin{align}
        \rho &= (1.39181/r_s)^3\,,\\
        \rho_D &= (1.75313/r_s)^3=1.9985 \,\rho\,,
    \end{align}
    where $\rho$ is given in \unit{\gram\per\cubic\centi\meter}.
    \item The fractions of bound states are determined by temperature and density in units of the binding energy, 
    \begin{align}\label{eq:atom-bind-energy}
    E_B &=\frac{m_r}{2\hbar^2}\left(\frac{e^2}{\epsilon_r}\right)^2= \SI{13.59}{\eV}\,,\\\label{eq:bohr-radius}
    a_B &= \frac{\hbar^2}{m_r}\frac{\epsilon_r}{e^2} = \SI{0.529}{\angstrom}\,,\\\label{eq:mol-bind-energy}
    E_B^M &= \SI{4.52}{\eV} \sim \SI{52400}{\kelvin}\,,\\\label{eq:mol-distance}
    a^M_{pp} & \approx 1.4\,a_B\,,\\
    \label{eq:ionization-temp}
     T^{\rm ion} & \approx \frac{E_B}{k_B}\approx \SI{157600}{\kelvin}\,,
    \end{align}
    where $m_r$ is the reduced mass, $m_r=m_e\frac{M}{M+1}$, with the mass ratio $M=m_p/m_e$, and for hydrogen $\epsilon_r=1$. In Eqs.~(\ref{eq:mol-bind-energy}) and (\ref{eq:mol-distance}) we indicated the values of the binding energy of hydrogen molecules and the associated equilibrium proton distance.
    Complete thermal ionization of atoms occurs for temperatures $T\gtrsim T^{\rm ion}$, Eq.~(\ref{eq:ionization-temp}), corresponding to the binding energy. 
    \item \textbf{Pressure ionization. Mott effect. Insulator-metal transition (IMT).}
    In a many-particle system, ionization can occur also at zero temperature. In the case of a sustained overlap of neighboring atoms (molecules), Coulomb correlations, quantum, and exchange effects lead to a shift of the of individual bound state energies and a lowering of the ionization (dissociation) energy which eventually vanishes at a critical density, which is sometimes called ``Mott density'' in the plasma physics literature \cite{red-book,green-book}. PIMC simulations predicted a critical coupling parameter 
    \begin{align}\label{eq:mott-density}
    r_s^{\rm Mott} &\approx 1.2\,,    
    \end{align}
     for pressure ionization of hydrogen atoms at low temperature \cite{bonitz_prl_5}. At finite temperature, pressure ionization occurs at reduced densities. The topic of pressure ionization and ionization potential depression will be discussed in more detail in Sec.~\ref{sssec:ion-pot}.
     Note that the above picture of pressure ionization is essentially a single-atom picture, where an isolated atom is embedded into a surrounding medium which is often appropriate in the high-temperature plasma phase.

     At low temperatures, in the condensed phase, the atoms or molecules of hydrogen form an ordered lattice. Then pressure ionization and dissociation is a many-particle effect. Compression of the crystal leads to a change of the band structure and, eventually, to a closure of the energy gap between the highest occupied and the lowest unoccupied band: an insulator to metal transition (IMT). This effect was first studied by  E.~Wigner and H.B.~Huntington~\cite{wigner-huntington_35} predicting that molecular hydrogen (an insulator) would transition to metallic atomic hydrogen because of band filling.  N.F.~Mott~\cite{Mott_49, mott_rmp_68} further considered the effects of electron correlation and computed critical values for the electron density beyond which metallization occurs. These values strongly depend on the material and the crystal symmetry, and the corresponding values for the coupling parameter are significantly higher than the hydrogen plasma result (\ref{eq:mott-density}), for a recent overview, see Ref.~\cite{imada_rmp_98}. The IMT in hydrogen will be discussed in detail in Sec.~\ref{sssec:hydrogen-lowt}.
    \item The free electron properties in a partially ionized hydrogen plasma can be estimated from the jellium parameters, by a simple rescaling of the density, $n_e \to n_e^*$ \cite{hamann_prr_23}, the corresponding parameters will be denoted by an asterisk:
    \begin{align}\label{eq:rs*}
        r_s^*(\alpha) &= \frac{a^*}{a_B} = \alpha^{-1/3}\cdot r_s \,,\\\label{eq:qf*}
        q_F^*(\alpha) &= (3 \pi^2 n_e^*)^{1/3} = \alpha^{1/3}\cdot q_F \,,\\\label{eq:theta*}
        \Theta^*(\alpha) &= \frac{k_B T}{E^*_F} =\alpha^{-2/3}\cdot \Theta   \,,\\
        \chi^*_e(\alpha) &= \alpha \chi_e\,,\\
        \label{eq:omplasma*}
        \omega_p^*(\alpha) &=  [(4\pi n^*_e e^2)/(m_e)  ]^{1/2} = \alpha^{1/2}\cdot \omega_p\,.
    \end{align}
    For a given total density $n$ and temperature $T$ these quantities depend on the degree of ionization, $\alpha(n,T)$, Eq.~(\ref{eq:alpha-def}).
    This estimate of electronic properties applies the ``chemical picture'', cf. Sec.~\ref{subsec:chem-model} and assumes that the influence of the bound states on the free electron properties is negligible.
\end{itemize}

\subsection{Phases of hydrogen\label{sssec:phases-overview}}
%
%\textcolor{red}{David, Burkhard, Michael, Ronald}

\begin{figure}[h]\centering
\includegraphics[width=0.5\textwidth]{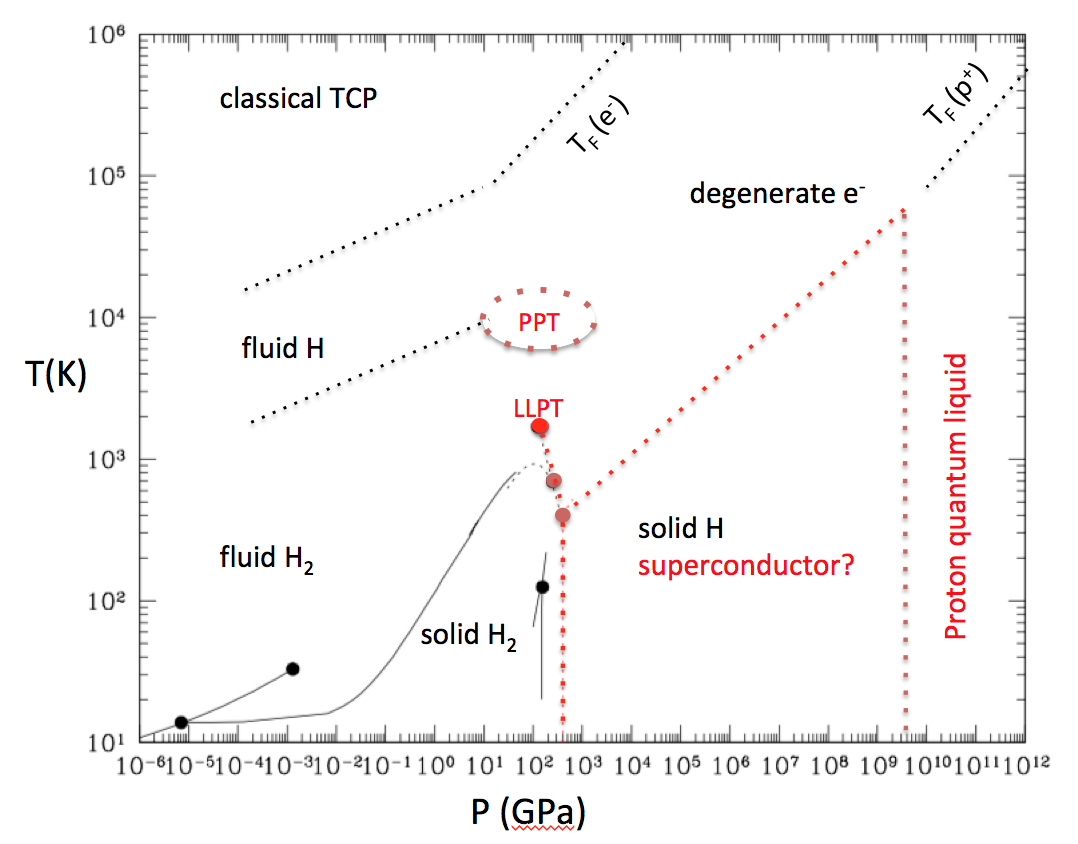}
\caption{ Hydrogen phase diagram. Solid black lines show the boundaries between the gas, liquid, and solid phases as measured in static experiments. The solid circles show the location of critical or triple points  (black: observed, red: predicted). The black dashed lines are crossovers between the classical behavior of electrons and protons at high temperatures. $T_{\rm F} (e^-)$ and $T_{\rm F} (p^+)$ are the Fermi temperatures of electrons and protons, respectively.  
 ``PPT'' indicates the critical region of the hypothetical Plasma Phase Transition, Sec.~\ref{sssec:hydrogen-hight}, ``LPPT'' indicates the Liquid-Liquid Phase Transition. Thermal ionization occurs at $T \sim T^{\rm ion}=157\,000$K, Eq.~(\ref{eq:ionization-temp}).
 The red dotted line shows the estimated extent of solid atomic hydrogen.
 }
\label{fig:h-phase-diag} 
\end{figure} 
Hydrogen has a very rich phase diagram, many details of which are under active investigation.  Fig.~\ref{fig:h-phase-diag} shows a simplified overview.  
At high temperatures and low pressures (top left corner) hydrogen behaves as a classical gas of charged particles -- a two-component electron-ion plasma (TCP). Upon cooling, below the binding energy, $T\le T^{\rm ion}$, hydrogen atoms and molecules form, the fraction of which increases as the temperature decreases. 

At higher pressures, cooling potentially gives rise to a second fluid-fluid phase transition in partially ionized hydrogen. There have been extensive discussions in the literature about whether there are two independent phase transitions. L.D.~Landau and Ya.B.~Zeldovich~\cite{zeldovich-landau_44}, K.H.~Schramm~\cite{schramm_zphys_61} and G.~Norman and A.~Starostin~\cite{norman-starostin_68} predicted that ionization and recombination of the plasma proceeds via a ``plasma phase transition'' (PPT), see Fig.~\ref{fig:h-phase-diag}. We discuss theoretical and experimental predictions in Sec.~\ref{sssec:hydrogen-hight} and conclude that there is no evidence for the PPT. A second liquid-liquid phase transition (LLPT) at megabar pressures and temperatures around \SI{1000}{\kelvin} likely occurs between molecular hydrogen and metallic fluid hydrogen \cite{weir_prl_96}.

Lowering the temperature further, hydrogen freezes. We discuss the location of the melting line in Sec.~\ref{sssec:hydrogen-lowt}. There, we also consider phenomena predicted to occur upon compression, in particular, the insulator-to-metal transition in solid hydrogen  (E.~Wigner and H.B.~Huntington \cite{wigner-huntington_35}). Whether this transition coincides with the transition from molecular hydrogen to atomic hydrogen is still an open question.   
N.~Ashcroft~\textit{et al.}~\cite{ashcroft_prl_68,ashcroft_prl_77} has suggested that the solid atomic hydrogen phase will be a room-temperature superconductor because of the large phonon energies and electron-proton coupling. 
 Finally, in Sec.~\ref{sssec:hydrogen-proton-cryst} we concentrate on the compression of hydrogen gas up to TPa and PPa at relatively high temperatures, see Fig.~\ref{fig:h-phase-diag}. These pressure gives rise to the formation of an atomic (proton) crystal that is bounded by a classical proton fluid at lower pressures, and a quantum proton liquid at higher pressures.

\subsubsection{Partially ionized hydrogen below the Mott density. Hypothetical plasma phase transition (PPT)\label{sssec:hydrogen-hight}}
We start by considering partially ionized hydrogen at temperatures $\SI{10000}{\kelvin} \lesssim T \lesssim T^{\rm ion}$, Eq.~\eqref{eq:ionization-temp}, and 
 densities below the Mott density, $r_s \gtrsim r_s^{\rm Mott}$, Eq.~(\ref{eq:mott-density}). The degree of ionization $\alpha$ increases monotonically with temperature (thermal ionization) and density (pressure ionization) which is accompanied by a strong increase in electrical conductivity. This parameter range attracted particular attention in the 1970s to 1990s, due to the prediction of the so-called plasma phase transition (PPT) -- a hypothetical first-order phase transition that was thought to be analogous to the van der Waals gas-liquid transition. The idea was (K.H.~Schramm, Ref.~\cite{schramm_zphys_61}, G.~Norman and A.N.~Starostin, Ref.~\cite{norman-starostin_68}) that the combination of overall attractive Coulomb forces in the plasma and short-range electron repulsion (due to quantum effects), below a critical temperature may lead to a fluid with two phases of different degree of ionization.
Additional arguments for the existence of the PPT were similar effects in electrolytes, alkali metals, and electron-hole plasmas in semiconductors, e.g. \cite{red-book}. The critical point of the PPT was closely linked to the ionization potential of the hydrogen atom, and most predictions settled around $T^{\rm cr} \approx 0.1\,T^{\rm ion}\sim \SI{15800}{\kelvin}$, for a representative collection of results, see Tab.~\ref{tab:ppt}.
The majority of predictions of the PPT originated from chemical models (CM, the top part of Tab.~\ref{tab:ppt}),
for details, cf. Sec.~\ref{subsec:chem-model}. 
\begin{table}[]
    \centering
    \begin{tabular}{|c|c|c|c|c|}
    \hline
    Ref., year & $T^{\rm cr}$[K] & $p^{\rm cr}$[GPa] & $\alpha, r_s, n, \rho$ & method \\
    \hline
     \cite{schramm_zphys_61}, 1961 & 15\,920 &  &  & CM\\
     \cite{norman-starostin_68}, 1968 & $0.1 T^{\rm ion}$ &  & $n^{cr}=\left(\frac{0.1}{a_B}\right)^{3}$ & CM$^\dagger$\\
     \cite{norman-starostin_70}, 1970 & $ 10\,913$ &  & $n^{cr}=\left(\frac{0.096}{a_B}\right)^{3}$ & CM\\
     \cite{ebeling_ap_73}, 1973 & 12\,600 & 95.0 & $\rho^{cr}=0.95$ & CM\\
     \cite{franck_ap_80}, 1980 & $< 9000$ & & $\rho^{cr}=1.0$ & CM\\
     \cite{robnik-kundt_83}, 1983    & 19\,000 & 24.0 & $\alpha^{cr}=0.50$ &  CM\\
     \cite{ebeling-richert_85}, 1985 & 16\,500 & 22.8 & $\alpha^{cr}=0.32$ & CM\\
     \cite{saumon-chabrier_prl_89}, 1989 & 15\,000 & 64.6 & $\alpha^{cr}=0.2$ & CM\\ 
     \cite{saumon-chabrier_92}, 1992 & 15\,300 & 61.4 & $\alpha^{cr}=0.18$ &  CM\\
     \cite{saumon_aps_95}, 1995 & 15\,311 & 61.4 & $\alpha^{cr}=0.075$ &  CM\\
     \cite{schlanges-etal.95cpp}, 1995 & 14\,900 & 72.3  & $\alpha^{cr}=0.4$ &  CM\\
     \cite{reinholz_PRE_95}, 1995 & 15\,000 & 23 & $\rho^{cr}=0.13$ & CM \\
     \cite{ebeling_ppcf_96}, 1996 & 13\,000 && $n^{cr}=a_B^{-3}$ & CM \\
     \cite{beule_prb_99}, 1999 & $>$10\,000 & $\sim 100$  &  &  CM\\     \hline
     \cite{filinov-norman_75}, 1975 & $>$ 100\,000 &  &  & FPIMC \\
     \cite{magro_prl_96}, 1996 & 11\,000 & $\sim $48 & $r^{cr}_s=2.2$ & RPIMC$^f$\\
     \cite{militzer_pre_00}, 2000 & - & - & no PPT & RPIMC$^v$\\
     \cite{filinov_jetpl_01}, 2001 &  $>$10\,000 & & $n^{cr}>10^{22}$ & FPIMC \\
     \cite{filinov_cpp_03}, 2003 &  $>$10\,000 & & $n^{cr}>10^{22}$ & FPIMC \\\hline
     \cite{grigoriev_jeptl_72}, 1972 &  & $<$ 280 &  & Exp.$^1$\\
%     \hline
     \cite{weir_prl_96}, 1996 & - & - & no PPT & Exp.$^2$\\
     \cite{fortov_prl_07}, 2007 &&&& Exp.$^1$\\
     \hline
     \end{tabular}
    \caption{Results for the hypothetical plasma phase transition (top rows: chemical models, CM; middle rows: PIMC simulations; bottom rows: experiments, Exp.), including critical temperature, pressure, degree of ionization or density ($\rho$ is given in g$\,$cm$^{-3}$; $n$ is given in cm$^{-3}$). $^\dagger$: data are from a preprint of V.A.~Alekseev et al., Ref. 5 in \cite{norman-starostin_68}. FPIMC: fermionic PIMC; RPIMC: restricted PIMC. f: free-particle nodes; v: variational nodes; 1: explosives-driven shock wave; 2: laser-driven shock wave. 
    %The presented results are not confirmed by more accurate later simulations or by experiments.
    Predicted transitions in the liquid phase (LLPT) are listed separately, in Tab.~\ref{tab:llt}.}
    \label{tab:ppt}
\end{table}
\\

The PPT was also investigated with first-principle fermionic PIMC simulations \cite{filinov-norman_75,filinov_jetpl_01,filinov_cpp_03}. However, these simulations are severely hampered by the sign problem, cf. Sec.~\ref{sec:FSP} and could not achieve converged results~\footnote{In Ref.~\cite{filinov-norman_75}. The authors published ``preliminary estimates'' for the two-phase region that extends beyond $T=\SI{100000}{\kelvin}$, based on PIMC results with just $P=2$ high-temperature factors.  In Refs.~\cite{filinov_jetpl_01,filinov_cpp_03} V. Filinov \textit{et al.} used $P=20$ high-temperature factors, $N=50$ electrons and ions. At $T=\SI{10000}{\kelvin}$, in the density range of \SIrange{0.1}{1.5}{\gram\per\cubic\centi\meter}, they reported strong fluctuations and the formation of ``metallic clusters'' that they interpreted as evidence for the PPT, without providing evidence for convergence of the simulations.}.
On the other hand, restricted PIMC simulations with free-particle nodes reported a PPT~\cite{magro_prl_96} but were not confirmed by more accurate later simulations with variational nodes \cite{militzer_pre_00}, see also Refs.~\cite{bonev_prb_04, Militzer_PRE_2001}, cf. bottom part of Tab.~\ref{tab:ppt}.

Finally, on the experimental side [bottom of Tab.~\ref{tab:ppt}] hydrogen metallization and a density jump at high pressure were reported in shock compression experiments using explosives~\cite{grigoriev_jeptl_72}. 
On the other hand, S.T.~Weir \textit{et al.} \cite{weir_prl_96}  
    observed metallization of hydrogen and deuterium in shock experiments, around $p=\SI{140}{\giga\Pa}$ at temperatures where hydrogen is a fluid, but did not observe a phase transition.
    V.E.~Fortov \textit{et al.} also used explosives-driven shock compression of deuterium \cite{fortov_prl_07} and claimed proof of the PPT, based on observation of a ``density jump'' in the range of $p=\SIrange{127}{150}{\giga\Pa}$. However, the evidence is not convincing~\footnote{The analysis is based on the comparison to simple chemical models and to the PIMC simulations of Ref.~\cite{filinov_jetpl_01}. Also, the reported strong increase of the conductivity in the same density range is not proof of a phase transition.}, and the studied temperature range of \SIrange{2550}{4000}{\kelvin} (as deduced from models) is confined to the fluid phase. Thus, the observations could rather be an indication of the liquid-liquid transition, see Sec.~\ref{sssec:hydrogen-lowt}.
     \\ 
    
    The topic of the PPT has been frequently discussed in reviews and textbooks, see e.g.~\cite{redmer_phys-rep_97,ebeling-norman_03,mc-mahon_RevModPhys.84.1607}. Also,    Refs.~\cite{norman_cpp_19,norman_ufn_21} contain a recent overview of the history of the PPT, from the point of view of their Russian authors. 
    Table~\ref{tab:ppt} summarizes the extensive literature on the PPT. 
    Chemical models, the majority of which have predicted a PPT with a critical temperature $T^{\rm cr}\gtrsim \SI{10000}{\kelvin}$ are not reliable when the ionization fraction changes significantly, as is the case with the PPT. The PPT may be an artifact  ``built into'' these models \cite{chabrier_jpa_06,chabrier_ass_07}. For more details, see Sec.~\ref{subsec:chem-model}. State-of-the-art first-principles simulations in the physical picture show no indications of a PPT. 
    % RR: I have added the work of W. Lorenzen which appeared in parallel to Miguel's work 
    DFT-MD simulations of W.~Lorenzen \textit{et al.}~\cite{Lorenzen2010} and coupled electron-ion (CEIMC) simulations of M.A.~Morales \textit{et al}.~\cite{Morales2010, Pierleoni2016} have produced evidence of a phase transition at much lower temperatures in the fluid phase (LLPT), which differs qualitatively from the PPT, for more details, see Tab.~\ref{tab:llt} and Sec.~\ref{sssec:hydrogen-lowt}.

\begin{table}[h]
    \centering
    \begin{tabular}{|c|c|c|c|}
    \hline
    Reference, year & $T^{\rm cr}\,$[K] & $p^{\rm cr}\,$[GPa]  & method \\
    \hline
     \cite{weir_prl_96}, 1996    & $< 5\,200$ & no LLPT  &   shock experiment\\
     \cite{fortov_prl_07}, 2007 &&& shock experiment\\
     \hline
     \cite{scandolo_pnas_03}, 2003 &$\lesssim 2\,000$ & $\gtrsim 125 $ & Car-Parrinello\\
     \cite{bonev_prb_04}, 2004 & $< 4\,500$ && DFT, PBE\\
     \cite{delaney_prl_06}, 2006 & -- & no LLPT & CEIMC\\
     \cite{militzer1}, 2007 & -- & no LLPT & DFT\\
     \cite{Holst_PRB_2008}, 2008 & -- & no LLPT & DFT\\     
     \cite{Lorenzen2010}, 2010 & 1400 & 132 & DFT\\
     \cite{Pierleoni2016}, 2010-16 & $\lesssim 2\,000$ & $\gtrsim 120$ & CEIMC\\
      & $\lesssim 2\,000$ & $\gtrsim 97$ & DFT\\
      \hline
 \end{tabular}
    \caption{Experimental (above line) and simulation (below line) results for the first order liquid-liquid phase transition (LLPT) including temperature range (in 1000 K) and pressure range. PBE: PBE XC-functional; Predictions of phase transitions in the plasma phase (PPT) are listed separately in Tab.~\ref{tab:ppt}. For details see text and Fig.~\ref{fig:phase-diagram-low-t-optics}.}
    \label{tab:llt}
\end{table}

\subsubsection{Hydrogen phase diagram at low temperature. Metallization. Liquid-liquid phase transition (LLPT)\label{sssec:hydrogen-lowt}}
\begin{figure}
    \centering
    \includegraphics[width=0.52\textwidth]{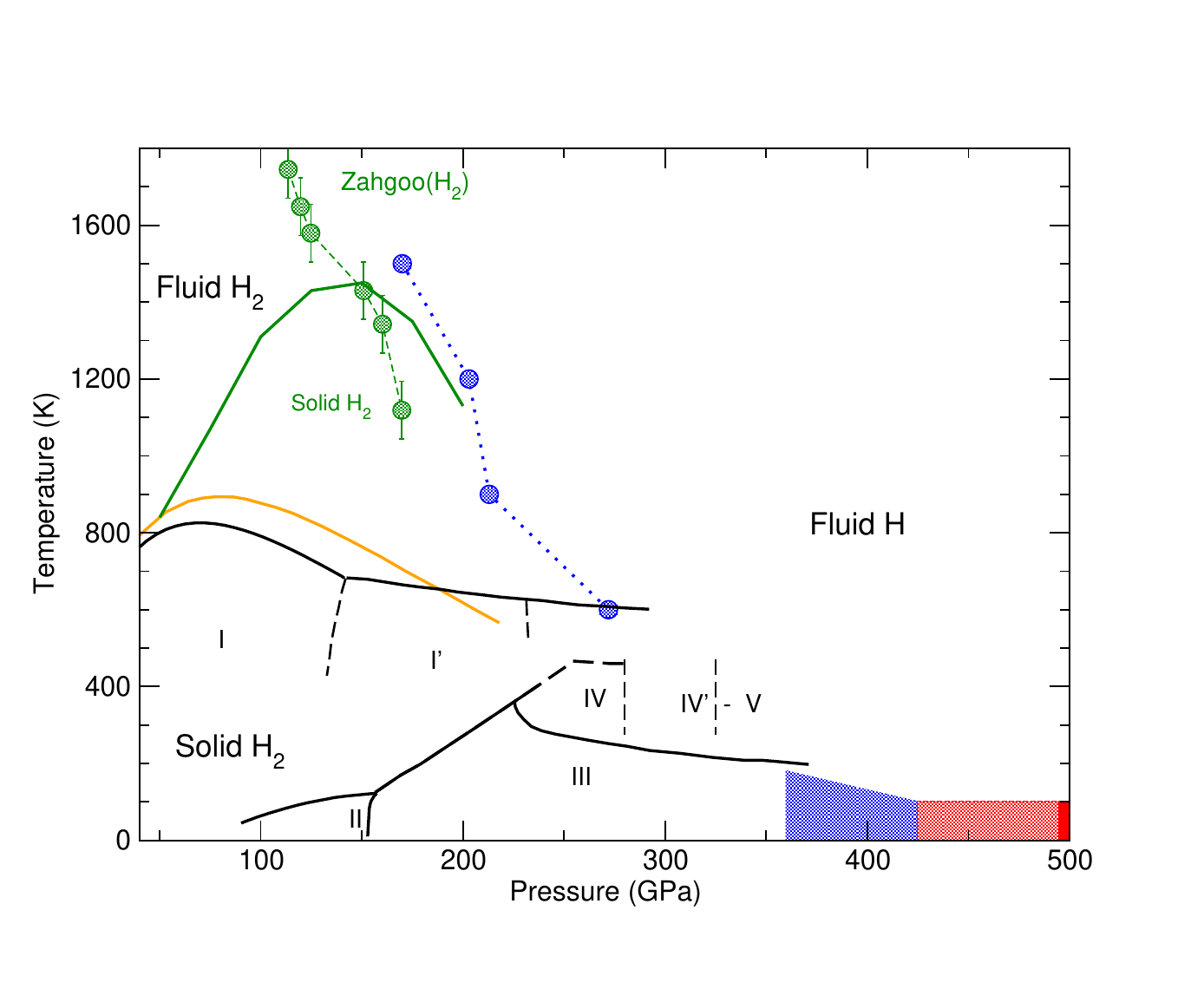}
    \caption{
    Experimentally inferred solid hydrogen phase diagram (solid and dashed black lines). Several non-metallic crystal phases have been detected. The melting line (black line) is reentrant and has been measured up to about \SI{300}{\giga\Pa}~\cite{Zha2017}. The low-temperature phase III is where metallization has been measured. The semimetallic state is entered at \SI{360}{\giga\Pa} \cite{Eremets2019} and indicated by the blue shaded area. It persists up to \SI{425}{\giga\Pa} where a sudden collapse of the direct gap is detected~\cite{loubeyre_nat_20} which is represented by the red shaded area. At \SI{495}{\giga\Pa} a reflective sample has been reported \cite{dias_science_17} (represented by the dark red vertical bar). Two theoretical melting lines are also reported, one obtained by free energy methods for classical nuclei within BOMD with DFT-PBE (orange line) \cite{Morales2010} and one recently obtained by the two-phase method for a system of quantum nuclei using a machine learning force field (DeePMD) trained on QMC energies and forces (green continuous line)~\cite{Niu2023}. Green circles: experimental data for warm hydrogen in DAC where signatures of a phase transition were detected~\cite{Zaghoo2016}. Blue circles: CEIMC predictions for the LLPT line for quantum protons \cite{Pierleoni2016}. 
    }
    \label{fig:phase-diagram-low-t}
\end{figure}

The phase diagram of hydrogen under extreme conditions is still uncertain, mainly because of experimental difficulties in producing and controlling the extreme physical conditions required 
\cite{gregoryanz_mre_20} 
and the limited information experiments obtain about the system. At low temperatures in the solid phase, the most important missing information is the crystal structure as a function of temperature and pressure. In Fig.~\ref{fig:phase-diagram-low-t} we report the present low-temperature phase diagram emerging from experimental information. Up to six different crystal structures have been detected but the transition lines are traced based on discontinuities and changes, resp., of the vibrational frequencies of the infrared (IR) and Raman spectra, while measurements of Bragg peaks are mostly missing. A notable exception is along the \SI{300}{\kelvin} isotherm where special X-ray spectroscopy has been developed and applied to confirm that the structure of phase I is m-HCP up to \SI{250}{\giga\Pa} %\CP{to some pressure} 
\cite{Ji2019,Ji2020}. Candidate structures have been inferred by the \textit{ab-initio} Random Structure Search (AIRSS) method \cite{Pickard2007} based on DFT calculations with phonon corrections. 

A relevant and still partially unanswered question concerns the mechanism by which solid hydrogen metallizes upon increasing pressure~\cite{gregoryanz_mre_20}. 

Important advances towards the metallization of solid hydrogen at low temperature were announced by different groups in recent years.
In 2017 the group at Harvard reported the observation of metallic hydrogen in diamond anvil cell (DAC) experiments at \SI{495}{\giga\Pa} below \SI{80}{\kelvin}~\cite{dias_science_17}. This conclusion was based on the sudden appearance of a reflective sample that was interpreted as evidence for the metallization of hydrogen. This interpretation has been criticized by others, \cite{gregoryanz_mre_20} and so far the results have not been reproduced. Almost simultaneously the group in Mainz reported evidence of the formation of a semi-metallic, still molecular, phase at around \SI{350}{\giga\Pa} and below \SI{100}{\kelvin}. They employed both optical probes and direct electric measurements in a DAC 
\cite{eremets_nat-phys_19}. In 2020, P.~Loubeyre \textit{et al.}~\cite{loubeyre_nat_20} reported results using a toroidal DAC with synchrotron radiation. They measured the IR absorption profile over a wide range of pressure and detected complete absorption at \SI{427}{\giga\Pa} and \SI{80}{\kelvin}, which was interpreted as a sudden closure of the direct energy gap, a strong indicator for the state of a ``good metal''.  Although sample visual inspection and reversibility of the transition upon pressure release suggest that this metal is still molecular, strong experimental evidence, such as the persistence of the vibron signal, is missing. Thus it is conceivable that the observed collapse of the infrared gap signals a metallization through molecular dissociation. 

\begin{figure}
    \centering
    \includegraphics[width=0.48\textwidth]{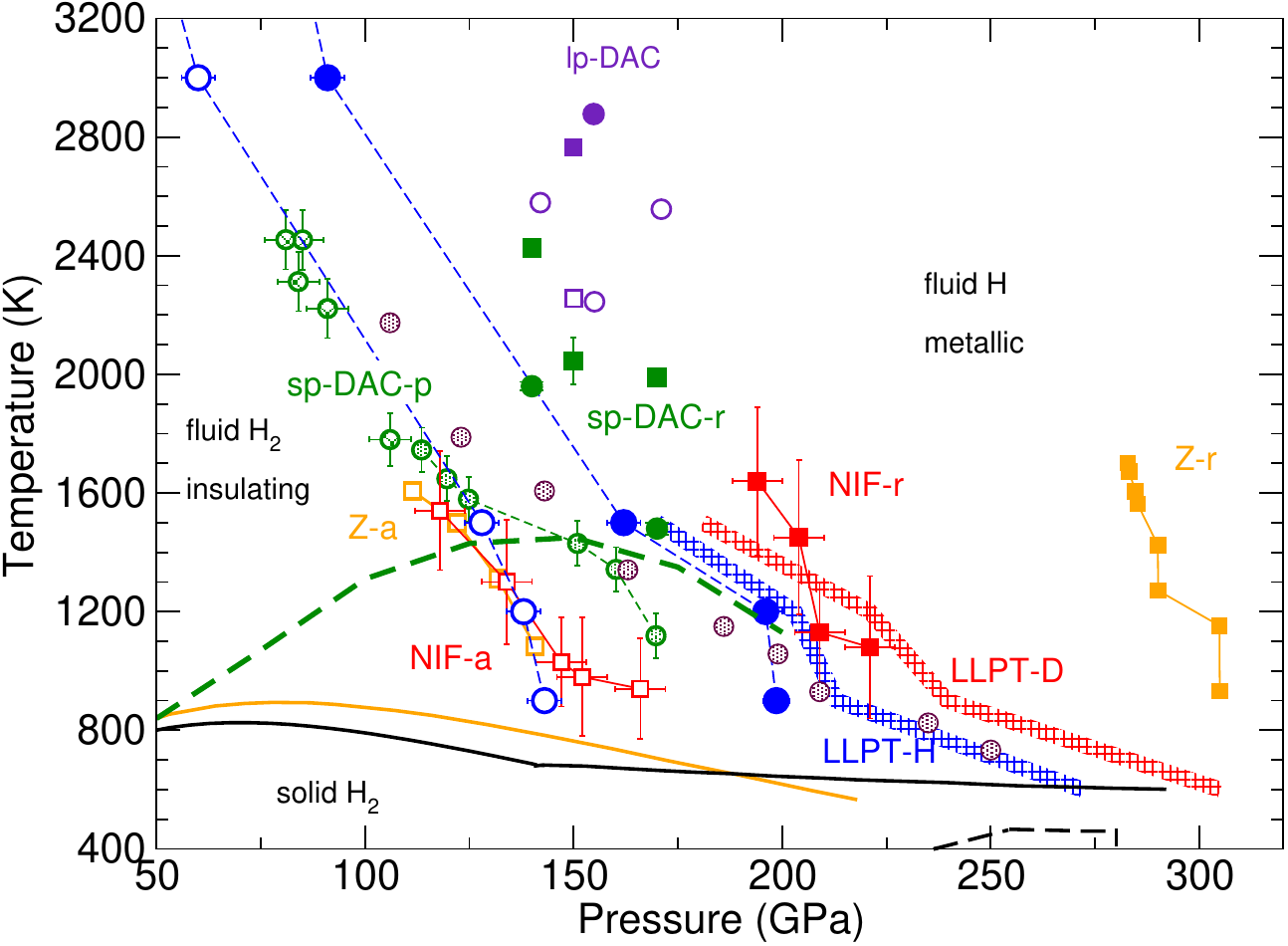}
    \caption{
%    \CP{I propose to put this figure to replace the previous one.} 
    Phase diagram of fluid hydrogen and deuterium around the LLPT line. Shaded lines (blue for hydrogen and red for deuterium) are the LLPT predicted by CEIMC~\cite{Pierleoni2016, Pierleoni2017}. Filled symbols: estimates of the LLPT from the reflectivity coefficient; open symbols indicate the inception of absorption. Squares correspond to deuterium, and circles to hydrogen. Shown are data from sp-DAC (green), Z-machine (orange), NIF (red), and lp-DAC methods (purple). DAC-p, data from sp-DAC corresponding to the temperature plateau from Refs.~\cite{Dzyabura_PNAS_2013,Zaghoo2016} (T $\leq 1\,700~$K) and from Ref.~\cite{Ohta2015} (T $\geq 1\,700~$K); DAC-r, data from sp-DAC at R = 0.3; lp-DAC \cite{Jiang2020}, filled purple points are conducting conditions, and open purple points are nonconducting conditions (for both hydrogen and deuterium); NIF-a, data from NIF when the absorption coefficient $\mu \simeq \SI{1}{\micro\meter^{-1}}$; NIF-r, data from NIF at R = 0.3 \cite{Celliers_Science_2018}; Z-a, data from Z-machine when the sample becomes dark; Z-r, data from Z-machine at the observed discontinuity in reflectivity \cite{Knudson_Science_2015}. 
    Blue points: theoretical predictions from Ref.~\cite{Rillo2019}: Filled circles show when R=0.3 for H/vacuum interface; open circles show when $\mu=\SI{1}{\micro\meter^{-1}}$. 
    Brown shaded circles: recent predictions for a system of quantum protons from PIMD simulation with the DFT-SCAN-rvv10 approximation \cite{Hinz2020}. The points correspond to a conductivity $\sigma=\SI{2000}{\siemens\per\centi\meter}$. Melting lines are as in Fig.~\ref{fig:phase-diagram-low-t}. Figure adapted from Ref.~\cite{Rillo2019}.}
    \label{fig:phase-diagram-low-t-optics}
\end{figure}
In Fig.~\ref{fig:phase-diagram-low-t}, we sketch the current low-temperature phase diagram up to the IMT region.  Below \SI{200}{\kelvin} in phase III, we report the transition to the semimetallic state at \SI{350}{\giga\Pa}~\cite{Eremets2019} as a blue shaded area and the later transition to the ``metallic'' state at \SI{425}{\giga\Pa} \cite{loubeyre_nat_20} as the red shaded area. At \SI{495}{\giga\Pa}, we also report the results of the Harvard group~\cite{Dias2017} as the red-filled area. No experimental information is available for metallization at temperatures higher than $\SI{\sim 100}{\kelvin}$. 
Candidate structures for the low temperature phase III, emerging from AIRSS~\cite{Pickard2007,Pickard2012} and a subsequent QMC analysis \cite{McMinis2015}, are C2/c-24 and Cmca-12. Gap closure of those structures has been detected with QMC methods \cite{Gorelov2020} finding that the C2/c-24 structure undergoes an indirect gap closure around \SI{360}{\giga\Pa}, in quantitative agreement with the experimental transition to the semimetallic phase. Moreover, it was shown that the direct gap of the C2/c-24 phase follows the experimental absorption edge behavior with pressures from Ref.~\cite{loubeyre_nat_20}. We emphasize that this agreement can only be obtained by taking into account the zero point motion of the protons~\cite{Gorelov2020}. Moreover, modeling of the correlated electrons requires going beyond common DFT approximations and using QMC methods \cite{Rillo2018, azadi_prb_19, Gorelov2020}.
A recent study \cite{Monacelli2020,monacelli_nat-phys_23}, based on the Stochastic Self-Consistent Harmonic Approximation (SSCHA) and QMC-corrected energies suggests that the observed collapse of the direct gap at \SI{425}{\giga\Pa} \cite{loubeyre_nat_20} is related to a structural transition from the C2/c-24 to the Cmca-12 metallic structure, in both hydrogen and deuterium, as verified by experiments~\cite{Loubeyre2022}.

Above \SI{200}{\kelvin} phase III transforms into a new phase IV when the molecular vibron line splits into two characteristic frequencies suggesting that different molecules may experience two different environments~\cite{Howie2012}. For this phase, theory predicted a layered structure in which regular molecular layers are intercalated with planar layers where molecules form almost regular hexagons, precursors of a transition to a hexagonal monoatomic layer at higher pressure~\cite{Pickard2012, Monserrat2018}. Not much is known about metallization in this higher temperature phase (see however Ref.~\cite{Dalladay-Simpson2016, Monserrat2018}). 

The black lines in Fig.~\ref{fig:phase-diagram-low-t} are experimental phase boundaries. However, experimental information about the crystalline structures of those phases are missing, and the experimental determination of the melting line is not based on the evidence of the vanishing of long-range order (Bragg peaks) above the melting line.
We also report the melting line from two different theories. The orange line is from BOMD simulations with the DFT-PBE approximation~\cite{Bonev2004,Morales2010}. The green line is a recent prediction from a Machine Learning force field trained on QMC energies and forces (ML-QMC)~\cite{Niu2023}. Within DFT, the location of the melting line and LLPT line strongly depends on the approximation for the exchange-correlation functional.
The DFT-PBE melting line seems to agree better with the experiments, although it was observed that, considering quantum nuclei within this theory, leads to unphysical results \cite{Morales2013, Pierleoni2016}. At pressures between \SI{50}{\giga\Pa} and \SI{200}{\giga\Pa}, ML-QMC  predicts a different solid structure that is much more stable than in PBE-PIMD calculations. This prediction has yet to be confirmed by experiments.

Above the melting line 
fluid hydrogen can be either molecular or atomic. First-principle simulations, both by Born-Oppenheimer Molecular Dynamics (BOMD) and by Coupled Electron-Ion Monte Carlo (CEIMC) \cite{scandolo_pnas_03, Lorenzen2010, Morales2010, Pierleoni2016, Rillo2019, karasiev2021,Bryk2020}, predict the existence of a weakly first order LLPT between the molecular insulating fluid and the mostly monoatomic metallic fluid, below a critical temperature estimated to be $T^{\rm cr}_{\rm LLPT}\sim \SI{2000}{\kelvin}$. This is signaled by a small discontinuity in the specific volume at a given pressure and temperature. 
The precise value of $T^{\rm cr}_{\rm LLPT}$ is not known but, above $T^{\rm cr}_{\rm LLPT}$, molecular dissociation and metallization increase progressively with pressure. In Fig.~\ref{fig:phase-diagram-low-t-optics}, we report CEIMC predictions (cf. blue and red shaded lines) for the LLPT line of hydrogen and deuterium between \SI{600}{\kelvin} and \SI{1500}{\kelvin} \cite{Pierleoni2016}. Other lines from BOMD are in qualitative agreement, but their precise location depends on the specific XC approximation adopted \cite{Pierleoni2016, Karasiev_PRB_2022}. Unequivocal experimental evidence of the occurrence of a first-order transition is presently missing.

Experiments on fluid hydrogen are performed either by static compression, with DAC \cite{Zaghoo2016}, or by dynamic compression, with shock wave techniques \cite{weir_prl_96,fortov_prl_07,Knudson_Science_2015,Celliers_Science_2018}, cf. Fig.~\ref{fig:h-phase-diag-exp}. During the pressure increase, the sample first becomes opaque when the electronic gap [cf. Sec.~\ref{ss:h-gap}] becomes comparable to the energy of the probe laser (absorption coefficient $\mu\geq \SI{1}{\micro\meter^{-1}}$), and later, at higher pressure, it turns reflective, signaling the occurrence of the metallic state (reflectivity $R=0.3$). 

In Fig.~\ref{fig:phase-diagram-low-t-optics}, we show static compression data at $R=0.3$ (DAC-r) from Refs.~\cite{Zaghoo2016,Zaghoo2018} (short-pulse DAC, sp-DAC) and Ref.~\cite{Jiang2020} (long pulse DAC, lp-DAC) on hydrogen and deuterium, together with shock wave data on deuterium from the Sandia group \cite{Knudson_Science_2015} and from the Livermore group \cite{Celliers2018}. The inception of absorption ($\mu\geq \SI{1}{\micro\meter^{-1}}$) in those experiments is represented by open symbols while closed symbols represent the occurrence of reflective samples ($R=0.3$). Also, data from sp-DAC show their temperature plateau (DAC-p). Together with experimental data, Fig.~\ref{fig:phase-diagram-low-t-optics} also reports theoretical predictions (obtained within the Kubo-Greenwood approach, cf. Sec.~\ref{sss:dft-kubo-greenwood}) for the inception of an opaque sample and inception of a metallic sample,  from Ref.~\cite{Rillo2019}. While we see an essential agreement between different experiments and with the theory, for the absorption threshold,  reflectivity measurements are not in agreement between the two different shock wave experiments. Note that in those experiments, temperature is not directly measured but it is inferred from a model EOS, and predictions are sensitive to the adopted model. Theoretical predictions for the reflectivity threshold are closer to the NIF-r data and are also in agreement with sp-DAC-r measurements.

In Ref.~\cite{Rillo2019}, it was shown that $R=0.3$ corresponds roughly to a conductivity value of $\sigma=\SI{2000}{\siemens\per\centi\meter}$, although the correspondence depends on the refractive index of the considered interface (see Ref.~\cite{Rillo2019} for details). This value of the conductivity of metallic hydrogen was observed as saturation value in early shock wave experiments~\cite{weir_prl_96}. In Fig.~\ref{fig:phase-diagram-low-t-optics} we also report theoretical predictions of the occurrence of metallic hydrogen ($\sigma=\SI{2000}{\siemens\per\centi\meter}$) from PIMD with a modern XC approximation (SCAN+rVV10)\cite{Hinz2020}. While below the critical point, those predictions are in agreement with other estimates, at higher temperatures this behavior deviates towards the absorption threshold. 

Before closing this section we should note that the new ML-QMC melting line has a potentially large impact on interpreting and understanding this part of the phase diagram. For instance, the thermodynamic path followed during the compression in the shock wave experiments might be influenced by the occurrence of crystallization and subsequent re-melting of the sample. We observe that, except for the $\mu=\SI{1}{\micro\meter^{-1}}$ data, all other data show a  ``kink'' as the temperature is lowered including that of the sp-DAC-p data. The relative location of the new melting line and the LLPT line from CEIMC suggests that this kink might be the signature of a reentrant melting transition -- a speculation that needs confirmation. Also, we note that a higher melting line restricts the LLPT region to a much smaller portion of the phase diagram.

\subsubsection{Hydrogen phase diagram in the high-density plasma phase. Proton crystal (atomic solid hydrogen)\label{sssec:hydrogen-proton-cryst}}
Finally, we consider temperatures exceeding \SI{10000}{\kelvin} and densities exceeding the Mott density, $r_s \lesssim r_s^{\rm Mott}$, where the hydrogen plasma is fully ionized. With increasing density,  the electrons eventually approach an ideal Fermi gas; however, the protonic order can change. Due to the large proton to electron mass ratio, the proton degeneracy is smaller by almost five orders of magnitude, $\chi_e/\chi_p \approx 78\,700$. Consequently, there exists a density range given by [cf. the definitions (\ref{eq:chi}) and (\ref{eq:lambda-debroglie})]
\begin{align}\label{eq:tcp-crystal-range}
\frac{1}{\Lambda_p^{3}}\gg n \gg \frac{1}{\Lambda_e^{3}}\,,     
\end{align}
corresponding to a temperature range $T_{\rm FP}\ll T \ll T_{\rm F}$, where the electrons are strongly degenerate but the protons are still classical, cf.~Fig.~\ref{fig:h-phase-diag}.
There, the system is reasonably well described by a classical one-component plasma (OCP) model of ions embedded into a uniform neutralizing electron background. Depending on the density and temperature, protons will exhibit gas-like, liquid-like or solid-like behavior. The range of the proton crystal (solid atomic hydrogen in a b.c.c structure) is indicated in Fig.~\ref{fig:h-phase-diag}). Screening effects at high density reduce the melting temperature by a factor of about 2~\cite{Liberatore}. The liquid-solid phase boundary at high temperature is approximately given by the classical proton coupling parameter, Eq.~\eqref{eq:gamma}, $\Gamma\approx 175$, whereas the transition to a quantum proton liquid occurs towards high density, around $r_{\rm sp}=\frac{r_{\rm s}}{M}\approx 100$, as a result of proton quantum effects. Finally, the gas-liquid transition is observed in the range $\Gamma \sim 10\dots 20$, e.g. \cite{bonitz_rpp_10}.

\begin{figure}
    \centering
    \includegraphics[width=0.45\textwidth]{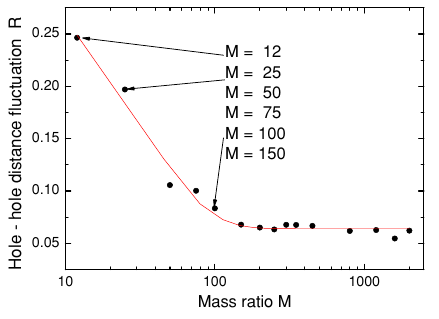}
    \caption{Mean square relative distance fluctuations (in units of the mean interparticle distance) of the heavy particles (ions, holes), Eq.~\eqref{eq:lindemann}, as a function of the mass ratio $M$, showing the role of heavy particle quantum effects. Two-component PIMC simulations for $k_BT=\frac{2}{3}E_B$ and $r_s=0.63$, with the binding energy $E_B$, Eq.~\eqref{eq:atom-bind-energy}.  Taken from Ref.~\cite{bonitz_jpa_06} with the permission of the authors.}
    \label{fig:hole-distance-fluct}
\end{figure}

This behavior is not restricted to hydrogen but is observed also for other dense electron-ion systems.
For a given temperature and density, the liquid-solid boundary differs for different plasmas, as it depends on the ion to electron mass ratio $M$. Results of two-component PIMC simulations are shown in Fig.~\ref{fig:hole-distance-fluct}. There we plot the relative interparticle distance fluctuations
\begin{align}\label{eq:lindemann}
    R = \frac{1}{d}\sqrt{\left\langle r^2_{ij}\right\rangle}\,,
\end{align}
i.e. the mean fluctuations of the distance of neighboring particles in units of the mean interparticle distance, $d$, which provide a very approximate criterion for a solid-liquid transition (modified Lindemann criterion), e.g. \cite{boening_prl_8}: whereas in the solid phase, particles are localized around their lattice positions and distance fluctuations are below a threshold value on the order of $0.05\dots 0.15$, $R$ increases rapidly in the liquid phase. The figure shows that this transition occurs (at fixed temperature and density) when $M$ is reduced, around a value of $M=80$. 

In Ref.~\cite{bonitz_prl_5} also analytical estimates for the existence of an ion crystal were provided. The density range, $[n^{(1)},n^{(2)}]$, the maximum temperature $T^*$ as well as the critical mass ratio were found to be given by
\begin{align}\label{eq:icryst-n1}
  n^{(1)} &= \frac{3}{4\pi}\left( \frac{1}{r_s^{\rm Mott}} \right)^3    \,,\\\label{eq:icryst-n2}
  n^{(2)} &= n^{(1)} \left( \frac{M+1}{M^{\rm cr}+1} \right)^3\,,\\\label{eq:icryst-t}
  T^* &= \frac{4 E_B}{k_B}\frac{Z^2(M+1)}{\Gamma^{\rm cr}\,r_s^{\rm cr}}\,,\\\label{eq:icryst-mcrit}
  M^{\rm cr} &= \frac{r_s^{\rm cr}}{Z^{4/3}\,r_s^{\rm Mott}} -1\,,
\end{align}
where $Z$ is the ion to electron charge ratio. For hydrogen, it follows $n^{(1)} = \SI{0.9d24}{\centi\meter^{-3}}$, $n^{(2)} = \SI{e28}{\centi\meter^{-3}}$, and $T^* = \SI{66000}{\kelvin}$.
The approximate range of the proton crystal is indicated in Fig.~\ref{fig:h-phase-diag} by the red dotted line and denoted as ``solid (atomic) hydrogen''.
Note that, for more accurate estimates of the solid-liquid transition, one has to take into account screening of the ion-ion interaction by the electrons \cite{Liberatore,Militzer_Graham_2006}, e.g. via an effective ion-ion potential, an  effective screening dependent coupling parameter or based on the shape of the pair distribution function, e.g. \cite{hamaguchi_jcp_96,ott_pop_11,ott_pop_14}.

%\item 
As noted above, the crystallization of heavy charged particles is also relevant for other systems. Prominent examples are white dwarf stars where crystallization of carbon ions ($C^{6+}$) and oxygen ions ($O^{8+}$) is expected to occur in the core. Due to the increased ion mass the inequality (\ref{eq:tcp-crystal-range}) leads to a much broader density range and higher maximum temperature.
Equations~(\ref{eq:icryst-n1})--(\ref{eq:icryst-t}) yield, for carbon [oxygen], $n^{(1)} = \SI{2d26}{\centi\meter^{-3}}$ [\SI{6.6d26}{\centi\meter^{-3}}], $n^{(2)} = \SI{3.7d33}{\centi\meter^{-3}}$ [\SI{2.7d34}{\centi\meter^{-3}}], and $T^*= \SI{e9}{\kelvin}$ [\SI{4.2d9}{\kelvin}].
%
%\item 
As a final note, liquid or crystal formation is also relevant for certain semiconductors. In fact, hole crystallization was predicted, among others, by A.~Abrikosov \cite{abrikosov_jlcm_78,abrikosov_88} who predicted a critical hole to electron mass ratio of $M^{\rm cr}\approx 100$ for CuCl. Fermionic PIMC simulations for electron-hole plasmas yielded \cite{bonitz_prl_5,bonitz_jpa_06,filinov_cpp12_2} $M^{\rm cr}\approx 83$, in 3D, and $M^{\rm cr}\approx 60$, in 2D, cf. Eq.~(\ref{eq:icryst-mcrit}). Such mass ratios are feasible in intermediate valence semiconductors, such as Tm[Se, Te] \cite{wachter_prb_04}.
%\end{itemize}

%----------

\subsection{Open questions. Challenges for experiment and simulations}\label{sssec:phase-diagram-opn}
We conclude the analysis of the phase diagram of dense hydrogen by outlining open problems and indicating possible solutions. As we have seen before, the phase diagram of hydrogen is surprisingly complex and identification of phases and phase transitions or crossovers very sensitively depends on the analysis tools. As we have seen in the example of the shock Hugoniot, cf. Sec.~\ref{sec:introduction} and Fig.~\ref{fig:hugoniot}, various experiments and simulation methods may lead to strongly differing predictions which requires a careful analysis.

\begin{itemize}
    \item The high-pressure phases of condensed hydrogen and the LLPT are particularly complex, with many competing possible orders. The simulation results presented in Sec.~\ref{sssec:hydrogen-lowt} include extensive predictions that still await experimental tests.
    \item Accurate experiments and first principle simulations are needed for definitive findings. Reliable simulation results are of great value for benchmarking other less fundamental methods. Examples are fermionic PIMC simulations for the jellium model \cite{schoof_prl15,dornheim_prl16, dornheim_physrep_18}, as well as for hydrogen \cite{filinov_pre_23} as discussed in Sec.~\ref{sec:td}.
    \item The DFT and QMC simulations are currently limited to treating a few hundred atoms. While pressure and internal energy seem to converge fairly quickly with increasing system size, this limitation may lead to biases in the vicinity of phase transitions. Extrapolation to the thermodynamic limit [cf. Sec.~\ref{ss:models-particles}] is necessary to establish an accurate long-range order and critical exponents. Methods that can treat larger system sizes are helpful. A promising technique is to combine DFT or QMC results with machine learning techniques, e.g.~\cite{Fiedler_PRM_2022,fiedler_npj_23,Niu2023}, but the accuracy and reliability to reproduce in general the underlying DFT or QMC simulations still needs to be established.
    \item 
    The high-temperature and high-pressure range is of particular interest for ICF, cf. Fig.~\ref{fig:overview}. This wide range of material conditions places a severe set of constraints on computational techniques used to compute equations of state and transport properties. The radiation-hydrodynamic codes, cf. Sec.~\ref{sss:qhd}, use physics models such as the equation of state, electrical and thermal conductivity, and plasma viscosity as input to the Navier-Stokes equations. It is important that the phenomena are well understood theoretically and modeled accurately so that the ICF design simulations are as accurate as possible. 
\end{itemize}

A major challenge for experiments with matter at extreme densities, temperatures, and pressures, in general, and dense hydrogen, in particular, is given by accurate diagnostics. Indeed, very often, basic parameters such as density and temperature cannot be directly measured and have to be inferred from other observations and theories.
The velocity interferometer for any reflector (VISAR) diagnostics is a standard tool to obtain pressure and density in a shocked state of the sample~\cite{VISAR,Falk_HEDP_2012}. If the pressure standard is known to have good accuracy and the shock is planar enough, it will deliver reliable results. Streaked optical pyrometry (SOP) is a diagnostic used to extract the temperature of the sample~\cite{SOP,Falk_HEDP_2012}. However, it usually measures (near) surface temperatures and the results depend on the (model-dependent and often unknown) reflectivities of the warm dense matter state in a grey body model.

A potentially particularly powerful method of diagnostics that overcomes problems inherent in the standard diagnostics fielded in WDM experiments is given by X-ray Thomson scattering (XRTS)~\cite{siegfried_review,sheffield2010plasma}. XRTS probes the electronic dynamic structure factor (DSF), $S_{ee}(\mathbf{q},\omega)$, convolved with the combined source-and-instrument function $R(\omega_s)$~\cite{Dornheim_T_follow_up}
\begin{eqnarray}\label{eq:convolution}
    I(\mathbf{q},\omega_s) = S_{ee}(\mathbf{q},\omega) \circledast R(\omega_s) \, .
\end{eqnarray}
The momentum transfer $\mathbf{q}$ is determined by the scattering geometry, $\omega=\omega_0-\omega_s$ denotes the energy loss of the scattered photon, and $\omega_0$ is the beam energy.
In principle, the DSF gives one detailed insight into the microphysics of the probed sample and the measured scattering intensity can be used to infer key parameters such as $T$, $\rho$, and the effective charge $Z$~\cite{glenzer_PRL_2007,Falk_HEDP_2012,Falk_PRL_2014,kraus_xrts,Tilo_Nature_2023,dornheim2024unraveling}.

In practice, however, XRTS experiments with hydrogen are notoriously difficult. Hydrogen has a small scattering cross-section, leading to a low photon count and, consequently, a noisy intensity signal, e.g.~\cite{Fletcher_RevSci_2016}.
This is particularly problematic for XRTS experiments with fusion plasmas, which have a very low repetition rate (e.g., one to three shots per day at the NIF~\cite{Moses_NIF}). 
XRTS experiments with hydrogen jets~\cite{Zastrau,Fletcher_Frontiers_2022} at modern X-ray free electron laser (XFEL) facilities, such as LCLS in the USA or the European XFEL in Germany, on the other hand, allow one to average over thousands of shots. Here, a crucial problem is given by the uncertain degree of inhomogeneity and nonequilibrium of the sample~\cite{Vorberger_PLA_2024}.

Let us conclude by outlining two additional, general challenges of XRTS experiments with warm dense matter. On the experimental side, the accurate interpretation of the measured intensity requires detailed knowledge of the source-and-instrument function $R(\omega_s)$, which is a nontrivial task for both backlighter sources~\cite{MacDonald_POP_2022} and XFEL facilities~\cite{kraus_xrts}.
From the theoretical perspective, understanding the measured signal usually requires model or simulation results for $S_{ee}(\mathbf{q},\omega)$, which is a difficult task. In our opinion, the most flexible tool for this purpose is given by time-dependent DFT (TDDFT), which is discussed in Secs.~\ref{sss:tddft} and \ref{sec:TDDFT} below.
An alternative strategy has recently been suggested by T.~Dornheim \textit{et al.}~\cite{Dornheim_T_2022,Dornheim2023,Dornheim_MRE_2023}, who have proposed to evaluate Eq.~(\ref{eq:convolution}) in the imaginary-time domain, see Sec.~\ref{sec:ITCF}. 
Remarkably, this allows for model-free diagnostics of parameters such as the temperature~\cite{Dornheim_T_2022,Dornheim_T_follow_up} and the static structure factor $S_{ee}(\mathbf{q})$~\cite{dornheim2023xray}, and allows for direct comparisons with quasi-exact PIMC simulations~\cite{dornheim2024unraveling,Dornheim_LFC_2024}, see also Sec.~\ref{sec:full_hydrogen_response}.

Given these recent developments, we are confident that future XRTS experiments with hydrogen will give valuable new insights into the equation of state components and other important properties.

\section{Theoretical concepts for warm dense hydrogen \label{sec:theory}}

\subsection{Basic equations. Models and first-principle computer experiments}\label{ss:simulations-intro}
\subsubsection{Basic equations}\label{sss:basic-equations}
The theoretical description of hydrogen starts with the Hamiltonian of $N_e=N_p=N$ electrons and protons with coordinates $(\textbf{r}_1,\dots \textbf{r}_N)$ and $(\textbf{R}_1,\dots \textbf{R}_N)$, respectively,
\begin{align}
    \hat H &= \hat H_e + \hat H_p + \hat W_{ep} \equiv \hat K + \hat V\,,\label{eq:hamiltonian-h}\\
    \hat H_e &= -\frac{\hbar^2}{2m_e}\sum_{i=1}^N \nabla_i^2 + \frac{1}{2}\sum_{i\ne j}^N w(\textbf{r}_i-\textbf{r}_j)\,,\label{eq:hamiltonian-e}\\
    \hat H_p &= -\frac{\hbar^2}{2m_p}\sum_{i=1}^N \nabla_i^2 + \frac{1}{2}\sum_{i\ne j}^N w(\textbf{R}_i-\textbf{R}_j)\,,\label{eq:hamiltonian-p}\\
    \hat W_{ep} &= - \sum_{i, j=1}^N w(\textbf{r}_i-\textbf{R}_j)\,,\label{eq:hamiltonian-ep}
\end{align}
where $w(r)=q^2/r$ is the repulsive Coulomb potential of two electrons (or protons) with $q= \mp e_0$. The total Hamiltonian $\hat H$, Eq.~\eqref{eq:hamiltonian-h}, can be decomposed in various ways, e.g. into contributions of electrons, $\hat H_e$, protons, $\hat H_p$, and electron-proton interactions, $\hat W_{ep}$, or into kinetic, $\hat K=\hat K_e + \hat K_p$, and interaction energy, $\hat V = \hat W_{ee} + \hat W_{pp} + \hat W_{ep}$. For simulations of dense hydrogen, we will assume a macroscopic,  spatially uniform system without external potentials. Exceptions (additional terms in the hamiltonian) are Sec.~\ref{sec:dynamics}, where we compute the density response of hydrogen by studying the reaction to a harmonic field, and, Sec.~\ref{sss:stopping-g1-g2}, where the time evolution is studied that follows the impact of an energetic particle beam.\\

There exist a few methods that allow us to solve the many-body problem with the Hamiltonian (\ref{eq:hamiltonian-h}) exactly (i.e. without approximations or statistical noise from sampling). These include exact diagonalization (configuration interaction, CI) and density matrix renormalization group (DMRG) approaches which are, however, limited to small particle numbers $N$ and/or small number of basis functions and have not been used for dense hydrogen. They are valuable for benchmarking of other methods in model cases that might be relevant to dense hydrogen.

Before discussing simulation approaches that are being applied to dense hydrogen, we briefly summarize the main many-particle properties that are of interest, both for comparison with experiments and for basic physical understanding.

\subsubsection{Observables of interest}\label{sss:observables-overview}

The main quantities of interest can be classified into the following groups for which we also indicate the sections where they are discussed in this paper.
\begin{description}
\item[I.] The first relevant set of observables is given by the \textit{thermodynamic properties} of dense hydrogen, cf.~the top segment in Tab.~\ref{tab:observables} below. This includes the equation of state (pressure) [Sec.~\ref{sss:rpimc-vs-fpimc}], total energy, free energy etc. More detailed information on the structural properties and interaction effects is obtained from the species-resolved static structure factors $S_{ab}(\mathbf{q})$ and pair distribution functions $g_{ab}(\mathbf{r})$ which will be studied in Sec.~\ref{sssec:pdf}.
Complementary microscopic information is contained in both the electronic and ionic momentum distributions $n_e(\mathbf{q})$ and $n_i(\mathbf{q})$, [Sec.~\ref{sec:n(k)}], which are important, e.g.~for the characterization of the effective charge state or, in the case of the ions, for the estimation of ion impact reaction or nuclear fusion rates, e.g.~\cite{starostin_ppr_05,starostin_jetp_17}.
\item[II.] The second group of important observables describe \textit{electronic dynamic  properties}. These include the dynamic structure factor, $S_{ee}(\mathbf{q},\omega)$, which is the key property in XRTS experiments, cf.~Eq.~(\ref{eq:convolution}), and the imaginary-time correlation function (ITCF) $F_{ee}(\mathbf{q},\tau)$, which is connected to the former via a two-sided Laplace transform, Eq.~(\ref{eq:Laplace}).
Interestingly, the ITCF has emerged as an important observable for XRTS diagnostics in its own right~\cite{Dornheim_T_2022,Dornheim_T_follow_up,dornheim2023xray,dornheim2024unraveling,Schoerner_PRE_2023}, as it gives one straightforward and model-free access e.g.~to the temperature. 
\item[III.] The third group of observables are \textit{electronic transport and optical properties}, such as the conductivity $\sigma(\omega)$, the heat conductivity $\lambda(\omega)$, cf. Sec.~\ref{sss:results-sigma}, and the opacity $\kappa(\omega)$, Sec.~\ref{sss:results-opacity}. The latter is a key property in optical transmission experiments, e.g.~at the NIF~\cite{Lutgert_POP_2022}.
\item[IV.] A related fourth group of properties includes \textit{spectral information}, such as the density of states (DOS), as well as the single-particle  spectral function $A(\mathbf{q},\omega)$ and the Matsubara or nonequilibrium Green functions, Sec.~\ref{sss:g1-g2}. Note that the spectral function can be directly measured in photoemission experiments.
Closely related properties are the energy gap that is important to characterize the low-temperature phases of hydrogen, cf. Sec.~\ref{ss:h-gap}, and the ionization potential depression, Sec.~\ref{ss:ipd-overview}, which characterizes the ionization of high density hydrogen in the gas phase.
\item[V.] The fifth set of relevant observables is given by the \textit{ion dynamic properties}, that have traditionally been estimated on the basis of molecular dynamics simulation. In particular, we consider the ionic structure factor,  $S_{ii}(\mathbf{q},\omega)$, which is important for an analysis of the collective modes (ion acoustic modes) and sound speed of the heavy particles [Sec.~\ref{sssec:i-acoustic_H}]. Other important observables are the diffusivity, $D_\textnormal{ion}$,  viscosity, $\eta_\textnormal{ion}$, and the ionic thermal conductivity $\lambda_\textnormal{ion}$.
\item[VI.] The sixth group of observables are related to important \textit{non-equilibrium properties} which characterize the equilibration of hydrogen after an external excitation. In the bottom segment of Tab.~\ref{tab:observables} we list three examples: the relaxation time, $t_\textnormal{rel}$, during which the electron momentum distribution, $n_e(\textbf{q})$, thermalizes to a Fermi function (in general, to a correlated equilibrium distribution),  the equilibration time between electrons and protons, $t_\textnormal{eq}$, and the stopping power, $S_x$, Sec.~\ref{sss:stopping-g1-g2} (and related energy loss properties).

\end{description}

The above list does not pretend to be complete but concentrates on quantities that have been in the focus recently.
Other observables that are potentially interesting for future investigations, but which are not considered here, include various nonlinear response functions~\cite{Dornheim2020,Dornheim_PRR_2022,Tolias_2023,Dornheim_JPSJ_2021,Dornheim_PRE_2021}, effective forces~\cite{Dornheim_JCP_2022} and potentials~\cite{Kukkonen_PRB_1979,Kukkonen_PRB_2021}.

Now, obvious questions arise: which simulation approaches are available to compute the above quantities? And, in case a computation is possible, how accurate and reliable are the results? How can the accuracy be estimated or verified? Finally, if several methods are available -- which one is preferable? In the following, we attempt answers to these questions, with a focus on applications to dense hydrogen. A summary of the relation between methods and observables is presented in Tab.~\ref{tab:observables}. A comparison of accuracy and limitations of various methods is presented in Tabs.~\ref{tab:methods}--\ref{tab:methods-noneq}.

\subsubsection{Hierarchy of simulation approaches}\label{sss:simulations-hierarchy}
For dense hydrogen in the \textit{ground state}, the most accurate results, so far, have been obtained using diffusion or variational Monte Carlo (DMC, VMC). Ground state properties are not the focus of the present paper, for an overview, see Ref.~\cite{mc-mahon_RevModPhys.84.1607}.
For \textit{finite temperature} properties, one has to switch to a mixed ensemble description based on the $N$-particle density operator $\hat\rho$. Here, the most accurate approach is provided by fermionic path integral Monte Carlo (FPIMC) which is free of systematic errors, and convergence to the exact thermodynamic limit can be achieved by proper extrapolation with respect to the particle number $N$ and the number of high-temperature factors $P$, see Sec.~\ref{subsec:jellium}. FPIMC is the only method which is capable of producing unbiased accurate predictions without requiring benchmarks by other simulations. The limiting factor of FPIMC is the fermion sign problem (FSP, cf. Sec.~\ref{sec:FSP}) manifesting itself by an exponentially small signal to noise ratio, which currently limits simulations to temperatures $T\gtrsim 0.5 T_F$. There are various concepts to avoid this limitation by formulating PIMC in second quantization (CPIMC), see e.g. Sec.~\ref{sec:PIMC}, but the accessible parameter range remains restricted by the FSP.

To extend the simulations of hydrogen to the parameters of interest (or to more complex systems), as well as to increase the computational efficiency, a large arsenal of approximate methods has been developed, both for equilibrium and nonequilibrium situations. An overview is presented in Tabs~\ref{tab:methods}--\ref{tab:methods-noneq}, and a more detailed discussion is given in the sections below. The purpose of these tables is to provide a compact comparison of important methods which includes their applicability range and the quantities that they are capable of providing. The methods are listed in the order of their accuracy that is \textit{generally expected}, based on their construction and the involved basic approximation schemes.
Note that the actual accuracy of different methods, when applied to dense hydrogen, may differ from these expectations.
While the ultimate tests are, of course, experiments, they are extremely challenging and complex in the case of dense hydrogen and currently afflicted by large uncertainties or disagreements between different experimental groups and techniques.  Therefore, we base our tests on computer experiments 
for which the accuracy has been convincingly established and which serve as benchmarks. These are fermionic PIMC (FPIMC) results for the uniform electron gas, e.g.~\cite{dornheim_pop17,dornheim_physrep_18}, and hydrogen~\cite{filinov_pre_23}. Presently only a few such benchmarks are available, therefore, in the present paper we produce additional comparisons which are all listed in Tabs~\ref{tab:methods} and~\ref{tab:methods2}.
 
Note that these tests refer to specific parameter ranges of dense plasmas and they are performed for a special choice of approximation (such as self-energy or exchange-correlation functional in the case of Green functions or DFT, respectively).
Therefore, the results of the benchmarks cannot be directly generalized to other parameters or approximations. Nevertheless, they provide valuable guidelines for the behavior of different methods and approximations, when applied to dense hydrogen.
\begin{table*}[th]
%\label{tab:methods}
%
\caption{First-principles simulation methods for dense hydrogen in thermodynamic equilibrium at finite temperature ordered by their expected accuracy. Numbers I.--V. in the column ``Observables'' refer to the list in Sec.~\ref{sss:observables-overview}.
 $G^M(q,\tau)$: Matsubara Green function of ``imaginary time'' $\tau$. The column ``benchmarks'' lists available relevant benchmarks of pressure ($p$), interaction energy ($V$), degree of ionization ($\alpha$), for new ones provided in this paper the relevant figures are indicated, for details see footnotes and main text. For a more complete list of observables, see Tab.~\ref{tab:observables}. \\}
\begin{tabular}{|c|c|c|c|c|c|}
   \hline 
   \textbf{Method} & \textbf{Observables} & \textbf{approximations} & \textbf{main limitations} & \textbf{benchmarks} & \textbf{Section}\\
   \hline
   Fermionic PIMC &  I., II. & number of particles $N$ & fermion sign problem &  & \\
   (FPIMC) & $G^M(q,\tau)$ & number of high-T factors $P$ & $\Theta \gtrsim 0.5$  & N/A & \ref{subsec:jellium}\\
   & nonlinear response & & statistical errors  & & \\
          \hline
   Restricted PIMC    &  &  &  & $p,  \alpha$ & \\
   (RPIMC) & I. & fixed node approximation & $\Theta \gtrsim 0.1$ & based on FPIMC & \ref{sec:alternatives}\\
   & & & & Figs.~\ref{fig:rpimc-pimc-FVT2-pressure}, \ref{fig:ion-degree-mil-fil} & \\
\hline
     &  I., IV. & BO approximation & &  & \\
   CEIMC   & DOS, energy gap & Electrons in ground state &  $\Theta \lesssim 0.1$ & N/A & \ref{subse:ceimc}\\
   &  & using VMC or RQMC & & &\\
\hline\hline
       &  I., II., III., IV.  &  &  moderate  & interaction energy & \\
    Green functions  & $A(q,\omega)$ & Selfenergy $\Sigma$   & coupling strength & based on FPIMC\footnote{Benchmarks of potential energy of Ref.~\cite{dornheim_physrep_18} for jellium.} &\ref{sss:g1-g2} \\
       & DOS & &  &  model systems  & \\
\hline  
\hline
   Kohn-Sham-  & I., II. & BO approx., XC functional  & & $p$ for LDA, PBE & \\
   DFT-MD       & III.  & Kubo-Greenwood relation  & no electron collisions &  and KDT16 & \ref{subsec:dft}\\
                &  V. & XC kernel  & & Fig.~\ref{fig:press_dftmd2} &\\
          \hline
   Orbital free    &  &   &  &  & \\
   DFT-MD       &  I., V. & BO approx., XC functional & no electron collisions  & N/A & \ref{subsec:dft}\\
                &  &   &  &  &\\
          \hline
\end{tabular}
\label{tab:methods}
\end{table*}

In the following, we summarize the results of the available benchmarks, starting with Tab.~\ref{tab:methods}. 1.) Among the approximate methods, RPIMC (even with free-particle nodes) is the most accurate one, achieving an accuracy of the equation of state of better than $2\%$, in a broad parameter range of partially ionized hydrogen (no tests are possible for high densities, $r_s \lesssim 3\dots 4$). Only at the lowest temperature, $T\sim \SI{15000}{\kelvin}$, deviations reach $6\%$ \cite{filinov_pre_23}, cf. Figs.~\ref{fig:rpimc-pimc-FVT2-pressure}, \ref{fig:ion-degree-mil-fil} in Sec.~\ref{sss:rpimc-vs-fpimc}. 2.) Equilibrium Green functions (EGF) with weak coupling self-energies achieve an accuracy better than $10\%$, for the thermodynamic properties of the model UEG, for $r_s \lesssim 1$. With the cumulant approach J.J.~Kas and J.J.~Rehr \cite{Kas_PRL_2017} achieved similar accuracy for a significantly broader range of $r_s$-values. 3.) Kohn-Sham DFT with PBE exchange--correlation functionals exhibits deviations of up to  $7\%$ for the pressure of hydrogen in the density-temperature range corresponding to $r_s\gtrsim 4$ and $T\gtrsim \SI{30000}{\kelvin}$. When finite-temperature functionals are being used, the accuracy improves substantially, to better than $2\%$, for $T\sim \SI{60000}{\kelvin}$, and larger deviations for $T\sim \SI{30000}{\kelvin}$, for details, see Fig.~\ref{fig:press_dftmd2} and Sec.~\ref{subsec:results-finitet-xc}.
Orbital-free DFT is generally expected to be significantly less accurate than KS-DFT, but no benchmarks are available so far.

%%%%%%%%%%%%%%%
A second group of methods that involve stronger approximations are listed
in Tab.~\ref{tab:methods2}. They are expected to be of lower accuracy than the methods in Tab.~\ref{tab:methods}. A few benchmarks are performed in this paper for semiclassical MD with quantum potentials, cf. Fig.~\ref{fig:rpimc-pimc-FVT2-pressure}, and for chemical models, cf. Figs.~\ref{fig:rpimc-pimc-FVT2-pressure} and~\ref{fig:Nfrac}. For example, semiclassical MD with the improved Kelbg potential is able to reproduce the equation of state for temperatures above \SI{60000}{\kelvin} below a critical density (that varies with temperature) with an accuracy of the order of $1\dots 3\%$, see the discussion of Fig.~\ref{fig:rpimc-pimc-FVT2-pressure}. On the other hand,
fluid variational theory (FVT) reproduces the equation of state at low temperatures within approximately $20\%$ but fails at temperatures exceeding \SI{20000}{\kelvin}.

%%%%%%%%%%%%%%%

\begin{table*}[ht]
%\label{tab:methods2}
%
\caption{Further simulation methods for dense hydrogen in thermodynamic equilibrium at finite temperature, ordered by their expected accuracy. Numbers I.--VI. in the column ``Observables'' refer to the list in Sec.~\ref{sss:observables-overview}. $\alpha$: degree of ionization, $x_A$: degree of dissocation (atom fraction). For a more complete list of observables, see Tab.~\ref{tab:observables}.
\\}
\begin{tabular}{|c|c|c|c|c|c|}
   \hline 
   \textbf{Method} &  \textbf{Observables}  &\textbf{approximations} &\textbf{main limitations} & \textbf{benchmarks} & \textbf{Section}\\
   \hline
   Average atom   &  I., II., III.  &  XC functional, & Surrounding atoms  & PDF  & \\
    model    & IV., V.  &  single ion center & not treated explicitly & based on FPIMC & \ref{ss:aa-models} \\
    &&&& Fig.~\ref{fig:PDFT62P} &\\
    \hline\hline
   & &  & wave function modeled   &  & \\
    Wavepacket MD & I., II. & classical dynamics  & by one or few Gaussians & N/A & \ref{sssec:wpmd}\\
          & V., VI. & pair potentials & Approximations needed to  && \\
          &&  & prevent uncontrolled spreading && \\
\hline
    &  & classical dynamics & & pressure $p$,   & \\
   Semiclassical MD & I., II. & quantum pair potentials &  $k_BT \gtrsim 0.5\,$Ry & based on FPIMC &\ref{ssubsec:pair-pot-md}\\
             & V., VI. & e.g. improved Kelbg pot. & only pair exchange & Fig. \ref{fig:rpimc-pimc-FVT2-pressure} & \\
\hline\hline
   Chemical models &  I.  & equilibrium $n(k)$& approximations for interaction & $p, x_A$ & \\
   e.g. FVT  & part of thermo-  & $\alpha, x_A$ & spatial homogeneity & based on FPIMC  & \ref{subsec:chem-model}\\
            & dynamic functions & & & Figs.~\ref{fig:rpimc-pimc-FVT2-pressure}, \ref{fig:Nfrac}& \\
          \hline
\end{tabular}
\label{tab:methods2}
\end{table*}

%%%%%%%%%%%%%%%
Important simulation methods that allow one to access nonequilibrium properties are summarized in Tab.~\ref{tab:methods-noneq}. So far they have been tested much less and not for dense hydrogen, for which currently no benchmarks exist. The most accurate approach is nonequilibrium Green functions (NEGF) and the associated quantum kinetic equations (QKE) for which a few tests for lattice systems against DMRG exist. They indicate that, for weak to moderate coupling, time-dependent observables can be computed with errors below $10\%$, provided the proper self-energies (third order approximation or T-matrix) are being used. The remaining nonequilibrium approaches listed in Tab.~\ref{tab:methods-noneq} involve significantly more restrictive approximations than NEGF, but presently no accuracy tests are available. Benchmarks could be produced in the future based on NEGF simulations by designing well-defined model cases for which all methods of interest are feasible.

%%%%%%%%%%%%%%%

\begin{table*}[ht]
%\label{tab:methods-noneq}
%
\caption{Nonequilibrium simulation methods for dense hydrogen ordered by their expected accuracy. QKT: quantum kinetic theory. Numbers I.--VI. in the column ``Observables'' refer to the list in Sec.~\ref{sss:observables-overview}. $n(\textbf{r},t), \textbf{u}(\textbf{r},t)$: density and velocity field. $n^{\rm eq}(\textbf{k})$: equilibrium momentum distribution; $\omega(\textbf{q})$: plasmon dispersion. nl excitations: nonlinear excitations. For a more complete list of observables, see Tab.~\ref{tab:observables}.
\\
}
\begin{tabular}{|c|c|c|c|c|c|}
   \hline 
   \textbf{Method} & \textbf{Observables} & \textbf{approximations} & \textbf{main limitations} &  \textbf{benchmarks} & \textbf{Section}\\
   \hline
   Nonequilibrium    & I.--IV., VI. & & moderate & $n_i(t)$  & \\
   Green functions  & nonequilibrium DOS & Selfenergy $\Sigma$ & coupling strength & based on DMRG\footnote{Benchmarks of site occupations, $n_i(t)$, of Ref.~\cite{schluenzen_prb17} for Hubbard model.} & \ref{sss:qke}\\
   (NEGF, QKT)  & & & model systems &    &\\
\hline
   &   & XC potential & accuracy of electron &  & \\
   Real-time   TDDFT  & II., III., VI. & adiabatic approximation  & collisions unclear  & N/A & \ref{sss:tddft}\\
            &&&  &&\\
            \hline\hline
   Quantum  & $n(\textbf{r},t), \textbf{u}(\textbf{r},t)$ & $n(\textbf{k})\equiv n^{\rm eq}(\textbf{k})$ & $\Theta \gtrsim 1$, $\Gamma \ll 1$ & forces and potentials & \\
   Hydrodynamics & $\omega(\textbf{q})$ & no exchange effects & limited length \& & based on KS-DFT\footnote{Benchmarks of Ref.~\cite{moldabekov_scipost_22}. }  & \ref{sss:qhd}\\
   & nl excitations, shocks &  & time resolution& &\\
\hline
    & $n(\textbf{r},t), \textbf{u}(\textbf{r},t)$ & $n(\textbf{k})\equiv n^{\rm eq}(\textbf{k})$ & $\Theta \gtrsim 1$, $\Gamma \ll 1$ & & \\
   Hydrodynamics & $\omega(\textbf{q})$ & no quantum effects & limited length \& & N/A  & \ref{sss:qhd}\\
   & nl excitations, shocks & no exchange effects  & time resolution& &\\
\hline
\end{tabular}

\label{tab:methods-noneq}
\end{table*}

In the following, we give a brief overview of the methods presented in this article which can be loosely grouped into particle-based simulations and continuum models.

%23, 138, 8
\definecolor{mygreen}{rgb}{0.09, 0.54, 0.03}
\definecolor{myorange}{rgb}{0.99, 0.69, 0.1}

\begin{table*}[t]
\caption{Summary of relevant observables and simulation methods available to compute them. EOS: equation of state (including energy,  pressure and other thermodynamic variables); $S_{ee}(\mathbf{q})$, $S_{ii}(\mathbf{q})$, $S_{ei}(\mathbf{q)}$: static structure factors; $n_e(\mathbf{q})$: electronic momentum distribution function; $S_{ee}(\mathbf{q},\omega)$: electronic dynamic structure factor; $F_{ee}(\mathbf{q},\tau)$: ITCF [Eq.~(\ref{eq:Laplace})]; $\sigma(\omega)$: electric conductivity in the optical limit; $\kappa(\omega)$: opacity; DOS: density of states; IPD: ionization potential depression; $S_{ii}(\mathbf{q},\omega)$: ion dynamic structure factor; $D_i$: diffusion coefficient; $\eta_i$: viscosity; $\lambda_i$: ion thermal conductivity; $t_\textnormal{rel}$: relaxation time of the electron momentum distribution; $t_\textnormal{eq}$ electron-ion equilibration time; $S_x$: stopping power. \textcolor{mygreen}{Yes} indicates a demonstrated capability, \textcolor{myorange}{Yes} indicates the potential capability and/or some significant limitation that is explained in a footnote, and \textcolor{red}{No} indicates that estimating the observable in question is fundamentally not possible. Note that we do not compare the accuracy of different methods and do not list limitations of the parameter range that is accessible by different methods; this information is provided in Tabs.~\ref{tab:methods}--\ref{tab:methods-noneq}.
}

\begin{center}
  \begin{tabular}{  c | c | c | c || c | c | c || c | c | c || c || c  }
  \hline\hline
     & FPIMC & RPIMC & CEIMC & KS-DFT & OF-DFT &  AA & WP-MD & SC MD & CM & RT-TDDFT & QKT, GF\footnote{\label{foota}Quantum kinetic theory (QKT) and Green functions (GF, including equilibrium and nonequilibrium Green functions) simulations for dense hydrogen are currently limited to full ionization and moderate coupling.}\footnote{Extension to partially ionized hydrogen appears realistic by using Kohn-Sham orbitals as input}  \\ \hline
    \hline
    EOS & \textcolor{mygreen}{Yes} & \textcolor{mygreen}{Yes} & \textcolor{mygreen}{Yes} & \textcolor{mygreen}{Yes} & \textcolor{mygreen}{Yes} & \textcolor{mygreen}{Yes} & \textcolor{mygreen}{Yes} & \textcolor{mygreen}{Yes} & \textcolor{mygreen}{Yes} & \textcolor{red}{No}  &  \textcolor{myorange}{Yes} 
    \\
        $S_{ee}(\mathbf{q})$ & \textcolor{mygreen}{Yes} & \textcolor{mygreen}{Yes} & \textcolor{mygreen}{Yes} & \textcolor{myorange}{Yes\footnote{By integrating over linear-response TDDFT results for $S_{ee}(\mathbf{q},\omega)$ involving additional uncontrolled approximations such as $K_\textnormal{xc}(\mathbf{q},\omega)$.}} & \textcolor{red}{No} & \textcolor{myorange}{Yes\footnote{By integrating over $S_{ee}(\mathbf{q},\omega)$ computed from an approximate dynamic collision frequency in the optical limit.}} & \textcolor{mygreen}{Yes} & \textcolor{mygreen}{Yes} & \textcolor{red}{No} & \textcolor{myorange}{Yes\footnote{Indirectly by integrating over $S_{ee}(\mathbf{q},\omega)$.}} &  \textcolor{myorange}{Yes} \\
        $S_{ii}(\mathbf{q})$ & \textcolor{orange}{Yes} & \textcolor{mygreen}{Yes} & \textcolor{mygreen}{Yes} & \textcolor{mygreen}{Yes} & \textcolor{mygreen}{Yes} & \textcolor{mygreen}{Yes} & \textcolor{mygreen}{Yes} & \textcolor{mygreen}{Yes} & \textcolor{red}{No} & \textcolor{red}{No} & \textcolor{red}{No}  \\
        $S_{ei}(\mathbf{q})$ & \textcolor{mygreen}{Yes} & \textcolor{mygreen}{Yes} & \textcolor{mygreen}{Yes} & \textcolor{mygreen}{Yes} & \textcolor{mygreen}{Yes} & \textcolor{mygreen}{Yes} & \textcolor{mygreen}{Yes} & \textcolor{mygreen}{Yes} & \textcolor{red}{No} & \textcolor{red}{No} & \textcolor{red}{No}  \\
        $n_e(\mathbf{q})$ & \textcolor{mygreen}{Yes} & \textcolor{mygreen}{Yes} & \textcolor{mygreen}{Yes} & \textcolor{myorange}{Yes\footnote{In principle, one can compute $n_e(\mathbf{q})$ by integrating over the spectral function e.g.~in the GW approximation~\cite{PhysRevB.86.195123}.}} & \textcolor{red}{No} & \textcolor{red}{No} & \textcolor{myorange}{Yes\footnote{Using fully anti-symmetrized models non-classical momentum distributions are obtained~\cite{feldmeier2000molecular}, however, the wave packet approximation will influence the result when $\mathbf{q}^{-1}$ is comparable to and smaller than the size of the wave packet}} & \textcolor{red}{No} & \textcolor{red}{No} & \textcolor{red}{No} & \textcolor{myorange}{Yes} 
        \\\hline
                $S_{ee}(\mathbf{q},\omega)$ & \textcolor{myorange}{Yes\footnote{Via a difficult analytic continuation of $F_{ee}(\mathbf{q},\tau)$ [cf.~Eq.~(\ref{eq:Laplace})], see, e.g., Ref.~\cite{Dornheim2018b} for the UEG model.}} & \textcolor{red}{No} & \textcolor{red}{No} & \textcolor{mygreen}{Yes} & \textcolor{red}{No} & \textcolor{mygreen}{Yes} & \textcolor{myorange}{Yes\footnote{\label{h}The limited functional form is noticeable for $\mathbf{q}^{-1}$ comparable to packet size and the computations has previously been computed in a semi-classical manner~\cite{zwicknagel2006wpmd} but quantum formulations are under development.}} & \textcolor{myorange}{Yes} & \textcolor{red}{No} & \textcolor{mygreen}{Yes} &  \textcolor{myorange}{Yes} \\
                $F_{ee}(\mathbf{q},\tau)$ & \textcolor{mygreen}{Yes} & \textcolor{red}{No} & \textcolor{red}{No} & \textcolor{mygreen}{Yes} & \textcolor{red}{No} & \textcolor{mygreen}{Yes} & \textcolor{myorange}{Yes$^{\textnormal{\ref{h}}}$} & \textcolor{red}{No} & \textcolor{red}{No} & \textcolor{mygreen}{Yes} &  \textcolor{myorange}{Yes} \\
                $\sigma(\omega)$ & \textcolor{red}{No} & \textcolor{red}{No} & \textcolor{red}{No} & \textcolor{mygreen}{Yes} & \textcolor{red}{No} & \textcolor{mygreen}{Yes} & \textcolor{red}{No} & \textcolor{red}{No} & \textcolor{red}{No} & \textcolor{mygreen}{Yes} &  \textcolor{myorange}{Yes} \\
                $\kappa(\omega)$ & \textcolor{red}{No} & \textcolor{red}{No} & \textcolor{red}{No} & \textcolor{mygreen}{Yes} & \textcolor{red}{No} & \textcolor{mygreen}{Yes} & \textcolor{red}{No} & \textcolor{red}{No} & \textcolor{red}{No} & \textcolor{mygreen}{Yes} & \textcolor{myorange}{Yes} \\
                DOS & \textcolor{myorange}{Yes}\footnote{While not having been demonstrated, it is in principle possible to compute the Matsubara Green function $G_\textnormal{M}(\mathbf{q},\tau)$~\cite{boninsegni2}, use it as the basis for a reconstruction of the single-particle spectral function $A(\mathbf{q},\tau)$~\cite{JARRELL1996133}, and then integrate the latter over the momentum $\mathbf{q}$.} & \textcolor{red}{No} & \textcolor{red}{No} & \textcolor{mygreen}{Yes} & \textcolor{red}{No} & \textcolor{mygreen}{Yes} & \textcolor{red}{No} & \textcolor{red}{No} & \textcolor{red}{No} & \textcolor{red}{No} &  \textcolor{myorange}{Yes}  \\
                IPD & \textcolor{myorange}{Yes\footnote{Despite the absence of orbitals in PIMC, information about IPD is encoded in $F_{ee}(\mathbf{q},\tau)$, but this has not yet been explored for hydrogen. An alternative approach to the IPD is presented in Sec.~\ref{sss:ipd-qmc}}} & \textcolor{red}{No} & \textcolor{red}{No} & \textcolor{mygreen}{Yes} & \textcolor{red}{No} & \textcolor{mygreen}{Yes} & \textcolor{red}{No} & \textcolor{red}{No} & \textcolor{mygreen}{Yes} & \textcolor{mygreen}{Yes} &  \textcolor{myorange}{Yes}   \\\hline
                
                $S_{ii}(\mathbf{q},\omega)$ & \textcolor{red}{No} & \textcolor{red}{No} & \textcolor{red}{No} & \textcolor{mygreen}{Yes} & \textcolor{mygreen}{Yes} & \textcolor{mygreen}{Yes} & \textcolor{mygreen}{Yes} & \textcolor{mygreen}{Yes} & \textcolor{red}{No} & \textcolor{red}{No} & \textcolor{red}{No}\\
                $D_{\rm ion}$ & \textcolor{red}{No} & \textcolor{red}{No} & \textcolor{red}{No} & \textcolor{mygreen}{Yes} & \textcolor{mygreen}{Yes} & \textcolor{mygreen}{Yes} & \textcolor{mygreen}{Yes} & \textcolor{mygreen}{Yes} & \textcolor{red}{No} & \textcolor{red}{No} & \textcolor{red}{No}\\
                $\eta_{\rm ion}$ & \textcolor{red}{No} & \textcolor{red}{No} & \textcolor{red}{No} & \textcolor{mygreen}{Yes} & \textcolor{mygreen}{Yes} & \textcolor{mygreen}{Yes} & \textcolor{mygreen}{Yes} & \textcolor{mygreen}{Yes} & \textcolor{red}{No} & \textcolor{red}{No} & \textcolor{red}{No}\\
                $\lambda_{\rm ion}$ & \textcolor{red}{No} & \textcolor{red}{No} & \textcolor{red}{No} & \textcolor{mygreen}{Yes} & \textcolor{mygreen}{Yes} & \textcolor{mygreen}{Yes} & \textcolor{mygreen}{Yes} & \textcolor{mygreen}{Yes} & \textcolor{red}{No} & \textcolor{red}{No} & \textcolor{red}{No}
                \\\hline

                $t_\textnormal{rel}$ & \textcolor{red}{No} & \textcolor{red}{No} & \textcolor{red}{No} & \textcolor{red}{No} & \textcolor{red}{No} & \textcolor{red}{No} & \textcolor{red}{No} & \textcolor{red}{No} & \textcolor{red}{No} & \textcolor{myorange}{Yes\footnote{Computing $t_\textnormal{rel}$ is possible in principle, but would require a fully dynamic XC-potential, which is unknown in practice.}} & \textcolor{myorange}{Yes}  \\

                $t_\textnormal{eq}$ & \textcolor{red}{No} & \textcolor{red}{No} & \textcolor{red}{No} & \textcolor{myorange}{Yes \footnote{Within linear response, using KS-DFT obtained electron-phonon matrix elements or electron-ion friction coefficients~\cite{Akhmetov_2023,Simoni_2019}}} & \textcolor{myorange}{Yes \footnote{Using $S_{ii}(\vec q)$ in approximate models for electron-ion collisions (e.g., see the discussions in Ref. \cite{Moldabekov2019}) and combining it with the Mermin model for the dielectric function of electrons \cite{Reinholz_pre_2000}.}} & \textcolor{myorange}{Yes \footnote{Using an effective ion potential in approximate models \cite{Gericke_pre_2002, Kodanova_MRE_2017}. }} & \textcolor{mygreen}{Yes} & \textcolor{mygreen}{Yes} & \textcolor{red}{No} & \textcolor{mygreen}{Yes} & \textcolor{myorange}{Yes}  \\

%                \begin{tabular}{@{}c@{}}Stopping\\ power\end{tabular} 
                $S_x$
                & \textcolor{red}{No} & \textcolor{red}{No} & \textcolor{red}{No} & \textcolor{myorange}{Yes\footnote{Electronic stopping power is computed on the linear response level.}} & \textcolor{myorange}{Yes \footnote{Via the Mermin dielectric function by using $S_{ii}(\vec q)$ in models for the calculation of the electron-ion collision frequency (e.g.~in the Ziman formula \cite{Ziman_1961}).}} & \textcolor{myorange}{Yes \footnote{Within the linear response regime \cite{Hentschel_pop_2023}. }} & \textcolor{mygreen}{Yes} & \textcolor{mygreen}{Yes} & \textcolor{red}{No} & \textcolor{mygreen}{Yes} & \textcolor{myorange}{Yes}    \\\hline
  \end{tabular}
\end{center}
\label{tab:observables}
\end{table*}

\subsubsection{Particle-based simulations}\label{ss:models-particles}
Simulations using particles are well known from classical physics, the most important ones being thermodynamic Monte Carlo (MC) and molecular dynamics (MD). These simulations are potentially exact, i.e. they are free of approximations and systematic errors (based on ``first principles'') if the pair interactions are accurately known. For quantum systems, such as dense hydrogen, 
a true ``first principles'' approach is fermionic PIMC (FPIMC) which was discussed above. To extend FPIMC to a broader parameter range and eliminate the fermion sign problem , several approximate methods have been proposed. The most important ones, based on the so-called fixed node approximation, are restricted PIMC (RPIMC, Sec.~\ref{sec:alternatives}) and coupled electron-ion PIMC (CEIMC, Sec.~\ref{subse:ceimc}) that have been very successfully applied to dense hydrogen and other WDM systems. While the accuracy of the fixed node approximation is not known \textit{a priori}, benchmarks against FPIMC simulations for dense hydrogen at $\Theta \gtrsim 0.5$ confirmed that the relative error of RPIMC in the equation of state and the total energy does not exceed a few percent, cf. Tab.~\ref{tab:methods} and Sec.~\ref{sss:rpimc-vs-fpimc}.
Moreover, the fixed node approximation, both in the ground state and at finite temperature, is variational in nature which provides an internal consistency check among different nodal restrictions and the possibility of systematic improvement by more flexible reference density matrices (RPIMC) or electronic ground state wave functions (CEIMC).\\

The second successful particle-based quantum simulation approach is the combination of Kohn-Sham DFT with MD for the ions, cf. Sec.~\ref{subsec:dft}. It uses a much more severe set of approximations compared to RPIMC: the first is the Born-Oppenheimer approximation to decouple the electron and ion dynamics. The second is a many-body approximation to account for interaction effects which is formulated in terms of the exchange-correlation (XC) functional, see Tab.~\ref{tab:methods}. There exist a number of approximate functionals but their applicability range and accuracy are not known \textit{a priori}; they have to be established by benchmarks for model systems or against FPIMC simulations, which will be done in Sec.~\ref{sec:td}.

%``ab initio''  \\

Particle-based simulations use a finite simulation cell with a small to moderate particle number, $N\sim10-10^3$, and the results are repeated for an increasing number $N$. Scaling versus $N^{-1}$ is performed to establish convergence to the thermodynamic limit (TDL) for all quantities of interest, see e.g. Ref.~\cite{Chiesa_PRL_2006,Drummond_PRB_2008,Holzmann_PRB_2016,filinov_pre_23,dornheim_prl16,dornheim_physrep_18,Dornheim_JCP_2021}. If the scaling of certain observables with the particle number is known analytically, explicit finite size corrections can be derived and the transition to the TDL be performed for any finite value of $N$.\\

A prominent example for the latter case is the interaction energy of the UEG, which is readily expressed as an integral over the static structure factor $S(\mathbf{q})$. This allows for two potential sources of finite-size errors: a) the explicit dependence of $S^N(\mathbf{q})$ on the system size, and b) the approximation of the continuous $\mathbf{q}$-integral by a discrete sum over lattice vectors in the finite simulation cell. In the ground state, S.~Chiesa \textit{et al.}~\cite{Chiesa_PRL_2006} found that it holds $S^N(\mathbf{q})\approx S^\infty(\mathbf{q})$, to a remarkable degree. Moreover, they proposed to correct the discretization error in leading order based on the $q\to0$ limit of $S(\mathbf{q})$, which is known for the UEG~\cite{kugler_bounds}; the resulting finite-size correction scales as $\sim 1/N$ and is often sufficient for applications in the ground state.
While the idea is easily extended to finite temperatures~\cite{Brown_PRL_2013}, this first-order correction becomes insufficient at high densities and temperatures~\cite{dornheim_prl16,dornheim_pop17,dornheim_physrep_18}. General strategies to correct finite size errors beyond leading order have been discussed \cite{Holzmann_PRB_2016} and successfully applied to hydrogen systems at zero electronic temperatures. At finite temperature, T.~Dornheim \textit{et al.}~\cite{dornheim_prl16} introduced a beyond leading order correction based on a suitable trial function for $S(\mathbf{q})$ that, in practice, can be computed on the level of the RPA.  Finally, we note a recent idea by T.~Dornheim and J.~Vorberger~\cite{Dornheim_JCP_2021}, 
how to remove the small finite-size errors from $S^N(\mathbf{q})$ in the high-density regime.

\subsubsection{Continuum models}\label{ss:models-continuum}
Here we briefly discuss models that are formulated in the thermodynamic limit, i.e. the limits $N\to \infty$ and $V\to \infty$ have been performed while the density $n=N/V$ has been kept constant. These models, in one way or the other, use approximations for the interaction part, $\hat V$, of the Hamiltonian \eqref{eq:hamiltonian-h} which are often weak coupling approximations  (perturbation expansions). Alternatively, particular relevant physical effects can be taken into account selectively, such as bound states or dynamical screening. Examples of the latter approximation schemes are Feynman diagram expansions of many-body self-energies \cite{stefanucci2013nonequilibrium, schluenzen_jpcm_19} or decoupling approximations of the BBGKY hierarchy of reduced density operators \cite{bonitz_98teubner}. 

The most accurate, but also the most complex of these approximation schemes, in the case of thermodynamic equilibrium, is  
Equilibrium (or imaginary time or Matsubara) Green functions. The method depends on a single input quantity -- the selfenergy $\Sigma$ -- and would be exact, would $\Sigma$ be known. In practice, of course, one has to use approximations for $\Sigma$, and the accuracy of the results is not known \textit{a priori}, cf. Tab.~\ref{tab:methods}. Here the central quantity is the Matsubara Green function, $G^M(i\omega_n,\textbf{k})$, which, besides statistical information (the k-dependence), also contains spectral information (spectral function and density of states) via the dependence on the Matsubara frequencies $\omega_n$. The accuracy of various selfenergy approximations is unknown \textit{a priori}, and only a few tests are available, for the UEG \cite{Chen2019,Haule22}, for more information, see Sec.~\ref{sss:g1-g2}. They confirm good accuracy of  EGF, if self-energies within their expected range of validity are being used, see Tab.~\ref{tab:methods}.

A less accurate scheme, compared to Green functions theory, is DFT. It uses a simpler quantity as the basis, compared to the Green function -- the density, $n(\textbf{r})$, which depends only on the coordinate, but not the momentum (or frequency). There exist various approximation schemes for the interaction energy as a functional of $n(\textbf{r})$ which avoid Kohn-Sham orbitals (Orbital-free DFT), for details, see Sec.~\ref{subsec:dft}. Further simplified approaches are [cf. Tab.~\ref{tab:methods2}] average atom models, Sec.~\ref{ss:aa-models} and chemical models, Sec.~\ref{subsec:chem-model}. The latter are used
mainly to compute the mean chemical composition of partially ionized plasmas without any spatial or temporal resolution. \\

Finally, in Tab.~\ref{tab:methods-noneq} we list
nonequilibrium models that capture time-dependent properties and, in part, the relaxation behavior of dense hydrogen. Here the most accurate approach is Nonequilibrium Green functions (NEGF) theory and the associated quantum kinetic equations, see  Sec.~\ref{sss:g1-g2}. Other approaches, again in a sequence of decreasing accuracy, are Real-time TDDFT [Sec.~\ref{ss:basics}], wavepacket MD [Sec.~\ref{sssec:wpmd}], and semiclassical MD [Sec.~\ref{ssubsec:pair-pot-md}]. Lastly, additional coarse graining leads to equations for macroscopic space-dependent observables, such as densities, velocity fields, energy density, and so on, which obey hydrodynamic equations, cf. Sec.~\ref{sss:qhd}. 

Using simpler models with more approximations, in general, allows one to reduce the computational effort and access larger length scales, longer simulation times, or more complex systems. Finding the optimal compromise between accuracy and computational cost is an important issue to which we return in Sec.~\ref{ss:outlook}.

%\newpage

\subsection{Chemical models of partially ionized hydrogen}\label{subsec:chem-model}

\subsubsection{Physical vs. chemical picture}\label{sssub:chem-phys-pic}
First-principles many-particle approaches start with physical particles, such as electrons and protons, as ``fundamental'' ingredients. This can be regarded as the ``physical picture'' and it is at the heart of analytical approaches, such as the many-particle Schrödinger equation, density matrix methods, or nonequilibrium Green functions, as well as first-principle simulations, such as quantum Monte Carlo, cf. Sec.~\ref{sec:PIMC}, or density functional theory, Sec.~\ref{subsec:dft}. Of course, the electron states in an isolated hydrogen atom can be rigorously subdivided into bound and scattering states, based on the sign of the electron energy (relative energy of the electron-ion pair). The situation changes immediately at finite temperatures, where thermal energy allows for the excitation of bound electrons into the continuum (ionization). Similarly, in a system of many atoms under high pressure, the electronic states are renormalized by the surrounding particles leading to the formation of bound states of several atoms (e.g. molecules) or the break up of atoms due to pressure ionization (Mott effect), cf. Sec.~\ref{sec:wd-hydrogen}. In these cases, a clear separation between bound and free particles does not exist anymore. Nevertheless, it is often, both intuitive and technically advantageous, to introduce such a subdivision, as it allows for mapping of some of the complex properties of the many-particle system onto a significantly simpler ``chemical picture''. 

There exist various criteria for how to ``identify bound states'' in a first principles approach. This includes the definition of bound state wave functions or of bound contributions of two-particle density matrices and Green functions, e.g., Refs.~\cite{red-book,green-book}. In QMC simulations, one can use criteria based on the electron-proton or proton-proton pair distribution function \cite{filinov_pla_00} or on the extension of the electron path in the vicinity of a proton, e.g., Ref.~\cite{filinov_pre_23}, for QMC applications, see Sec.~\ref{sssec:alpha}. On the other hand, in DFT, one might use electronic density gradients~\cite{moldabekov2023bound}, as in the electron localization functions (ELF)~\cite{Becke_1990}, or maximally localized Wannier functions~\cite{Marzari_2012} to obtain information about the nature of bound states. In DFT-AA models [cf. Sec.~\ref{ss:aa-models}], there are several plausible definitions of the average ionization \cite{murillo2013partial}.
Obviously, the results for the fractions of free and bound particles may depend on the criterion, which requires special care if such an analysis is performed. This is the case, in particular, for densities and temperatures around which the degree of ionization or dissociation changes rapidly (cf. Sec.~\ref{sssec:alpha}).

Once such a subdivision has been performed, a transition to chemical models can be made.
The key to all ``chemical picture'' models is the assumption that electrons, protons, hydrogen atoms, and molecules can be regarded as distinct ``chemical'' species. This can be extended to treating additional bound states such as molecular ions or different excited atomic states as separate ``species''. An extended discussion of why fundamental and composite particles ``must be treated in a \textit{democratic} way'' (i.e. on equal footing) can be found in the text book of W.~Ebeling, W.D.~Kraeft and D.~Kremp \cite{red-book}. The main advantage of chemical models is that one can apply standard tools of thermodynamics or chemical kinetics to compute equilibrium states of dense plasmas including the fractions of free electrons, atoms, and molecules as well as the conditions for phase transitions, cf. Secs.~\ref{sssec:hydrogen-hight} and \ref{sssec:hydrogen-lowt}.

\subsubsection{Chemical equilibrium. Saha equation}\label{sssec:saha}
% done by RR, April 7
Within the chemical picture, using standard thermodynamics of chemically reacting systems, the equilibrium composition of a hydrogen plasma follows either from minimizing the derivatives of the proper thermodynamic potential (e.g. free energy) with respect to the particle number or by equating the chemical potentials of the species for each ``reaction'' or balance equation. In case of hydrogen, there are two main ``reactions'': $e+p \rightleftarrows A$ and $A+A \rightleftarrows M$
leading to the thermodynamic stability conditions
\begin{align}\label{eq:saha-a}
    \mu_e + \mu_p &= \mu_A\,,\\
    2 \mu_A &= \mu_M\,.
    \label{eq:saha-m}
\end{align}
Note that each chemical potential contains an ideal part, which is known, and an interaction contribution, $\mu_a=\mu^{\rm id}_a+\Delta\mu_a, \; a=e, p, A, M$.  Chemical models are formally exact, would $\Delta\mu_a$ be known \footnote{This implies, in particular, that the atomic contribution, $\Delta \mu_A$, includes contributions from all bound states and that all interactions appearing in the Hamiltonian are mapped uniquely on the interaction parts of the chemical potentials.}. This can be exploited by mapping QMC data on a chemical model and extracting unknown parameters, such as the ionization potential depression, cf. Sec.~\ref{sss:ipd-qmc}.

In practice, however, approximations are used which we discuss below. But first, we recall the ideal chemical potentials for free (a=e, p) and bound (B=A, M) particles:
\begin{align}
    \beta\mu^{\rm id}_a &=  \ln \chi_a \,,\label{eq:mu-free-nondegenerate}\\
        \beta\mu^{\rm id, F}_e &= I_{1/2}^{-1}(\chi_e^*/2)\,,\label{eq:mu-free-degenerate}\\
    \beta\mu^{\rm id}_B &= \ln \chi_B -  \ln Z^{\rm int}_B\,,\label{eq:mu-bound}
\end{align}
where $I^{-1}_{1/2}$ is the inverse of the Fermi integral of order $1/2$, $\chi^*_e=\alpha\chi_e$, with $\chi_e$ being the degeneracy parameter of electrons, Eq.~\eqref{eq:chi}, and
$Z^{\rm int}_B$ is the internal partition function for atoms or molecules. Here Eq.~\eqref{eq:mu-free-nondegenerate} and Eq.~\eqref{eq:mu-free-degenerate} refer to nondegenerate and degenerate (fermions) particles, respectively.

We first obtain the two Saha equations (ionization and dissociation equilibria) for the case that all particles are nondegenerate.  
Inserting the results \eqref{eq:mu-free-nondegenerate} and \eqref{eq:mu-bound}
into Eqs.~\eqref{eq:saha-a} and \eqref{eq:saha-m}, we obtain
\begin{align}
    \frac{x_A}{\alpha^2}\bigg|_{\rm class} &= \chi_e(n_e) Z^{\rm int}_A(T)e^{\beta(\Delta\mu_e+\Delta\mu_p-\Delta\mu_A)}\,,\label{eq:saha-h-nondegenerate}\\
     \frac{x_M}{x_A^2} &= \chi_A(n_A) Z^{\rm int}_M(T)e^{\beta(2\Delta\mu_A-\Delta\mu_M)}\,,\label{eq:saha-h2-nondegenerate}
\end{align}
whereas for quantum degenerate electrons, from Eq.~\eqref{eq:mu-free-degenerate} it follows for the ionization equilibrium instead
\begin{align}
    \frac{x_A}{\alpha}\bigg|_{\rm Fermi} &= 2 Z^{\rm int}_A(T) e^{\beta \mu^{\rm id, F}_e(\chi_e^*) }e^{\beta(\Delta\mu_e+\Delta\mu_p-\Delta\mu_A)}\,.\label{eq:saha-h-degenerate}
\end{align}
For any given density and temperature, the two Saha equations can be solved for $\alpha$ and $x_A$ whereas $x_M$ follows from Eq.~\eqref{eq:xm-equation}. These equations still require input for the partition functions and for the interaction parts of the chemical potentials. Before discussing these questions, we illustrate the physical meaning of the ionization equilibrium for the case of low temperatures where the atomic partition function can be approximated by the ground state, $Z^{\rm int}_A(T) \approx e^{-\beta E_1}=e^{\beta I}$, where $I$ is the (positive) ionization potential. Then the remaining terms on the r.h.s. of Eq.~\eqref{eq:saha-h-degenerate} can be understood as contributions $\Delta I$ that cause a reduction of the ionization potential. Rewriting Eq.~\eqref{eq:saha-h-degenerate} in the form of the classical case, Eq.~\eqref{eq:saha-h-nondegenerate}:
\begin{align}
    \frac{x_A}{\alpha^2}\bigg|_{\rm Fermi} &= \chi_e(n_e) e^{\beta(I +\Delta I + \Delta I_{\rm Fermi}})\,,\label{eq:saha-h-degenerate_2}\\
     \Delta I &= \Delta\mu_e+\Delta\mu_p-
     \Delta\mu_A\,,\label{eq:deltai}\\
          \Delta I_{\rm Fermi} &= \mu_e^{\rm id,F}(\chi_e^*) - k_BT\ln \frac{\chi^*_e(n_e^*,T)}{2}\,.
\end{align}
Here $\Delta I$ arises from all interactions in the system, whereas $\Delta I_{\rm Fermi}$ is an additional energy shift due to the spin statistics given by the difference of the chemical potentials of an ideal Fermi gas and ideal classical gas.

Let us now return to the partition function. Since the canonical partition function of Coulomb bound states, $\sum_n n^2 e^{-\beta E_n}$, is diverging, e.g. \cite{green-book}, improved versions that take into account also the spectrum of scattering states have been discussed. Applying higher-order Levinson theorems when performing integration by parts, the leading divergent terms can be projected into the contribution of scattering states so that the well-known Planck-Larkin partition function follows,
\begin{align}
    Z^{\rm int}_A(T) &= \sum_{n=1}^\infty n^2 \{ e^{-\beta E_n} - 1- \beta E_n\}\,,
    \label{eq:z-planck-larkin}
\end{align}
which is convergent; for details, see~\cite{reinholz_PRE_95}. For the molecular partition function, we refer to Refs.~\cite{schlanges-etal.95cpp,reinholz_PRE_95} and references therein. Let us briefly discuss the approximations of the interaction contribution to the chemical potentials; for details, see~\cite{reinholz_PRE_95,redmer_phys-rep_97}.

\begin{itemize}
    \item The contribution of the interaction between charged particles to the chemical potential is known analytically in the limiting cases of low (Debye-H\"uckel) and high densities (Thomas-Fermi). Efficient Padé formulas that are applicable for arbitrary densities and temperatures were constructed by W.~Ebeling and co-workers~\cite{ebeling-richert_85}, recently benchmarked~\cite{dornheim_pop17} and possibly improved by QMC results, see Sec.~\ref{subsec:jellium}.
    \item The (repulsive) interaction between neutrals is often treated within the hard-sphere model for which accurate formulas were derived by N.F.~Carnahan and K.E.~Starling~\cite{Carnahan_1969} and G.A.~Mansoori \textit{et al.}~\cite{Mansoori_1971}. Improved approaches such as Fluid Variational Theory (FVT, see Ref.~\cite{ross_jcp_83}) consider also attractive van-der-Waals-like contributions, see~\cite{juranek_jcp_00,juranek_jcp_02}. 
    \item The interaction between charged and neutral particles is of particular importance in partially ionized plasmas. The respective contribution to the chemical potential can be described by calculating the second virial coefficient with respect to a screened polarization potential as derived by R.~Redmer \textit{et al.}~\cite{Redmer_1987}. 
\end{itemize}
Detailed calculations for the thermodynamic properties of partially ionized hydrogen plasma within the general scheme outlined above were performed by many groups using slightly different approximations for each of the three interaction parts. We would like to mention here the work of W.~D\"appen \textit{et al.}~\cite{daeppen_aj_88} on the solar atmosphere, of D.~Saumon and G.~Chabrier~\cite{saumon-chabrier_prl_89,saumon-chabrier_92} on a wide-range EOS of dense hydrogen plasma, and the FVT of H.~Juranek \textit{et al.}~\cite{juranek_jcp_00,juranek_jcp_02} applied for the calculation of the EOS of hydrogen and, in particular, the Hugoniot curve. 

The dissociation degree (fraction of atoms), as predicted by an advanced chemical model (FVT), is shown below, in Fig.~\ref{fig:FVT2-alpha} \cite{juranek_jcp_02}. At low temperatures, good agreement with tight-binding molecular dynamics (TB-MD) simulations~\cite{lenosky_pre_99} is achieved. With increasing temperature, deviations from QMC simulations are increasing, as is discussed in more detail in Sec.~\ref{sssec:alpha}. Also, the thermodynamic quantities of dense hydrogen calculated within FVT are in good agreement with quantum Monte Carlo results, at low temperature \cite{juranek_jcp_02}. Illustrations are shown 
for the equation of state in Fig.~\ref{fig:rpimc-pimc-FVT2-pressure} below, a discussion is given in Sec.~\ref{sss:rpimc-vs-fpimc}.

\subsubsection{Chemical models for dynamic properties. Chihara decomposition}\label{sssub:chihara}
The chemical picture concept to distinguish between bound and free electrons has also been extended to dynamical (frequency-dependent) quantities. In particular, this assumption
leads to a decomposition of the total dynamic electron structure factor, 
$S_{ee}(\qv,\omega)$, according to~\cite{Chihara_1987,Chihara_2000,siegfried_review,Wuensch_2011}
\begin{align}
\MoveEqLeft \bar{Z}_A S_{ee}(\qv,\omega)=\nonumber\\
&\sum_{\alpha,\beta}\sqrt{x_{\alpha}x_{\beta}}
\left[F_{\alpha}(\qv)+g_{\alpha}(\qv)\right]
\left[F_{\beta}(\qv)+g_{\beta}(\qv)\right]S_{\alpha\beta}(\qv,\omega)\nonumber\\
&+Z_fS_{ee}^0(\qv,\omega)\nonumber\\
&+\sum_{\alpha}x_{\alpha}Z_{\alpha}^b\int d\omega' \tilde{S}_{\alpha}^{ce}(\qv,\omega-\omega')S_{\alpha}^s(\qv,\omega')\,.
\label{eq_chihara}
\end{align}
Here $\bar{Z}_A$ is the average atomic number of the ion species $\alpha=\beta=\{\ldots\}$ such that $N_e=\bar{Z}_A N_i$. The first line describes the contribution of the bound electrons of various ion species represented by the form factor $F_{\alpha}$ and the contribution of the free electrons that screen the ionic charges in screening clouds $g_{\alpha}$. These electronic densities are convolved with the ion structure factors $S_{\alpha\beta}$ and are weighted according to their concentrations $x_{\alpha}=n_{\alpha}/\sum_{\beta}n_{\beta}$. The 2nd line contains the dynamic structure factor of the free electrons in the system, $S_{ee}^0(\qv,\omega)$, with $N_e=Z_f\sum_{\alpha}N_{\alpha}+\sum_{\alpha}Z_{\alpha}N_{\alpha}$. The third line describes transitions of electrons between free and bound states. Here, $\tilde{S}_{\alpha}^{ce}$ describes the correlations of the bound electrons of species $\alpha$ whereas $S_{\alpha}^s$ is the ionic self structure factor.

Several methods exist to calculate the necessary input quantities for the Chihara formula~(\ref{eq_chihara}). If the ion dynamics can be neglected, as is often the case [putting $S_{ii}(\qv,\omega)=S_{ii}(q)\delta(\omega)$], then the ion structure can be obtained from HNC calculations~\cite{Wuensch_2009,Fletcher_Frontiers_2022}, from DFT-MD~\cite{Vorberger_PRE_2015}, or even from PIMC simulations~\cite{dornheim_2024ab}. The form factor $F$ is from an isolated atom or ion and can either be approximated by hydrogenic wave functions or be calculated directly from DFT~\cite{Wuensch_2008b}. The screening cloud $g$ can be taken from linear response theory and may even include closed electron shell effects~\cite{Gericke_2010}. The dynamic structure of the free electrons $S_{ee}^0$ is described using the extended Mermin dielectric function featuring electron-electron local field corrections~\cite{Fortmann_PRE_2010}. The contributions of bound-free transitions and also free-bound transitions to the dynamic structure factor are usually obtained using the impulse approximation in an isolated atom picture in which the medium effects are inserted via effective ionization potentials, i.e., via atomic level shifts using ionization potential depression (IPD) models~\cite{Mattern_2012,Mattern_2013,stewart-pyatt_66}, see Sec.~\ref{sssec:ion-pot}.

Several improvements on the most basic models have been suggested and tested against experiments. DFT and DFT-MD methods can be applied to calculate, both the ion-ion and electron-ion structure, $\left[F(q)+g(q)\right]=S_{ei}(q)/S_{ii}(q)$, so that the first line of Eq.~(\ref{eq_chihara}) can be determined from quantum mechanical simulations alone~\cite{Vorberger_PRE_2015}. Similarly, calculations of the dynamic optical dielectric function using the Kubo-Greenwood formula~\cite{kubo1957statistical,greenwood1958boltzmann}, cf. Sec.~\ref{sss:dft-kubo-greenwood}, based on DFT Kohn-Sham orbitals can be used as improved collision frequency entering the Mermin dielectric function (\ref{eq:mermin_df}) for the free electron structure factor~\cite{plagemann2012dynamic,Sperling_PRL_2015,Witte_PRL_2017}. The combination of all these concepts, including also the DFT determination of the bound-free part, allows one to almost overcome the Chihara picture. The full inelastic spectrum is then obtained using the KG approach based on DFT extended by a Mermin ansatz and the elastic part can be directly sampled from DFT-MD~\cite{Bethkenhagen_2020,Schoerner_PRE_2023,PhysRevB.102.195127}.
Latest efforts utilize the linear response time-dependent DFT (LR-TDDFT) for the inelastic part of the spectrum together with the usual DFT-MD for the elastic part
~\cite{Schoerner_PRE_2023,PhysRevB.102.195127}.

\subsection{Density functional theory simulations}\label{subsec:dft}
DFT-based approaches such as Kohn-Sham (KS)-DFT and orbital-free (OF)-DFT have become the most commonly used first-principle simulation methods for WDM \cite{karasiev_importance_2016}. This is due to a combination of acceptable accuracy over the desired parameter space with a manageable computational cost.

\subsubsection{Basic concepts}\label{ss:basics}
KS-DFT and OF-DFT treat the electronic component quantum mechanically while considering ions as classical particles.
Both flavors of DFT rely on a one-to-one correspondence between the equilibrium electronic density $n_0(\vec r)$ and the external potential, including the one of the ions, $V_{ei}(\vec r)$.
For the time-independent case, the proof of this correspondence is provided by the Hohenberg-Kohn theorems~\cite{hohenberg_inhomogeneous_1964}, for the ground state, which was extended to finite temperatures by D.~Mermin \cite{Mermin_1965, Eschrig_2010}.  
In the time-dependent case, E.~Runge and E.K.U.~Gross showed that a one-to-one mapping exists between the time-dependent single-particle density and the single-particle potential  \cite{Runge_Gross_1984}.
Consequently, one can consider a system of non-interacting fermions in a local effective potential ${v}_{\rm KS}(\vec r)$ (KS potential) that generates the same single-particle electronic density as of a real (interacting) system. Moreover, physical properties of electrons, such as free energy and pressure, can be computed as functionals of the electron density $n(\vec r)$.

In the Kohn-Sham (KS) formulation of DFT, the electronic subsystem is described by the single-particle Schr\"odinger-type equations of auxiliary non-interacting fermions:
\begin{eqnarray}\label{eq:psi_KS}
    i\hbar\partial_t \psi_{\uline{\vec R}}^{j}\left(\vec r, t\right)=\left( -\frac{\hbar^2}{2m_e}\vec{\nabla}^2+{v}_{\rm KS}(\vec r)\right)\psi_{\uline{\vec R}}^{j}\left(\vec r, t\right), 
\end{eqnarray}
where the orbitals $\psi_{\uline{\vec R}}^{j}$ depend parametrically on the ionic positions $\uline{\vec R}=\left(\vec R_1,...,\vec R_{N_{ion}}\right)$ [$N_{ion}$ is the total number of ions].
Equation~(\ref{eq:psi_KS}) provides a framework for the simulation of the time evolution of the electronic density $n(\vec r,t)=\sum_j f_j\left|\psi_{\uline{\vec R}}^{j}(\vec r,t)\right|^2$ in terms of the single-particle KS orbitals $\psi_{\uline{\vec R}}^{j}$, with $f_j$ being the occupation numbers according to the Fermi-Dirac distribution, and the sum is performed over all orbitals.
In Eq. (\ref{eq:psi_KS}), the KS potential is defined as the sum of the classical electrostatic (Hartree) potential, ${v}_{\rm H}(\vec r)$, the exchange-correlation (XC) potential, ${v}_{\rm xc}(\vec r)$, and an external potential that contains the potential of the ions, ${v}_{\rm ext}(\vec r)$, i.e.,   ${v}_{\rm KS}(\vec r)={v}_{\rm H}(\vec r)+{v}_{\rm XC}(\vec r)+{v}_{\rm ext}(\vec r)$. The XC potential plays a key role in compensating for the differences between the auxiliary non-interacting fermions in the effective field and a real system.
The exact form of ${v}_{\rm xc}(\vec r)$ is not known in advance and has to be approximated or provided by a higher-level quasi-exact method such as PIMC. 
In this way, as soon as the XC potential is defined, the KS approach substantially simplifies the computational task of describing quantum-many particle systems.

In practice, because of the large difference in masses of electrons and ions, an additional approximation that is often used for WDM simulations is the Born-Oppenheimer (BO) approximation \cite{Born1927}, in which electrons are assumed to be in equilibrium in the field of a given configuration of ions. Therefore, the BO approximation reduces the KS-DFT to the time-independent situation in the absence of a time-dependent external potential. From Eq. (\ref{eq:psi_KS}) then follows
\begin{eqnarray}\label{eq:ground_KS}
    \left( -\frac{\hbar^2}{2m_e}\vec{\nabla}^2+{v}_{\rm KS}(\vec r)\right)\phi_{\uline{\vec R}}^{j}\left(\vec r\right)=\epsilon^j \phi_{\uline{\vec R}}^{j}\left(\vec r\right), 
\end{eqnarray}
where $\phi_{\uline{\vec R}}^{j}\left(\vec r\right)$ results from the factorization \\
$\psi_{\uline{\vec R}}^{j}\left(\vec r, t\right)=\phi_{\uline{\vec R}}^{j}\left(\vec r\right)\exp(-i\epsilon^jt/\hbar)$, and for the density we have $n(\vec r)=\sum_j f_j\left|\phi_{\uline{\vec R}}^{j}(\vec r)\right|^2$.

Since an equilibrium system is considered, the solution of Eq. (\ref{eq:ground_KS}) is combined with the minimization of the free energy of the system:
\begin{align}
\mathcal F_{\uline{\vec R}}[n] = \mathcal E^{kin}_{\mathrm{s},\uline{\vec R}}[n] - k_B T \mathcal S_{\mathrm{s},\uline{\vec R}}[n] +\mathcal U_{\uline{\vec R}}[n] + \mathcal F^{\rm xc}_{\uline{\vec R}}[n] + \mathcal V^{ei}_{\uline{\vec R}}[n] , \label{eq:F[n]}
\end{align}
where $\mathcal E^{kin}_{{\mathrm{s},\uline{\vec R}}}[n]$ and $\mathcal S_{\mathrm{s},\uline{\vec R}}[n]$ are the kinetic energy and entropy of the auxiliary system of non-interacting fermions computed using KS orbitals and eigenenergies, respectively. Further, in Eq. (\ref{eq:F[n]}), $\mathcal U_{\uline{\vec R}}[n]$ is the Hartree mean-field potential energy,  and  $\mathcal V^{ei}_{\uline{\vec R}}[n]$ is the potential energy due to electron-ion interaction. The functional derivative of the XC functional  $\mathcal F^{\rm xc}_{\uline{\vec R}}[n]$ with respect to density defines the XC potential ${v}_{\rm xc}(\vec r)$ of the effective potential in KS-DFT. For WDM applications, we discuss the various approximations of the  XC functional in Sec. \ref{sss:functionals}.

 Neglecting dynamic electronic polarization effects, the BO approximation allows one to drive the dynamics of ions using forces from the ground state (equilibrium) DFT density of electrons. The BO approximation is expected to be sufficiently accurate for most of the ensemble-averaged macroscopic properties, such as the EOS.

There are a number of packages available that solve Eq. (\ref{eq:ground_KS}) using a variety of numerical methods, libraries, and programming languages. For example, freely available open source codes are {\sc GPAW} \cite{GPAW1, GPAW2, ase-paper, ase-paper2}, {\sc Quantum- Espresso} \cite{Giannozzi_2009, Giannozzi_2017}, and {\sc Abinit}  \cite{Gonze2020, Romero2020, Gonze2016, Gonze2009, Gonze2005, Gonze2002}. A commercial alternative is for example VASP~\cite{Kresse_1993,Kresse_1994,kresse1996efficient,kresse1996efficiency}. Being massively parallel and rigorously tested, these codes with numerous postprocessing capabilities make KS-DFT a powerful tool for first-principle simulation of quantum many-particle systems.

Within DFT, an alternative route is based on the direct minimization of the free energy, Eq.~(\ref{eq:F[n]}), expressed as a functional of the density, without employing KS orbitals. This is the strategy of orbital-free DFT (OF-DFT).
Besides the approximation for the XC term, this requires an explicit formula for the non-interacting free energy, $\mathcal F_{\mathrm{s}}[n]=\mathcal E^{kin}_{\mathrm{s}}[n]-k_BT \mathcal S_{\mathrm{s}}[n]$, as a functional of the density.  Since the exact form of this functional is not known, $\mathcal F_{\mathrm{s}}[n]$ is approximated based on theoretical constraints and information about limiting cases, such as the UEG or an isolated hydrogen atom 
\cite{karasiev_prb_12,karasiev_prb_13,Sjostrom_prb_2013, Luo_prb_2020, Moldabekov_prb_2023}. While being less accurate than KS-DFT, on the other hand, OF-DFT has a much lower computational cost. This enables the simulation of tens of thousands of particles at elevated temperatures \cite{MIHAYLOV2024108931}. For certain materials, the active development of various approximations for $\mathcal F_{\mathrm{s}}[n]$ allows one to perform simulations with sufficient accuracy (e.g., for molecular dynamics) in both ambient \cite{Michele_Chem_Rev} and extreme conditions \cite{Karasiev_cpc_2014}.
Because of the high degree of ionization (i.e., weak electron-ion coupling) and reduced relevance of quantum effects (e.g., compared to the case with $\Theta\lesssim 1$ for hydrogen), OF-DFT is particularly valuable in the limit of high temperatures with $\Theta\gg1$ \cite{MIHAYLOV2024108931, Fiedler_PRR_2022}, e.g., for EOS calculations \cite{Mihaylov_prb_2021}. In this way, OF-DFT is a valuable complement to KS-DFT. Note that there are several open source and freely available codes for OF-DFT-based MD simulations at high temperatures (see e.g. Refs.~\cite{Karasiev_cpc_2014,MIHAYLOV2024108931, DFTpy}).

The main reason for the increase in the computational cost of  KS-DFT is the need for a much larger number of orbitals as compared to the number of particles in the simulation.
This is the result of the significant smearing of the occupation numbers beyond the Fermi energy when $\Theta\gtrsim1$.
To have accurate results at WDM conditions, in addition to other parameters, convergence with respect to the number of used orbitals (i.e., the smallest occupation number) is mandatory \cite{Blanchet_pop_2020}.  For example, keeping the smallest occupation number at about $10^{-6}$, for 108 hydrogen atoms, it was shown using
the highly efficient GPU implementation of VASP \cite{vaspgpu1, vaspgpu2},  that the computation cost scales as $\sim T^2$, at $T\gtrsim 10~{\rm eV}$ and $r_s=2$ \cite{Fiedler_PRR_2022}. Such a quadratic scaling becomes intractable when the number of atoms increases to thousands.
Therefore, new numerical methods are required to push the computational capabilities of  KS-DFT to high temperatures.
To meet the demands of WDM research, S.~Zhang \textit{et al.} \cite{Zhang_POP_2016} introduced the so-called extended KS-DFT where high-energy orbitals are approximated using the analytical formula for the density of states of the free electron gas. The energy threshold $E_c$ above which states are treated analytically is a convergence parameter in simulations using the extended KS-DFT. We note that the extended KS-DFT scheme is implemented in \textit{Abinit}~\cite{Blanchet_cpc_2022}. Another similar hybrid approach was suggested by P.~Hollebon and T.~Sjostrom \cite{Hollebon_PRB_2022}, where the density contribution due to orbitals above $E_c$ is approximated using the Thomas-Fermi model.
In addition, the most recent development in the adaptation of the KS-DFT to meet the needs of high-temperature applications is the stochastic formulation of the KS-DFT \cite{Cytter_Rabani_prb2018, Baer_prl_2013, fabian_WIRE_2019, White_prl_2020,10.1063/5.0163303}.

\subsubsection{Exchange-Correlation Functionals}\label{sss:functionals}

The accuracy of the KS-DFT results depends, to a large degree, on the form of the used XC functional. 
A local density approximation (LDA) is arguably the simplest form one can use for the XC functional  $\mathcal{F}^{\rm xc}_{\uline{\vec R}}[n]$. To simplify the notation, we omit the parametric dependence on the ionic positions and write $\mathcal{F}^{\rm xc}_{\uline{\vec R}}[n]=\mathcal{F}_{\rm xc}[n]$.

For extended systems, the most successful XC-functionals on the LDA-level are based on the UEG model:
\begin{eqnarray}
\mathcal{F}_{\rm xc}[n]&=&\int {\mathrm{d}\vec r}~n(\vec r)f_{\rm xc}[n(\vec r)] \nonumber\\
& \approx& \int {\mathrm{d}\vec r}~n(\vec r)f^{\rm UEG}_{\rm xc}[n(\vec r)]=\mathcal{F}_{\rm xc}^{\rm LDA}[n],
\end{eqnarray}
where $f^{\rm UEG}_{\rm xc}[n(\vec r)]$ is the XC-free energy per particle of the UEG at a given temperature and the local value of the density. 

In practice, the quasi-exact QMC results for the UEG are parametrized (fitted) for KS-DFT applications. 
For example, widely used parametrizations of the ground state QMC results of D.~Ceperley and B.~Alder \cite{Ceperley_Alder_PRL_1980} were provided by J.P.~Perdew and A.~Zunger \cite{Perdew_Zunger_PRB_1981}, and by S.H.~Vosko \textit{et al.} \cite{vwn}.

The inclusion of the corrections due to density inhomogeneity, in the first order, leads to the additional functional dependence on density gradients $\nabla n(\vec r)$.
The most often used and universal XC functionals on this second ``rung'' beyond the LDA are referred to as generalized gradient approximations (GGA). The GGA level functionals, such as PBE \cite{Perdew_PRL_1996} and PBEsol \cite{Perdew_prl_2008} for $T=0$, are designed using various physical (e.g., the QMC data for the UEG limit) and mathematical constraints (e.g.,  Levy's uniform scaling condition \cite{Levy_1989}).
The extension beyond the GGA leads to the meta-GGA rung of functionals like SCAN, where the functional dependencies on the Laplacian of the density $\Delta n(\vec r)$ and on the kinetic energy density $\tau=\sum_j f_j \left|\nabla \phi_{j}\left(\vec r\right)\right|^2$ are added \cite{Sun_prl_2015}. Further advanced inclusion of the non-locality into the XC-functional results in higher rung classes like the hybrid-GGA HSE functional that features exact exchange~\cite{HSE_2003}. This hierarchy of the XC functionals is often referred to as Jacob's ladder, following J.P.~Perdew and K.~Schmidt \cite{Perdew_AIP_2001}. Numerous XC-functionals are available through the Libxc library of XC-functionals \cite{libxc_2012, libxc_2018}.
Similar to the ground state applications, accurate QMC data for the UEG at finite temperatures allowed the development of the XC functionals on different rungs of Jacob's ladder for warm dense matter applications (see Sec. \ref{sss:finit-t-xc}).

\subsubsection{Finite-temperature effects in the functionals}\label{sss:finit-t-xc}

Currently, the vast majority of DFT simulations of WDM use a ground state 
XC-functional without explicit temperature dependence, that is 
${\cal F}_{\mathrm{xc}}[n,T] \approx  E_{\mathrm{xc}}[n]$ -- an approach 
known as the ground-state approximation (GSA). Ground-state functionals 
beyond the LDA rung take into account inhomogeneity effects, but completely miss thermal XC effects. The simplest LDA XC-free energy 
is based on the highly accurate QMC data for the uniform electron gas (UEG) at finite temperature of T.~Dornheim \textit{et al.} \cite{groth_prl17},
and is presented by two equivalent parameterizations already mentioned above, 
the KSDT~\cite{ksdt} and GDSMFB~\cite{groth_prl17} functionals.
 
A ground state GGA functional 
with an additive thermal correction at the LDA level was presented in 
Ref.~\cite{Sjostrom_2014}. J.~Kozlowski \textit{et. al.} exploited a similar idea 
presenting a ground-state PBE functional with a multiplicative 
LDA-level thermal correction~\cite{kozlowski2023generalized}. These two approximations take 
into account the inhomogeneity effects only at the ground-state level, 
i.e. depend on the ground state reduced density gradient variables without explicit 
temperature dependence. Another drawback of these two functionals is that they
are not consistent with the XC-finite-temperature gradient expansion -- one of 
the known and important constraints for the XC-free energy. 
 
V.~Karasiev \textit{et al.} developed a non-empirical fully thermal GGA-level XC-functional, 
KDT16~\cite{Karasiev_PRL_2018}, based on rigorous constraints 
with proper temperature-dependent reduced density gradient 
variables. The KDT16 XC reduces to 
the ground state PBE in the zero-temperature limit, such that it can be used 
across the entire temperature range. 
Recently developed thermal functionals based on the de-orbitalized 
Laplacian-dependent SCAN and its regularized-restored (R2)
version, TSCANL and TR2SCANL respectively, are based on a universal 
thermal additive correction at the GGA level using a perturbative-like self-consistent approach \cite{Karasiev_PRB_2022}.
Semi-local functionals suffer from a fundamental drawback -- underestimating the electronic band gap.
Hybrid XC functionals, such as the global PBE0 \cite{PBE0} and the range separated HSE \cite{HSE_2003} are known to predict 
qualitatively correct band gap values. The recently developed thermal KDT0 hybrid~\cite{PhysRevB.101.245141} is based 
on a mixture of finite-temperature Hartree-Fock (HF) exchange and thermal KDT16 GGA XC.
KDT0 hybrid provides significant improvements to calculations of the entire band structure at temperatures
within the WDM regime, such that the accuracy of optical properties is improved as well upon accounting
thermal XC effects at the hybrid (HF plus density functional approximation) level~\cite{Goshadze.PRB.107.155116.2023}. In Sec.~\ref{subsec:results-finitet-xc} we test the relevance of temperature effects on the functionals for dense partially ionized hydrogen.

\subsection{Average-Atom Models (DFT-AA)}\label{ss:aa-models}
Like the DFT-MD models described in the previous sections, DFT-based average-atom models (DFT-AA) treat electrons quantum mechanically, determining an equilibrium electronic density that is self-consistent with an electron-ion potential. Like DFT-MD models, DFT-AA models can describe the electrons using either orbital-free approximations \cite{thomas1927,Fermi1928} or Kohn-Sham orbitals \cite{Liberman1979,Dharma-wardana-1982,Wilson2006}, and are sensitive to the choice of the exchange-correlation functional. Unlike DFT-MD models, however, DFT-AA models do not determine a full three-dimensional electronic density that varies with the positions of multiple ions; instead they model only a single ion,  implicitly averaging both the electronic density $n_e(r)$ and the screened potential, $V_{ei}(r)$ into spherically symmetric quantities around a single ion center. The effects of changing ion density are incorporated through changes in the boundary conditions at the ionic Wigner-Seitz radius $d_i$, Eq.~(\ref{eq:ws-radius}): typically the potential is forced to vanish at $d_i$ (although other types of boundary conditions have been explored \cite{Liberman1979,Callow2022}. 

DFT-AA models can also predict ionic properties, modeling correlations in the external environment by computing the self-consistent response of electrons in a cavity absent a central nuclear charge \cite{Dharma-wardana_PRE_2012,starrett2012fully,starrett2014hedp}. This external potential can be used to define a neutral pseudo-atom (NPA) charge density and an ion-ion potential that determines the static ion-ion structure factor and radial ion distribution function.

Averaging over both electronic and ionic configurations makes all-electron DFT-AA models computationally highly efficient over a very wide range of densities and up to very high temperatures. As a consequence they provide electronic contributions to many modern equation-of-state tables such as SESAME ~\cite{sesame}. However, averaging over ion ensembles and enforced spherical symmetry make DFT-AA models less reliable than DFT-MD at low temperatures where molecular and lattice structures become important.

Many extensions to basic DFT-AA models have extended their functionality beyond equations of state. The Kohn-Sham electronic orbitals can be used to compute transition matrix elements needed for static \cite{Faussurier2010,sterne,Burrill2016} or dynamic \cite{Johnson2006Optical} conductivities and opacities, and for the collisions and dynamic structure factors relevant to X-ray Thomson Scattering \cite{Johnson_pre_2012,Souza2014,Hentschel_pop_2023} and stopping powers \cite{Faussurier2010,Hentschel_pop_2023}. Expansions of these orbitals into real, integer-occupied electronic configurations can provide detailed X-ray opacities \cite{FaussurierConfigs2018,hansen2023}. Examples of opacities from our DFT-AA model are shown in Fig.~\ref{opacity_D2}
in Sec.~\ref{sss:results-opacity}.

\subsection{Path integral Monte Carlo (PIMC)\label{sec:PIMC}}
\begin{figure}[ht]\centering
\includegraphics[width=0.499\textwidth]{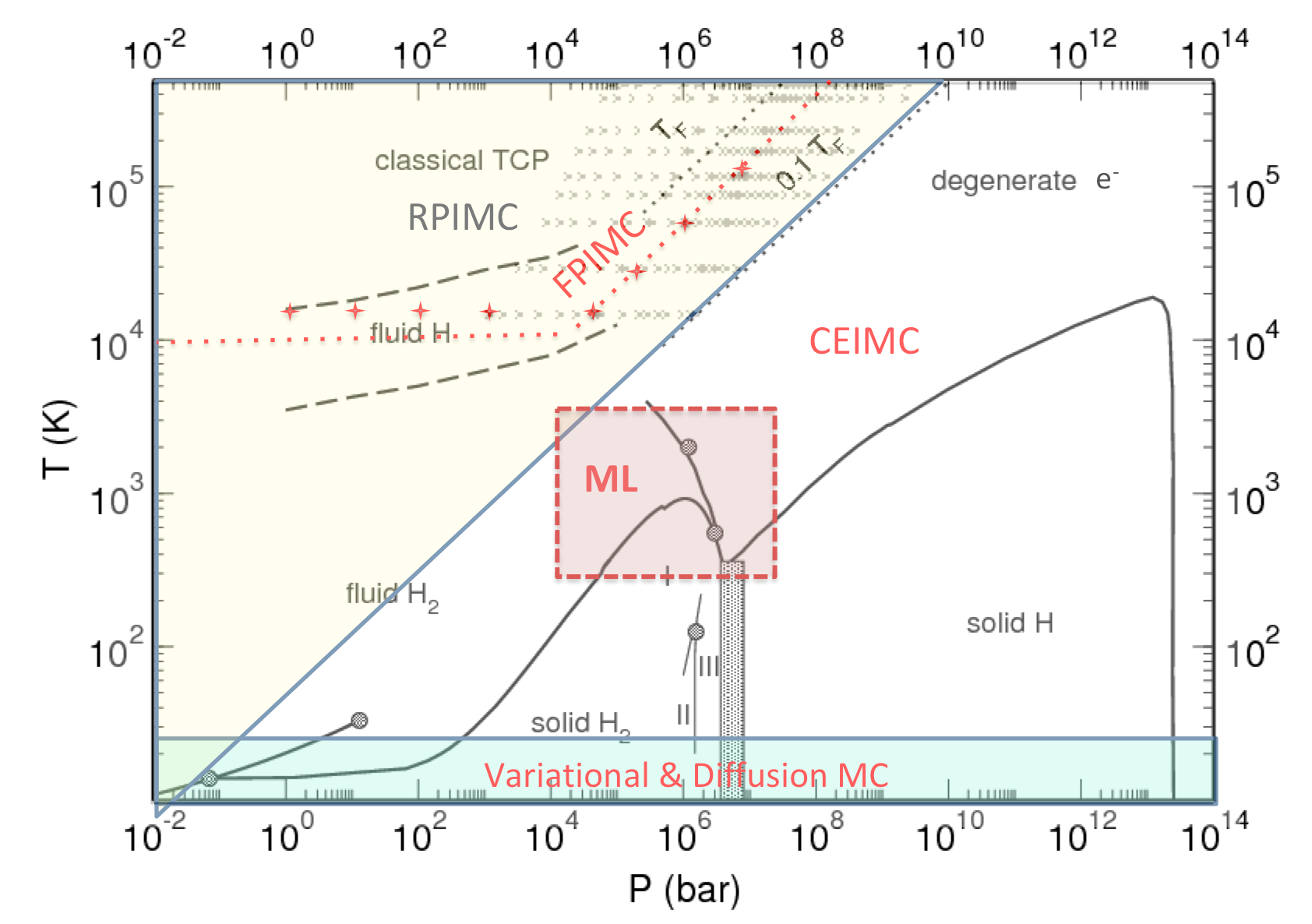}
\caption{ Applicability range of different simulation methods across the hydrogen phase diagram. RPIMC: restricted PIMC, extends to about $0.1\,T_{\rm F}$, Sec.~\ref{sec:alternatives}, CEIMC: coupled electron-ion Monte Carlo, Sec.~\ref{subse:ceimc}, FPIMC: fermionic PIMC, extends to about $0.5\,T_{\rm F}$ and was applied to temperatures above $\SI{10000}{\kelvin}$, Sec.~\ref{sec:PIMC_hydrogen}.  ML denotes region fit by QMC-based machine learning force fields in Ref.~\cite{Niu2023} and also the region where CEIMC has been applied.
Red pluses show  FPIMC simulation points from Ref.~\cite{filinov_pre_23} that mark the border of what is feasible today.
Light black crosses show regions covered by RPIMC database \cite{hu_ICF}. The other lines are introduced in Fig.~\ref{fig:h-phase-diag}.}
\label{fig:h-phase-diag-qmc} 
\end{figure}

Originally introduced for the simulation of low temperature liquid $^4$He~\cite{Fosdick_PR_1966,Jordan_PR_1968}, the PIMC approach~\cite{cep,Berne_JCP_1982,Takahashi_Imada_PIMC_1984,boninsegni1} has emerged as one of the most powerful finite-temperature methods in statistical physics and quantum chemistry. PIMC simulations have been successfully applied to warm dense hydrogen, where a number of different variants have been developed which include restricted (fixed nodes) PIMC (RPIMC), fermionic (direct) PIMC (FPIMC), configuration PIMC (CPIMC), and coupled electron-ion Monte Carlo (CEIMC). In Fig.~\ref{fig:h-phase-diag-qmc}, we give an overview of the parameter range where the different finite temperature methods apply for dense hydrogen. In the limit of vanishing temperatures, variational and diffusion Monte Carlo methods directly address ground state properties \cite{Ceperley1987}. In this section, we focus on FPIMC, Secs.~\ref{subsec:jellium} and \ref{sec:PIMC_hydrogen}, RPIMC, Sec.~\ref{sec:alternatives}, CPIMC, Sec.~\ref{sec:cpimc}, and CEIMC, Sec.~\ref{subse:ceimc}.

Before discussing applications to hydrogen, 
 we introduce the basic idea behind the PIMC method for the Uniform Electron Gas (UEG) model~\cite{dornheim_physrep_18,loos}---the archetypical model system of interacting quantum electrons~\cite{quantum_theory}---in which the ions are replaced by a homogeneous neutralizing positive background. 
Subsequently, we will generalize the PIMC approach to actual electron--ion systems such as hydrogen and discuss how to avoid path collapse due to the singular Coulomb attraction in an efficient way~\cite{MILITZER201688}.
Finally, we will touch upon a variety of possible concepts to avoid the fermion sign problem~\cite{troyer,Dornheim2019a}, which otherwise renders direct PIMC simulations of hydrogen computationally unfeasible for degenerate electrons.

\subsubsection{PIMC simulation of the uniform electron gas}\label{subsec:jellium}

The UEG constitutes the quantum mechanical analog of the classical one-component plasma (OCP). Its Hamiltonian simplifies Eq.~(\ref{eq:hamiltonian-h}) and is given by (using Hartree atomic units) ~\cite{dornheim_physrep_18}
\begin{eqnarray}\label{eq:Hamiltonian_UEG}
    \hat{H}_\textnormal{UEG} = -\frac{1}{2}\sum_{l=1}^N \nabla_l^2 + \frac{1}{2}\sum_{l\neq k}^N W_\textnormal{E}(\hat{\mathbf{r}}_l,\hat{\mathbf{r}}_k) + \frac{N}{2}\xi_\textnormal{M}\ ,
\end{eqnarray}
where $W_\textnormal{E}(\hat{\mathbf{r}}_l,\hat{\mathbf{r}}_k)$ is the Ewald pair potential that a) takes into account the interaction also with the infinite periodic array of images and b) with the neutralizing uniform positive background. The final term in Eq.~(\ref{eq:Hamiltonian_UEG}) is given by the Madelung constant that accounts for the self-interaction between a charge and its own array of images.
 A detailed discussion of the Ewald potential has been presented by L.M.~Fraser \textit{et al.}~\cite{Fraser_PRB_1996}.

To derive the PIMC approach, we consider a system of $N$ spin-unpolarized (i.e., $N^\uparrow=N^\downarrow=N/2$) electrons in the canonical ensemble where the volume $\Omega=L^3$, number density $n=N/\Omega$ and inverse temperature $\beta=1/k_\textnormal{B}T$ are fixed. The partition function in coordinate space then reads
\begin{widetext}
\begin{eqnarray}\label{eq:Z}
    Z_{N,\Omega,\beta} = \frac{1}{N^\uparrow!N^\downarrow}\sum_{\sigma^\uparrow\in S_{N^\uparrow}}\sum_{\sigma^\downarrow\in S_{N^\downarrow}} (-1)^{N_\textnormal{pp}} \int_\Omega \textnormal{d}\mathbf{R}\ \bra{\mathbf{R}} e^{-\beta\hat{H}} \ket{\hat{\pi}_{\sigma_{N^\uparrow}}\hat{\pi}_{\sigma_{N^\downarrow}}\mathbf{R}}\ ,
\end{eqnarray}
\end{widetext}
where $\mathbf{R}=(\mathbf{r}_1,\dots\mathbf{r}_N)^T$ represents the coordinates of both spin-up and spin-down electrons. Note that we have to explicitly take into account all possible permutation elements $\sigma^i$ of the respective permutation group $S_{N^i}$ (with $\hat{\pi}_{\sigma_{N^i}}$ being the corresponding permutation operator, $i=\uparrow,\downarrow$). Evidently, the partition function is given by a sum over contributions that change their sign as a function of the number of pair permutations $N_{pp}$. This is the root cause of the fermion sign problem~\cite{Dornheim2019a} that is discussed in more detail below.

At this point, the main practical problem is the evaluation of the matrix elements of the density operator $\hat{\rho}=e^{-\beta\hat{H}}$ since the kinetic and potential contributions to the full Hamiltonian $\hat{H}=\hat{K}+\hat{V}$ do not commute. To overcome this obstacle, one can make use of the exact semi-group property of the density operator
\begin{eqnarray}\label{eq:semi_group}
    \hat{\rho} = \prod_{\alpha=0}^{P-1} e^{-\epsilon\hat{H}}\ ,
\end{eqnarray}
where the reduced inverse temperature is defined as $\epsilon=\beta/P$. Inserting $P-1$ unity operators of the form $\hat{1}=\int\textnormal{d}\mathbf{R}_\alpha \ket{\mathbf{R}_\alpha}\bra{\mathbf{R}_\alpha}$ then leads to the familiar expression~\cite{dornheim_physrep_18}
\begin{widetext}
\begin{eqnarray}\label{eq:Z_modified}
    Z_{N,\Omega,\beta} = \frac{1}{N^\uparrow!N^\downarrow}\sum_{\sigma^\uparrow\in S_{N^\uparrow}}\sum_{\sigma^\downarrow\in S_{N^\downarrow}} (-1)^{N_\textnormal{pp}} \int_\Omega \textnormal{d}\mathbf{R}_0\dots\textnormal{d}\mathbf{R}_{P-1}\ \bra{\mathbf{R}_0} e^{-\epsilon\hat{H}} \ket{\mathbf{R}_1} \dots
    \bra{\mathbf{R}_{P-1}}e^{-\epsilon\hat{H}} 
    \ket{\hat{\pi}_{\sigma_{N^\uparrow}}\hat{\pi}_{\sigma_{N^\downarrow}}\mathbf{R}_P}\ ,
\end{eqnarray}
\end{widetext}
with $\mathbf{R}_0=\mathbf{R}_{P}$. The original Eq.~(\ref{eq:Z}) has thus been re-cast as an integral over $P$ sets of particle coordinates where the corresponding density matrix has to be evaluated at $P$ times the original temperature.
For a sufficiently large $P$, one may introduce a suitable high-temperature factorization such as the \textit{primitive approximation} $e^{-\epsilon\hat{H}}\approx e^{-\epsilon\hat{K}}e^{-\epsilon\hat{V}}$; convergence of the latter is ensured by the well-known Trotter formula~\cite{trotter}
\begin{eqnarray}\label{eq:trotter}
    e^{-P\epsilon(\hat{K}+\hat{V})} = \lim_{P\to\infty} \left( e^{-\epsilon\hat{K}}e^{-\epsilon\hat{V}} \right)^P\ .
\end{eqnarray}
It is important to note that Eq.~(\ref{eq:trotter}) only holds for operators that are bounded from below~\cite{kleinert2009path}. This is not the case for hydrogen due to the diverging Coulomb attraction between electrons and ions, which will require special care, cf.~Sec.~\ref{sec:PIMC_hydrogen}.
In addition, we note that more efficient factorization schemes have been discussed in the literature~\cite{sakkos_JCP_2009,Zillich_JCP_2010,Takahashi_Imada_PIMC_1984}, which is of particular interest for the PIMC simulation of fermions~\cite{Chin_PRE_2015,Dornheim2015a,dornheim_cpp_19,filinov_pre_23} as we
elaborate in Sec.~\ref{sec:alternatives} below.

Finally, the partition function can be expressed as
\begin{eqnarray}\label{eq:Z_final}
    Z_{N,\Omega,\beta} = \SumInt\textnormal{d}\mathbf{X}\ W(\mathbf{X})\ ,
\end{eqnarray}
i.e., as an integral over the high-dimensional meta-variable $\mathbf{X}=(\mathbf{R}_0,\dots,\mathbf{R}_{P-1})^T$ where the symbolic notation $\SumInt\textnormal{d}\mathbf{X}$ contains all integrals as well as the summation over all permutations.
Each particular $\mathbf{X}$ can be interpreted as a path configuration through the \textit{imaginary time} $t=-i\hbar\tau$ with $\tau\in[0,\beta]$, where each particle is represented on each of the $P$ discrete imaginary time slices. Further, each path contributes proportionally to its weight $W(\mathbf{X})$, which is a function that can be readily evaluated in practice.

\subsubsection{The fermion sign problem\label{sec:FSP}}

Eq.~(\ref{eq:Z_final}) is a $3PN$-dimensional integral, with $N\sim\mathcal{O}\left(10-10^2\right)$ and $P\sim\mathcal{O}\left(10^2\right)$ as well as the summation over $N !$ permutations. Its numerical evaluation using standard quadrature methods is thus unfeasible in practice. % due to the well-known curse of dimensionality~\textcolor{red}{[cite]}. 
However, a stochastic estimation  based on the Metropolis algorithm~\cite{metropolis} overcomes this bottleneck, as its efficiency depends only weakly on the dimensionality~\cite{bonitz_rinton}. The task at hand is thus generating a Markov chain of \textit{configurations} $\mathbf{X}$, which are distributed according to $P(\mathbf{X})=W(\mathbf{X})/Z_{N,\Omega,\beta}$. Indeed, path sampling schemes~\cite{cep,boninsegni1,boninsegni2,Dornheim_PRB_nk_2021} allow for quasi-exact simulations of $N\sim\mathcal{O}\left(10^3-10^4\right)$ bosons (such as superfluid $^4$He~\cite{cep,Ferre_PRB_2016,ultracold2}) or boltzmannons (i.e., hypothetical distinguishable quantum particles, e.g., Refs.~\cite{Clark_PRL_2009,dornheim_cpp16}), giving unprecedented insights into important phenomena such as superfluidity~\cite{cep,ultracold2,Dornheim_PRA_2020}.
For fermions, on the other hand, the $(-1)^{N_\textnormal{pp}}$ term leads to both positive and negative weights $W(\mathbf{X})$, thus preventing the interpretation of $P(\mathbf{X})$ as a probability distribution.

Formally, this obstacle is easily avoided by switching to a modified configuration space that is defined by
\begin{eqnarray}\label{eq:modified_configuration_space}
    W'(\mathbf{X}) &=& \left| W(\mathbf{X})\right|\ , \quad P'(\mathbf{X}) = \frac{W'(\mathbf{X})}{Z'_{N,\Omega,\beta}}\ , \\ \nonumber Z'_{N,\Omega,\beta}&=&\SumInt\textnormal{d}\mathbf{X}\ W'(\mathbf{X}) \ .
\end{eqnarray}
The expectation value of an arbitrary observable $\hat{A}$ is then computed as
\begin{eqnarray}\label{eq:ratio}
\braket{\hat A} = \frac{\braket{\hat{A}\hat{S}}'}{\braket{\hat{S}}'}   \ , 
\end{eqnarray}
with $S(\mathbf{X})=W(\mathbf{X})/W'(\mathbf{X})$ and
\begin{eqnarray}
    \braket{\hat{A}}' &=& \frac{1}{Z'_{N,\Omega,\beta}} \SumInt\textnormal{d}\mathbf{X}\ W'(\mathbf{X}) A(\mathbf{X}) \ ,
\end{eqnarray}
where $A(\mathbf{X})$ is the \textit{estimator} of $A$ in the path-integral representation.
The denominator in Eq.~(\ref{eq:ratio})  the \textit{average sign} $S\equiv\braket{S}'$ ~\cite{Dornheim2019a} and measures the cancellation between positive and negative contributions to the fermionic partition function.

The relative error of a Monte Carlo estimate for $\braket{\hat{A}}$ assuming $N_\textnormal{MC}$ statistically independent samples is then given by~\cite{Ceperley1996}
\begin{eqnarray}\label{eq:relative_error}
    \frac{\Delta A}{A} = \frac{\sigma_A}{S\sqrt{N_\textnormal{MC}}} = \frac{e^{(f'-f)N\beta} \sigma_A}{\sqrt{N_\textnormal{MC}}} \ .
\end{eqnarray}
Here $\sigma_A$ is the intrinsic variance of the estimator $A(\mathbf{X})$, and $f'$ and $f$ are the free-energy densities of $Z'_{N,\Omega,\beta}$ and $Z_{N,\Omega,\beta}$ with $f'\geq f$. 
Eq.~(\ref{eq:relative_error}) thus directly implies that the statistical error of a fermionic PIMC simulation increases exponentially with increasing system size and decreasing temperature (or, equivalently, increasing $\beta$). 
This exponential wall can only be compensated by increasing the number of Monte-Carlo samples as $1/\sqrt{N_\textnormal{MC}}$, which quickly becomes computationally unfeasible.
This is the \textit{fermion sign problem}~\cite{Dornheim2019a}, that prevents the direct application of the PIMC method over substantial parts of the relevant WDM parameter space, see also Ref.~\cite{dornheim_pop17} for a recent overview article.
This unsatisfactory situation has sparked a remarkable surge of developments in the field of fermionic QMC simulations over the last decade, e.g., Refs.~\cite{Driver_PRL_2012,Brown_PRL_2013,schoof_cpp15,schoof_prl15,Dornheim2015a,Blunt_PRB_2014,dornheim_jcp15,groth_prb16,dornheim_prb16,Malone_JCP_2015,Malone_PRL_2016,yilmaz_jcp_20,Rubenstein_JCTC_2020,Hirshberg_JCP_2020,Dornheim_Bogoliubov_2020,Joonho_JCP_2021,filinov_pre_23,Dornheim_JCP_xi_2023,Dornheim_JPCL_2024}; a selection of methods that are particularly relevant for the simulation of hydrogen is shown in Secs.~\ref{sec:alternatives} and \ref{ssub:fsp-reduction} below. Another idea is to develop PIMC simulations in different quantum mechanical representations where the FSP may appear in different parameter regions. This concept is discussed in Sec.~\ref{sec:cpimc}.

\subsubsection{Configuration PIMC (CPIMC) simulation of the UEG\label{sec:cpimc}}
The physical origin of the fermion sign problem in FPIMC simulations are quantum exchange effects of fermions. The path integral concept maps the simulation of quantum systems at a temperature $T$ on that of classical systems at a $P$ times higher temperature, cf. Eq.~\eqref{eq:semi_group}. Thus, it is not surprising that fermionic PIMC simulations become increasingly difficult with increasing quantum degeneracy, i.e. for increasing $\chi$ or decreasing $\Theta$, as indicated in Fig.~\ref{fig:h-phase-diag-qmc}. On the other hand, for the limit of strong degeneracy, i.e. low $\Theta$, or that of a nearly ideal Fermi gas, i.e. $r_s \ll 1$, well-known concepts exist. Here we will not consider the low-temperature case where ground state methods such as diffusion or variational Monte Carlo are appropriate, see Fig.~\ref{fig:h-phase-diag-qmc}, but concentrate on finite temperatures and low $r_s$-values. In fact,  for $r_s \to 0$, the ideal Fermi gas limit is recovered where $N$-particle states are given by Slater determinants of one-particle orbitals. The $N$-particle density operator $\hat \rho$, as discussed in Sec.~\ref{subsec:jellium}, can now be transformed into a new representation, in addition to the previously considered coordinate representation that we reproduce in the first line, we sketch the idea of configuration PIMC (CPIMC) developed by M.~Bonitz, T.~Schoof and S.~Groth, in the second quantization~\cite{schoof_cpp11},
\begin{align}\nonumber
    e^{-\beta \hat H} & \longrightarrow \sum_{\sigma^\uparrow}\sum_{\sigma^\downarrow} (-1)^{N_\textnormal{pp}}\langle \textbf{R} |e^{-\beta(\hat K + \hat V)}|\hat{\pi}_{\sigma_{N^\uparrow}}\hat{\pi}_{\sigma_{N^\downarrow}}\textbf{R}'\rangle\,,\\
    & \longrightarrow \langle \{n\}| e^{-\beta(\hat K + \hat V)} |\{n'\}\rangle\,,\qquad \text{CPIMC} \,.   
\end{align}
In the coordinate representation, one uses $N$-particle coordinate states $|\textbf{R}\rangle$ which are (non-antisymmetrized) product states of one particle states, $|\textbf{r}_i\rangle$, with $i=1\dots N$ and the antisymmetrization being implemented via a sum over all different permutations, $\sigma^\uparrow$ and $\sigma^\downarrow$, giving rise to sign-alternating terms that cause the FSP. In CPIMC (the second line), one proceeds differently: no antisymmetrization of the density operator is performed but, instead, the matrix elements are computed with antisymmetric $N$-particle states, $|\{n\}\rangle = |n_1n_2\dots \rangle$, where $n_k$ are the occupation numbers of all single-particle orbitals (occupation number representation). While fermionic PIMC in coordinate representation becomes trivial 
%exact for $\hat K \to 0$, i.e.  %%% TD: I changed this; PIMC is always formally exact
in the classical limit (large $r_s$), CPIMC becomes trivial for $\hat V \to 0$, i.e. for an ideal Fermi gas. In both cases, the density matrix is diagonal. Departure from these limits leads to an increasing fermion sign problem: for coordinate space simulations if $r_s$ is lowered and for CPIMC, if $r_s$ is increased. 
The behavior of CPIMC has been studied in detail for various systems, see Refs.~\cite{schoof_cpp11,schoof_cpp15,schoof_prl15,groth_prb16,dornheim_prb16} where these general considerations are confirmed.
Further results are available for the static structure factor \cite{groth_jcp17} and for the momentum distribution function \cite{hunger_pre_21}, for overviews, see Refs.~\cite{cpimc_springer_14,dornheim_physrep_18}.

\begin{figure}
    \centering
    \includegraphics[width=0.45\textwidth]{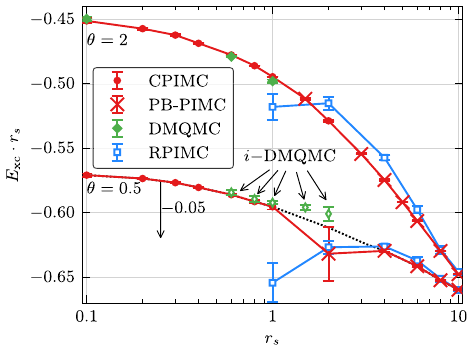}
    \caption{Exchange-correlation energy of $N=33$ spin-polarized electrons in jellium for two dimensionless temperatures: $\Theta=2$ and $\Theta=0.5$. First principle CPIMC and PB-PIMC data can be connected exactly. RPIMC data from Ref.~\cite{Brown_PRL_2013}, DMQMC from Ref.~\cite{Malone_PRL_2016}. In the limit $r_s\to 0$ the Hartree-Fock result is approached and the curves should be horizontal, as observed in CPIMC and DMQMC. $\Theta=0.5$ is the lowest temperature for which CPIMC and PB-PIMC can be smoothly connected, as PB-PIMC is afflicted with increasing errors at $r_s \lesssim 2$.
    For $\Theta=0.5$ all data have been shifted by $0.05~$Ha. Reproduced from Ref.~\cite{dornheim_pop17} with permission of the authors.}
    \label{fig:cpimc-pbpimc-jellium}
\end{figure}

Results for the thermodynamic properties of the  uniform electron gas (the single-particle orbitals generating the Slater determinants are plane waves)
were presented in Refs.~\cite{schoof_prl15, schoof_cpp15}, and we show the exchange-correlation energy of the UEG 
 in Fig.~\ref{fig:cpimc-pbpimc-jellium}. While coordinate space simulations (RPIMC, see Sec.~\ref{sec:alternatives}, and PB-PIMC \cite{Dornheim2015a}, see Sec.~\ref{ssub:fsp-reduction}) are 
 %severely 
 hampered by the FSP when approaching $r_s=1$ \textit{from above}, CPIMC easily fills the gap at small $r_s$-values. At the same time, CPIMC is afflicted by an FSP when $r_s$ approaches $0.5\dots 1$, \textit{from below}.

Thus, an interesting complementarity of the two representations is observed, and a combination of the two allows one to achieve
 first principle PIMC results, for all densities, thereby effectively avoiding the sign problem, as is demonstrated in Fig.~\ref{fig:cpimc-pbpimc-jellium}. A combination of both approaches without gaps is possible for temperatures $\Theta \gtrsim 0.5$, for $N=33$ particles.
 The figure also contains a comparison to RPIMC simulations by E.~W.~Brown \textit{et al.} \cite{Brown_PRL_2013,Brown_PRB_2013} which became increasingly inaccurate when $r_s$ was reduced towards unity. Finally, we also included in Fig.~\ref{fig:cpimc-pbpimc-jellium} data
from density matrix QMC (DMQMC). This approach is similar to CPIMC and was developed by 
 N.S.~Blunt \textit{et al.}~\cite{Blunt_PRB_2014}, as a finite-temperature extension of full-CI-QMC (FCIQMC)~\cite{Shepherd2012,RuggeriAlavi2018}. DMQMC was subsequently applied to the UEG by F.~Malone \textit{et al}.~\cite{Malone_JCP_2015,Malone_PRL_2016}, their data is shown by the green diamonds. 

To conclude this section, we have shown that a combination of two fermionic PIMC methods in complementary representations is a very promising approach to avoid the FSP, not only for jellium but also for hydrogen. However, so far no CPIMC results for hydrogen are available yet. Other strategies to alleviate the fermion sign problem will be discussed in Secs.~\ref{sec:alternatives} and \ref{ssub:fsp-reduction}.

\subsubsection{The Restricted Path Integral Monte Carlo method (RPIMC)\label{sec:alternatives}}

  As discussed above, the direct PIMC method requires sampling both paths and permutations of paths.  For fermions, this requires a minus sign whenever an odd permutation is encountered as R.P.~Feynman and A.R.~Hibbs noted in 1965 \cite{Feynman_Hibbs}.    As explained in Ref.~\cite{Ceperley1996}, this leads to an exponential increase in the computational effort, cf.~Eq.~(\ref{eq:relative_error}).

  In ground-state QMC methods, the fixed-node (FN) method has proved useful to get beyond this limitation \cite{Ceperley_Alder_PRL_1980,anderson2007quantum}.   Here one posits a good many-body trial function (typically from a mean field calculations)  and constructs a solution with the same sign. A stochastic process can be used to find the optimal magnitude.  If the nodes are correct, one obtains the exact energy.  Otherwise, the computed energy lies below the variational Monte Carlo energy but above the exact energy. This rigorous upper bound allows one to judge the relative accuracy of the nodal surfaces even in the absence of experimental data or other methods. If one could completely parameterize nodal surfaces, one could use this to arrive at the exact surface.  The FN approximation is particularly accurate for hydrogenic systems since, in the molecular phase, the nodes are between the molecules and have less influence on the computed properties; in the atomic phase, the wave function is weakly correlated so nodes from DFT orbitals are accurate.

  It is possible to generalize the FN method to non-zero temperature using the restricted path integral method~\cite{Ceperley1991}  (RPIMC). One posits a many-body fermion density matrix  $\rho_T (\mathbf{R},\mathbf{R}'; \tau) $ for all $0 \leq \tau \leq \beta $. Then one can show that the exact density matrix satisfies the identity: %\textcolor{red}{TD: I do not understand this equation; one would have to integrate over sth like $\mathbf{R}_*$, no?}
  \begin{equation}
   \rho_F (\mathbf{R}_{\beta} ,\mathbf{R}_*;\beta) = \SumInt\textnormal{d}\mathbf{X}\ W(\mathbf{X})\
%\sum_P 
%\oint_{\in \Upsilon (R_*)} dR_t e^{-S[R_t] }   
  \end{equation}
where the paths go from  $P\mathbf{R}^* \rightarrow \mathbf{R}_{\beta}$ without crossing a node. That is:  %(restricted to the domain $\Upsilon$)
  \begin{equation}
   \rho_T (\mathbf{R}_\tau, \mathbf{R}_* ; \tau) >  0\, .  
   \label{node}
  \end{equation}
The many-body position that sets the nodes, $\mathbf{R}_*$, is called the reference point. The contribution of the paths $W(\mathbf{X})$ is the same as in Eq. \ref{eq:Z_final}.
This is an identity as long as the sign of the trial density matrix is correct. To obtain observables such as the internal energies, and the pair correlation functions or the momentum distributions one samples the reference point $R_*$ and the permutation, $P$, over their allowed values.  For observables diagonal in coordinate space (e.g. the partition function and the electron density) $\mathbf{R}_{\beta}=\mathbf{R}^*$ and only even permutations contribute, all contributions are non-negative and can be sampled without a sign problem. When off-diagonal elements of the density matrix are sampled to compute, e.g., the momentum distribution, negative contributions still enter the computed averaged even though all paths obey the nodal restriction in Eq.~(\ref{node})~\cite{Militzer2019}.

There remain two main challenges. First, the exact fermionic density matrix is only known in a few cases, e.g., for noninteracting particles. So one is required to employ a trial density matrix, which introduces an uncontrolled approximation into the calculations. For temperatures at or above the degeneracy temperature, one can show that the nodes approach exponentially (in $T$) those of the free particle density matrix,
\begin{equation}
   \rho_T(\mathbf{R},\mathbf{R}^*;\tau) \sim \det_{ij} \left\{ \exp\left[ -m(\mathbf{r}_i-\mathbf{r}^*_j)^2/(2\tau \hbar^2)\right] \right\}\, .
\end{equation}

On the other hand, in the low-temperature limit, if the ground state is unique (non-degenerate), the fermion density matrix will factor into the product $\rho(\mathbf{R},\mathbf{R}'; \beta) \rightarrow \Psi^*(\mathbf{R}) \Psi(\mathbf{R}')$, and RPIMC reduces to the ground state FN method. 
To determine the nodal structure between these two limits, two approximations have been derived. Variational nodes have been developed~\cite{militzer_pre_00} and then applied to dense hydrogen~\cite{militzer_prl_00,Mi01,MC01}. Furthermore, free-particle nodes and bound states around the nuclei have been combined into one trial density matrix to extend the applicability range of RPIMC simulations to lower temperatures~\cite{Militzer_PRL_2015}. 

The second challenge is to sample the space of path coordinates and permutations efficiently with Monte Carlo methods in the presence of the nodal restriction. This poses a different challenge than in the bosonic case because the restriction singles out the reference point as being special and the nodal restriction places a non-local restriction on Monte Carlo updates.  As the temperature is lowered the paths and the reference point become difficult to move. It is not known if this is a fundamental problem. See Refs.~\cite{Ce91,Ceperley1996} for details of the RPIMC method.

Beginning with Ref.~\cite{PC94} in 1994, the RPIMC method has been applied to study hydrogen numerous times, which will be discussed in the next section. Starting with helium~\cite{Mi06,Mi09}, RPIMC with free-particle nodes has been applied to study many heavier elements up to neon~\cite{Driver_PRL_2012,Zhang2018,DriverNitrogen2016,Driver2015b,Driver2015,ZhangCH2017}. The inclusion of bound states into the nodal structure allowed one to simulate elements and compounds as heavy as silicon~\cite{ZhangSodium2017,Driver2018,Gonzalez2020,gonzalez2020equation}. For each material, the RPIMC results were combined with predictions from density functional molecular dynamics simulations to obtain one consistent equation of state (EOS) table that covers a wide range of density-temperature conditions so that shock Hugoniot curves can be inferred. These EOS tables were combined into one first principles EOS (FPEOS) database~\cite{FPEOS,militzer2020nonideal}. RPIMC has also been used to construct an EOS for the uniform electron gas~\cite{Brown_PRB_2013,Brown_PRL_2013,ksdt}.

\subsubsection{PIMC simulations of hydrogen\label{sec:PIMC_hydrogen}}

The main difference between PIMC simulations of the UEG, cf.~Sec.~\ref{subsec:jellium}, and hydrogen is given by the Coulomb attraction between electrons and protons. Since the latter is not bounded from below, it prevents the direct utilization of the primitive approximation in the Trotter formula, Eq.~\eqref{eq:trotter}.
To overcome this obstacle,  one can replace the primitive approximation with the pair action as described in \cite{Po88,Ceperley1995,MILITZER201688} 
\begin{eqnarray}\label{eq:pair}
e^{-\varepsilon\Phi_\textnormal{pair}(\mathbf{R},\mathbf{R}';\varepsilon)} = e^{-\varepsilon \sum_{l<k}^N \phi(\mathbf{r}_{lk},\mathbf{r}_{lk}';\varepsilon)}\ ,
\end{eqnarray}
where the kinetic part has been omitted for simplicity. Here $\phi$ is the logarithm of the exact density matrix for two charged particles: electron-electron or electron-proton. 
A further simplification has been proposed by G.~Kelbg~\cite{kelbg_ap_63,kelbg_ap_63_2}, based on a first-order perturbation expansion of Eq.~(\ref{eq:pair}) with respect to the coupling strength. 
%{\bf (We should probably explain that this potential is still being used to dynamical simulation of ions and electrons. Cite Graziani? For nondynamical simulations, it is obsolete.)} 
This result was extended to strong coupling in Ref.~\cite{filinov_jpa03}, leading to the ``improved Kelbg potential'', IKP, see Sec.~\ref{ssubsec:pair-pot-md}.
Although the resulting effective quantum pair potential~\cite{filinov_pre04} is easy to implement and avoids the path collapse in PIMC simulations due to the divergence of the bare Coulomb attraction, its convergence is substantially slower compared to the full pair approximation~\cite{filinov_pre04,bonitz_cpp_23, Bohme_PRE_2023,Bohme_PRL_2022}, making the latter the preferred approach for PIMC simulations.
For completeness, we note that such effective potentials are still being used in approximate dynamic simulations of electron--ion systems, see e.g.~Refs.~\cite{glosli2008,GRAZIANI2012105} and Sec.~\ref{sssec:i-acoustic_H}.

\begin{figure}
    \centering
    \includegraphics[width=0.47\textwidth]{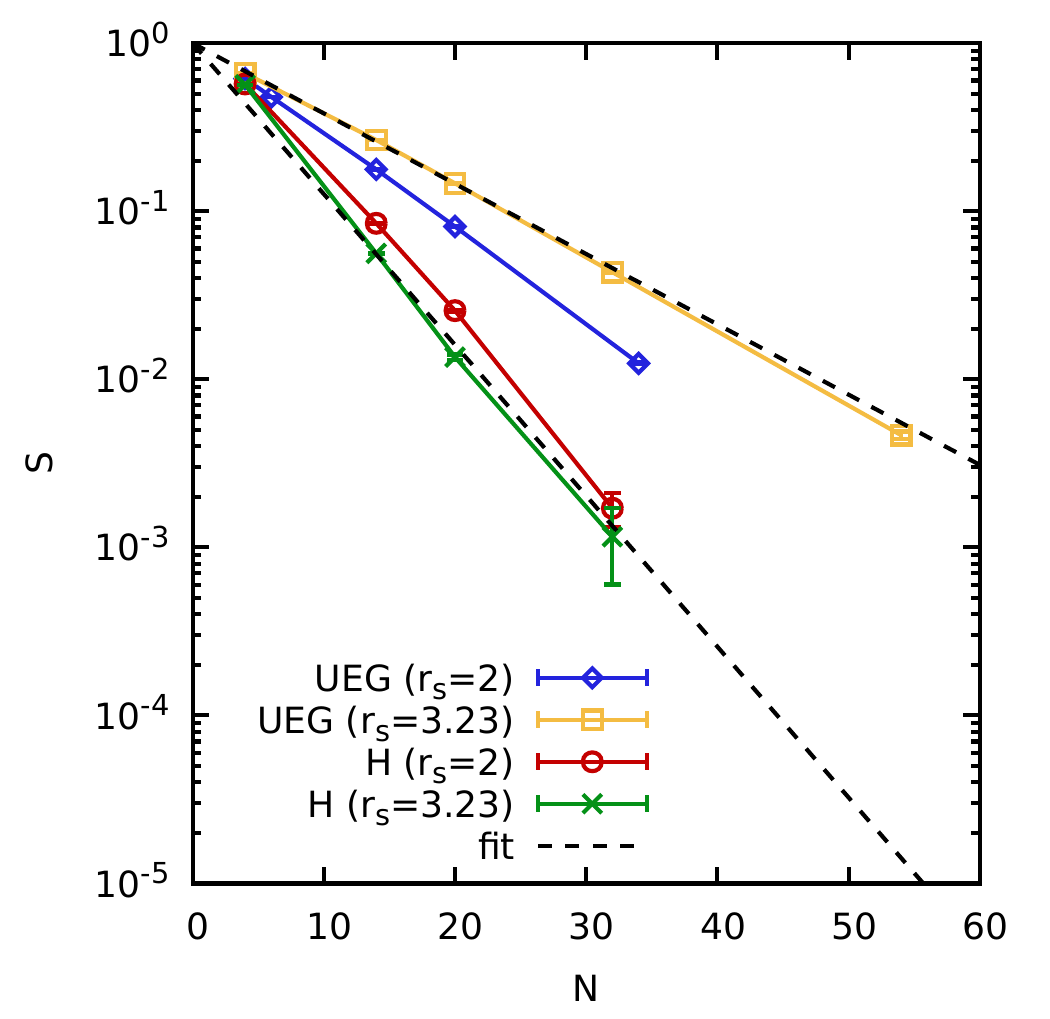}
    \caption{Average sign $S$ as a function of the number of electrons $N$ at the electronic Fermi temperature $\Theta=1$, for $r_s=2$ ($\rho=\SI{0.34}{\gram\per\cubic\centi\meter}$, $T=\SI{12.53}{\eV}$) and $r_s=3.23$ ($\rho=\SI{0.08}{\gram\per\cubic\centi\meter}$, $T=\SI{4.80}{\eV}$). The dashed black lines show exponential fits to the PIMC data, cf.~Eq.~(\ref{eq:relative_error}).
    Taken from Ref.~\cite{Dornheim_JCP_2024} with the permission of the authors. 
    \label{fig:sign}}
\end{figure}

A second difference between PIMC simulations of the UEG and hydrogen is the impact of fermionic exchange effects and their manifestation with respect to the fermion sign problem. In Fig.~\ref{fig:sign}, we show the average sign from direct PIMC simulations at the electronic Fermi temperature $\Theta=1$ for $r_s=2$ ($\rho=\SI{0.34}{\gram\per\cubic\centi\meter}$, $T=\SI{12.53}{\eV}$) and $r_s=3.23$ ($\rho=\SI{0.08}{\gram\per\cubic\centi\meter}$, $T=\SI{4.80}{\eV}$) as a function of the number of electrons $N$. For all depicted data sets, we find an exponential decrease of $S$ with increasing system size, cf.~Eq.~(\ref{eq:relative_error}).
For the UEG model (blue diamonds and yellow squares), decreasing the density (at constant degeneracy) leads to a less severe sign problem as the formation of permutation cycles~\cite{Dornheim2019c} is effectively suppressed by the stronger Coulomb repulsion between the electrons.
Interestingly, the opposite trend has been reported for hydrogen~\cite{Dornheim_JCP_2024}. First, the sign problem is overall more severe compared to the UEG under the same conditions. This is a direct consequence of the increased degree of inhomogeneity due to the presence of the protons around which the electrons tend to cluster even, if they are effectively \textit{unbound}.
Second, we find a reduced sign for hydrogen at the lower density, as the inhomogeneity induced formation of permutation cycles of electrons located around the protons predominates over the separation of electrons due to the Coulomb repulsion observed for the UEG. 
We note that the Pauli principle ensures the stability of a system of protons and electrons when the temperature is lowered, whereas a corresponding system of hypothetical distinguishable quantum particles would collapse~\cite{Lieb_RMP_1976}.

To deal with the more pronounced sign problem in PIMC simulations of hydrogen, three strategies have been successfully employed: i) B.~Militzer, D.M.~Ceperley and others~\cite{militzer_pre_00} have used the restricted PIMC method, see Sec.~\ref{sec:alternatives}; ii) A.~Filinov and M.~Bonitz~\cite{filinov_pre_23} have employed antisymmetric imaginary-time propagators (i.e., determinants). By grouping together positive and negative terms in the determinants, the sign problem is substantially alleviated~\cite{Dornheim2015a,dornheim_jcp15}, allowing for simulations at lower temperature compared to direct PIMC; and iii) the $\xi$-extrapolation method introduced by Y.~Xiong and H.~Xiong~\cite{Xiong_JCP_2022}, which removes the exponential scaling with respect to the system size at moderate temperatures~\cite{Dornheim_JCP_xi_2023,Dornheim_JPCL_2024,dornheim2024unraveling,Dornheim_JCP_2024,Dornheim_LFC_2024}. These concepts are discussed in more detail in Sec.~\ref{ssub:fsp-reduction}.

An additional difference between hydrogen and the UEG arises for either the CEIMC method, described below (Sec.~\ref{subse:ceimc}), or in the RPIMC method (Sec.~\ref{sec:alternatives}), because the trial density matrix or electron trial function, has an important dependence on the protonic configuration.  Accurate antisymmetric trial functions may
give rise to additional computational effort.

\subsubsection{Concepts to alleviate the Fermion sign problem}\label{ssub:fsp-reduction}

Over the last few years, a number of concepts to alleviate the fermion sign problem have been suggested. Here, we focus on two ideas that have recently been used for the PIMC simulation of warm dense hydrogen, as discussed in Sec.~\ref{sec:PIMC_hydrogen}: i) the utilization of antisymmetrized imaginary-time propagators~\cite{filinov_pre_23} and ii) the controlled extrapolation over the continuous spin-variable $\xi$~\cite{Dornheim_JCP_2024,Dornheim_LFC_2024}.

As mentioned in Sec.~\ref{sec:FSP}, the sign problem comes from the cancellation of positive and negative terms due to the $(-1)^{N_\textnormal{pp}}$ term in Eq.~(\ref{eq:Z_modified}).
Using the idempotency property of the antisymetrization operator and the definition of the determinant, one can rewrite the partition function as~\cite{zamalin_77,Takahashi_Imada_PIMC_1984,dornheim_physrep_18,Chin_PRE_2015,Dornheim2015a,Lyubartsev_2005}
\begin{widetext}
\begin{eqnarray}\label{eq:Z_determinant}
    Z_{N,\Omega,\beta} = \int\textnormal{d}\mathbf{X}\ \prod_{\alpha=0}^{P-1} e^{-\epsilon V(\textbf{R}_\alpha)} \textnormal{det}\left(\rho_0(\mathbf{R}_\alpha^\uparrow,\mathbf{R}_{\alpha+1}^\uparrow,\epsilon) \right) \textnormal{det}\left(\rho_0(\mathbf{R}_\alpha^\downarrow,\mathbf{R}_{\alpha+1}^\downarrow,\epsilon) \right)\ ,
\end{eqnarray}
\end{widetext}
where $V(\mathbf{R}_\alpha)$ includes all potential energy contributions from time-slice $\alpha$ and $\rho_0(\dots)$ is the noninteracting kinetic density matrix. Comparing Eqs.~(\ref{eq:Z_modified}) and (\ref{eq:Z_determinant}), it becomes evident that a large number of positive and negative terms have been grouped together in the determinants. While the determinants can still be both positive and negative, this can lead to a considerable reduction of the sign problem since at least parts of the cancellation are carried out analytically.

\begin{figure}
    \centering
    \includegraphics[width=0.37\textwidth]{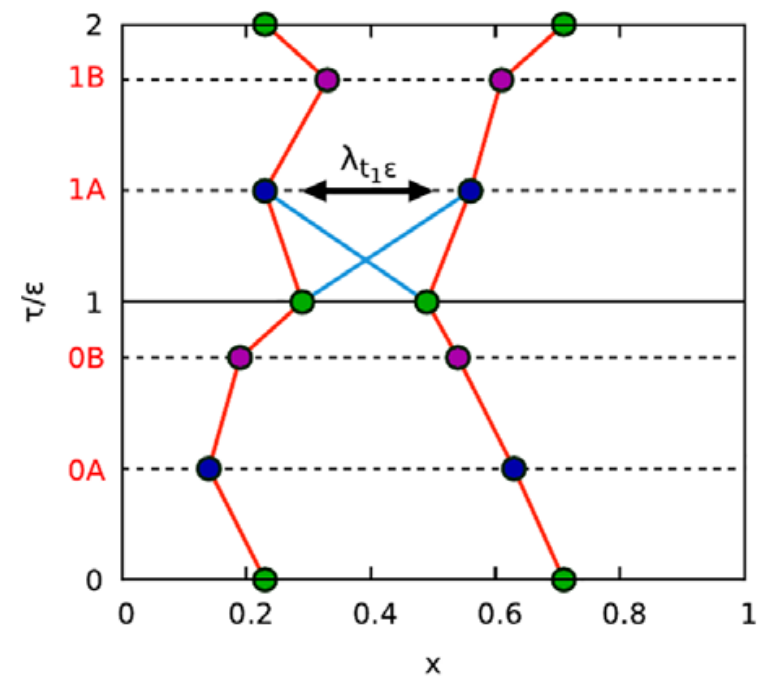}
    \caption{Schematic illustration of the grouping of diagonal (positive) and off-diagonal (negative) terms in an antisymmetrized imaginary-time propagator. Using the fourth-order propagator from Ref.~\cite{Chin_JCP_2002}, every imaginary-time step $\epsilon$ is further divided into three sub-intervals of unequal length.
   % Taken from Ref.~\cite{review}. 
   Reprinted from \textit{Physics Reports} \textbf{744}, T.~Dornheim, S.~Groth, and M.~Bonitz, "The uniform electron gas at warm dense matter conditions" 1-86, copyright 2018, with permission from Elsevier.
    \label{fig:determinant}}
\end{figure}
The basic idea behind this strategy is illustrated in Fig.~\ref{fig:determinant}, where we show a configuration of $N=2$ identical fermions (e.g.~electrons) in the $\tau$-$x$-plane. Within the direct PIMC approach, one would either consider the diagonal (red) connections that are associated with a positive sign of the configuration weight, or the off-diagonal (blue) connections corresponding to a pair exchange with a negative sign. By using the determinants in Eq.~(\ref{eq:Z_determinant}), one can evaluate both terms at the same time, which reduces the amount of cancellations in the average sign $S$.
In practice, this strategy only works when the diagonal and off-diagonal contributions have a comparable weight; this is the case when the distance between two beads on the same time slice is comparable to the associated thermal wavelength $\lambda_\epsilon=\sqrt{2\pi\epsilon}$, see the black arrow in Fig.~\ref{fig:determinant}.
Since $\lambda_\epsilon$ decreases with increasing number of time slices $P$, the effect of the determinants vanishes, and one recovers the full sign problem of the direct PIMC method for $P\to\infty$~\cite{dornheim_cpp_19}.

To avoid this conundrum, it has been suggested~\cite{Dornheim2015a,Chin_PRE_2015} to combine the determinants with a higher-order factorization of the density operator that allows for sufficient accuracy with a small number of high-temperature factors $P$. This is the basic idea of the permutation blocking PIMC (PB-PIMC) method~\cite{Dornheim2015a,dornheim_jcp15,dornheim_pre17,dornheim_cpp_19,dornheim_physrep_18,dornheim_pop17}, which has recently been extended to the grand canonical ensemble~\cite{filinov_cpp_21} and applied to the simulation of warm dense hydrogen by A.~Filinov and M.~Bonitz~\cite{filinov_pre_23}.

\begin{figure}\centering
\includegraphics[width=0.45\textwidth]{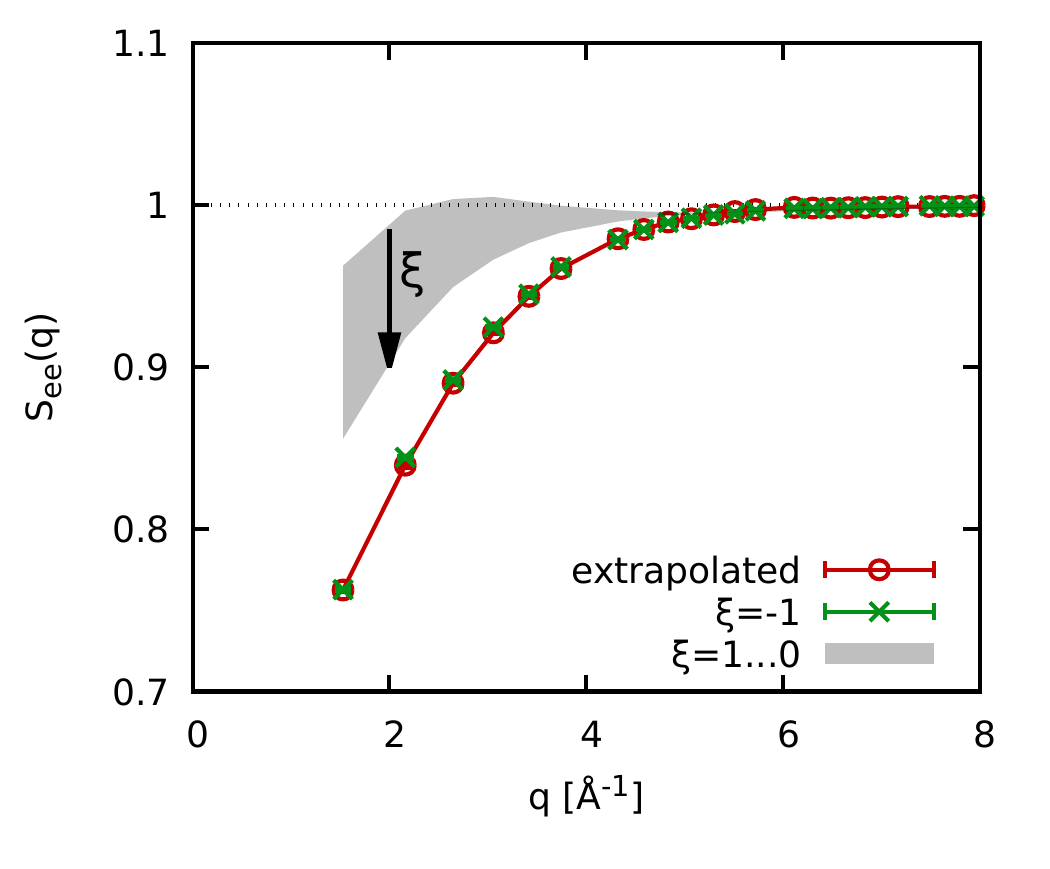}
\caption{\label{fig:xi_extrapolation} Electron--electron static structure factor of warm dense hydrogen for $N=14$ atoms at $r_s=2$ ($\rho=\SI{0.34}{\gram\per\cubic\centi\meter}$) and $\Theta=1$ ($T=\SI{12.53}{\eV}$). Green crosses: exact direct fermionic PIMC results (i.e., $\xi=-1$); red circles: extrapolation from the sign-problem free domain of $\xi\in[0,1]$ via Eq.~(\ref{eq:xi_fit}); grey area: PIMC results for $0\leq\xi\leq1$. Taken from Ref.~\cite{Dornheim_JCP_2024} with the permission of the authors.
}
\end{figure}

A different approach has recently been introduced by Y.~Xiong and H.~Xiong~\cite{Xiong_JCP_2022}, who have suggested to replace the factor $(-1)^{N_\textnormal{pp}}$ in Eq.~(\ref{eq:Z_modified}) by $\xi^{N_\textnormal{pp}}$, where $\xi\in[-1,1]$ is a continuous variable.
The physically meaningful cases of Bose-Einstein, Maxwell-Boltzmann, and Fermi-Dirac statistics are recovered for $\xi=1$, $\xi=0$, and $\xi=-1$, respectively.
The basic strategy to avoid the sign problem is then to carry out simulations in the sign-problem free domain of $\xi\in[0,1]$ and subsequently extrapolate to the fermionic limit based on the empirical relation~\cite{Xiong_JCP_2022}
\begin{eqnarray}\label{eq:xi_fit}
    A(\xi) = a_0 + a_1\xi + a_2\xi^2\ ,
\end{eqnarray}
where $a_0$, $a_1$, and $a_2$ are the free fit parameters. 
Having originally been introduced for path integral MD, this idea was applied to PIMC simulations of electrons in a harmonic trap and the warm dense UEG in Ref.~\cite{Dornheim_JCP_xi_2023}.
In practice, Eq.~(\ref{eq:xi_fit}) works remarkably well for weak to moderate levels of quantum degeneracy, where it is capable of delivering quasi-exact results with a typical relative error bar of $\sim0.1\%$. However, the idea breaks down for low temperatures, when the bosonic and fermionic systems become too dissimilar~\cite{Dornheim_JCP_xi_2023}.
In practice, the applicability of the extrapolation can be checked for small systems with $N\sim10$ electrons where exact fermionic PIMC simulations (i.e., for $\xi=-1$) are still feasible. Its reliability for larger systems is then ensured by the local nature of fermionic exchange effects at moderate to high temperatures~\cite{Kohn_PNAS_2005}.
As a result, one gets a non-empirical PIMC approach for the simulation of fermions without the exponential bottleneck with respect to the system size. This has recently allowed for simulations of the UEG with up to $N=1000$ electrons~\cite{Dornheim_JPCL_2024}.

Subsequently, the $\xi$-extrapolation method has been successfully applied to study the static density response~\cite{dornheim_2024ab}, imaginary-time structure, and structural properties~\cite{Dornheim_JCP_2024} of warm dense hydrogen; for completeness, there are further results for strongly compressed beryllium in Ref.~\cite{dornheim2024unraveling}.
As an example, we show results for the electron--electron static structure factor $S_{ee}(\mathbf{q})$ for $N=14$ hydrogen atoms at $r_s=2$ and $\Theta=1$ in Fig.~\ref{fig:xi_extrapolation}. The green crosses show exact fermionic PIMC results, and the red circles have been obtained by extrapolating from the sign-problem free sector $0\leq\xi\leq1$ (shaded grey area). The two data sets are in excellent agreement over the entire $q$-range, which demonstrates the reliability of the $\xi$-extrapolation for these parameters.

Let us conclude by comparing a few strengths and weaknesses of the RPIMC, permutation blocking PIMC, and $\xi$-extrapolation approaches.
The RPIMC method is available over the broadest range of parameters (in particular with respect to low temperatures), cf. Fig.~\ref{fig:h-phase-diag-qmc}. Moreover, it works for a variety of materials~\cite{Militzer_PRE_2021}, including second-row elements~\cite{Militzer_PRL_2015}. These strengths come at the cost of an approximation that is very difficult to check in practice. Additionally, the straightforward access of direct PIMC to different ITCFs is lost.
The PB-PIMC method, on the other hand, provides a higher degree of certainty, as the convergence with the number of high-temperature factors $P$ can, in principle, be checked, even though this can be difficult in practice.
Moreover, PB-PIMC does provide access to all ITCFs, but the imaginary-time grid on which these can be resolved is often very small due to the requirement of a small $P$ for the alleviation of the sign problem.
Finally, the $\xi$-extrapolation method appears to be as limited with respect to the temperature as the direct PIMC method (although new concepts are continually being developed~\cite{Xiong_PRE_2023}), i.e., to a region of the phase space where the impact of Fermi statistics is not too profound. Instead, its key strength is the removal of the exponential scaling with the number of electrons at parameters where PIMC simulations are generally possible, at least for small systems.
At the same time, it does not require any external input such as the nodal structure in RPIMC, and it gives one full access to all ITCFs on an arbitrarily dense $\tau$-grid.

\subsection{Coupled electron-ion Monte Carlo (CEIMC)}
\label{subse:ceimc}
Within the adiabatic approximation, solutions to the full Schr{\"o}dinger equation of the coupled electron-ion system
are expanded in exact eigenfunctions of the electronic Hamiltonian for fixed (given) nuclear positions.
The coupling of different adiabatic electronic eigenstates only occurs via the ionic kinetic energy operator and is suppressed
by the large mass ratio, $M=m_I/m_e \gg 1$. Within the Born-Oppenheimer  approximation (BOA) \cite{Born1927},
the wave function of the coupled electron-ion system factorizes into an electronic and nuclear
part which can be regarded as the starting point of an expansion in powers of $(1/M)^{1/4}$
for the total energy \cite{Born1988}. Far below the electronic Fermi temperature, electrons are frozen in the adiabatic ground state, and the nuclear distribution can be obtained from its
Born-Oppenheimer energy surface, e.g. the electronic ground state energy 
for a static nuclear configuration.
By employing the BOA, we have completely decoupled the electronic problem. Calculations of the BO potential of the nuclei can now be done by any suitable method, independently from sampling of the nuclear degrees of freedom. A frequent choice is the use of
DFT to determine electronic energies and forces within classical molecular dynamics simulations of the nuclei (BOMD) \cite{MARX2009}.  

Within coupled electron-ion Monte Carlo (CEIMC) \cite{Pierleoni2004} the electronic BO
energy is determined via ground state Monte Carlo methods, e.g. variational (VMC) or reptation (RMC) Monte Carlo, whereas the nuclear configurations at finite temperature are sampled 
either classically, according to their Boltzmann weight, or quantum mechanically
using Path-Integral Monte Carlo methods. This decoupling of electronic and nuclear degrees of freedom 
enables QMC simulation for temperatures below those applicable to RPIMC or FPIMC calculations
(as shown in the figure).
However, since VMC or RMC calculations of the electronic energies have stochastic errors, to guarantee unbiased Monte Carlo sampling of the nuclear configurations
the nuclear Monte Carlo process uses the penalty method \cite{Ceperley99}. A detailed description of CEIMC can be found in Refs.~\cite{Pierleoni2006,mc-mahon_RevModPhys.84.1607}.

The accuracy of CEIMC depends on the choice of the trial wave function underlying the electronic QMC calculations. Typically one uses a Slater-Jastrow wave function augmented by backflow and three-body correlations \cite{Holzmann2003} with orbitals in the Slater determinant taken from DFT \cite{Pierleoni2008}. Any residual dependence of QMC energies on parameters of the
trial wave functions as well as the choice of the DFT functional used for the orbitals can be
quantified and controlled by minimizing the QMC energies.  The possibility to quantify the accuracy of different QMC calculations based on the variational principle represents one of the main advantages compared to using DFT energies where one usually relies on experimental input to justify the choice of the functional.

Diffusion (DMC) \cite{Foulkes_RMP_2001} or reptation Monte Carlo (RMC) \cite{Baroni1999} stochastically improves any trial wave function via
imaginary time propagation. Similar to PIMC described above, direct sampling suffers from a strong sign problem
which can be circumvented by the use of the fixed-node~(FN)  approximation. %, obtained by electronic RPIMC approaching zero temperature. 
Assuming  the BOA, the quality of the trial wave function and its residual fixed node error can be studied separately on snapshots of nuclear configurations; this
greatly simplifies the search for accurate 
electronic wave functions, e.g. compared to improving the nodal surface underlying RPIMC.  
In practice, the influence of the nodal surface can be studied by the influence of the underlying DFT orbitals \cite{Pierleoni2016}, optimizing localized atomic orbitals \cite{Mazzola2018}, 
multi-determinant wave functions \cite{Clay2019},
or backflow transformations \cite{Holzmann2003}.

However, the simplifications occurring in the BOA, due to the electronic description at zero temperature, come with one major drawback,
since electronic correlations are no longer limited by the thermal wavelength; electronic coherence may extend over the whole simulation cell. The resulting sensitivity to boundary conditions can result in an important sensitivity to the size of the supercell; in other words to the number of particles in the simulation.
Those finite-size errors can be drastically reduced by employing twist averaged boundary conditions
\cite{Lin2001,Holzmann2016} in the QMC calculations of the BO energy. Leading order finite-size effects can be estimated directly from the properties of the underlying trial wave function \cite{Chiesa_PRL_2006,Holzmann2016} without the need for extrapolations based on simulating different system sizes. This is particularly important for CEIMC where numerical extrapolations are computationally expensive or impossible. 

Since finite size considerations, based on the analytical structure of the wave functions \cite{Holzmann2016}, guarantee size consistency,
more recently developed iterative and neural backflow wave functions \cite{Taddei2015,Holzmann2019,Holzmann_PRL_2020,Wilson2023,Xie_PRL_2023,Casella2023,Pescia2023}, 
as well as FCIQMC \cite{Shepherd2012,RuggeriAlavi2018} and coupled cluster methods \cite{Shepherd2013}
offer the possibility to systematically study and
reduce the fixed-node error for systems with a small number of electrons. So far, most of these studies have been done on the homogeneous electron gas.

Beyond structural properties of the nuclei including quantum and thermal motion beyond the harmonic
approximation, CEIMC also provides important insights into electronic properties. The many-body BO wave function of the
electrons yields direct access to off-diagonal matrix elements of the reduced (single-electron) density matrix which
encodes the localized (insulating) or extended (Fermi-liquid) behavior of the electrons \cite{Pierleoni2016,Pierleoni2018,Gorelov2020a}, cf.~Sec.~\ref{ssec:n(k)-h}. Further, electronic excitation gaps can be determined within QMC accuracy,
by variations of the electronic chemical potential within the grand-canonical ensemble \cite{Yang2020}, cf. Sec.~\ref{ss:h-gap}.

Within CEIMC, 
excitation gaps including nuclear quantum and thermal effects can be consistently determined \cite{gorelov_prl_20,gorelov2023electronic}. The closure of the fundamental gap pinpoints the transition to metallic hydrogen.  
In solid hydrogen, the character of the electronic wavefunction at different chemical potential contains information
on the band structure of added or removed electrons \cite{Gorelov2020b} and indicates the closure of the indirect gap
at \SIrange{330}{380}{\giga\Pa} into a bad metal phase (blue shaded area in Fig.~\ref{fig:phase-diagram-low-t}) with a direct gap closing at higher pressures \SIrange{\sim 450}{500}{\giga\Pa} (red shaded area in Fig.~\ref{fig:phase-diagram-low-t}) \cite{gorelov_prl_20}. Concerning the LLPT, CEIMC found an abrupt closure of the gap at the molecular-atomic transition (see Figs~\ref{fig:phase-diagram-low-t} and \ref{fig:phase-diagram-low-t-optics})
whereas a cross-over between insulating and metallic liquid is observed at around \SI{3000}{\kelvin}, above the critical temperature \cite{Gorelov2020a}.

Due to the BO decoupling of electronic and nuclear degrees of freedom, CEIMC naturally connects with many electronic structure methods used in material science. On one hand, it provides QMC based results of fundamental quantities which can be used to validate DFT based methods as well as GW-Bethe Salpeter equation (BSE) approaches where experimental results are missing or difficult to interpret as is the case in high pressure hydrogen.  On the
other hand, it allows one to investigate approximations such as BO, by comparison with PIMC or zero temperature DMC for conditions of temperature and pressure where several methods are applicable.

Sampling classical nuclear degrees of freedom by Langevin dynamics based on QMC forces within the
BO approximation has been developed in Refs.~\cite{Attaccalite2008,Mazzola2014,Mazzola2015,Mazzola2017}. Comparable  results to CEIMC are found \cite{Mazzola2018}, once
finite size and basis set errors of previous results \cite{Mazzola2014,Mazzola2015,Mazzola2017} 
are accounted for.

\subsection{Semiclassical and quantum molecular dynamics}\label{subsec:md}
There exist a variety of concepts to extend classical molecular dynamics to quantum systems. This includes Wigner function QMD \cite{filinov_prb_2,vfilinov_jpa_03}, path integral molecular dynamics (PIMD) \cite{cao_jcp_94, polyakov_jcp_10} or wave packet molecular dynamics (WPMD) \cite{heller1975time}. In addition, many attempts were made to employ classical MD and account for quantum and spin effects by proper modification of the pair interaction, see Sec.~\ref{ssubsec:pair-pot-md}.

\subsubsection{Wave packet molecular dynamics (WPMD)}\label{sssec:wpmd}
Wave packet molecular dynamics was originally proposed by E.\ J.\ Heller \cite{heller1975time} after the observation that a Gaussian wave function remains a Gaussian in a quadratic potential and its time evolution follows classical equations. More generally, equations of motion can be derived via the time-dependent variational principle by considering variations of the action \cite{feldmeier2000molecular},
\begin{equation}
    \mathcal{S} = \int dt\; \bra{Q} i\hbar \frac{d}{dt} - \Hat{H} \ket{Q}\,,
\end{equation}
where the state $\ket{Q} = \ket{Q(Q_{\mu})}$ is parametrised by a complex set of parameters $Q_{\mu}$. The equations of motion for the parameters readily follow,
\begin{equation}
    i\hbar \sum_{\nu} \mathcal{C}_{\mu\nu} \frac{dQ_{\nu}}{dt} = \frac{\partial \mathcal{H}}{\partial Q_{\mu}^*} \,,
\end{equation}
with the hamiltonian function $\mathcal{H} = \bra{Q}\Hat{H}\ket{Q} / \braket{Q|Q}$ and the norm matrix 
\begin{equation}
    \mathcal{C}_{\mu\nu} = \frac{\partial^2}{\partial Q_{\mu}^* \partial Q_{\nu}} \ln \braket{Q|Q}\,,
\end{equation}
where we have used the convention $\ket{Q} = \ket{Q(Q_\mu)}$ and $\bra{Q} = \bra{Q(Q_\mu^*)}$. 

Commonly, a restricted state $\ket{Q}$ that does not span the full Hilbert space is used to aid computational performance and model electron and ion dynamics simultaneously. However, such an approximation limits the dynamics to a sub-manifold, and the approximation is closely related to the McLachlan and Dirac-Frenkel variational principles \cite{lasser2022various}. The actual choice of the trial state is therefore important for the model and a variety of suggestions has been put forward, see Ref.\ \cite{grabowski2014review} and references therein. For the study of hydrogen systems, a Gaussian with a time-dependent width has been used most frequently for single-electron states, even if recent extensions have been suggested \cite{angermeier2021investigation,svensson2023development}. This limited functional has a ground-state binding energy of the hydrogen atom $E_0 \approx -11.5\,\text{eV}$, but some attempts have been made to explicitly include the 1s state in the model \cite{ebeling1997quantum}. 

D.~Klakow \textit{et al.} suggested an approximation to exchange effects, by considering a pairwise anti-symmetrisation of the kinetic energy \cite{klakow1994hydrogen,klakow1994semiclassical}, an idea that was extended by the electron force field (eFF) model by the introduction of fitting parameters to better match electron-ions bounding in low-Z elements \cite{su2007excited,su2009dynamics,jaramillo2011large}. These types of pairwise Pauli potentials have been primarily used to study dynamic processes in hydrogen, e.g.\ proton transport \cite{yao2021reduced,angermeier2023disentangling}, electron-proton temperature equilibration \cite{ma2019extremely}, plasma oscillations \cite{zwicknagel2006wpmd} and electron stopping power \cite{liu2021molecular} but also constructed a variational approximation to the imaginary time density matrix~\cite{MP00}.

For lower temperature conditions, a complete Slater determinant has been used as a trial state, producing a correctly antisymmetrized state, including effects on energies and the norm matrix \cite{knaup2003wave,jakob2007wave,jakob2009wave}. A low-temperature insulator-metal phase transition has been predicted for temperatures $T < \SI{4000}{\kelvin}$, below $\SI{150}{\giga\Pa}$ \cite{jakob2007wave,jakob2009wave}.

Y.~Lavrinenko \textit{et al.} have suggested including the effect of exchange and correlations in the wave packet model by evaluating DFT functionals based on the density profiles from the wave packets \cite{lavrinenko2019wave,Lavrinenko_2021,lavrinenko2021high}, instead of manipulating the state directly. This has the greatest effect at high densities, $n_e > \SI{e23}{\centi\meter^{-3}}$, and is seen to reduce the maximum compression along the Hugoniot compared to the explicitly anti-symmetrized models, more in accordance with experimental evidence \cite{Lavrinenko_2021}.

It is well documented \cite{knaup1999wave,knaup2001wave,knaup2003wave,morozov2009localization} that at sufficiently high temperatures the wave packets expand indefinitely, something that needs to be regularized. This expansion is arguably due to an inability to localize an electron on multiple ions \cite{grabowski2013wave}. Multiple different suggestions have been made to limit the expansion \cite{knaup1999wave,Ebeling2006,lavrinenko2016reflecting,ma2019extremely}, but the effect of this regularisation must be tested \cite{morozov2009localization}.

\subsubsection{Effective quantum pair potentials}\label{ssubsec:pair-pot-md}
The idea to ``correct'' the pair interaction for quantum diffraction and exchange effects was advanced by many authors, including G.~Kelbg and co-workers~\cite{kelbg_ap_63,kelbg_ap_63_2,hoffmann_ap_66} and  C.~Deutsch and co-workers~\cite{deutsch_pla_72,deutsch_pla_78, deutsch_pra_81}.
G.~Kelbg considered the canonical two-particle density matrix at a finite temperature $k_BT=1/\beta$ which obeys the Bloch equation
\begin{equation}
 \frac{\partial \rho_{ab}}{\partial \beta} = \frac{\hbar}{2m_{ab}} \Delta \rho_{ab} - V_{ab} \rho_{ab} \;.
\end{equation}
For weak coupling the result for the off-diagonal density matrix is
\begin{eqnarray}
&\rho_{ab} (r_a,r_a',r_b,r_b') = C \exp \Big[- \frac{m_a}{2 \beta \hbar^2} (r_a - r_a')^2  \Big] \nonumber\\
& \times\exp \Big[- \frac{m_b}{2 \beta \hbar^2} (r_b - r_b')^2  \Big] \exp[ - \beta \Phi(r_a,r_a',r_b,r_b')]\,,
\nonumber
%\label{OFFD1}
\end{eqnarray}
and involves kinetic energy parts (the Gaussian factors) and an effective ``quantum potential'' $\Phi$. The complicated coordinate dependence is often simplified to a diagonal potential, $U^{\rm IK}$, that depends only on the distance of the pair,
\begin{eqnarray}
 U^{\rm IK}_{ij}(r)=\frac{q_i q_j}{r} \left[1- e^{-\frac{r^2}{\lambda_{ij}^2}} +\frac{\sqrt{\pi} r}{\lambda_{ij} \gamma_{ij}}\left(1-\rm{erf} \left[\frac{\gamma_{ij} r}{\lambda_{ij}}\right]\right) \right]\,, \label{eq:improved-Kelbg}
\end{eqnarray} 
where erf denotes the error function, and we introduced the de Broglie wavelength with the reduced mass of the pair, $\lambda_{ij}=h(2\pi m_{ij}k_BT)^{-1/2}$.
Consider first the case $\gamma_{ij}\to 1$ for which one recovers the potential $U^{\rm K}$ derived originally by G.~Kelbg \cite{kelbg_ap_63,kelbg_ap_63_2}, for a recent overview, see Ref.~\cite{bonitz_cpp_23}.
In the limit $T\to \infty$ the 
potential approaches the classical Coulomb potential whereas, for decreasing $T$, the potential exhibits systematically increasing deviations from the Coulomb potential, due to quantum diffraction effects and approaches a finite value at zero separation, $U^{\rm K}_{ij}(0)=\frac{q_i q_j}{\lambda_{ij}}$. At the same time, it is known that, while the derivative $U^{\rm K'}_{ij}(0)$ is correct, the absolute value is not, if the weak coupling approximation is violated \cite{filinov_jpa03}. The correct value can be restored by fitting to the exact solution of the pair problem, with a single additional parameter, $\gamma_{ij}(T)$, and the corresponding results have been termed ``Improved Kelbg potential (IKP)'' \cite{filinov_jpa03,filinov_pre04}. Its value at zero pair separation is 
$$U^{\rm IK}_{ij}(0;T)=\frac{q_i q_j}{\gamma_{ij}(T)\lambda_{ij}(T)}\,,$$
which gives the fit parameter a simple interpretation: as a result of pair interaction effects the ``extension'' of a quantum particle becomes $\gamma_{ij}(T)\lambda_{ij}(T)$ compared to the ``ideal extension'' $\lambda_{ij}(T)$. While e-e repulsion increases the extension, electron-proton attraction reduces it, ultimately to the Bohr radius.
\begin{figure}
    \centering
    \includegraphics[width=0.4\textwidth]{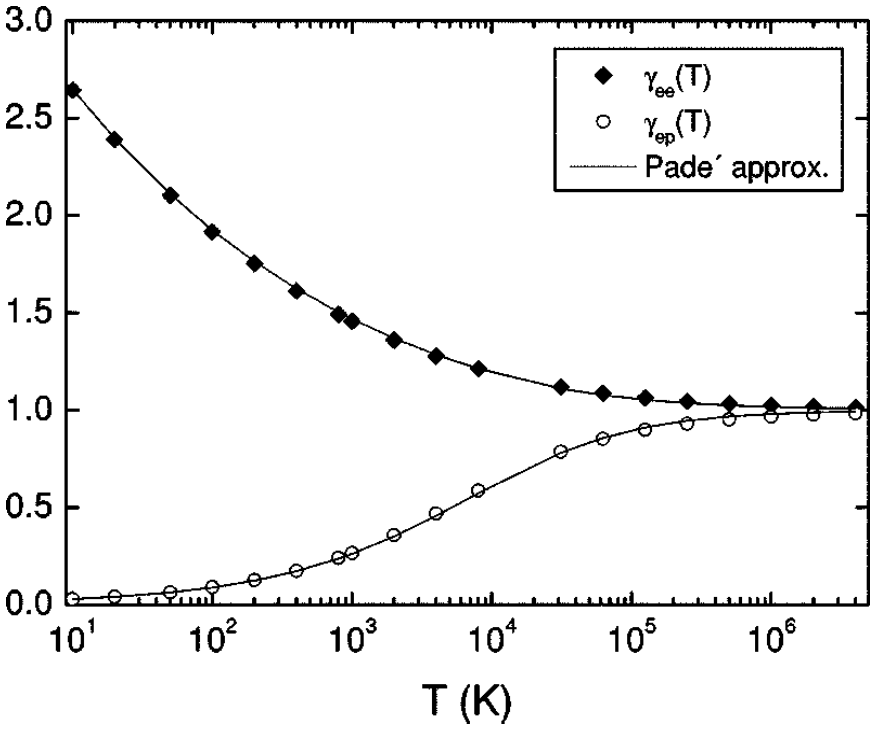}
    \caption{Fit parameters of the improved Kelbg potential (\ref{eq:improved-Kelbg}) for e-p and e-e (without exchange) interaction. $\gamma_{ee}$ and $\gamma_{ep}$ can be understood as modifications of the de Broglie wavelength due to e-e and e-p interaction, respectively, see text. Reprinted from Ref.~\cite{filinov_pre04} with permission of the authors.}
    \label{fig:ikp-gama(t)}
\end{figure}
The parameters $\gamma_{ee}(T)$ and $\gamma_{ep}(T)$ have a simple Padé representation given in Ref.~\cite{filinov_pre04}
and are shown in Fig.~\ref{fig:ikp-gama(t)}.

The Kelbg potential was used in PIMC simulations of hydrogen and electron-hole plasmas by V. Filinov \textit{et al.}~\cite{filinov_jetpl_00,filinov_jetpl_01,filinov_ppcf_01,filinov_cpp_03} and also in simulations of quark-gluon plasmas \cite{filinov_prc_13,filinov_cpp09,filinov_cpp12}. The improved Kelbg potential was used by A. Filinov \textit{et al.} in recent PIMC simulations of hydrogen~\cite{filinov_pre_23, bonitz_cpp_23}. There it was found that, at low temperatures, the convergence (with respect to the number of high-temperature factors $P$) with the Kelbg potential is very slow and faster with the IKP. Thus, the use of the exact pair density matrix is advantageous for such conditions. With increasing temperature the convergence improves rapidly for the IKP, see also Sec.~\ref{sec:PIMC_hydrogen}.\\

Aside from PIMC, the Kelbg potential was also used in semiclassical MD simulations of partially ionized hydrogen for the computation of the density correlation function and the plasmon spectrum \cite{golubnichiy_cpp_02,golubnychiy_pre_1,bonitz_jpa_03}.
MD simulations with the IKP of the thermodynamic properties of partially ionized hydrogen were reported in Ref.~\cite{filinov_pre04} and allowed to accurately reproduce the equation of state for $T\gtrsim \SI{50000}{\kelvin}$. They break down, however,  when molecules form because the IKP does not prevent cluster formation of more than two hydrogen atoms, due to the missing exchange of more than two electrons. 
Novel MD results with the IKP will be presented below in Sec.~\ref{sssec:i-acoustic_H} where we compute the ion-acoustic plasma mode in dense hydrogen.\\

The use of the IKP in classical MD simulations can be understood as a simple case of force fields. Further improvements of this concept have recently been achieved via machine learning approaches which we discuss in Sec.~\ref{ssubsec:ml-md}.

Note that the quantum potentials discussed above are important in order to reproduce short-distance phenomena in semiclassical simulations. On the other hand, to capture screening and collective effects of quantum particles inside a plasma, the Coulomb potential has to be renormalized at large distances. Examples are the Debye potential or the dynamically screened Coulomb potential, Eq.~(\ref{eq:vs-definition}). 

Finally, the appropriate potential to compute the interaction between two electrons in a nonideal plasma is the  Kukkonen-Overhauser potential \cite{Kukkonen_PRB_1979}. It has been tested in PIMC simulations in Ref.~\cite{Dornheim_JCP_2022} and was recently used to explain the roton minimum observed in the plasmon dispersion of the correlated UEG \cite{dornheim_comphys_22}, see Sec.~\ref{sec:roton}.

\subsubsection{Effective ion-ion potentials. Machine-learning concepts
%[Jan, Tobias, Mandy]
}\label{ssubsec:ml-md}
Methods to extract effective ion-ion potentials from first principle simulations have been studied for a number of years. 
They are expected to close the gap between cheap but heuristic pair potentials and computationally expensive ab-initio calculations, needed to better explore all parts of the phase diagram of hydrogen. 
Forces or potentials from DFT-MD can be matched to a functional form of the potential or can be fit by a freely varying function~\cite{Izvekov_2004,Toth_2007,Akin-Ojo_2008,Vorberger_effect_2013}. At the same time, model potentials have been fit to structural or thermodynamic data~\cite{lcbopii_2005}.

In this way, machine-learned (ML) effective potentials between molecules, atoms, and ions are a new class of potentials that are promising to be more flexible and transferable than effective potentials that are obtained in the more traditional ways. 

In particular, the part of the phase diagram where the Born-Oppenheimer
approximation can be applied seems to be predisposed to replacing
\textit{ab-initio} DFT or QMC based BO energies by approximate ML models
developed over recent years \cite{Behler2007,Bartok2010,Zhang2018,Chmiela2017,Jinnouchi_2019,Fiedler_PRM_2022,fiedler_npj_23}.
The red box in Fig.~\ref{fig:h-phase-diag-qmc} indicates the phase-space region where ML effective potentials have been created and applied to liquid and solid hydrogen \cite{Zong2020,Cheng2020,Tirelli2022,Niu2023}.

A ML model for the BO potential surface is trained on energies and forces of $10^3-10^5$ static nuclear configurations calculated by DFT or QMC methods, e.g. from
\textit{ab-initio} MD, PIMD, or CEIMC runs. ML-based methods seem to perform equally well
for QMC-based energies or forces despite the intrinsic stochastic errors  \cite{Niu2023,Ceperley_2024}.

By running MD or PIMD simulations with effective ion potentials, one can then afford calculations on much bigger time and length scales.
Although ML model potentials in particular are built to ensure transferability, there is
limited practical experience for hydrogen so far. The smooth cross-over between the molecular and atomic limit found in Ref.~\cite{Cheng2020,Cheng2021} from a ML model of PBE-BO energies has not been confirmed later by direct \textit{ab-initio} PBE-MD~\cite{karasiev2021}. The absence of the liquid-liquid phase transition in the ML simulations might be due to the too coarse grid of initial BO-DFT-MD simulations. 

A ML potential study \cite{Niu2023} including nuclear quantum effects and QMC BO-energies (as well as PBE and vdW-DF1) predicted melting of the solid at considerably higher temperatures for the ML QMC potential energy surfaces
(see green line in Fig.~\ref{fig:phase-diagram-low-t}) than the one obtained within PBE, with ML vdW lying in between,
as well as a possible structural phase transition to a Fmmm-4 phase at higher temperature. The higher melting line prediction from ML models is in qualitative agreement with direct simulation by DFT-MD with vdW-DF functional and by CEIMC \cite{Niu2023}.

An alternative, though closely related strategy to achieve size transferability based on machine-learned DFT results has recently been proposed by J.A.~Ellis \textit{et al.}~\cite{Ellis_PRB_2021}. Instead of constructing an effective potential, their strategy is to learn the local density of states (LDOS) and to parametrize it based on local SNAP descriptors~\cite{THOMPSON2015316}. The LDOS then gives one access to standard DFT observables such as the energy or the electronic density, which, in principle, can be evaluated for very large numbers of ions; results for \num{131072} Be atoms have been presented in Ref.~\cite{fiedler_npj_23}.
The application of this method to dense hydrogen constitutes an interesting topic for future work.

\subsection{Time-dependent simulations}\label{subsec:noneq}
Dense plasmas exposed to external fields may be driven far from equilibrium and undergo a cascade of fast to slow relaxation processes, extending from femtosecond to nanosecond scales, before they return to equilibrium. Knowledge of the nonequilibrium behavior is important for understanding many experiments with dense hydrogen and deuterium, including inertial confinement fusion, for a recent discussion, see Ref.~\cite{Vorberger_PLA_2024}.

While truly first principles methods such as QMC [Sec.~\ref{sec:cpimc}] exist for the ground state and thermodynamic properties of dense quantum plasmas, no approaches of similar accuracy and capability have been developed for nonequilibrium situations yet. Even though there exist real-time QMC methods, such as continuous time QMC that are being used in condensed matter and cold atom physics to study impurity models, e.g.~\cite{gull_rmp_11}, they are afflicted with an additional phase problem. This limits their application to very short simulation times and, to our knowledge, has ruled out their application to dense plasmas so far. Similarly, extensions of path integral methods to nonequilibrium are available; various versions of path integral molecular dynamics (PIMD) have been proposed, e.g.~\cite{cao_jcp_94,polyakov_jcp_10,filinov_prb_2}, Applications to dense hydrogen have concentrated so far on \textit{equilibrium} situations to capture nuclear quantum effects \cite{Hinz2020}. On the other hand,  the accuracy and applicability range of these methods for dense plasmas out of equilibrium remains to be fully explored.  

Thus, the main method to simulate dense plasmas in nonequilibrium is quantum kinetic theory, cf. Sec.~\ref{sss:qke}, extensions of DFT to nonequilibrium, cf.  Sec.~\ref{sss:tddft} and hydrodynamics, cf. Sec.~\ref{sss:qhd}.

\subsubsection{Quantum kinetic equations (QKE)}\label{sss:qke}
Even though the equations of motion of quantum many-particle systems are known -- the time-dependent N-particle Schrödinger equation, $|\Psi_N(t)\rangle$, in case of pure states, and the von Neumann equation for the density operator $\hat \rho_N(t)$, for a mixed state -- their solution for plasmas is neither possible nor necessary. To compute time-dependent macroscopic observables, such as transport, optical or dielectric properties, collision rates, or reaction cross sections, knowledge of far simpler quantities is sufficient. The proper link between the microscopic quantum-mechanical equations and the macroscopic quantities and their equations of motion is provided by quantum kinetic theory \cite{bonitz_qkt} -- the proper generalization of Boltzmann's kinetic equation,
\begin{align}    
    \left(
    \partial_t + \textbf{v} \cdot \nabla_\textbf{r}
     + 
    \textbf{F}_a(\textbf{r},t) \cdot \nabla_\textbf{p}
    \right) 
    f_a(\textbf{r},\textbf{p},t) = 
    I_a(\textbf{r},\textbf{p},t)\,,
    \label{eq:boltzmann-eq}
    \\
    \mathbf{F}_a(\mathbf{r},t)  =  q_a \textbf{E}(\mathbf{r},t) + \frac{q_a}{c} \textbf{v}\times \textbf{B}(\mathbf{r},t)\,.
    \label{eq:lorentz-force}
\end{align}
Here $f_a(\textbf{r},\textbf{p},t)$ is the phase space distribution function of particle species ``a'' which evolves in time due to spatial inhomogeneities, all forces, $\textbf{F}_a$, as well as scattering processes described by the collision integral $I_a$. For the case of a plasma, $\textbf{F}_a$ is the Lorentz force, Eq.~(\ref{eq:lorentz-force}), which involves the electromagnetic field that obeys Maxwell's equations. Since the charged particles induce an electromagnetic field themselves, the associated ``mean field'' (Vlasov or Hartree field) is fully accounted for by this force.
The distribution function is normalized to the total particle number, $N_a$, and gives access to all single-particle observable in general nonequilibrium situations, including the particle density, $n_a$ and the mean velocity, $\textbf{u}_a$,
\begin{align}\nonumber
      N_a(t) & = \int \frac{d^3r \,d^3p}{(2\pi\hbar)^3} \,f_a(\textbf{r},\textbf{p},t)\,,\\
      \label{eq:f-n}
      n_a(\textbf{r},t) & = \int \frac{d^3p}{(2\pi\hbar)^3}\, f_a(\textbf{r},\textbf{p},t)\,,\\
      \textbf{u}_a(\textbf{r},t) & = \int \frac{d^3p}{(2\pi\hbar)^3} \,\textbf{v} f_a(\textbf{r},\textbf{p},t)\,,\nonumber
\end{align}
which provides the starting point for deriving hydrodynamic equations, cf. Sec.~\ref{sss:qhd}. Furthermore, the direct relation between distribution function and density, Eq.~\eqref{eq:f-n}, allows one to establish connections between quantum kinetic equations and time-dependent DFT, cf. Sec.~\ref{sss:tddft}.

Let us now turn to the r.h.s. of Eq.~(\ref{eq:boltzmann-eq}) which describes all scattering processes involving two or more particles, i.e.
\begin{equation}\nonumber
    I_a = \sum_b I_{ab} + \sum_{bc} I_{abc}\,.
\end{equation}
The three-particle integrals describe a large variety of scattering processes including inelastic processes, such as excitation/de-excitation or ionization of atoms and molecules, e.g. \cite{blue-book}. On the other hand, the integrals $I_{ab}$ describe binary elastic scattering processes and are crucial for correctly describing correlation and thermalization effects in the plasma. Here we concentrate on the two-particle integrals and, for illustration, present the \textit{Balescu-Lenard integral} \cite{balescu60,lenard60} which contains other collision integrals as limiting cases (see below) and plays a central role in plasma kinetic theory \cite{bonitz_qkt}
\begin{align}\label{eq:lb-col-int}
    I_a(\textbf{p}) = & \frac{1}{8\pi^2}\sum_b \int \mathrm{d}\mathbf{p}'\mathrm{d}\mathbf{k}\,\left[V^s_{ab}(k,\mathbf{k}\cdot\mathbf{v})\right]^2\delta[\mathbf{k}\cdot (\mathbf{v}- \mathbf{v}')]\cdot \\\nonumber
    & \times (\mathbf{k}\cdot\nabla_\mathbf{p})  \mathbf{k}\cdot \left[\nabla_\mathbf{p}f_a(\mathbf{p})f_b(\mathbf{p}')-\nabla_{\mathbf{p}'}f_b(\mathbf{p}')f_a(\mathbf{p})\right] \,.\\
    & V^s_{ab}(k,\omega) = \frac{V_{ab}(k)}{|\epsilon(k,\omega)|}\,,\quad 
    V_{ab}(k) = \frac{4\pi q_aq_b}{k^2}
    \,.\label{eq:vs-definition}
\end{align}
\textbf{Discussion of the collision integral. Special cases}. The integral $I_a$ describes the temporal change of the distribution $f_a(\textbf{r}, \textbf{p},t)$ due to binary (or more complex) collisions with particles of all types in all possible momentum states $\textbf{p}'$ which are occupied with probability $f_b(\textbf{p}',t)$. The collision integral (\ref{eq:lb-col-int}) is the lowest order perturbation theory result, it is proportional to the square of the interaction potential (second order in the coupling parameter $\Gamma$, second Born approximation, SOA) and contains a number of limiting cases which we briefly recall. i) for $\epsilon \to 1$, $I_a$ reduces to the Landau collision integral which contains the square of the Fourier transform of the Coulomb potential. In that case the $k$-integral diverges logarithmically which is ``fixed'' by introducing finite minimum and maximum values for the k-integration, resulting in the so-called ``Coulomb logarithm'' (for a discussion of its use in classical plasma kinetic theory see Ref.~\cite{braginskii_65}), 
\begin{align}
\ln\Lambda_c=\ln \frac{k_{\rm max}}{k_{\rm min}}= \ln \frac{b_{\rm min}}{b_{\rm max}}\,.    \label{eq:coulomb-log}
\end{align}
The existence of a finite maximum wave number, $k_{\rm max}$ (or minimum distance, $b_{\rm min}$), is usually justified by the finite extension of particles which is of the order of the De Broglie wavelength $\Lambda_a$ [cf. Sec.~\ref{ssubsec:pair-pot-md}], leading to the choice $k_{\rm max}\to \Lambda_a^{-1}$. The common choice of $k_{\rm min}=1/b_{\rm max}$ is $1/r_{\rm D}$ -- the inverse of the Debye screening length, Eq.~(\ref{eq:debye-length}). This is motivated by the screening of the long-range Coulomb interaction in a plasma, giving rise to the effective replacement of the Coulomb potential by the Debye potential, $\frac{q_aq_b}{r} \to \frac{q_aq_b}{r}e^{-r/r_{\rm D}}$. In degenerate quantum systems, the Debye length is replaced by the Thomas-Fermi length, $r_{\rm TF}$, Eq.~(\ref{eq:tf-length}), or by a proper interpolation between the two limits.

Use of the Debye potential leads to ii) the statically screened second Born approximation where, in Eq.~(\ref{eq:vs-definition}), $\epsilon(k)=1+(k r_D)^{-2}$. iii) This result is further improved by taking into account the dynamics of the screening cloud via the dynamic Vlasov dielectric function, $\epsilon(k,\omega)$, of an ideal plasma giving rise to the dynamically screened pair potential, $V^s(k,\omega)$, Eq.~(\ref{eq:vs-definition}). This corresponds to the Balescu-Lenard kinetic equation which selfconsistently includes screening effects (no phenomenological cut-off $k_{\rm min}$ is needed) and collective excitations (plasmons) in the two-particle scattering process, for a derivation see Ref.~\cite{schroedter_cpp_24}. In the many-body theory, this approximation is directly related to the GW approximation (see below). Note that, for dense plasmas, quantum generalizations of the kinetic equation (\ref{eq:boltzmann-eq}) and of the collision integral are necessary which are straightforward and will be discussed below in the context of the G1--G2 scheme.\\
\textbf{Relaxation Time approximation (RTA).} An important and popular special case of the collision integral (\ref{eq:lb-col-int}) is obtained by linearization with respect to small deviations from the asymptotic equilibrium distribution, $f_a^{\rm eq}$  \cite{bgk_pr_54,rostoker_pfl_60},
\begin{align}
    I_a(\textbf{p},t)|_{\rm RTA} = -\frac{1}{\tau(n,T)}\left\{f_a(\textbf{p},t) -  f_a^{\rm eq}(\textbf{p}) \right\}\,,
    \label{eq:i-rta}
\end{align}
where for $f_a^{\rm eq}(\textbf{p};n,T)$ a Maxwell, Bose or Fermi distribution is used. 
Furthermore, $\tau$ denotes a characteristic time during which $f_a(\textbf{p},t)$ approaches $f_a^{\rm eq}$ which is computed for a plasma in thermodynamic equilibrium. The explicit expression for $\tau$ depends on the choice of the collision integral which is then linearized around $f_a=f_a^{\rm eq}$. The integral (\ref{eq:i-rta}) is sometimes called BGK (Bhatnagar, Gross, Krook) collision integral and is the starting point for the derivation of linear response quantities from kinetic equations, e.g.~\cite{ludwig_jpcs_10}. An example is dielectric and optical properties that can be derived from the kinetic equation (\ref{eq:boltzmann-eq}). While setting $I_a \to 0$ leads to the mean field (Vlasov or RPA) results for the dielectric function, using $I_a(\textbf{p},t)|_{\rm RTA}$ allows one to include collision effects in a sum-rule preserving way, see the discussion of the Mermin dielectric function, Eq.~(\ref{eq:mermin_df}), in Sec.~\ref{sec:roton_H}.

Aside from linear response theory applications, both, the Landau and Balescu-Lenard equation have been frequently solved for electron-hole plasmas and dense quantum plasmas that are driven far away from equilibrium, e.g.~\cite{scott_prl_92,binder_prb_92,kosse-etal.97,scullard_pop_16} to study the relaxation towards thermodynamic equilibrium~\cite{Vorberger_pop_2009}. However, the proper choice of the collision integral is often not clear, and, therefore, the resulting time evolution of $f_a(\textbf{r},\textbf{p},t)$ is only qualitatively correct. \\
\textbf{Properties and failures of the kinetic equation (\ref{eq:boltzmann-eq}) with the collision integral (\ref{eq:lb-col-int}).} 
It easily verified \cite{bonitz_qkt} that the kinetic equations (\ref{eq:boltzmann-eq}, \ref{eq:lb-col-int}) 
\begin{enumerate}
    \item obey conservation laws of particle number and mean momentum;
    \item conserve the mean kinetic energy, $\langle p^2/2m \rangle$ where averaging is with $f_a(t)$. This, however, violates the correct conservation law of an interacting system where total energy, i.e. the sum of kinetic and interaction energy, is conserved. 
    \item have an asymptotic solution, $f_a^{\rm eq}(\textbf{p})$, which is given by the Fermi or Bose distribution (if quantum generalizations for fermions or bosons have been performed, e.g. \cite{bonitz_qkt}). However, in an interacting quantum system, the equilibrium distribution is different, as we will show explicitly in Sec.~\ref{sec:n(k)}.
    \item Numerical solutions of the quantum Balescu-Lenard equation (\ref{eq:boltzmann-eq}, \ref{eq:lb-col-int}) revealed an unphysically rapid thermalization, indicating that the collision integral (\ref{eq:lb-col-int}) is not applicable in situations far from equilibrium, at short times.
\end{enumerate}
Before discussing how to overcome these limitations of Markovian kinetic equations we briefly comment on the status of equilibrium Green functions (EGF) theory. 

\subsubsection{Equilibrium Green functions (EGF)}\label{sss:egf}
Nonequilibrium Green functions, $G_{ij}(t,t')$ which will be discussed in Sec.~\ref{sss:g1-g2} depend on two time arguments and are defined on the round trip Keldysh contour \cite{keldysh_contour}. The NEGF contains thermodynamic equilibrium as a limiting case where the Keldysh contour shrinks to its imaginary branch, e.g. ~\cite{stefanucci2013nonequilibrium,balzer-book} and $G_{ij}(t,t') \to G^M_{ij}(\omega)$, where the frequency $\omega$ is the Fourier adjoint of the time difference, $t-t'$. This limit coincides with the Matsubara Green functions that have been actively investigated since the late 1950s, e.g. \cite{matsubara55,abrikosov-etal.62}. This technique has been extensively applied to dense partially ionized plasmas in the 1970s and 1980s by the Rostock group, e.g. \cite{red-book, green-book}. The strength of equilibrium and nonequilibrium Green functions theory is the systematic approach (via Feynman diagrams) to incorporate correlation effects into the theory. At the same time, the set of available self-energies is limited and their validity range and accuracy are not known \textit{a priori}. A few tests have been performed against FPIMC simulations for the uniform electron gas using the dynamic second order Born approximation (Montroll-Ward and e$^4$ self-energies) \cite{schoof_prl15,dornheim_physrep_18}. Within their expected range of validity, $r_s\lesssim 1$, the relative error of the interaction energy $V$ reaches $10\%$ \cite{dornheim_physrep_18} which is the value shown in Tab.~\ref{tab:methods}. J.J.~Kas \textit{et al.} presented improved EGF results that are based on a cumulant expansion and were also applied to the strong coupling regime, $r_s\lesssim 40$~\cite{Kas_PRL_2017,Kas_PRB_2019}. They reported overall good agreement with FPIMC-based fits (errors of the order of $10\%$ for the interaction parts of the thermodynamic functions), however, significantly larger errors and unphysical temperature dependence were observed at intermediate densities \cite{dornheim_physrep_18}. Note that these errors translate into much smaller errors for the total thermodynamic quantities, which is remarkable, considering the large density and temperature range.

A major advantage of Green functions methods is the direct access -- via the frequency dependence of $G^M$ -- to spectral properties such as the spectral function or density of states. This is achieved via an iterative solution of the Dyson equation, 
\begin{align}
    \label{eq:dyson-gm}
    G^M(\omega) = G^M_0(\omega) + G^M_0(\omega)\Sigma(\omega)G^M(\omega)\,.
\end{align}
This equation would be exact, would the selfenergy be known. In practice, approximations for the functional dependence, $\Sigma[G^M]$ are being used. For a discussion of the solution of the Dyson equation, see Ref.~\cite{joost_cpp_21}. The second approach to the spectral properties consists of solving the Bethe-Salpeter equation (BSE) -- the equation of motion of the two-particle Green function, e.g.~\cite{blue-book,stefanucci2013nonequilibrium}.

\textbf{Ab initio EGF simulations}. 
The large computational effort of Green functions theory arises from the dependence of the Green function on two basis indices, i.e. the quantities in Eq.~(\ref{eq:dyson-gm}) have to be understood as matrices, $G^M(\omega) \to G_{ij}^M(\omega)$. 
In the case of spatially uniform systems, such as dense plasmas or jellium, a plane wave basis in terms of momentum eigenstates is appropriate, leading to diagonal matrices, $G^M_{p,p'} \sim \delta_{p,p'}$ and $\Sigma_{p,p'}\sim \delta_{p,p'}$. However, even in that case, the dimension of the matrices can be very large, since $p$ are vectors, for details Sec.~\ref{sss:g1-g2}.

A very promising concept to reduce the computational effort and increase the accuracy of the results is to use, instead of plane waves, a basis of Kohn-Sham orbitals obtained from an independent KS-DFT simulation for the same system. This leads to a powerful combination of DFT with Green functions known as \textit{ab initio} BSE or \textit{ab initio} GW simulations and was successfully applied to various systems in the ground state, e.g. \cite{PhysRevB.86.195123}. This has also been extended to excited state and optical properties, e.g. with the yambo code \cite{marini_yambo_09}.

\subsubsection{Nonequilibrium Green functions and the G1-G2 scheme}\label{sss:g1-g2}
The deficiencies of standard kinetic equations that were listed above led in the  1990s to the development of generalized quantum kinetic equations by D.~Kremp, M.~Bonitz, and co-workers, who used nonequilibrium Green functions (NEGF)~\cite{keldysh_contour,bonitz_pss_19_keldysh,kadanoff-baym, blue-book,balzer-book} and reduced density operators \cite{bonitz_qkt}. The resulting quantum kinetic equations are free of the aforementioned problems, e.g.~\cite{bonitz-etal.96jpcm,bonitz-etal.96pla,binder-etal.97prb} and were applied, among others, to quantum plasmas in strong laser fields \cite{kremp_99_pre,bonitz_99_cpp,haberland_01_pre} and to the correlated dielectric function of the uniform electron gas \cite{kwong_prl_00}. In recent years more advanced collision integrals could be implemented that include strong coupling effects (T-matrix approximation). Moreover, extensive tests of the accuracy were performed by comparison to cold atom experiments and exact diagonalization and density matrix renormalization group (DMRG) calculations \cite{schluenzen_prb16, schluenzen_prb17,schluenzen_jpcm_19}. Even though these tests were possible only for lattice models, they have allowed one, for the first time, to rigorously benchmark the accuracy of quantum kinetic theory simulations with different collision integrals (self-energies). This has made quantum kinetic theory a predictive tool for nonequilibrium applications and a possible benchmark for other time-dependent approaches for dense quantum plasmas such as time-dependent DFT and quantum hydrodynamics. \\

\textbf{The G1-G2 scheme.}\\
The main limitation of NEGF simulations is their high computational load: nonequilibrium simulations scale cubically with the number of time steps, $N_\textnormal{t}$, which is due to the two-time dependence of the NEGF and the time nonlocal structure of the collision integrals (memory integration). This has changed dramatically when 
N.~Schlünzen, J.-P.~Joost, and M.~Bonitz were able to
reformulate the NEGF equations as two coupled time-local equations 
for the one- and two-particle Green functions by invoking the generalized Kadanoff-Baym ansatz (GKBA) \cite{lipavsky_generalized_1986}: this allowed them to reduce the scaling to first order in $N_\textnormal{t}$ \cite{schluenzen_prl_20,joost_prb_20}. Remarkably, this scaling is achieved not only for simple perturbation theory (SOA, Landau collision integral) but also for advanced self-energies, including T-matrix and GW [generalization of the Balescu-Lenard approximation, Eq.~(\ref{eq:lb-col-int})] and combinations thereof resulting in the dynamically screened ladder (DSL) approximation \cite{joost_prb_22}. This has already triggered a large number of applications for lattice systems and 2D quantum materials. Below, we present the equations in a form suitable for dense quantum plasmas including fully ionized hydrogen.

\begin{widetext}
    Assuming a spatially uniform system consisting of multiple components (labeled $\alpha$ and $\beta$ with charge numbers $Z_\alpha$ and $Z_\beta$), the equations of motion, in the G1--G2 scheme, are conveniently rewritten in the momentum representation (to simplify the notation, we omit the time dependencies of $G^\gtrless$ and $V_q$ in some places) \cite{bonitz_pssb23}
\begin{align}\label{eq:g1-equation}
 &  \mathrm{i}\hbar\frac{\text{d}}{\mathrm{d}t}G^<_{\mathbf{p}\alpha}(t) =[I+I^\dagger]_{\mathbf{p}\alpha}(t),\qquad I_{\mathbf{p}\alpha}(t)=\pm\mathrm{i}\hbar  Z_\alpha\sum\limits_{\mathbf{kq},\beta}Z_\beta V_{\mathbf{q}}(t)\,\mathcal{G}^{\beta\alpha}_{\mathbf{kpq}}(t)\,,\\
 & G^<_{\mathbf{p}\alpha}(0) = G^{<0}_{\mathbf{p}\alpha}\,,\quad
  \mathcal{G}^{\alpha\beta}_{\mathbf{kpq}}(0) = \mathcal{G}^{\alpha\beta \,0}_{\mathbf{kpq}}\,,\quad V_q(t)=f^{\rm AS}(t)V_q\,,\quad \lim_{t\to -\infty}f^{\rm AS}(t) = 0\,, \quad f^{\rm AS}(t\ge 0) = 1\,,
  \label{eq:g1-g2-inicond}\\
    \label{eq:g2-equation}
   & \mathrm{i}\hbar \frac{\text{d}}{\text{d}t} \mathcal{G}^{\alpha\beta}_{\mathbf{kpq}}(t) -
\mathcal{G}^{\alpha\beta}_{\mathbf{kpq}}(t)\left[h^\text{HF}_{\mathbf{k}-\mathbf{q},\alpha}(t)+h^\text{HF}_{\mathbf{p}+\mathbf{q},\beta}(t)-h^\text{HF}_{\mathbf{k},\alpha}(t)-h^\text{HF}_{\mathbf{p},\beta}(t)\right]
    =
    \hat\Psi^{\pm,\alpha\beta}_{\mathbf{kpq}}(t)+
    \Pi^{\alpha\beta}_{\mathbf{kpq}}(t)\,,
\\
\nonumber
    \hat\Psi^{\pm,\alpha\beta}_{\mathbf{kpq}} & =(\mathrm{i}\hbar)^2\left[V_{|\mathbf{q}|}^{\alpha\beta} \pm \delta_{\alpha\beta}V^{\alpha\alpha}_{|\mathbf{k}-\mathbf{p}-\mathbf{q}|}\right]\cdot\left[G^>_{\mathbf{k}-\mathbf{q},\alpha}\,G^>_{\mathbf{p}+\mathbf{q},\beta}\,G^<_{\mathbf{k},\alpha}\,G^<_{\mathbf{p},\beta}-G^<_{\mathbf{k}-\mathbf{q},\alpha}\,G^<_{\mathbf{p}+\mathbf{q},\beta}\,G^>_{\mathbf{k},\alpha}(t)\,G^>_{\mathbf{p},\beta}\right]\,,\\
    \\
    \Pi^{\alpha\beta}_{\mathbf{kpq}} &=\pi_{\mathbf{kpq}}^{\alpha\beta}-\left[\pi_{\mathbf{p}+\mathbf{q},\mathbf{k}-\mathbf{q},\mathbf{q}}^{\beta\alpha}\right]^*,\quad\text{where}\quad \pi_{\mathbf{kpq}}^{\alpha\beta}=(\pm)_\beta(\mathrm{i}\hbar)^2\left[G^>_{\mathbf{p}+\mathbf{q},\beta}(t)\,G^<_{\mathbf{p},\beta}-G^<_{\mathbf{p}+\mathbf{q},\beta}\,G^>_{\mathbf{p},\beta}\right]\sum\limits_{\mathbf{p}'\gamma}V_{|\mathbf{q}|}^{\alpha\gamma}\,\mathcal{G}^{\alpha\gamma}_{\mathbf{k}\mathbf{p}'\mathbf{q}}\,.
    \nonumber
\end{align}
Here, the single-particle Green function is related to the distribution function by $-i\hbar G^<_\alpha(\textbf{p},t)=f_\alpha(\textbf{p},t)$, whereas $i\hbar G^>_\alpha(\textbf{p},t)=1 - f_\alpha(\textbf{p},t)$. The two-particle function entering the collision integral in the kinetic equation~(\ref{eq:g1-equation}) denotes the correlated part of the two-particle Green function, $\mathcal{G}=G^{(2)}-G^{(2)}_{\rm HF}$, where $G^{(2)}$ is related to the two-particle distribution function by $(i\hbar)^2G^{(2)}_{\alpha\beta}=f^{(2)}_{\alpha\beta} $, and the Hartree-Fock hamiltonian is given by $h^\text{HF}_{\mathbf{p},\alpha}(t)= h_{\mathbf{p},\alpha}(t)+\sum_\textbf{q} V_{\textbf{p}-\textbf{q}}G^<_\alpha(\textbf{q},t)$. Note that the G1--G2 scheme is complemented with initial conditions (\ref{eq:g1-g2-inicond}) for the one- and two-particle Green function at time $t=0$. In contrast to the Boltzmann equation, the initial state may be correlated \cite{semkat_00_jmp,semkat_99_pre}, i.e. $\mathcal{G}(0)\ne 0$, which is achieved by slowly turning on the pair interaction (``adiabatic switching'') during the solution of the equations, for a finite period of time $t<0$, e.g. Refs.~\cite{bonitz_cpp13, hermanns_prb14, schluenzen_jpcm_19}.
\end{widetext}

Equation (\ref{eq:g1-equation}) is the generalized quantum kinetic equation which, due to the assumed homogeneity and absence of external fields does not contain additional terms on the l.h.s. (for generalizations to plasmas in strong fields, see Refs.~\cite{kremp_99_pre,bonitz_99_cpp,haberland_01_pre}). The collision integral $I$ generalizes the Balescu-Lenard integral (\ref{eq:lb-col-int}) to the case of arbitrary correlations the properties of which are defined by the two-particle correlation function $\mathcal{G}$. It obeys a separate equation of motion, Eq.~(\ref{eq:g2-equation}) which is the nonequilibrium generalization of the time-diagonal Bethe-Salpeter equation (BSE) when the GKBA has been applied \cite{schroedter_pssb23}. The terms on the r.h.s. specify the many-body approximation (selfenergy): if $\Pi$ is neglected, Eq.~(\ref{eq:g2-equation}) corresponds to the static second Born approximation (generalization of the Landau equation). The $\Psi$ term describes the two-particles scattering process including exchange and Pauli blocking.
On the other hand, if the polarization term $\Pi$ is included, the treatment of correlations is significantly improved to the dynamically screened second Born approximation, i.e. to the 
 GW approximation (non-Markovian quantum Lenard-Balescu equation).
 Dyson equation for $W$). It is well known that the polarization term $\Pi$ gives rise to a Dyson equation for the inverse dielectric function or the dynamically screened potential $V^s(k,t,t')$ \cite{bonitz_qkt} -- the nonequilibrium generalization of the potential (\ref{eq:vs-definition}), 
 \begin{align}\label{eq:vs-dyson}
     V^s(\textbf{k},t,t') =& V(\textbf{k})\delta(t-t') + \\
     &+ V(\textbf{k})\int_{t'}^t d\bar t \,P(\textbf{k},t,\bar t)V^s(\textbf{k},\bar t,t')\,,\nonumber
 \end{align}
 where $P$ is the retarded longitudinal polarization function. Equation (\ref{eq:vs-definition})
 yields the nonequilibrium inverse dielectric function,
 \begin{align}\label{eq:epsm1-noneq}
     \epsilon^{-1}(\textbf{q},t,t') = \frac{V^s(\textbf{q},t,t')}{V(\textbf{q})}
 \end{align}
 and the complete plasmon spectrum, including its time evolution in nonequilibrium.

For a stationary system, the dependence on the center of mass time, $T=(t+t')/2$ vanishes, and Fourier transformation with respect to $\tau=t-t'$ yields, applying the convolution theorem, the frequency dependence of all quantities
\begin{align}\label{eq:vs-def}
    V^s(\textbf{k},\omega) = \frac{V(\textbf{k})}{1 - V(\textbf{k})P(\textbf{k},\omega)} = 
    \frac{V(\textbf{k})}{\epsilon(\textbf{k},\omega)}\,.
\end{align}
This leads the dielectric function of an ideal system as well as the inverse dielectric function
\begin{align}\label{eq:eps-def}
    \epsilon(\textbf{k},\omega) &= 1 - V(\textbf{k})P(\textbf{k},\omega)\,,\\
    \epsilon^{-1}(\textbf{k},\omega) &= 1 + V(\textbf{k})\chi(\textbf{k},\omega)\, ,
\label{eq:epsm1-def}
\end{align}
where the latter is expressed in terms of the density response function $\chi$, for the definition, see Eq.~(\ref{eq:density-response}). 
Using Eqs.~(\ref{eq:eps-def}) and (\ref{eq:epsm1-def}), one finds that the density response function obeys the equation
\begin{align}\label{eq:chi-def}
    \chi(\textbf{k},\omega) = \frac{P(\textbf{k},\omega)}{1 - V(\textbf{k})P(\textbf{k},\omega)}\,,
\end{align}
a relation that is also being used in time-dependent DFT, see Sec.~\ref{sss:tddft}.
The RPA result for $\chi$ follows when the polarization function of an ideal Fermi gas (Lindhard polarization, $P \to \chi_0$) is being used. One way to go beyond the RPA result for $\chi$ is to introduce the dynamic local field correction, $G(\textbf{k},\omega)$, leading to 
\begin{align}\label{eq:chi-correlated-def}
    \chi(\textbf{k},\omega) = \frac{\chi_0(\textbf{k},\omega)}{1 - V(\textbf{k})\chi_0(\textbf{k},\omega)[1-G(\textbf{k},\omega)]}\,,
\end{align}
The function $G$ is discussed in more detail in Sec.~\ref{sec:dynamics}. 

 The second way to incorporate correlation effects is to further upgrade the BSE (\ref{eq:g2-equation}) which leads to the dynamically screened ladder approximation which takes into account plasmon effects beyond the RPA as well as bound states in the plasma environment (ionization potential depression), cf. Sec.~\ref{sssec:ion-pot}.
  
 So far, the G1-G2 equations have been solved for dense quasi-1D plasmas \cite{makait_cpp_23}, and 
 new results will be presented in Sec.~\ref{sss:stopping-g1-g2}.
 Extensions of the G1--G2 scheme to 2D and 3D plasmas are straightforward, in principle, but suffer from a new bottleneck: the large memory consumption when storing the correlation function $\mathcal{G}^{\alpha\beta}_{\mathbf{kpq}}$. This problem is expected to be solved in the near future by applying 
 a novel quantum fluctuations approach developed by E.~Schroedter, M.~Bonitz and co-workers \cite{schroedter_cmp_22,schroedter_23,schroedter_pssb23}.

\subsubsection{Time-dependent DFT (TDDFT)}
\label{sss:tddft}

Real-time time-dependent Kohn-Sham Density Functional Theory (RT-TDDFT) is based on Eqs.~(\ref{eq:psi_KS}), which provides the time-dependent one-particle density $n(\vec r,t)=\sum_j f_j\left|\psi_{\uline{\vec R}}^{j}(\vec r,t)\right|^2$.
Although RT-TDDFT can be formulated to be formally exact, in practice several approximations are used.
Most importantly for WDM applications, the XC functional is often utilized in a static (adiabatic) approximation, meaning that $v_{xc}$ depends only on the density value at a given time moment and does not have an explicit dependence on time; memory effects in the XC potential are thus neglected. This is not critical for systems such as WDM and dense plasmas if the excitation spectrum is qualitatively similar to that of the ideal electron gas. 
Examples of such features are plasmons at small wavenumbers (where collective oscillations are not strongly damped) and features of single-particle oscillations at large wavenumbers (where the kinetic energy of an electron,  $\hbar^2q^2/2m_e$, dominates over interaction energy terms). In addition, in the same way, for the same reasons, and with the same quality as in the equilibrium KS-DFT, RT-TDDFT with an adiabatic XC functional provides information about the energy levels of the orbitals localized around ions. 

The drawback of using the static XC functional in RT-TDDFT is that it leads to time-independent occupation numbers \cite{Appel_Gross_2010}.
Second, for extended systems in general and WDM simulations in particular,  the initial state of the system is usually prepared using an equilibrium KS-DFT calculation with corresponding occupation numbers according to the Fermi-Dirac distribution. This in combination with the static approximation for  $v_{xc}$ prevents the legitimate application of RT-TDDFT for the simulation of non-equilibrium effects on a time scale shorter than the relaxation time [Sec.~\ref{sss:qke}], where the occupation numbers change significantly. Furthermore, the adiabatic approximation in $v_{xc}$ fails at perturbation frequencies (energies) that are outside the spectrum of the equilibrium KS-DFT state \cite{Thiele_Gross_prl_2008}.

RT-TDDFT does not require an explicit condition of weak coupling between the perturbing external field and the electrons. This allows one to use it for the simulation of processes that are beyond the linear response approximation, e.g., in the simulation of the stopping power~\cite{Kononov2023}.  In the context of ICF,  the latter application is of particular importance for the alpha particle energy loss in warm dense hydrogen \cite{kononov_2024, moldabekov_pre_20}. 

An alternative and formally equivalent formulation of TDDFT in the linear response regime is referred to as linear-response TDDFT (LR-TDDFT).
LR-TDDFT does not require an explicit time-dependent propagation of the wave functions since all needed information about the dynamic properties of the equilibrium system is already contained in the equilibrium state wavefunctions and eigenenergies. Therefore, the derivation of LR-TDDFT is based on a standard perturbative approach starting from the ideal (non-interacting) lowest-order approximation to the dynamic density response function. Further, the addition of density inhomogeneity  effects and corrections due to exchange and correlations results in a generalization of Eq.~(\ref{eq:epsm1-def}) for the inverse dielectric function to non-uniform systems,
\begin{equation}\label{eq:d_f}
      \varepsilon^{-1}_{\scriptscriptstyle \vec G,\vec G^{\prime}}(\vec k,\omega)=\delta_{\scriptscriptstyle \vec G,\vec G^{\prime}}+\frac{4\pi}{\left|\vec k+\vec G\right|^{2}}  \chi_{\scriptscriptstyle \vec G,\vec G^{\prime}} (\vec k,\omega),
\end{equation}
where  $\vec k$ is a wave vector restricted to the first Brillouin zone, and  $\vec G$ and $\vec G^{\prime}$ are reciprocal lattice vectors.  The dielectric function of a homogeneous system follows from $\vec G=\vec G^{\prime}$ components of the matrix $\varepsilon^{-1}_{\scriptscriptstyle \vec G,\vec G^{\prime}}(\vec k,\omega)$.

Similar as in Sec.~\ref{sss:g1-g2}, this equation can be transformed into
a Dyson equation for the microscopic electronic density response function 
$ \chi_{\scriptscriptstyle \vec G,\vec G^{\prime}}(\vec k,\omega)$ \cite{Byun_2020, martin_reining_ceperley_2016}:
\begin{equation}\label{eq:Dyson}
\begin{split}
\chi_{\scriptscriptstyle \mathbf G \mathbf G^{\prime}}(\mathbf k, \omega)
&= \chi^0_{\scriptscriptstyle \mathbf G \mathbf G^{\prime}}(\mathbf k, \omega)+ \displaystyle\smashoperator{\sum_{\scriptscriptstyle \mathbf G_1 \mathbf G_2}} \chi^0_{\scriptscriptstyle \mathbf G \mathbf G_1}(\mathbf k, \omega) \big[ v_{\scriptscriptstyle \mathbf G1}(\vec k)\delta_{\scriptscriptstyle \mathbf G_1 \mathbf G_2} \\
&+ K^{\rm xc}_{\scriptscriptstyle \mathbf G_1 \mathbf G_2}(\mathbf k, \omega) \big]\chi_{\scriptscriptstyle \mathbf G_2 \mathbf G^{\prime}}(\mathbf k, \omega),
\end{split}
\end{equation}
where   $\chi^{~0}_{\scriptscriptstyle \vec G,\vec G^{\prime}}(\vec k,\omega)$ is an ideal (non-interacting) density response function computed using Kohn-Sham orbitals, $v_{\scriptscriptstyle \mathbf G1}(\vec k)={4\pi}/{|\mathbf k+\mathbf G_1|^2}$ is the Coulomb potential in reciprocal space, and $ K^{\rm xc}_{\scriptscriptstyle \vec G_1,\vec G_2}(\vec k, \omega)$ is the exchange-correlation (XC) kernel defined as functional derivative of the exchange-correlation potential \cite{dynamic1}. Eq. (\ref{eq:Dyson}) represents a microscopic density response to an external perturbation with the frequency $\omega$ and the wavenumber $\vec q=\vec G+ \vec k$, where $\vec G$ is a reciprocal lattice vector, and thus constitutes a generalization of  Eq.~(\ref{eq:dynamic_density_response}) below.

The dependence on two different wavenumbers of the microscopic electronic density response function is due to the inhomogeneity of the system. To see this, it is illustrative to write the connection between the perturbing field $\delta v(\vec r, t)$ and the resulting density perturbation $\delta n(\vec r,t)$ in real space:
\begin{equation}
 \delta n (\vec r, t)=\int dt^{\prime} \int d \vec r' \, \chi(\vec r, t, \vec r^{\prime}, t^{\prime})\delta v(\vec r', t^{\prime})\,.   
\label{eq:density-response}
\end{equation}
For an equilibrium system perturbed by a weak field, we have $\chi(\vec r, t, \vec r^{\prime}, t^{\prime})=\chi(\vec r, t, \vec r^{\prime}, t^{\prime})=\chi(\vec r, \vec r^{\prime}, t-t^{\prime})$. The introduction of periodic boundary conditions allows one to rewrite $\chi(\vec r, t, \vec r^{\prime}, t^{\prime})=\chi(\vec r, \vec r^{\prime}, t-t^{\prime})$ in Fourier space in terms of $\vec G$ and $\vec G^{\prime}$ as $ \chi_{\scriptscriptstyle \vec G,\vec G^{\prime}}(\vec k,\omega)$ \cite{quantum_theory}.

For disordered (melted) extended systems such as warm dense matter, the system properties are homogeneous on average (over time or configurations). 
Therefore, to connect LR-TDDFT results with an experimental observable, such as XRTS, one needs to compute the averaged macroscopic density response function.
This has been explained in detail for the example of warm dense hydrogen in Refs. \cite{Moldabekov_PRR_2023, Moldabekov_JCP_averaging_2023}. 

We also point out that the Dyson equation (\ref{eq:Dyson}) is the equilibrium limit of the two-time Dyson equation (\ref{eq:vs-dyson}) that is studied in quantum kinetic theory, cf. Sec.~\ref{sss:qke}. This also provides the opportunity to establish direct links between approximations for $K_{\rm xc}$ and for the self-energy, as well as between the results of the two approaches.

\subsubsection{Classical and quantum hydrodynamics for ICF modeling}\label{sss:qhd}

Radiation-hydrodynamics (RH) codes such as Hydra \cite{marinak01, kritcher2022} and FLASH \cite{fryxell2000, saupe2023}, are the main computational tools used to design Inertial Confinement Fusion (ICF) experiments.  The codes are multi-dimensional, multi-physics (coupled radiation, hydrodynamics, and thermonuclear burn), and run on large parallel computing machines. Behind every ICF experiment are thousands and thousands of runs using the RH codes. This design element is critical to an experiment's success, due to the great expense of the targets and diagnostics.  Hence, the accuracy of RH codes is paramount. The process that each RH code goes through to ensure this accuracy is called verification and validation (VandV) \cite{ryanm-book}
%. VandV can be thought of as a type of quality control. Validation and verification of RH codes relies on 
and has to answer two questions: (1) Is the RH code solving the correct equations (validation)? (2) Is the RH code solving the equations correctly (verification)? The former question is typically addressed through a comparison of the RH code with experimental results of varying complexity. This might involve comparisons of Rayleigh-Taylor (RT) instability growth rates with experimental data coming from linear electric motor experiments \cite{banerjee2000}.  The latter question is typically addressed by a comparison of the RH code with analytic and semi-analytic results, including e.g. the Sod shock tube problem and the Marshak wave \cite{castor2004}. 

RH codes are tested on a large suite of validation and verification test problems before a code is used to simulate an ICF capsule. The trajectory [cf. Fig.~\ref{fig:overview}] of the fuel burning region in an ICF capsule traverses the warm dense to hot dense matter regime. This wide range of physical conditions places a severe demand on the accuracy of RH codes. The hydrodynamics usually involves some form of the Navier-Stokes equations, and the algorithms are designed to capture shock formation and propagation. A description of turbulent mixing is also needed in ICF simulations due to the prevalence of Rayleigh-Taylor, Richtmyer-Meshkov, and Kelvin-Helmholtz instabilities. Radiation is frequently treated with grey or multi-group diffusion \cite{castor2004}.  RH codes are made up of hundreds of input parameters if not thousands. The equations require as input numerous physics quantities, including EOS, fusion reaction rates, opacities, and electronic and ionic transport coefficients, such as thermal and electrical conductivity and viscosity. EOS models, in particular, are a focus of study in ICF implosions since they can affect shock timing and material compressibility and thereby determine instability growth rates and hydrodynamic coupling to the DT fuel.

The challenge users and code developers of the RH codes face is: what choices should they should make for an ICF simulation? For example, what is the best choice for the EOS for the fuel or ablator material? In many cases, high-quality experimental physics data that could inform users of the RH codes about what choices to make does not exist. The reason is that it is difficult to obtain data for a single physics model in regimes where the temperatures are in excess of 100 eV and pressures are far in excess of 1 Mbar. Most experiments in this regime are integrated, involving many different physics models. Computational physicists therefore must rely on the results for EOS, opacities, transport coefficients, etc., coming from fundamental physics codes like MD, atomic kinetics, kinetic theory, QMC, and TDDFT. 

Recently, a comparison of EOS models \cite{gaffney2018} and transport coefficients \cite{grabowski2020,stanek_pop_24} in regimes relevant to ICF and based on fundamental physics codes was documented from three code comparison workshops to which the high energy density physics community were invited to participate. A brief discussion of the results and open questions will be given in Secs.~\ref{sss:previous-comparisons-eos} and \ref{sss:previous-comparisons-transport}. One conclusion is that there is a strong demand for accurate simulation data for hydrogen that have predictive capability. 
In this paper, we present novel data for the equation of state and transport properties of dense hydrogen in the relevant parameter range that are based on first principles simulations. We also perform accuracy tests of various models, cf. Sec.~\ref{sss:rpimc-vs-fpimc} and \ref{subsec:results-finitet-xc} for dense hydrogen. This should help to reduce uncertainties of existing model predictions for parameters relevant to ICF.

At the same time, first principles-quality modeling of the entire ICF explosion is still out of reach.
QMC simulations, so far, only describe thermodynamic equilibrium situations, including static and dynamic properties, see Secs.~\ref{sec:PIMC} and \ref{sec:ITCF}. On the other hand, first principles time-dependent approaches, such as quantum kinetic methods [Sec.~\ref{sss:qke}] and time-dependent DFT (TD-DFT) [Sec.~\ref{sss:tddft}] only capture relatively short time periods, on the femtosecond to picosecond scale. At the same time, phenomena such as the hydrodynamic implosion of an ICF capsule or the shock propagation and various instabilities (such as the Rayleigh-Taylor instability) take place at significantly larger length and time scales that are currently inaccessible to the aforementioned \textit{ab initio} methods. Nevertheless, these first principle approaches could be very useful for benchmarks of the hydrodynamic models in well defined test cases.

A promising compromise between first principles simulations and hydrodynamics could be Quantum hydrodynamics (QHD) -- 
 an approach that captures the dynamics of the quantum many-body system in terms of hydrodynamic field variables such as density and velocity, directly extending classical hydrodynamics. A version of QHD that extends the picture of E.~Madelung and D.~Bohm~\cite{madelung_zp_27,bohm_pr_52} of one-electron quantum mechanics to many-particle systems
 was proposed by G.~Manfredi and F.~Haas \cite{manfredi_prb_21} and became very popular in quantum plasma simulations leading, however, also to poorly controlled applications, for discussions, see Refs.~\cite{Vranjes_2012, bonitz_pre_13,bonitz_pre_13_rep,bonitz_pop_19}. At the same time, this model does not 
 reproduce the correct plasmon dispersion. It was demonstrated by Zh. Moldabekov \textit{et al.} how this problem can be fixed and, moreover, how one can 
 correctly account for exchange and correlation effects (missing in the original formulation) e.g. by using Local field Corrections from QMC simulations \cite{zhandos_pop18,bonitz_pop_19} or how one can use input from DFT simulations \cite{moldabekov_scipost_22}. In fact, the importance of quantum effects for shock wave propagation was demonstrated recently \cite{graziani_cpp_21} which means that QHD could become a valuable tool also for ICF modeling. We will return to these questions in Sec.~\ref{ss:outlook}.

\section{Simulation Results}\label{sec:results}
In this section, we present state-of-the-art simulation results for the thermodynamic properties of dense hydrogen. This includes, in Sec.~\ref{sec:td}, the equation of state, pair distribution functions, degree of ionization, and ionization potential depression (IPD). In Sec.~\ref{sec:n(k)} we consider the momentum distribution function and, in Sec.~\ref{sec:dynamics}, static and dynamic density response properties. In Sec.~\ref{sec:transport} we summarize transport and optical properties, such as the electrical and thermal conductivity and the opacity. The results are a combination of existing data and novel simulations. At the beginning of each section, the origin of the data is explained including necessary details to allow for reproducibility.

\subsection{Thermodynamic properties}\label{sec:td}
\subsubsection{Previous comparisons}\label{sss:previous-comparisons-eos}
Comparisons of the equation of state results for dense hydrogen from different models have been performed at various places. A recent comparison by J.A.~Gaffney\textit{ et al.} \cite{gaffney2018} focused on ICF parameters and included extensive deuterium data
that resulted from a code comparison workshop to which the high energy density physics community was invited. The models present included CEIMC, KS-DFT, OF-DFT, AA-DFT, and various combined EOS tables.
The conclusions can be summarized as follows: (1) $5–10\%$ model-model variations exist throughout the relevant parameter space and can be much larger in regions where ionization and dissociation are occurring, (2) the deuterium EOS is particularly uncertain, with no single model able to match the available experimental data, and this drives similar uncertainties in the CH EOS, and (3) new experimental capabilities such as Hugoniot measurements around 100 Mbar and high-quality temperature measurements are essential to reducing EOS uncertainty. 

The reported large variations between models regarding the deuterium EOS is one of the motivations of the present paper. In the sections below we re-evaluate the hydrogen EOS in the difficult region of coexistence of atoms, molecules, and free charges for $T\gtrsim 15\,000$K and $r_s\gtrsim 3$. As we will show, in this parameter range, our benchmark comparisons allow for the appropriate choice of approximations in the first-principles approaches and significantly reduce the uncertainties in the EOS.

\subsubsection{Origin of simulation data}\label{sssec:data-td}
The quantum Monte Carlo data presented in this section which serve as benchmark data have been published recently: the RPIMC data are due to B.~Militzer \textit{et al.}, published in Ref.~\cite{Militzer_PRE_2021}. The fermionic PIMC data are due to A.~Filinov and M.~Bonitz, published in Ref.~\cite{filinov_pre_23}. \\

These PIMC data are compared to extensive new DFT simulation results. They have been obtained using the code {\sc VASP}~\cite{Kresse_1993,Kresse_1994,kresse1996efficient,kresse1996efficiency}. Standard issue PAW pseudopotentials were used~\cite{blochl1994projector,kresse1999ultrasoft}. The Mermin formalism of DFT was used to include temperature effects with the appropriate Fermi smearing of the occupation of the bands~\cite{Mermin_1965}. The XC functionals used include ground state LDA~\cite{Perdew_Zunger_PRB_1981}, PBE-GGA~\cite{Perdew_PRL_1996}, and the temperature-dependent KDT16 functional~\cite{Karasiev_PRL_2018}, for details see Sec.~\ref{subsec:dft}. The number of protons in the simulation box varies from $N=512$, for the highest densities, to $N=64$, for the lowest densities shown in the figures. We also ran DFT-MD simulations with different particle numbers and confirmed that finite size effects are negligible. The MD time step size was generally chosen to be $\Delta t = 0.1-0.2$~fs. 
We monitored the temperature, energy, and pressure fluctuations in the DFT-MD runs within the NVT ensemble for proper sampling and adjusted the thermostat parameter Nosè mass accordingly~\cite{Nose_1984,Hoover_1985}.
The number of bands needed to converge the DFT calculations (to within $0.2\%$) was adjusted such that the highest energy eigenvalue has an occupation not exceeding $5\times 10^{-5}$. This requires several thousand bands. Checks with a different number of k-points confirmed that most of the time the $\Gamma$-point was sufficient, and the Baldereschi mean value point was chosen for higher densities. The plane wave cutoff needed (for convergence to within $0.2\%$) was between $E_{\rm cut}=\SI{600}{\eV}$ for the standard PAW potential (used at lower densities) up to $E_{\rm cut}=\SI{1200}{\eV}$ for the hard GW PAW potentials used for the highest densities. These measures ensure that the pressure as obtained from DFT-MD is converged to better than $1\%$.

\subsubsection{Comparison of RPIMC, semiclassical MD, and FVT with fermionic PIMC simulations for the pressure}\label{sss:rpimc-vs-fpimc}
\begin{figure}\centering
\includegraphics[width=0.51\textwidth]{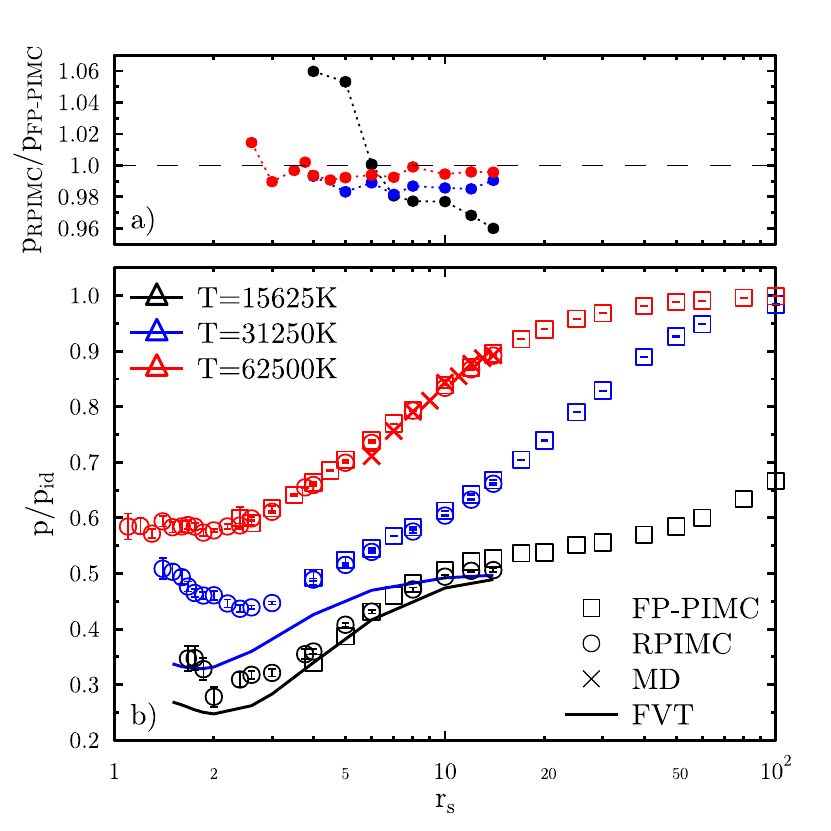}
\caption{\label{fig:rpimc-pimc-FVT2-pressure}
Panel b): Three isotherms of the hydrogen pressure (in units of the ideal Fermi pressure) -- comparison of FP-PIMC simulations~(fermionic propagator PIMC of Ref.~\cite{filinov_pre_23}, RPIMC simulations~\cite{Militzer_PRE_2021}, classical MD using the improved Kelbg potential [Sec.~\ref{ssubsec:pair-pot-md}], and from FVT~\cite{juranek_jcp_02}. Panel a): Ratio of the pressures from RPIMC and FP-PIMC for three different temperatures as a function of $r_s$, where results from both QMC simulations are available. }
\end{figure} 

We start the analysis of the thermodynamic properties of dense hydrogen by a comparison of the equation of state between available QMC results in Fig.~\ref{fig:rpimc-pimc-FVT2-pressure}. Using the recent fermionic PIMC results of Ref.~\cite{filinov_pre_23} as benchmarks allows us to judge the accuracy of the RPIMC simulations and the validity range of the used free-particle nodes. 
We also include new results from semiclassical MD simulations [Sec.~\ref{ssubsec:pair-pot-md}] and from a chemical model (Fluid Variational Theory, FVT) of H.~Juranek \textit{et al.}~\cite{juranek_jcp_02}, for details, see Sec.~\ref{subsec:chem-model}. It is quite obvious that, over a very broad range of densities and temperatures, FP-PIMC and RPIMC, which are completely independent simulations, agree to a remarkable degree. That validates the use of free particle nodes in RPIMC. For the two higher isotherms, the deviation is below $2\%$. At $T=\SI{15625}{\kelvin}$ we find a maximum deviation of $6\%$ in the region with a substantial amount of molecules. These differences are observed at the remarkably low value of $\Theta \approx 0.5$ which corresponds to the smallest $r_s$-value accessible in FP-PIMC due to the FSP where simulations are difficult to converge. We expect that these differences are due to the deteriorating quality of the nodal surfaces input in RPIMC. FVT, on the other hand,  can serve as a quick estimation of the EOS,  up to about $\SI{16000}{\kelvin}$ with up to $20\%$ deviation in the pressure. The molecular dissociation from $r_s=3$ to $r_s=14$ is described very well at $T=\SI{15625}{\kelvin}$ using FVT, which is also confirmed by Fig.~\ref{fig:Nfrac} below.

Finally, the comparison with semiclassical MD simulations that use the improved Kelbg potential [Sec.~\ref{ssubsec:pair-pot-md}] is presented for one isotherm of $T=62\,500$K (red crosses). The results are within $1\dots 3\%$ of the FPIMC data, for $r_s\gtrsim 7$, which is remarkable since the plasma contains a significant fraction of atoms. A similar agreement is observed for higher temperatures where the accessible density range increases with $T$. For example, for $T=95\,000$K ($T=125\,000$K) the density range grows to $r_s\gtrsim 5$ ($r_s\gtrsim 3$). For temperatures below the shown isotherm, the results are not reliable anymore because the current version of SC-MD does not describe molecules sufficiently accurately.

\subsubsection{Comparison of KS-DFT with fermionic PIMC }\label{subsec:results-finitet-xc}

\begin{figure*}[ht]
\centering
\includegraphics[width=\textwidth]{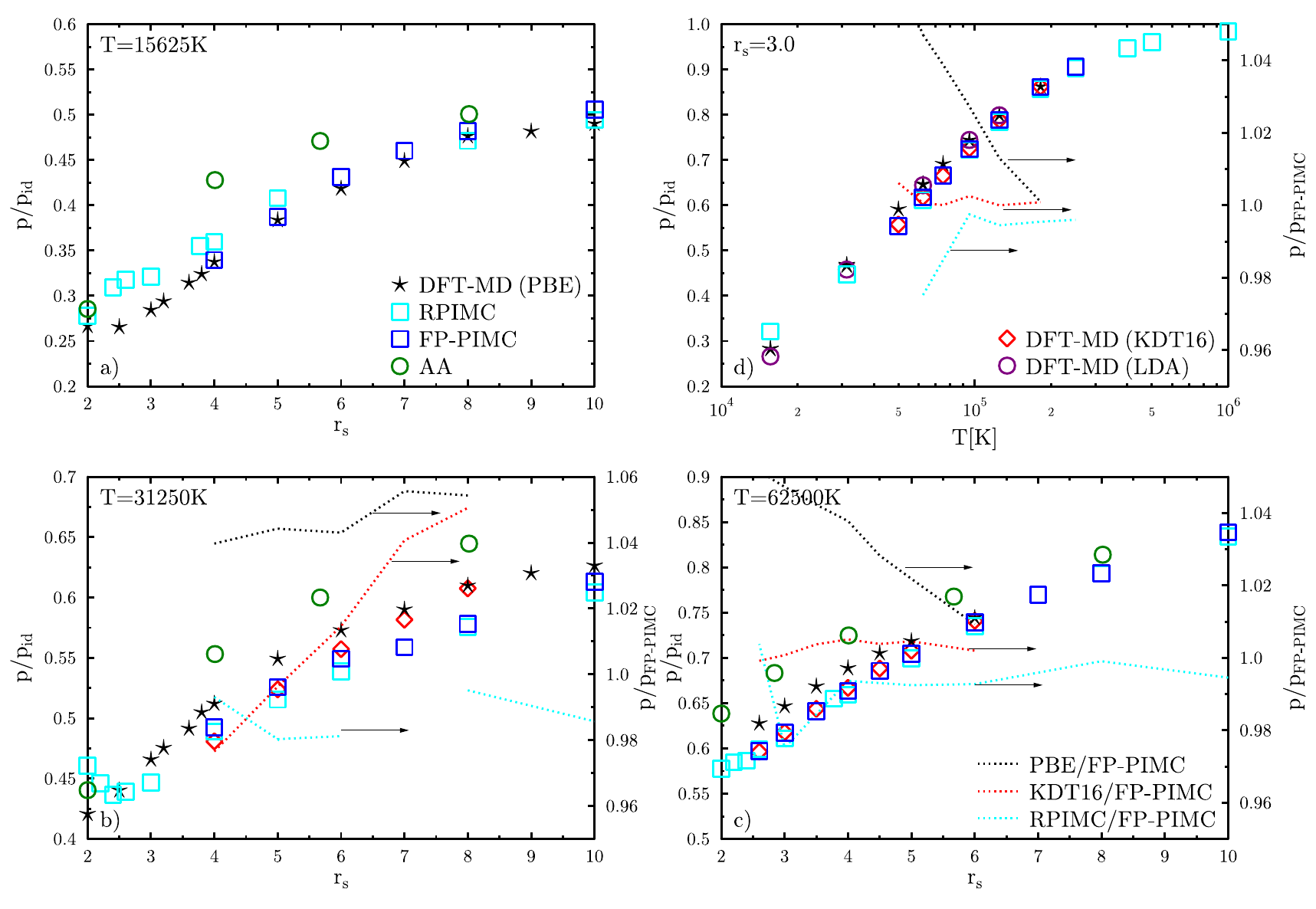}
\caption{\label{fig:press_dftmd2} 
Benchmarks of KS-DFT-MD results for the EOS of hydrogen (normalized to the pressure of an ideal Fermi gas). Panels a), b), and c) show three isotherms, and panel d) an isochore.  Panels c) and d) also show the ratio of various results to FP-PIMC (right hand side y-axis). We compare DFT-MD results with the zero-temperature LDA functional (violet circles), the PBE XC-functional (black stars), and using the temperature-dependent KDT16 functional (red diamonds)~\cite{Karasiev_PRL_2018}, respectively. RPIMC simulations~\cite{Militzer_PRE_2021} are shown by the cyan squares and FP-PIMC data~\cite{filinov_pre_23} by the blue squares. Also shown are average atom data (green circles), for a discussion, see text. The statistical errors of the data are within the size of the symbols.}
\end{figure*}

An in-depth comparison of state of the art DFT-MD simulations with quantum Monte Carlo results for the EOS of warm dense hydrogen is performed in Fig.~\ref{fig:press_dftmd2}. We show the same isotherms as in Fig.~\ref{fig:rpimc-pimc-FVT2-pressure} but, in addition, also an isochore for $r_s=3$. The comparison between the two different flavors of PIMC (at conditions where both RPIMC and FP-PIMC provide data) has already been established above. Here, we see clearly that RPIMC can provide data for high densities, $r_s\lesssim 3$, or low temperatures where the vanishing average sign makes this prohibitively expensive for FP-PIMC, even though the accuracy of RPIMC cannot be quantified.

Let us now compare 
Kohn-Sham DFT-MD simulations to the FPIMC results 
for densities larger than $r_s=4$. 
At $T=\SI{15625}{\kelvin}$, taken as an example for situations when a ground state XC-functional should be a good approximation, due to the still high electron degeneracy, we find excellent agreement for $r_s\ge4$ with FP-PIMC. For lower $r_s$ (higher densities), the RPIMC results consistently give higher pressures than DFT-MD indicating differences in the description of this system within which a substantial amount of molecules is present. There are two uncontrolled approximations here. Firstly, the XC-functional used in DFT seems to be sufficiently well approximated using PBE for $r_s\ge 4$. On the other hand, for $r_s<4$, no conclusion on the appropriate functional can be drawn yet, and it remains to investigate how other functionals, such as vdW, HSE, or SCAN behave. Secondly, the free-particle nodes used in RPIMC may lose validity. This is supported by the systematically too high pressures of RPIMC, already at $r_s=5$ and $r_s=4$.
%}

The situation is different for elevated temperatures of $T=\SI{31250}{\kelvin}$ and $T=\SI{62500}{\kelvin}$, see panels b) \& c) of Fig.~\ref{fig:press_dftmd2}. Again, for $r_s\ge 4$, PIMC shows a consistent picture whereas DFT-MD using a ground state PBE functional (but the temperature-dependent Mermin formulation of DFT) overestimates the pressure by up to $7\%$. On the other hand, taking into account temperature-dependent XC effects with the KDT16 GGA functional lowers the pressure such that the deviation from FP-PIMC is below $2\%$, for all conditions. It is interesting to observe that the main improvements, when using temperature-dependent XC-functionals, are found at intermediate temperatures of a few eV (up to \SI{20}{\eV}) such that the degeneracy is lifted but the ideal kinetic pressure is not dominant yet.
%}

The general conclusion is that, for the range of conditions presented here, that includes the molecular, atomic, and partially ionized fluid/plasma state of hydrogen and covers densities slightly higher than that of the solid to very low densities, and temperatures $T\gtrsim \SI{15000}{\kelvin}$,
the (temperature-dependent) GGA functional is accurate, and no meta, hybrid, exact exchange, or van der Waals functionals appear to be required. At the same time, larger differences are observed   for $T=31\,250$K, and the best choice of XC-functional remains open.

For completeness, in Fig.~\ref{fig:press_dftmd2} we also include data from a DFT-AA model using LDA exchange. Here, the ion pressure contribution is taken to be the classical ideal pressure at the given ion density and temperature, which tends to overestimate ion pressures at lower temperatures and densities where molecular bonding can occur. Electron pressures from DFT-AA models can be calculated using Virial theorem \cite{piron} or free-energy derivatives \cite{sterne, Wilson2006}, however, here the electron pressure has been calculated using the simple prescription of W.R.~Johnson \cite{johnson_pressure}.

%}

\subsubsection{Pair distribution functions and static structure factor}\label{sssec:pdf}
\begin{figure} [h] 
    \includegraphics[width=0.49\textwidth]{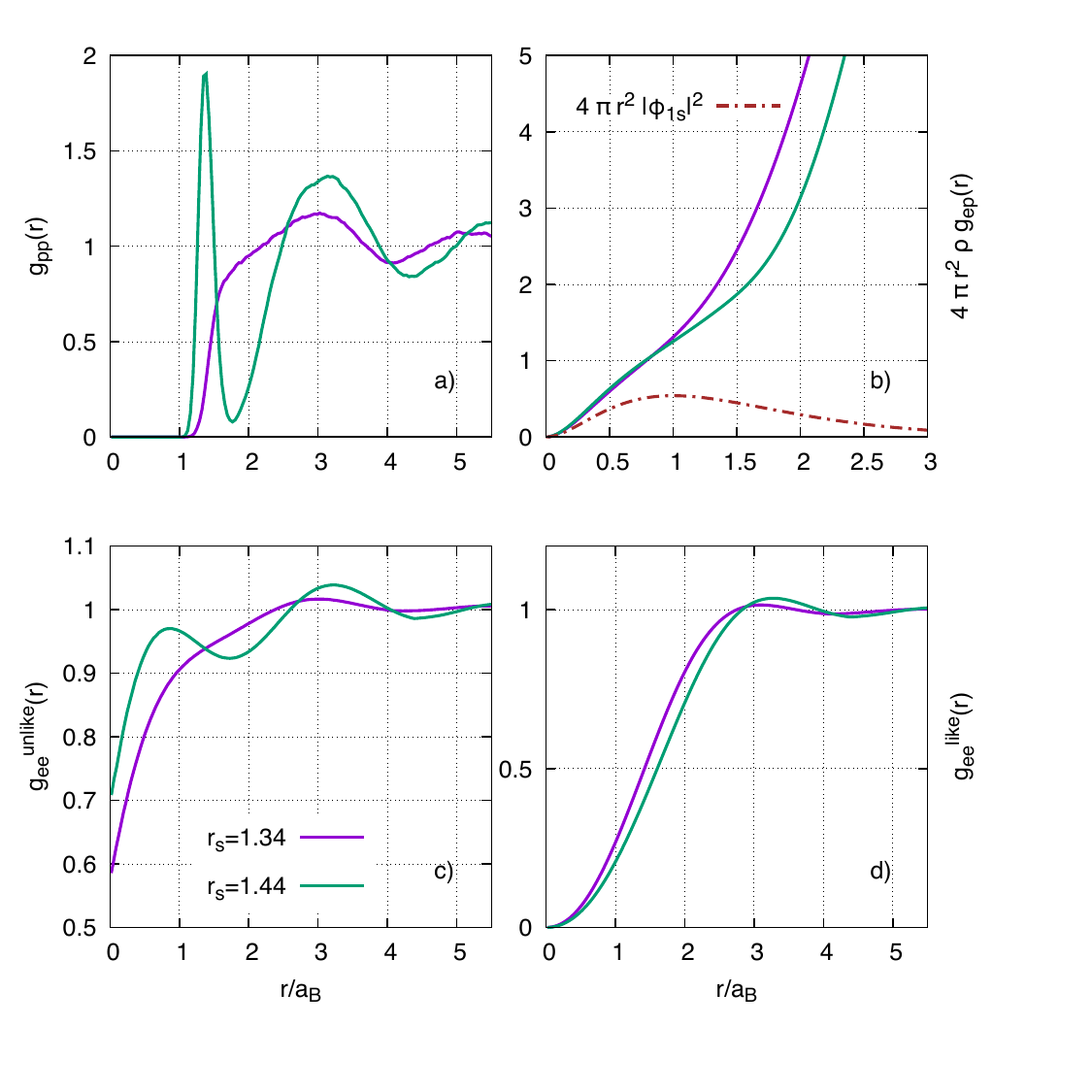}
    \caption{Radial distribution functions for hydrogen at $T=1\,200$K  for two densities across the LLPT. Panel a) proton-proton $g_{pp}(r)$, panel b): proton-electron $r^2 g_{ep}(r)$, panel c): spin-unlike electron $g^{\uparrow\downarrow}_{ee}(r)$, panel d): spin-like electrons $g^{\uparrow\uparrow}_{ee}(r)=g^{\downarrow\downarrow}_{ee}(r)$. Distances are in units of $a_B$. Red dash-dotted line: radial probability density of the hydrogen ground state. $r_s=1.44$ corresponds to the molecular fluid phase whereas $r_s=1.34$ corresponds to a metallic phase with mostly delocalized electrons, see also  Fig.~\ref{fig:nk_hydrogen_llpt}.}
    \label{fig:gr_T1200K_LLPT}
\end{figure}

The pair distribution functions (PDF or radial distributions) are a sensitive indicator of quantum and correlation effects in dense hydrogen. They also directly reflect the structural properties. Here, we illustrate this for several phases of hydrogen that were discussed in Sec.~\ref{sssec:phases-overview}, starting from low temperatures.
Figure~\ref{fig:gr_T1200K_LLPT} shows results for fluid hydrogen at $T=\SI{1200}{\kelvin}$, for two densities corresponding to $r_s=1.34$ and $r_s=1.44$. Those results have been obtained by CEIMC for classical protons and illustrate how correlations change across the LLPT. At the lower density ($r_s=1.44$) the system is molecular, as seen from the pronounced peak around $1.4 a_B$ in $g_{pp}$, followed by a nearly vanishing signal around $1.8 a_B$. Instead, at the higher density ($r_s=1.34$), the molecular peak disappears and the correlation function has a much smoother behavior signaling the absence of stable structures. In panel b) of the figure we report the proton-electron correlation $=4\pi \rho r^2 g_{ep}(r)$, where $\rho$ is the electron density, and we compare with the electronic density corresponding to the isolated atom ground state (red line). Instead of a peak at $r=1a_B$,  which would be due to hydrogen atoms, the simulations exhibit only a weak shoulder indicating that the vast majority of the electrons are delocalized, which is also confirmed by the results of Fig.~\ref{fig:nk_hydrogen_llpt}. 

In panels c) and d) we report the electron-electron pair distribution functions for spin-unlike and spin-like electrons, respectively. While little change with density is observed in the spin-like function, besides the trivial change in average distance, in the spin-unlike functions we can see the presence of molecules as a characteristic structure with a first maximum around $r\simeq 0.7 a_B$ induced by the charge accumulation between the protons to form the molecular bond, and a second maximum at about the distance between molecules. A more complete characterization of correlation in hydrogen around the LLPT, including a discussion of nuclear quantum effects obtained from CEIMC with nuclear path integrals is reported in Refs.~\cite{Pierleoni2016, Pierleoni2017, Pierleoni2018}.

\begin{figure}[ht]
\includegraphics[width=0.5\textwidth]{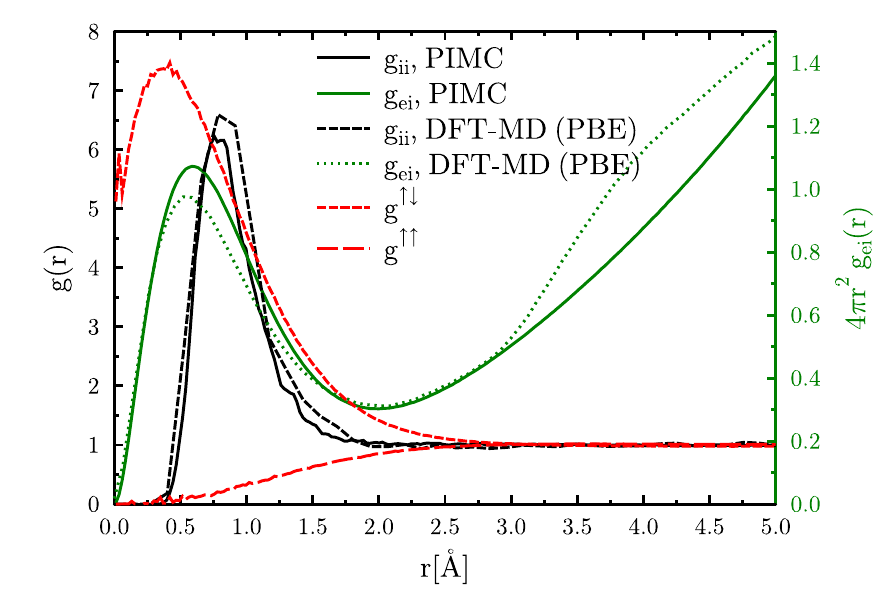}
\caption{Pair distribution functions of partially ionized hydrogen for $T=\SI{15640}{\kelvin}$ and $r_s=7$, from FP-PIMC simulations of Ref.~\cite{filinov_pre_23} using $P=96$ fourth order high-temperature factors and $N=34$. Green line: electron-ion PDF multiplied by $4\pi r^2$. Dotted lines: KS-DFT results with PBE-XC functional. 
}
\label{fig:PDFT15P}
\end{figure}

Next, we show results for the partially ionized plasma phase. Data from  fermionic PIMC simulations of Ref.~\cite{filinov_pre_23} are shown in Fig.~\ref{fig:PDFT15P} for $T=\SI{15640}{\kelvin}$ and $r_s=7$. The presence of hydrogen molecules can be identified from the peak in $g_{ii}(r)$ for ion-ion distances, $0.7\AA \leq r \leq 1.3 \AA$, as well as from  $g^{\uparrow\downarrow}(r)$, which indicates accumulation of electron pairs with different spins between two protons. The peak position of  $g^{ii}(r)$ agrees well with the ground state atom separation in a hydrogen molecule. 
The presence of atoms is reflected by the peak of the electron-ion function multiplied by $4\pi r^2$ which is close to $1a_B$. It is interesting to again compare the FPIMC data to KS-DFT results with the PBE functional (dotted lines). Both the e-i and i-i PDF are in reasonable agreement, indicating a similar number of molecules. 

\begin{figure}[ht]
\begin{center}
\hspace{-0.4cm}
\includegraphics[width=0.48\textwidth]{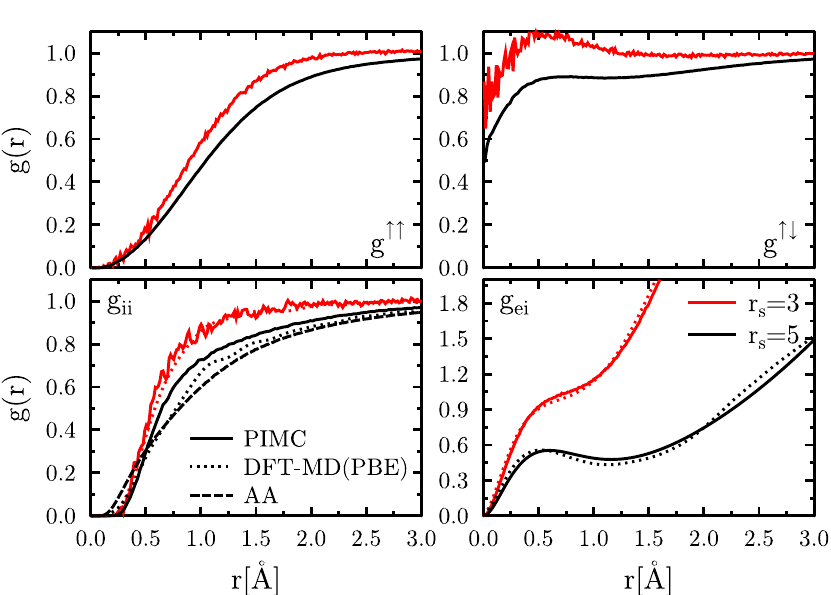}
\vspace{-0.3cm} 
\caption{Pair distribution functions of partially ionized hydrogen at $T=\SI{62500}{\kelvin}$ and $r_s=3$ and $r_s=5$. Top panels: e-e PDF with the indicated spin combinations. Bottom panels: i-i PDF (left) and $4\pi n_e r^2 g_{ei}(r)$ (right). Full lines: FP-PIMC results; dotted lines: KS-DFT with the PBE functional. Dashes: average atom model (AA).
}
\label{fig:PDFT62P}
\end{center}
\end{figure}

The pair distributions for a four times higher temperature and two densities, corresponding to $r_s=5$ and $r_s=3$, are shown in Fig.~\ref{fig:PDFT62P}. At this temperature, no molecules exist, as is confirmed by the monotonic ion-ion PDF. Also, there is only a small fraction, $x_A$, of atoms which is indicated by the broad peak (shoulder) of $r^2g_{ei}$, around one Bohr radius for $r_s=5$ ($r_s=3$), where $x_A$ increases when the density is lowered. The comparison between FPIMC and KS-DFT shows reasonable agreement for $g_{ii}$ where the latter predicts a slightly too strong proton-proton repulsion (too broad minimum at small distances). The electron-proton PDFs (bottom right figure) show excellent agreement with FPIMC.
The figure also allows for a comparison with average atom models (AA, dashed lines in the lower left figure). The results underestimate the i-i-repulsion leading to a significantly narrower minimum. 

\begin{figure} [h] 
    \includegraphics[width=0.43\textwidth]{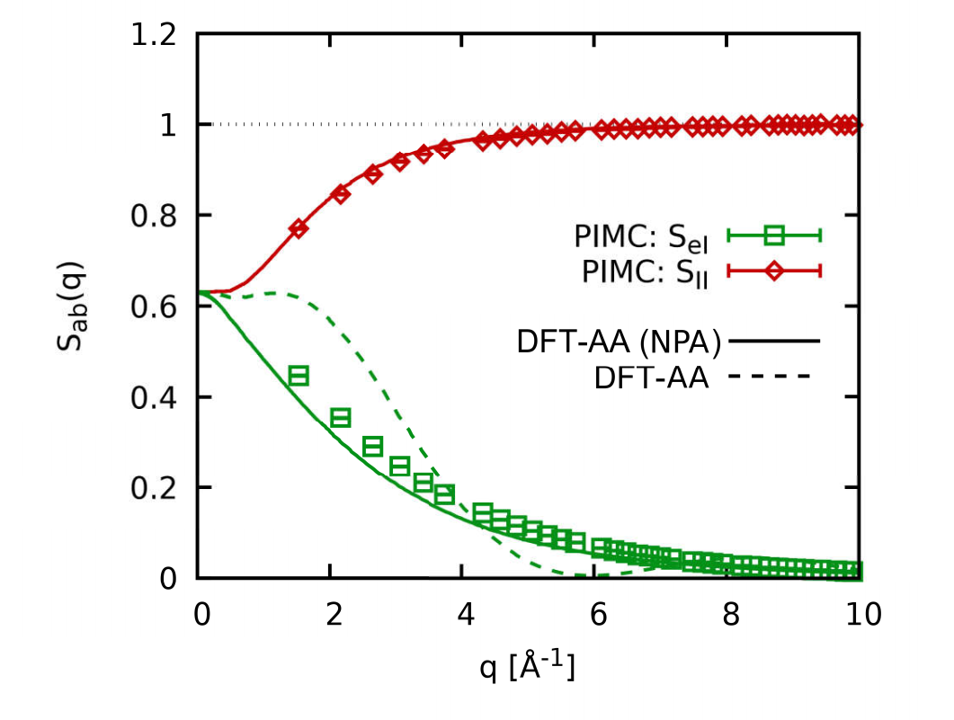}
    \caption{Static ion-ion and electron-ion structure factors for hydrogen at $T=\SI{145000}{\kelvin}$ and $r_s=2$. PIMC results from Ref.~\cite{dornheim_2024} are compared with two versions of the DFT-AA model, see text for details.}
    \label{fig:static_structure}
\end{figure}

We continue the analysis of the AA model, but now in Fourier space.
Figure \ref{fig:static_structure} shows a comparison of the electronic and ionic static structure factor,  between PIMC and DFT-AA, at a relatively high temperature (\SI{145000}{\kelvin}, or about \SI{12.5}{\eV}) and $r_s = 2$, where the plasma is fully ionized. There is excellent agreement in the static ion-ion structure factor $S_{ii}(q)$ predicted by the two approaches. For $S_{ei}(q)$, DFT-AA has two possible representations of the static electron-ion structure: one based on the self-consistent electron density within the Wigner-Seitz/ion-sphere cell about a single ion (labeled DFT-AA), and another one based on the neutral pseudo-atom electronic density (DFT-AA-NPA), which includes the effect of ion correlations and allows the electron density belonging to a single ion to extend beyond the ion-sphere radius \cite{Dharma-wardana_PRE_2012, starrett2014hedp}. The PIMC results are in much better agreement with the NPA results.

\subsubsection{Degree of ionization and dissociation}\label{sssec:alpha}

For chemical models, the degree of ionization, $\alpha(n,T)$, and the degree of dissociation, $\beta(n,T)$ [or, equivalently, the fraction of atoms $x_A$], follow from the solution of the coupled Saha equations, Eqs.~\eqref{eq:saha-h-degenerate} and \eqref{eq:saha-h2-nondegenerate}.
The accuracy of the results sensitively depends on the quality of the interaction contributions to the chemical potentials, as discussed in Sec.~\ref{subsec:chem-model} and is not known \textit{a priori}. A typical example will be discussed in Fig.~\ref{fig:FVT2-alpha} below. 

In contrast, in physical approaches such as quantum Monte Carlo or DFT, there is no strict subdivision of free and bound electrons. Only for an isolated atom such a subdivision could be introduced based on the sign of the energy eigenvalue -- negative (positive) for bound (free) electrons. However, for finite temperature this boundary is being ``washed out'' and, due to the additional thermal energy, highly excited atomic bound states will statistically contribute to the scattering spectrum. Moreover, in a plasma at high density, the interaction between atoms and the overlap of electronic orbitals from neighboring atoms may significantly modify the energy spectrum (shift of energy eigenvalues, lowering of ionization energy, Mott effect, etc.) which will be discussed in Sec.~\ref{sssec:ion-pot}.
In this section, we present first principle results for the degree of ionization and dissociation of hydrogen in different phases, that were discussed before. We start with the low-temperature case and the LLPT and then consider partially ionized hydrogen in the gas phase.

\begin{figure}
 \includegraphics[width=0.49\textwidth]{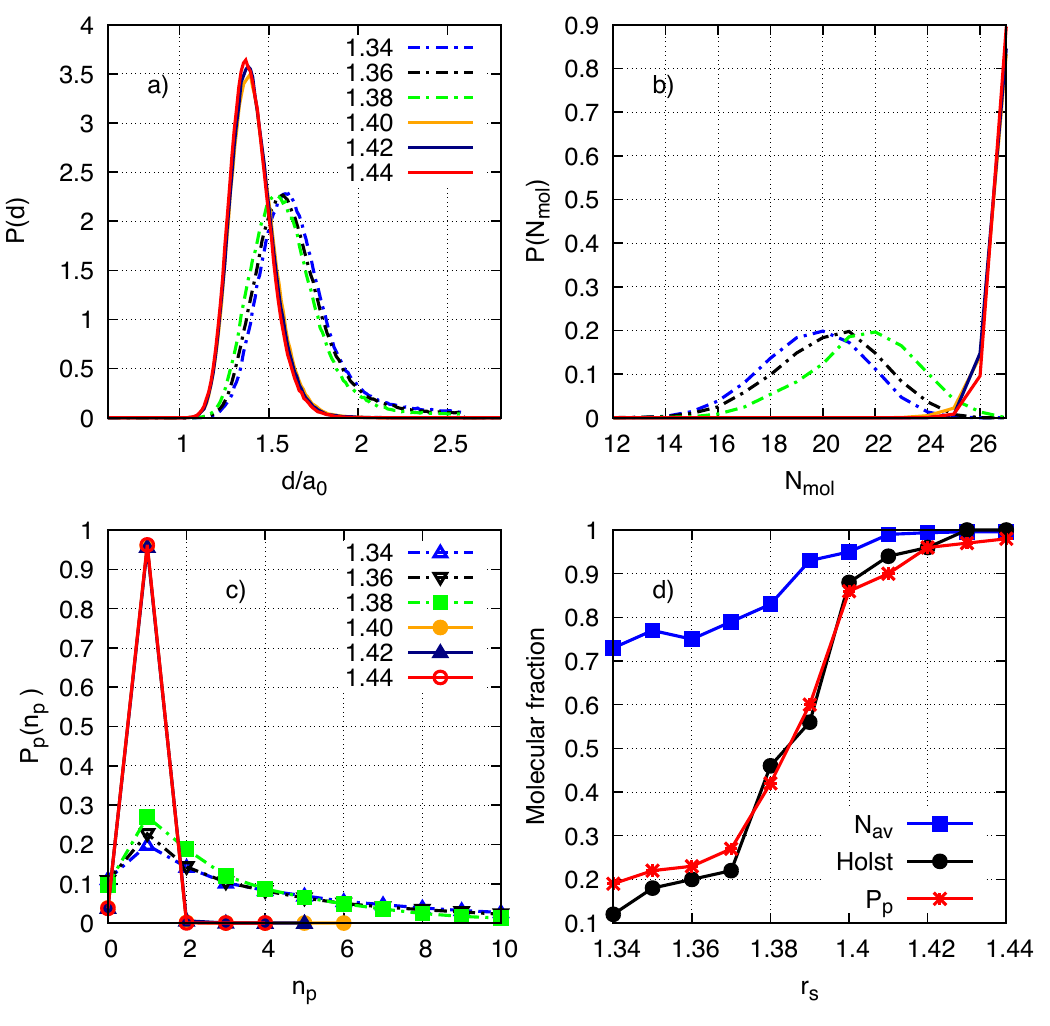}
    \caption{ Cluster analysis for classical protons at $T= 1\,200~$K.  (a): distribution of the bond distance in the pairs found by the cluster algorithm with a cut-off distance of $2.6a_B$. (b): distribution of the number of pairs with a cut-off distance of  $1.8a_B$.  (c) distribution of the number of different partners, for a given proton, in the pairs found with the large cut-off value. (d) molecular fraction from different estimators: $N_{\rm av}$ from the average number of molecules with the short cut-off value, %\textit{Holst} 
    from the coordination number at $r = 1.4a_B$, and $P_p$ from the values of the distribution in panel (c) for $n_p = 1$.}
    \label{fig:Mol_frac_1200K}
\end{figure}

The characterization of hydrogen across the LLPT in terms of the molecular fraction and the possibility of stable ionic composites like $H_2^+$ and $H_3^+$, based on CEIMC results, have been discussed in detail in Ref.~\cite{Pierleoni2017}. There, different possible definitions of the molecular fraction were discussed and compared. As an illustration,  in Fig.~\ref{fig:Mol_frac_1200K} we show the analysis for classical hydrogen along the $T=\SI{1200}{\kelvin}$ isotherm across the LLPT. As explained in Ref.~\cite{Pierleoni2017}, we assigned molecules for each configuration along the Monte Carlo trajectory by a cluster analysis and by a pair distance criterion. The ``bond length'' distribution at six different densities is shown in panel a) of Fig.~\ref{fig:Mol_frac_1200K}. A discontinuous behavior occurs at the LLPT: in the molecular phase (continuous lines) the distribution is narrow, strongly peaked around the molecular bond length ($\sim 1.4 a_B$), and independent of density. In the dissociated phase (dot-dashed lines) the distribution is wider, peaked at a distance larger than the molecular bond ($\sim 1.6a_B$), more asymmetric, with a detectable tail at large distances, and with a larger sensitivity to density. 

We also look at the distribution of the number of different neighbors a single proton experiences during the simulation, as shown in panel c) of Fig.~\ref{fig:Mol_frac_1200K}. This quantity is useful when nuclear dynamics is not available, as in CEIMC and PIMD, to emulate the persistence time criterion employed in molecular dynamics simulations of classical protons \cite{Vorberger2007a}. Again we observe a striking change at the LLPT: for densities up to $r_s=1.4$ very stable molecules are observed since the protons experience basically the same neighbor along the entire trajectory. For higher densities ($r_s < 1.4$), the distribution exhibits a long tail, indicating a strong attitude to being paired with many different neighbors during the sampling of the configurational space. One possible estimator of the molecular fraction is given by the maximum of this distribution and is reported in panel d) of the figure and denoted as $P_p$. Another possible estimator is obtained by computing the distribution of the number of distinct pairs within a cutoff distance of $1.8a_B$ corresponding to the first minimum of $g_{pp}(r)$, as shown in panel b) of the figure. Again an abrupt change of the distribution is observed at the LLPT: in the dissociated phase (dot-dashed lines) the distribution is rather broad and peaked around the value of 20, while in the molecular phase (continuous lines) we observe a strong peak at 27 which is the maximum number of pairs in our simulated system. The molecular fraction can then be defined as the average over those distributions, and it is denoted by $N_{\rm av}$ in the plot in panel c). 

Finally, a third definition of the molecular fraction has been proposed in Ref.~\cite{Holst_PRB_2008} as twice the coordination number obtained from $g_{pp}(r)$ at $r=1.4a_B$, i.e. at the distance of the molecular peak. This definition is also shown in panel d) and denoted as \textit{Holst}. For the present case, we see that the \textit{Holst} estimator and $P_p$ are in rather good agreement, whereas $N_{\rm av}$ is strongly overestimating the molecular fraction in the dissociated phase. This is the general trend also for other isotherms and for systems of quantum nuclei below the critical temperature of the LLPT \cite{Pierleoni2017}. Above the critical temperature, when molecular dissociation with pressure becomes a continuous process, the agreement between the \textit{Holst} estimator and $P_p$ is lost, and it is not clear which estimator is more reliable.

We now turn to higher temperatures which are in between the liquid phase studied in Fig.~\ref{fig:Mol_frac_1200K} and the partially ionized plasma phase above $T=\SI{15000}{\kelvin}$ that is considered below. This range is difficult to access by QMC simulations so we resort to an advanced chemical model (FVT), cf. Sec.~\ref{subsec:chem-model}.
\begin{figure}\centering
\includegraphics[width=0.35\textwidth]{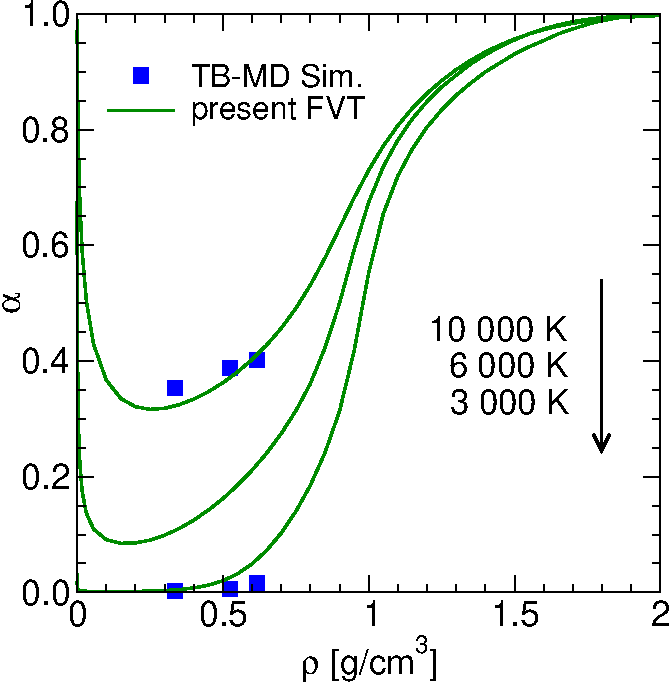}
\caption{\label{fig:FVT2-alpha}
Results for the dissociation degree $\alpha$ (fraction of atoms, $x_A$) of fluid hydrogen as a function of density for three temperatures. A comparison of the FVT version of Ref.~\cite{juranek_jcp_00} with tight-binding molecular dynamics simulations~\cite{lenosky_pre_99} is performed. The pressure-induced breakup of molecules occurs around $\rho=\SI{1}{\gram\per\cubic\centi\meter}$, corresponding to $r_s\approx 1.392$, which is consistent with the low-temperature behavior shown in Fig.~\ref{fig:Mol_frac_1200K}.d. Taken from H.~Juranek \emph{et al.}~\cite{juranek_jcp_02} with the permission of the authors.
}
%}
\end{figure} 
The dissociation degree, as predicted for hydrogen in the temperature range between \SI{3000}{\kelvin} and \SI{10000}{\kelvin}, is shown in Fig.~\ref{fig:FVT2-alpha} \cite{juranek_jcp_02}. Very good agreement with tight-binding molecular dynamics (TB-MD) simulations~\cite{lenosky_pre_99} is evident.
The comparison with Fig.~\ref{fig:Mol_frac_1200K}.d shows that, for the lowest temperature, the break up of molecules occurs nearly at the same density, $r_s\approx 1.39$, corresponding to the density $\rho\approx \SI{1}{\gram\per\cubic\centi\meter}$. At the same time, the density interval of the pressure-induced molecule dissociation is much larger than in the fluid phase where it occurred in the interval $r_s\in[1.37,1.40]$ corresponding to $\rho \in[0.982,1.049]$ gcm$^{-3}$. The comparison with Fig.~\ref{fig:FVT2-alpha} indicates that FVT predicts a much softer pressure dissociation. The comparison with TB-MD indicates that this could be due to limitations of the chemical model, so additional first principles simulations are needed to resolve this question.

\begin{figure}[h]
\hspace{-0.74cm}
\includegraphics[width=0.52\textwidth]{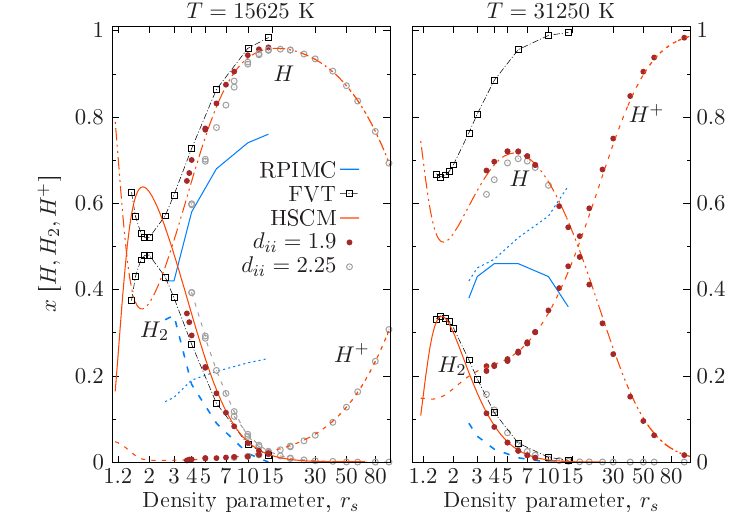}
\vspace{-0.4cm} 
\caption{Fraction of free electrons, atoms, and molecules, for two isotherms,   $T=\SI{15640}{\kelvin}$ (left) and $T=\SI{31250}{\kelvin}$ (right). FP-PIMC results from Ref.~\cite{filinov_pre_23} are plotted for $d^{cr}_H = 1.9 a_B$ (brown solid dots) [$d^{cr}_H = 2.25 a_B$, open gray circles], for details, see text. Blue lines: RPIMC data for $d^{cr}_H = 1.9 a_B$. Red lines: chemical model. 
Reproduced from Ref.~\cite{filinov_pre_23} with permission of the authors.}
\label{fig:Nfrac}
\end{figure}

We now turn to lower densities, $\rho \lesssim \SI{0.5}{\gram\per\cubic\centi\meter}$ where the fraction of atoms again increases. This is a statistical effect related to a decrease in the probability of two atoms to approach each other sufficiently closely in order to form a bond. This behavior is confirmed by fermionic PIMC simulations as well as by RPIMC simulations for $T=\SI{15625}{\kelvin}$ \cite{filinov_pre_23}, cf. left part of Fig.~\ref{fig:Nfrac}. There the two QMC simulations are compared to the results of  FVT and another chemical model. Interestingly,  we observe very good agreement between FP-PIMC and FVT, whereas the RPIMC results for the fraction of atoms differ significantly.
The behavior of FVT changes dramatically when the simulations move further into the gas phase, at $T=\SI{31250}{\kelvin}$. Here FVT yields good results for the fraction of molecules, however, it strongly overestimates the fraction of atoms. This is due to the ionization of atoms which is important for this temperature already at $r_s=3$, but missing in the FVT model indicating that it is restricted to lower temperatures. Note that a generalization of FVT to the case of partial ionization is possible; see, e.g.,~\cite{juranek_CPP_05}.
\begin{figure}
    \includegraphics[width=0.440\textwidth]{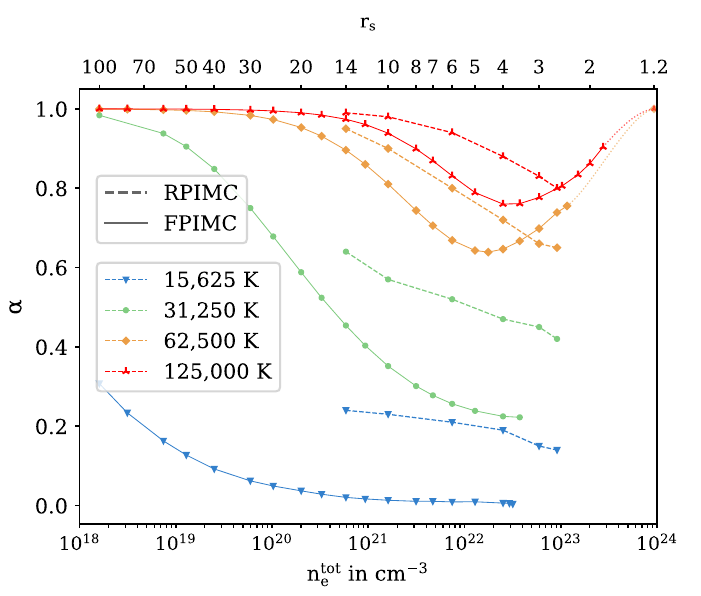}
    \caption{Four isotherms of the degree of ionization $\alpha$ of hydrogen vs. density or $r_s$. Comparison of RPIMC results of Ref.~\cite{Militzer_PRE_2001} and fermionic PIMC results of Ref.~\cite{filinov_pre_23}. Thin lines indicate the expected connection of the FPIMC data to the Mott density, $r_s=1.2$, where the FSP prohibits FPIMC simulations, for the two higher temperatures (for the lower temperatures, this is not possible). 
}
    \label{fig:ion-degree-mil-fil}
\end{figure}

In contrast to FVT, first-principles FPIMC simulations are possible for temperatures above $T=\SI{15000}{\kelvin}$ for a very broad range of densities, cf.\ Fig.~\ref{fig:Nfrac}. The fermion sign problem sets a lower limit for FPIMC of about $r_s\approx 4$ \cite{filinov_pre_23}. The criterion for the identification of molecules in FPIMC and RPIMC is based on a critical proton-proton distance, $d_{ii}$ that is indicated in the figure. By using two different values, the impact of the threshold can be quantified. The present criterion is similar to the one used in part b) of Fig.~\ref{fig:Mol_frac_1200K} for the analysis of hydrogen in the vicinity of the LLPT. On the other hand, the procedure to identify atoms in a fermionic PIMC simulation is based on a cluster analysis that counts the number of electrons inside a sphere around each proton, for details we refer to Ref.~\cite{filinov_pre_23}. The FPIMC simulations demonstrate that, for low densities corresponding to $r_s\gtrsim 15$ [$r_s\gtrsim 7$], at $T=\SI{15625}{\kelvin}$ [at $T=\SI{31250}{\kelvin}$] the atomic fraction decreases rapidly in favor of unbound electrons and protons. As in the case of the molecule break up discussed above, this is a statistics effect.

Results for the degree of ionization from FPIMC and RPIMC for four isotherms are shown separately in Fig.~\ref{fig:ion-degree-mil-fil}. The behavior that was discussed for the two temperatures shown in Fig.~\ref{fig:Nfrac} is here confirmed also for two higher temperatures. With increasing temperature, the fraction of atoms is, obviously, significantly lower, and their break up set in already at higher densities. Of particular interest is the minimum of the atom fraction which is observed around $r_s \sim 4.8$ [$r_s \sim 3.9$], for $T=\SI{62500}{\kelvin}$ [$T=\SI{125000}{\kelvin}$]. To the right of the minimum, i.e. for higher densities, the atom fraction, $x_A$, decreases due to pressure ionization, reaching zero at around $r_s=1.2$.

In conclusion, we stress again that the degree of ionization and the fractions of atoms and molecules are not physical observables. They cannot be rigorously computed, even by a first-principle simulation. Any result depends on the used criterion, as discussed above, even though the sensitivity to the chosen criterion can be verified. For completeness, we mention that other approaches use other criteria, in particular dynamical quantities to estimate the degree of ionization. For example, it was suggested to use the plasma frequency as an observable \cite{norman_cpp_15}.  In Ref.~\cite{Bethkenhagen_2020} it was demonstrated for carbon, how the degree of ionization can be derived from the dynamic conductivity in a DFT simulation using the Kubo-Greenwood formula [Sec.~\ref{sss:dft-kubo-greenwood}] and the Thomas-Reiche-Kuhn sum rule.

\subsection{Pressure and temperature-dependent ionization potentials}\label{sssec:ion-pot}
\subsubsection{Overview}\label{ss:ipd-overview}
The problem of a Coulomb bound state in a plasma medium has been studied for a long time. Physically, one expects that the ionization energy will be reduced by the medium, giving rise to an effective ionization potential, $I^{\rm eff}_{\rm ion}(n,T)$, that depends on density and temperature. 
Early works in that field are due to R.~Rompe and M.~Steenbeck~\cite{rompe-steenbeck_39} as well as G.~Ecker and W.~Kröll who considered non-degenerate electrons~\cite{ecker-kroell_63} and computed the ``ionization potential depression'' (IPD), i.e. the medium-induced lowering of the ionization potential. An improved model of IPD is due to J.C.~Stewart and K.D.~Pyatt,~Jr.~\cite{stewart-pyatt_66}. We also mention the computational analysis of F.~Rogers \textit{et al.}~\cite{rogers_pra_70} who solved the Schrödinger equation with a screened Coulomb potential and computed the reduction of the binding energy in dependence on the screening length. However, this does not fully describe the situation in a dense hydrogen plasma. A more comprehensive description is achieved within the  Bethe-Salpeter equation, which allows one to take into account dynamical screening, quantum, and Pauli blocking effects, e.g. \cite{zimmermann_pssb_78,red-book,haug_pqe_84}. Recent work by Massacrier \textit{et al.}~\cite{MassacrierBoehmeVorbergerSoubiranMilitzer2021} developed a new approach to describe the IPD that treats bound and free electronic states under a consistent set of assumptions and thereby removes the discrepancies in dense plasmas between first-principles calculations and average atom models that were reported in Ref.~\cite{Driver2018}. We also mention KS-DFT simulations of S.~Hu~\cite{hu_prl_17} for the shift of the K-edge in a strongly coupled fully degenerate carbon plasma that indicates significant differences from existing IPD models.

Before proceeding with a discussion of IPD, let us comment on the analogies with insulator-metal transition in the fluid phase. While in the gas phase, the pressure-induced transition from a non-conducting atomic (or molecular) phase to a conducting (fully ionized) plasma proceeds via the vanishing of the atomic binding energy (IPD), in the condensed phase, compression leads to band gap closure, resulting in a metallic state, for results, see Sec.~\ref{ss:h-gap}.
\\

\subsubsection{First-principles QMC results for IPD}\label{sss:ipd-qmc}
Based on the recently obtained first-principles QMC results for hydrogen \cite{filinov_pre_23}, here we present a novel approach to the ionization potential depression. The idea is to use, as input, the fractions of free electrons, $\alpha$, and atoms, $x_A$, in a partially ionized hydrogen plasma which depends on the effective binding energy of the electrons. However, what is not known is how different atomic (and molecular) orbitals are affected by interaction and quantum effects upon compression, i.e. spectral information is missing which is not available from the QMC simulations.

Here it is necessary to resort to alternative many-body theories. For example, it is known from the solution of the Bethe-Salpeter equation that all atomic levels are down-shifted in a plasma \cite{zimmermann_pssb_78,haug_pqe_84,red-book,seidel_pre_95}, i.e. \\ $E_n \to E_n(n,T) = \frac{E_1}{n^2}+\Delta E_n(n,T)$ and, simultaneously, the continuum edge (the zero energy level) shifts down as well, $0 \to \Delta I_0(n,T) \le 0$. Thus, the binding energies are renormalized according to 
\begin{align}
    E_n \to \tilde E_n(n,T) &= \frac{E_1}{n^2}+\Delta E_n(n,T) - \Delta I_0(n,T)\,,
    \label{eq:en-tilde}\\
    &= - I_n^{\rm eff}(n,T) \le 0\,,
    \label{eq:in-eff}
\end{align}
and directly yield the effective ionization potentials $I_n^{\rm eff}(n,T)$,
which vanish one by one, when the density is increased, as we demonstrate in the figures below. Vanishing of the ground state happens at the Mott density, 
$I_1^{\rm eff}(n^{\rm Mott},T) \to 0$
 corresponding to $r_s^{\rm Mott}\approx 1.2$, Eq.~\eqref{eq:mott-density}.
Note that, interaction effects, in general, remove the $l$-degeneracy of the hydrogen bound states, so $\Delta E_n$ follows from an average over all $l$.

Our first-principle approach to the renormalization of the binding energies is then to solve the Saha equation~\eqref{eq:saha-h-degenerate} for a single unknown quantity -- the continuum lowering, $\Delta I_0$, for given input values of $\alpha$ and $x_A$,

\begin{align}
    \frac{x_A}{\alpha} &= e^{\beta \mu_e^{\rm id,F}(\alpha\chi)} 2 \tilde Z_A^{\rm int}(T,n;\Delta I_0)\,. \label{eq:saha-fermi-ipd}
\end{align}
Here we assume that the partition function of the bound states in the plasma retains the same form as in vacuum, i.e. that of the Planck-Larkin partition function in which the energy eigenvalues are renormalized,
\begin{align}\label{eq:za-int-pl-n}
    \tilde Z_A^{\rm int}(T,n;\Delta I_0) & = \sum_{n=1}^\infty n^2 \left\{ e^{- \beta \tilde E_n} - 1 + \beta \tilde E_n\right\}\,,
\end{align}
with $\tilde E_n \to \text{min}(\tilde E_n, 0)$, i.e. eigenvalues that reach zero vanish from the sum. We solve the Saha equation (\ref{eq:saha-fermi-ipd}) iteratively for $\Delta I_0$ until the r.h.s. matches the given ratio $x_a/\alpha$, on the l.h.s.
\begin{figure}[h]\centering
\includegraphics[width=0.515\textwidth]{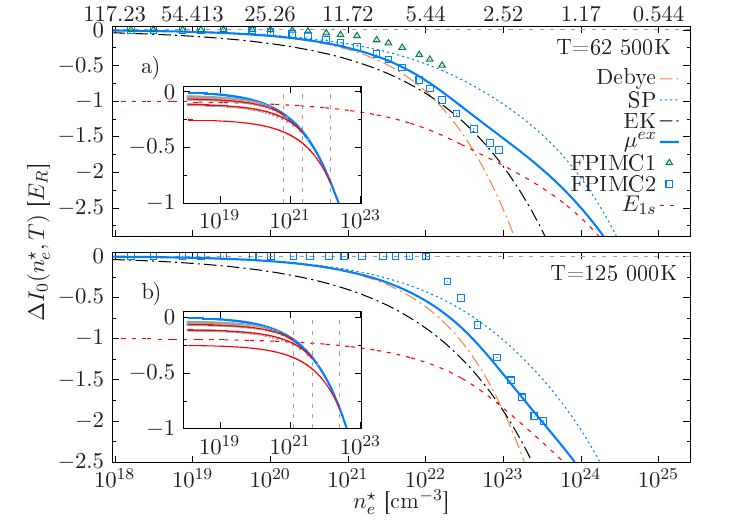}
\caption{\label{fig:detilde-hight} 
Shift of the continuum edge, $\Delta I_0$, versus free electron density $n_e^*=\alpha n_e$ (lower x-axis) and $r^*_s$ [Eq.~(\ref{eq:rs*}), upper x-axis], for temperatures $T=\SI{62500}{\kelvin}$ and \SI{125000}{\kelvin}. Lines with symbols: FPIMC results without (FPIMC1) and with (FPIMC2) renormalization of the bound state energies \cite{onate_jtap_16}. SP and EK denote the models of Stewart and Pyatt and Ecker and Kröll, respectively. $
    %\Delta \mu|_{\rm HSCM} 
    \mu^{ex}
    $ denotes the interaction part of the chemical potentials computed within the chemical model of Ref.~\cite{filinov_pre_23}. Insets show the renormalization of the 2s, 2p, 3s, 3p, 4s and 4p states. Vertical lines indicate the density where the 4s, 3s, and 2s levels merge into the continuum, according to the FPIMC2 simulation data. This happens around $r^*_s=13.59, 8.96$ and $4.86$, for $T=\SI{62500}{\kelvin}$, and $r^*_s=10.93, 7.26$ and $4.0$, for $T=\SI{125000}{\kelvin}$, respectively.
}
\end{figure} 
\begin{figure}[h]\centering
\includegraphics[width=0.515\textwidth]{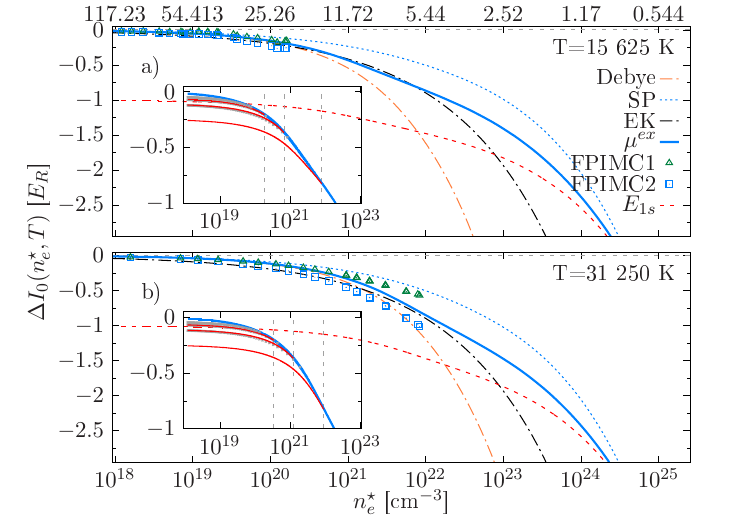}
\caption{\label{fig:detilde-lowt} 
Same as Fig.~\ref{fig:detilde-hight}, but for $T=\SI{15625}{\kelvin}$ and $\SI{31250}{\kelvin}$, respectively. The 4s, 3s and 2s levels vanish at $r^*_s=20.71, 13.48 $ and $5.93$, for \SI{15625}{\kelvin}, and at $r^*_s=16.81, 11.01 $ and $5.61$, for \SI{31250}{\kelvin}, respectively. These estimates are based on the FPIMC2 data, where available, and on the $\mu^{ex}$ data, otherwise.}
\end{figure} 

Fermionic PIMC results for the continuum edge are shown in Figs.~\ref{fig:detilde-hight} and ~\ref{fig:detilde-lowt} for two approximations for the level renormalization: the first (shown by the triangles and denoted ``FPIMC1'') entirely neglects the shift of the bound states, $\Delta E_n \to 0$, and only retains the continuum shift, $\Delta I_0$. The second takes into account the renormalization of the bound states in a consistent way, using approximate results from the solution of the Schrödinger equation with a Yukawa potential given in Ref.~\cite{onate_jtap_16}. Even though this neglects, among others, dynamic screening and spin statistics effect, this approximation should be appropriate for almost the entire density range for which the FPIMC results of Ref.~\cite{filinov_pre_23} are available. For the screening parameter, we take into account quantum effects. The results for several effective bound state energies, $\tilde E_n(n,T)$, used in approach 2 are also shown in Figs.~\ref{fig:detilde-hight} and ~\ref{fig:detilde-lowt} by the red dotted line, for the ground state, and the lowest excited states, in the insets. Vertical grey dashed lines indicate the free electron densities where the (renormalized) 4s, 3s, and 2s levels vanish.

Comparing the two approximations, we observe that both are in good agreement at low free electron densities, however, for $n^*_e \gtrsim \SI{e21}{\centi\meter^{-3}}$ differences increase rapidly. The renormalization of the bound state energies (FPIMC2) yields a substantially larger lowering of the continuum edge.
Our first-principles results also allow us to benchmark other models. While the Ecker-Kröll model (EK) \cite{ecker-kroell_63} is in reasonable agreement for low temperatures and up to moderate densities, the Stewart-Pyatt model (SP) \cite{stewart-pyatt_66} significantly underestimates the continuum lowering for most of the studied parameters.

\begin{figure}[h]\centering
\includegraphics[width=0.52\textwidth]{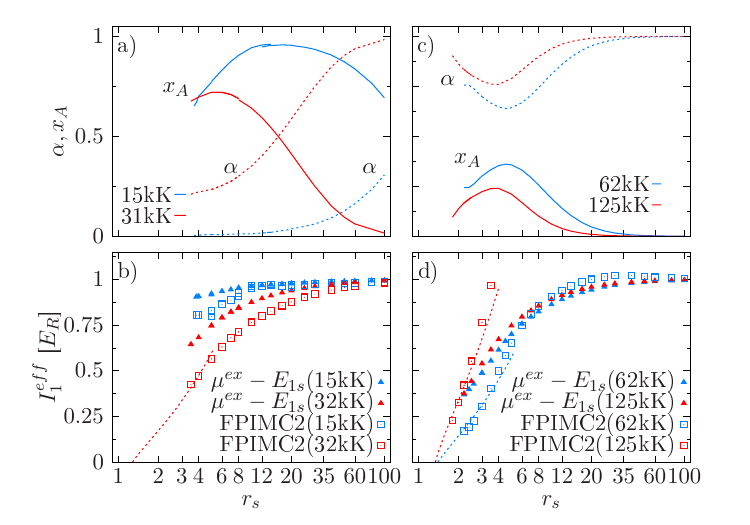}
\caption{\label{fig:ieff1} 
     The effective ionization energy of the ground state, $I^{\rm eff}_1$ (bottom) and a fraction of free electrons ($\alpha$, dots) and atoms (full line), cf. top figure,  versus total density parameter, $r_s=r_s^*\alpha^{-1/3}$, for \SI{31250}{\kelvin} and \SI{15625}{\kelvin} (left) and for \SI{62500}{\kelvin} and \SI{125000}{\kelvin} (right). Squares (triangles): FPIMC2 (chemical model) results using shifts of the continuum and ground state, as shown in Figs.~\ref{fig:detilde-hight} and \ref{fig:detilde-lowt}. Dotted line: extrapolation of the FPIMC2 results to the Mott density. }
\end{figure}

Finally, we present the effective ionization potential of the hydrogen ground state, i.e. the energy distance from the renormalized 1s-state energy to the lowered continuum, as a function of the total density parameter $r_s$ in Fig.~\ref{fig:ieff1}. Since the level shifts, [Figs.~\ref{fig:detilde-hight} and ~\ref{fig:detilde-lowt}] depend on the free electron density, the transition to $r_s$ invokes the FPIMC results for the degree of ionization, $\alpha$, which is reproduced, for reference, in the top part of the figure. 
The ionization potential $I_1^{\rm eff}$ remains close to $1E_R$ for low densities and up to approximately $r_s \sim 20$. In this parameter range (which is only of minor interest) the solution procedure sometimes does not converge (for FPIMC1) or it yields small fluctuations of the FPIMC2 results for the continuum edge [see the curves for $T=\SI{125000}{\kelvin}$ in Fig.~\ref{fig:detilde-hight}] which leads to reduced accuracy of $I_1^{\rm eff}$. This behavior is caused by the very low fraction $x_A$ of atoms, cf. Fig.~\ref{fig:ieff1}.c. where statistical errors become important.

For higher densities $r_s\lesssim 20$, $I_1^{\rm eff}$ starts to decrease monotonically, where the decrease becomes faster when the temperature increases. 
Comparing the two theoretical concepts where the bound state level shifts are neglected (FPIMC1) or taken into account (FPIMC2), we observe only small differences for the ground state ionization potential, in the studied parameter range. Therefore, we only show FPIMC2 data and compare them to the results for the interaction part of the chemical potential, $\mu^{ex}$, from the chemical model that is shown in Figs.~~\ref{fig:detilde-hight} and ~\ref{fig:detilde-lowt}. Interestingly, even though the FPIMC simulations are strongly hampered by the FSP and reach pressure ionization only for the two highest temperatures, cf. part c) of Fig.~\ref{fig:ieff1}, the smooth behavior of $I^{\rm eff}_1$ [cf. parts b) and d) of the figure] allow us to extrapolate the curves to zero, i.e. the Mott density, except for the lowest temperature.\\

To summarize this section, we presented a novel first principles approach to plasma-induced renormalization of hydrogen bound states that is quite general and equally applies to other materials and other theoretical methods, including RPIMC and DFT. The approach uses first principle data for the degree of ionization $\alpha$, as an input. Even though the definition of $\alpha$ depends on the chosen criterion, the influence of the chosen procedure can be easily tested and quantified. The main assumption going into the present model is the validity of the Planck-Larkin partition function (\ref{eq:za-int-pl-n}) also in a plasma environment. Further, the results indicated that it is important to accurately and consistently include the bound state level shifts (FPIMC2 results) which have to be provided as external input. Improved solutions to the two-particle problem in a medium, in particular at high density, will allow for further improvement in the results.

\subsection{Momentum distribution}\label{sec:n(k)}
The momentum distribution of electrons and protons is an important quantity of crucial importance for many properties of the system. This includes impact excitation and ionization rates,  chemical reaction rates, and fusion rates. In particular, it has been argued that the interplay of Coulomb correlations and quantum effects gives rise to an enhancement of fusion rates in dense plasmas, as compared to classical plasmas, e.g.~\cite{starostin_pa_02,starostin_ppr_05,starostin_jetp_17}. In fact, in classical plasmas, the momentum distribution, $n(k)$, always exhibits an exponential tail, in the high momentum limit, independently of the interaction strength, which is a consequence of the factorization of coordinate and momentum dependent terms in the classical partition function. In contrast, in quantum systems, an exponential tail is only observed in the ideal gas limit. In the presence of interactions however,
non-commutation of kinetic and interaction energy gives rise to a non-exponential power law decay of $n(k)$. This means analysis of the momentum tail allows for direct access to correlation effects in the plasma, in particular, to the so-called ``on top pair distribution'' -- $g(0)=\lim_{r\to 0} g^{\uparrow\downarrow}(r)$ -- the pair distribution of spin up and spin down particles at zero separation. 

Recently accurate results have been obtained for $n(k)$ and $g(0)$ of jellium which we discuss in Sec.~\ref{sec:n(k)}. After this, we briefly discuss the situation in hydrogen.

\subsubsection{Jellium}\label{ssec:n(k)-jellium}
\begin{figure}[h!]
    \centering    \includegraphics[width=0.44\textwidth]{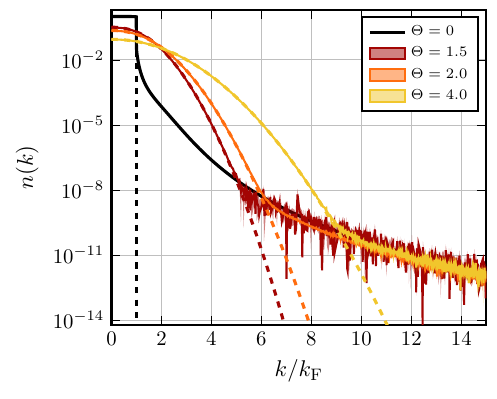}
    \caption{Nonexponential decay of the momentum distribution of moderately correlated and weakly degenerate electrons, $r_s=0.5$ and $T/T_F=1,5, 2.0, 4.0$. CPIMC results with $N=54$ particles are compared to the ground state (solid black, data of Ref.~\cite{gori_giorgi_short_range_2001}). For comparison,  the ideal Fermi distribution is shown by dashed lines of the same color as the interacting result.   Reproduced from Ref.~\cite{hunger_pre_21} with the permission of the authors.
}
    \label{fig:fullMDF_N54_rs05}
\end{figure}
\begin{figure}[h!]
    \centering
    \includegraphics[width=0.4\textwidth]{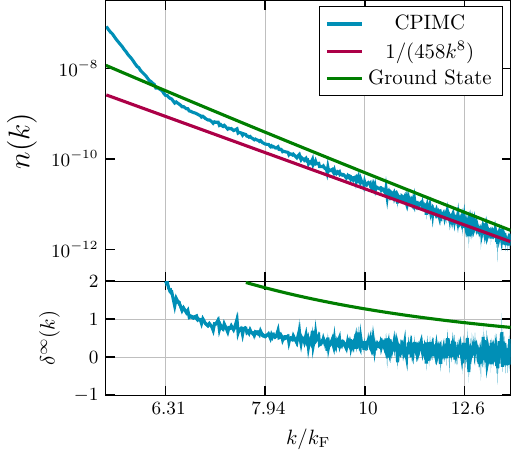}
    \caption{Tail region of the momentum distribution for $r_s=0.5$ and $\Theta=2$. Bottom figure shows the deviation from the asymptotic,     $\delta^\infty(k) = \frac{n(k)}{n^\infty(k)} - 1$. The $k^{-8}$ asymptotic is reached for $k\gtrsim 10k_F$ with a prefactor that is provided by the CPIMC simulations.
    Reproduced from Ref.~\cite{hunger_pre_21} with the permission of the authors.}
    \label{fig:tails_rs05_th2_N54}
\end{figure}
The existence of a non-exponential tail of $n(k)\sim k^{-8}$ for jellium in the ground state was demonstrated by E.~Daniel and S.H.~Vosko~\cite{daniel_vosko_momentum_1960}. Similar analytical predictions were due to V.M.~Galitskii \textit{et al.} \cite{galitskii_particle_1967} and J.C.~Kimball \cite{kimball_short_range_1975} whereas  H.~Yasuhara \textit{et al.}
established a direct relation between $n(k)$ and $g(0)$ for $T=0$, \cite{yasuhara_note_1976}
\begin{align}
    \lim_{k\to \infty} n(k)  = \frac{4}{9}\left(\frac{4}{9\pi}\right)^{2/3}\left( \frac{r_s}{\pi}\right)^2 \frac{k^8_F}{k^8}g^{\uparrow\downarrow}(0)\,,
      \label{eq:k8-g0}
\end{align}
that was extended to finite temperatures by J.~Hofmann \textit{et al.}~\cite{hofmann_short-distance_2013}. These are important results which, however, refer only to the occupation of the very high momentum states but do not contain information on the overall shape of $n(k)$, which requires simulations.\\

For the ground state, and moderate to strong coupling at $1\le r_s\le 10$, the entire momentum distribution has been computed by M.~Holzmann \textit{et al.}~\cite{Holzmann2011a} by reptation quantum Monte Carlo simulations. They particularly investigated the region around the Fermi momentum and determined the quasiparticle renormalization factor. 

For finite temperatures, $\Theta \gtrsim 1$, exact CPIMC results were presented recently by K.~Hunger \textit{et al.}~\cite{hunger_pre_21}, whereas T.~Dornheim \textit{et al.} presented direct PIMC results ~\cite{Dornheim_PRE_2021,Dornheim_PRB_nk_2021}. In  Fig.~\ref{fig:fullMDF_N54_rs05} we show 
CPIMC results for the momentum distribution $n(k)$ for a weakly coupled moderately degenerate electron gas and compare to the ground state and to an ideal Fermi gas. The figure clearly demonstrates the algebraic decay of the correlated momentum distribution (full lines) and its deviation from the exponential tail of the ideal system (dashed lines). The $k^{-8}$ asymptotic is confirmed in Fig.~\ref{fig:tails_rs05_th2_N54} and, in addition, the temperature and density dependence of the pre-factor (i.e. of $g(0)$) established \cite{hunger_pre_21}.
Note that, for $r_s \lesssim 1$ and temperatures above $0.5 T_F$, CPIMC proves to be highly efficient, being able to resolve $n(k)$ with an accuracy of ten decimal digits. Figure~\ref{fig:tails_rs05_th2_N54} also indicates that the power law asymptotic is reached only for large momenta where the occupation is already very low.

\begin{figure}
    \centering
    \includegraphics[width=0.48\textwidth]{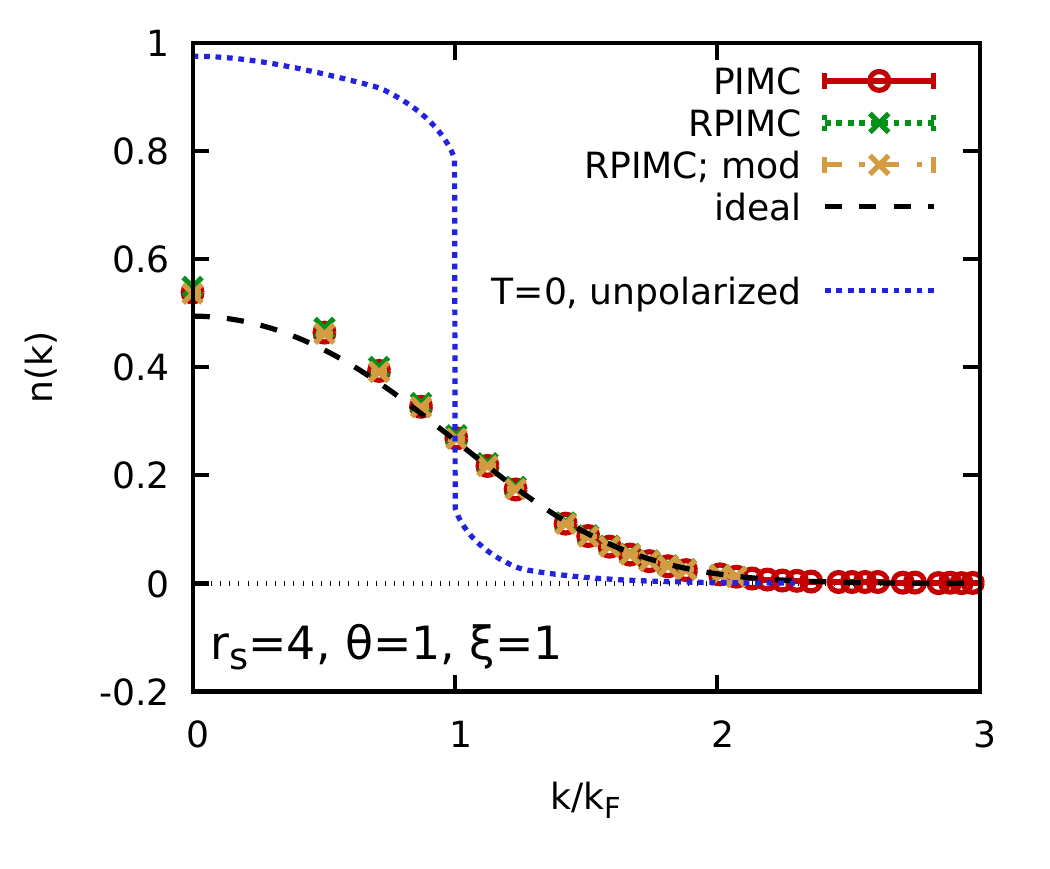}
    \caption{Momentum distribution $n(k)$ of the spin-polarized UEG at $r_s=4$ and $\Theta=1$ ($T=\SI{3.13}{\eV}$) for $N=33$ electrons. Red circles: direct PIMC; green crosses: RPIMC~\cite{Militzer2019}; yellow crosses: RPIMC with a modified normalization constant;
    dotted blue: ground-state QMC results \cite{Holzmann2011a} for the unpolarized UEG at the same density. Note that the Fermi momentum for both polarizations is different.
Adapted from Ref.~\cite{Dornheim_PRE_2021}. %
    \label{fig:nk_UEG} 
    }
\end{figure}

In Fig.~\ref{fig:nk_UEG}, we show complementary direct PIMC results (red circles) for $n(k)$ of a spin-polarized UEG with $N=33$ electrons at $r_s=4$ and $\Theta=1$. We note the larger occupation for $k\lesssim k_\textnormal{F}$ of the PIMC results compared to the ideal Fermi distribution (dashed black). This effect has first been reported by B.~Militzer and R.~Pollock~\cite{militzer_prl_02} and is connected to a lowering of the kinetic energy due to XC-effects. 
The green crosses show restricted PIMC results for the same conditions; they are in qualitative agreement with the exact PIMC data set but are systematically too high for all $k$.
Rescaling the RPIMC data set with a constant factor results in the yellow crosses, which are in good agreement with the direct PIMC data over the entire momentum range.
The observed discrepancy between PIMC and RPIMC has thus been attributed to a systematic error in the determination of the proper normalization in the latter implementation, whereas the direct PIMC results have been normalized automatically based on the extended ensemble approach introduced in Ref.~\cite{Dornheim_PRB_nk_2021}. At the same time, we note that no nodal error was found at these conditions~\cite{Dornheim_PRB_nk_2021,Dornheim_PRE_2021}.
A similar comparison for warm dense hydrogen constitutes an important topic for future works.
Finally, we include ground-state QMC results for the unpolarized UEG \cite{Holzmann2011a,PhysRevLett.105.086403} at the same density 
as the dotted blue curve. 
We note the substantially rounded edges; these features would be absent for an ideal Fermi gas where the ground-state momentum distribution is given by a simple step function and, therefore, originates from correlation effects due to the interacting electrons.

\subsubsection{Hydrogen}\label{ssec:n(k)-h}
Lowering the temperature below the electronic degeneracy temperature $T_F$,
the momentum 
distribution of jellium becomes steeper at the Fermi
momentum developing a discontinuity in the limit of $\Theta \to 0$. Thermal broadening
of the jump of the Fermi-Dirac occupation function
is of order $\Theta$, typically beyond the
experimental resolutions of synchrotron light-scattering experiments at ambient temperatures \cite{PhysRevLett.105.086403,PhysRevB.101.165124,PhysRevB.101.165125} in the regime where the Born-Oppenheimer approximation is expected to be applicable. 

Coulomb interactions between electrons lead
to the deviation from the ideal gas step-function
at zero temperature, reducing the magnitude
of the discontinuity at $k_F$, illustrated in Fig.~\ref{fig:nk_UEG} for $r_s=4$ corresponding 
to the density of valence electrons in crystalline metallic sodium at ambient pressure and temperature. However, anisotropies in the Fermi surface, e.g. due to band structure effects,
smear out the jump in $n(k)$ when spherically averaged \cite{PhysRevLett.105.086403}.
For an insulator, the electronic momentum distribution
is expected to remain continuous in any direction even at zero temperature. 

\begin{figure}
    \centering
    \includegraphics[width=0.5\textwidth]{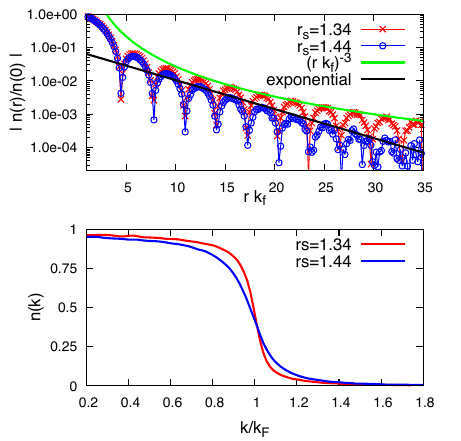}
    \caption{Electronic momentum distribution of hydrogen at two densities across the LLPT at $T=\SI{1200}{\kelvin}$, computed using a Slater-Jastrow backflow trial wavefunction with configurations from CEIMC. In the molecular phase (blue) the single-electron density matrix, $n(r)$, decreases exponentially with $r$, while in the dissociated phase (red) it decreases as $r^{-3}$ because the electrons are delocalized. 
    \label{fig:nk_hydrogen_llpt}}
\end{figure}

The electronic momentum distribution $n(k)$ of hydrogen has been computed by C.~Pierleoni \textit{et al.} using CEIMC simulations~\cite{Pierleoni2016,Pierleoni2018}, for details on the method, cf. Sec.~\ref{subse:ceimc}. In Fig.~\ref{fig:nk_hydrogen_llpt} we show results for $n(k)$ at $T=\SI{1200}{\kelvin}$ for two densities in the range of the LLPT, cf. Sec.~\ref{sssec:hydrogen-lowt}. 
For $r_s=1.44$ the system is in the molecular phase, whereas molecules are mostly dissociated at $r_s=1.34$, see Fig.~\ref{fig:Mol_frac_1200K} and the corresponding discussion in Sec.~\ref{sssec:alpha}. We, therefore, expect a noticeable change in the momentum distribution
reflecting metallic behavior with a sharp Fermi surface on the atomic side. Indeed, the width
of the distribution around $k_F$ is reduced by about a factor of two. 
This change occurs abruptly
at the LLPT transition whereas further density changes on the atomic or molecular side are much smoother \cite{Pierleoni2018}. The broadening of $n(k)$ around $k_F$ in the atomic phase is due to
anisotropies of the Fermi surface of the snapshots of different nuclear configurations in
the Born-Oppenheimer approximation. 

That these changes of the momentum distribution from $r_s=1.44$ to $r_s=1.34$ indicate a phase transition from insulating to metallic behavior is further supported by analyzing the
long distance behavior of the reduced single-particle density matrix $n(r)$, the Fourier transform of $n(k)$, shown in Fig.~\ref{fig:nk_hydrogen_llpt}.
Within the Born-Oppenheimer approximation, we have
$n(r)=\langle n(r|\uline{\vec R}) \rangle_{\uline{\vec R}}$,
with $\langle \dots \rangle_{\uline{\vec R}}$ indicating the averaging over the nuclear configurations
of
\begin{equation}
    n(r|\uline{\vec R})=\int  d \mathbf{r}_1 \cdots  d\mathbf{r}_N
    \Psi_0^*(  \mathbf{r}_1+  \mathbf{r}, \mathbf{r}_2, \dots \mathbf{r}_N)
    \Psi_0( \mathbf{r}_1, \mathbf{r}_2 \dots \mathbf{r}_N)
\end{equation}
where $\Psi_0(\cdots)\equiv \Psi_0(\cdots|\uline{\vec R})$ is the Born-Oppenheimer
electronic wave function which 
parametrically depends on the nuclear configuration $\uline{\vec R}$. 

From Fig.~\ref{fig:nk_hydrogen_llpt} we see that the envelope of the asymptotic long-range behavior changes
from an exponential to a power law, at the LLPT, confirming the change from the localized character of the
molecular liquid to a metallic state with a Fermi-liquid $r^{-3}$ decay, in the atomic phase.

\subsubsection{CEIMC results for the gap}\label{ss:h-gap}

Within CEIMC, it is further possible to calculate the fundamental electronic gap in the insulating phase \cite{Yang2020,gorelov_prl_20,Gorelov2020a}, the energy
gap corresponding to electron/hole doping. Assuming 
a strictly uniform background charge to assure charge neutrality,
the change of the free energy with respect to addition/removal of electrons reduces to different electronic Born-Oppenheimer excitation energies, averaged over nuclear configurations.
Since the additional electronic charge is necessarily smeared
out over the entire system due to the charge-compensating homogeneous background, modifications of the nuclear states
can be neglected in the thermodynamic limit \cite{gorelov_prl_20}. In the crystalline state,
electron addition and removal energies can be characterized by the symmetries of the crystal
structure, defined by the average charge density, even in the presence of
zero point and thermal nuclear motion \cite{Gorelov2020b} within the region of validity of the
BO approximation.

Experimentally, the electronic excitation spectrum is probed by optical absorption
corresponding to charge-neutral particle-hole excitations which can also be addressed with QMC-based methods 
\cite{Gorelov2023}.
Electron-hole binding and excitonic effects may lower the gap compared to the fundamental gap. Those effects are particularly important at lower pressures where
excitons are strongly bound \cite{gorelov2023electronic}. In addition, the localization of e-h 
excitations may also affect the BO energy surface of the nuclei in the vicinity of the excitons. Neutral electronic excitations averaged over nuclear trajectories of the ground state
BO energy surface remains an upper bound of the neutral excitation gap.

CEIMC calculations of the closing of the fundamental gap in solid hydrogen \cite{gorelov_prl_20} predicted that for \SIrange{\sim 370}{380}{\giga\Pa} an indirect gap closure occurs at \SI{200}{\kelvin}, 
for C2/c-24, and at \SI{\sim 340}{\giga\Pa}, for $Cmca-12$, whereas the direct gap remains open until \SI{\sim 450}{\giga\Pa} and 
\SI{\sim 500}{\giga\Pa}, respectively, suggesting a bad metal (or semimetal) scenario, qualitatively similar to experimental observations \cite{Eremets2017,Eremets2019,Dias2016,Dias2019}.
Predictions for the direct gap are in agreement with absorption measurements \cite{Loubeyre2002,loubeyre_nat_20} with a strong
reduction of \SI{\sim 2}{\eV} due to nuclear quantum effects, slightly
larger than the experimental extrapolation \cite{Loubeyre2022}
of \SI{\sim 1.5}{\eV} from $D_2$ to $H_2$ assuming a $m^{-1/2}$ 
isotope effect. Experiments have been performed at \SI{80}{\kelvin} so that the experimental value might be slightly biased by
residual finite temperature shifts.

\begin{figure}
    \centering
    \includegraphics[width=0.42\textwidth]{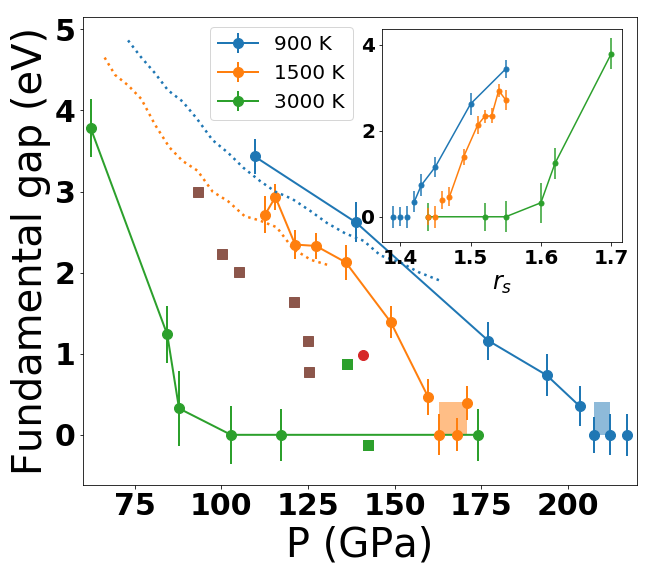}
    \caption{The fundamental energy gap of liquid hydrogen along the isotherms: $T=\qtylist{900;1500;3000}{\kelvin}$, as a function of pressure. Inset: the same gap as a function of $r_s$, a measure of density, Eq.~(\ref{eq:rs}). The lines connect the gap data only up to the molecular-atomic transition region. The colored rectangles show the coexistence region of the LLPT according to Ref.~\cite{Pierleoni2017}. The dotted lines are the gaps reported by P.M.~Cellier \textit{et al.} \cite{Celliers2018}. The brown and green squares are the results of W.J.~Nellis \textit{et al.} for temperatures of \SIrange{2000}{3000}{\kelvin} \cite{Nellis1999} reanalyzed in Ref.~\cite{Knudson2018}. The red circle is the gap reported by R.S.~McWilliams \textit{et al.} at \SI{2400}{\kelvin} \cite{McWilliams2016}. Adapted from Ref.~\cite{Gorelov2020a}.
    \label{fig:h2_gap}}
\end{figure}

\begin{figure}
    \centering
    \includegraphics[width=0.42\textwidth]{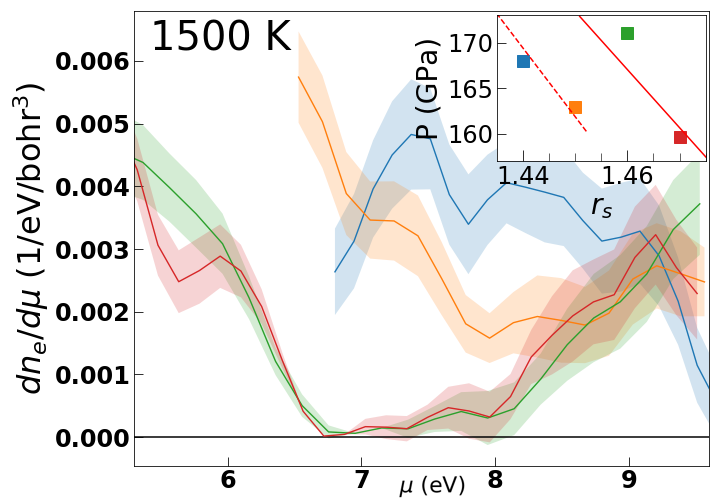}
    \caption{Density of states (DOS) of liquid hydrogen near the band edge at densities near the gap closure for the isotherm $T=\SI{1500}{\kelvin}$. The inset shows the equation of state, as reported in Ref.~\cite{Pierleoni2017}. The dashed and solid red lines indicate the atomic and molecular regions, respectively. The colors of the DOS match the colors of points in the insets. Adapted from Ref.~\cite{Gorelov2020a}.
    \label{fig:t1500_dos}}
\end{figure}

The gap closure of the molecular liquid has been studied at three different temperatures
around the LLPT \cite{Gorelov2020a}. Figure \ref{fig:h2_gap} shows that, below the critical temperature, the gap closure seems to occur abruptly, together with the molecular to atomic transition, whereas a smoother closure is observed at $T=\SI{3000}{\kelvin}$, supporting the existence of a cross-over between the two phases above the critical temperature. This picture is also supported by the electronic density of states, an example of which is reported in Fig.~\ref{fig:t1500_dos}. Experimental determinations of the gap are also reported in Fig.~\ref{fig:h2_gap} and are in reasonable agreement with QMC results, suggesting that the temperatures reached in the shock-wave experiments are in the same range as in the simulations.

To characterize the Fermi liquid behavior of metallic hydrogen, determination of the effective mass of single-particle excitations should be carried out.
However, since the effective mass of jellium is very close to the bare electron mass \cite{Haule22,Holzmann23}, one 
expects the excitation spectrum to be determined by (single-particle) band structure effects to a large extent.

\subsection{Density response properties}\label{sec:dynamics}

In this section, we will give an overview of a number of density response properties of warm dense hydrogen. As a starting point, we consider the modified Hamiltonian~\cite{Dornheim2023}
\begin{eqnarray}\label{eq:H_perturbed}
\hat{H}_{\mathbf{q},\omega,A} = \hat{H} + 2A\sum_{l=1}^N \textnormal{cos}\left(\mathbf{q}\cdot\hat{\mathbf{r}}-\omega t \right)\ ,
\end{eqnarray}
where $\hat{H}$ is the unperturbed Hamiltonian, Eq.~(\ref{eq:hamiltonian-h}), to which we add an external monochromatic perturbation of wave vector $\mathbf{q}$, frequency $\omega$, and perturbation amplitude $A$.
In the limit of infinitesimally weak perturbations, the induced density change with respect to the unperturbed system is described by [cf. Eq.~(\ref{eq:density-response})]
\begin{eqnarray}
    \Delta n(\mathbf{q},\omega) = \chi(\mathbf{q},\omega) A\ ;
\end{eqnarray}
it is thus a linear function of the perturbation strength, which is fully described by the dynamic linear density response function $\chi(\mathbf{q},\omega)$. We note that the latter is an equilibrium property of the unperturbed system.
From a theoretical perspective, it is very convenient to express the density response as [cf. Eq.~(\ref{eq:chi-correlated-def})]
\begin{eqnarray}\label{eq:dynamic_density_response}
    \chi(\mathbf{q},\omega) = \frac{\chi_0(\mathbf{q},\omega)}{1-\left[v_q + K_\textnormal{xc}(\mathbf{q},\omega)\right]\chi_0(\mathbf{q},\omega)}\ ,
\end{eqnarray}
where $\chi_0(\mathbf{q},\omega)$ denotes a known reference function. For the UEG, the latter is typically given by the temperature-dependent Lindhard function that describes the density response of an ideal Fermi gas at the same conditions~\cite{quantum_theory}. In that case, the complete information about electronic XC effects is included in the dynamic exchange-correlation kernel $K_\textnormal{xc}(\mathbf{q},\omega)$, which is related to the dynamic local field correction (LFC)~\cite{kugler1} by
\begin{eqnarray}
    K_\textnormal{xc}(\mathbf{q},\omega) =  - \frac{4\pi}{q^2}G(\mathbf{q},\omega)\ .
\end{eqnarray}
Indeed, setting $K_\textnormal{xc}(\mathbf{q},\omega)\equiv0$ in Eq.~(\ref{eq:dynamic_density_response}) corresponds to the well-known \emph{random phase approximation} (RPA) describing the electronic density response on a mean-field level. Unsurprisingly, $G(\mathbf{q},\omega)$ of the UEG constitutes key input for a host of practical applications such as the construction of nonlocal XC-functionals for DFT simulations~\cite{Lu_JCP_2014,Patrick_JCP_2015,pribram} or the interpretation of XRTS experiments~\cite{Fortmann_PRE_2010,Dornheim_PRL_2020_ESA,Zan_PRE_2021}. Therefore, the development of approximate expressions for $G(\mathbf{q},\omega)$ [or its static limit $G(\mathbf{q},0)$] has been a highly active topic over the last decades~\cite{IIT,stolzmann,stls_original,stls,stls2,vs_original,tanaka_hnc,Tanaka_CPP_2017,Tolias_JCP_2021,castello2021classical,dornheim_electron_liquid,Tolias_JCP_2023}.
Very recently, highly accurate results have become available both in the ground state~\cite{Chen2019} and at finite temperatures~\cite{Leblanc_PRL_2022,Hou_PRB_2022,dornheim_ML,Dornheim2018b,groth_prb_19,Dornheim_PRL_2020_ESA,Dornheim_PRB_ESA_2021,Dornheim_HEDP_2022,dornheim_HEDP,dornheim_electron_liquid,groth_jcp17,dornheim_pre17}. Among other things, these results have revealed a roton-type feature in the spectrum of density fluctuations~\cite{Dornheim2018b,dornheim_comphys_22,Dornheim_JCP_2022} for intermediate wavenumbers, which has 
subsequently been predicted to be even more pronounced in warm dense hydrogen~\cite{hamann_prr_23}, see Sec.~\ref{sec:roton} below.

For electron--ion systems such as hydrogen, the reference function in Eq.~(\ref{eq:dynamic_density_response}) is typically replaced by the Kohn-Sham response function $\chi_\textnormal{KS}(\mathbf{q},\omega)$, which depends on the utilized XC-functional, eventually leading to a Dyson's type equation, cf.~Eq.~(\ref{eq:Dyson}). Therefore, the clear-cut interpretation of $K_\textnormal{xc}(\mathbf{q},\omega)\equiv0$ as a mean-field description is no longer valid as both the kernel and the reference function contains some information about electronic XC-effects.
For hydrogen, first dependable results for $K_\textnormal{xc}(\mathbf{q},0)$ have been presented only recently~\cite{Bohme_PRL_2022,Bohme_PRE_2023,Dornheim2023,Moldabekov_JCTC_2023} and are discussed in Sec.~\ref{sec:static_density_response} below.

We also note that $\chi$ is directly related to the inverse dielectric function, cf. Eq.~(\ref{eq:epsm1-def}),
which describes the dynamical screening of the Coulomb interaction and also enters collision integrals of kinetic equations, cf. Sec.~\ref{sss:qke}.

A particularly important practical application of Eq.~(\ref{eq:dynamic_density_response}) is given by the fluctuation--dissipation theorem~\cite{quantum_theory}
\begin{eqnarray}\label{eq:FDT}
S_{ee}(\mathbf{q},\omega) = - \frac{\textnormal{Im}\chi(\mathbf{q},\omega)}{\pi n (1-e^{-\beta\omega})}\ ,
\end{eqnarray}
which relates the dynamic structure factor to the dynamic density response function. For example, Eq.~(\ref{eq:FDT}) constitutes the basis of LR-TDDFT simulations of XRTS measurements~\cite{PhysRevLett.120.205002,Ramakrishna_PRB_2021,frydrych2020demonstration,Voigt_POP_2021,Schoerner_PRE_2023,Moldabekov_PRR_2023,Dornheim2023,ranjan2023toward,dynamic2,gawne2024ultrahigh}.
In practice, XRTS measurements of hydrogen are generally challenging due to its comparably small scattering cross section~\cite{Zastrau,Fletcher_Frontiers_2022}.

\subsubsection{Static density response: snapshots\label{sec:static_density_response}}

In the static limit of $\omega\to0$, the electronic density response of hydrogen can be studied by carrying out a set of harmonically perturbed calculations that are governed by the Hamiltonian (\ref{eq:H_perturbed}) for a set of fixed proton coordinates.
Fully integrated PIMC simulations where both the electrons and protons are treated dynamically on the same level are discussed in detail in Sec.~\ref{sec:ITCF} below. Still, using PIMC to solve the electronic problem in the fixed external proton potential allows for direct comparisons with DFT simulations, which is interesting in its own right.

\begin{figure}\centering
\includegraphics[width=0.48\textwidth]{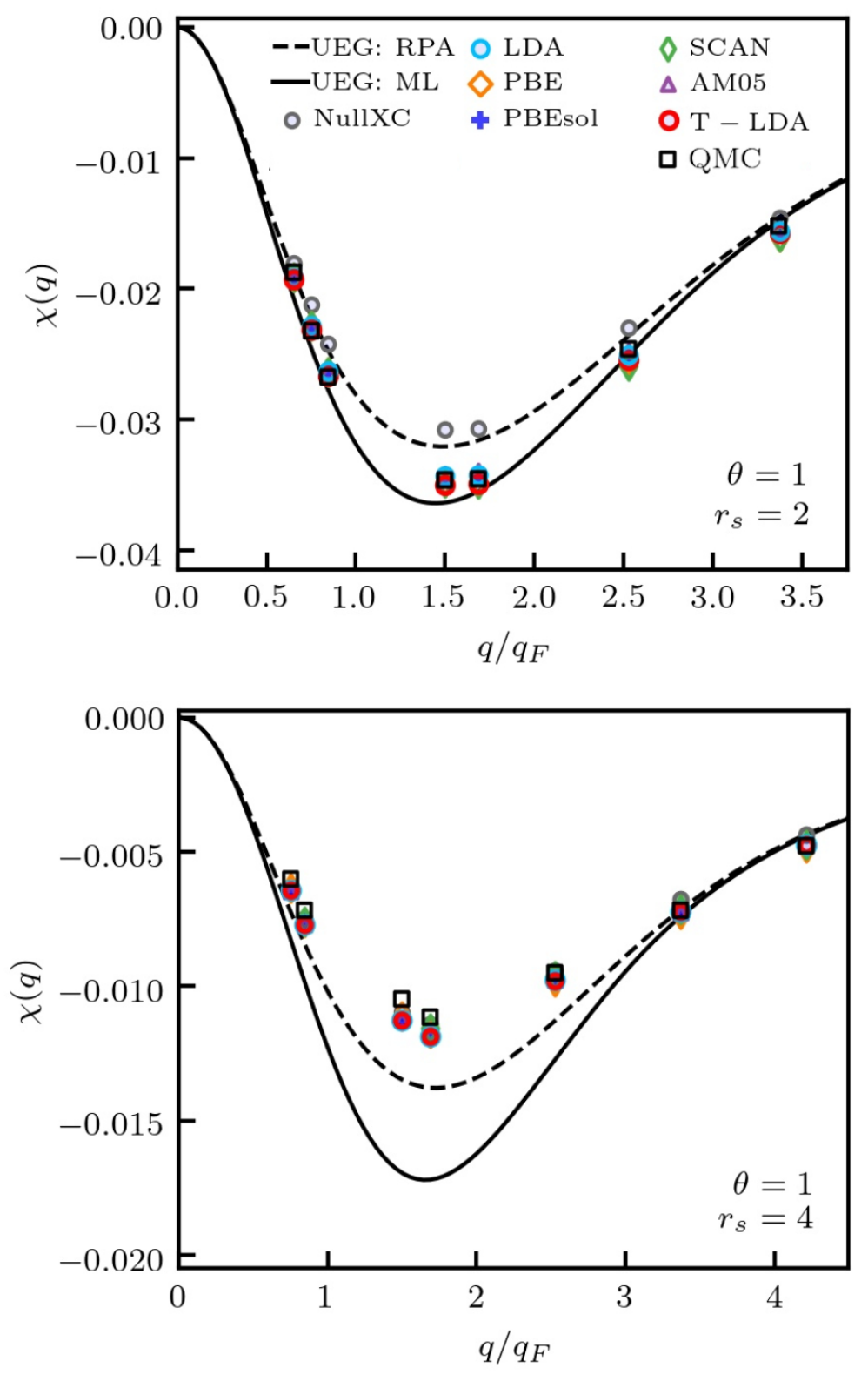}
\caption{\label{fig:Chi_static} Comparing PIMC and DFT results for the static linear density response function $\chi(\mathbf{q})$. Taken from Zh.~Moldabekov \emph{et al.}~\cite{Moldabekov_JCTC_2023} with the permisson of the authors.
}
\end{figure}

Following this idea, M.~B\"ohme \emph{et al.}~\cite{Bohme_PRL_2022} have presented the first accurate results for the static density response function $\chi(\mathbf{q})=\chi(\mathbf{q},0)$ [and, in this way, also the corresponding static XC-kernel $K_\textnormal{xc}(\mathbf{q})$]. 
The results are shown in Fig.~\ref{fig:Chi_static} as the black squares for $\Theta=1$, with the top and bottom panels corresponding to $r_s=2$ and $r_s=4$. In addition, we also show the exact static density response of the UEG~\cite{dornheim_ML} (solid black), as well as the RPA approximation to the latter (dashed black) at the same conditions.
For the higher density, most electrons are strongly delocalized throughout the system, and the density response of hydrogen closely follows the density response of the UEG over the entire wavenumber range. 
In stark contrast, the electrons are strongly localized around the protons for $r_s=4$, and the overall density response of hydrogen is substantially reduced compared to the UEG model at the same conditions.
Assuming a decomposition into effectively \emph{free} and \emph{bound} populations of electrons where only the former respond to the external perturbation, M.~B\"ohme \emph{et al.}~\cite{Bohme_PRL_2022} have reported a free-electron fraction (degree of ionization) of $\alpha=0.6$. This is in close agreement to the value of $\alpha=0.54$ reported in the RPIMC simulations of B.~Militzer and D.~Ceperley~\cite{Militzer_PRE_2001}, whereas more recent simulations by A.~Filinov and M.~Bonitz~\cite{filinov_pre_23} find a substantially lower value, $\alpha\lesssim0.3$, cf.~Fig.~\ref{fig:ion-degree-mil-fil}.

Subsequently, Zh.~Modabekov \emph{et al.}~\cite{Moldabekov_JCTC_2023} have used the same set-up for KS-DFT simulations of hydrogen, and the results for different XC-functionals are included as the colored symbols in Fig.~\ref{fig:Chi_static} for both densities. For $r_s=2$ all depicted functionals work very well, over the entire depicted $q$-range, although in particular the SCAN functional overestimates the true density response for $q>2q_\textnormal{F}$. In addition, the grey circles that have been computed without any XC-functional (``NullXC'') are in close agreement with the RPA result for the UEG.
For $r_s=4$, the situation is more complex. First and foremost, we find that here, too, all DFT results are in qualitative agreement with the exact PIMC reference data for all wavenumbers. Interestingly, DFT overestimates the magnitude of the true density response for $q\lesssim2q_\textnormal{F}$ independent of the utilized functional. This is likely a consequence of self-interaction effects~\cite{mori2014derivative}, which lead to an underestimation of the localization of the electrons around the ions. Since the latter depletes the ability of effectively bound electrons to react to the external perturbation, DFT does not capture the full reduction of $\chi(\mathbf{q})$ compared to the free UEG model. 
Second, we mention the 
increase in the density response compared to the UEG for the largest depicted value of $q$.
This has been explained in terms of isotropy breaking at certain length scales in the original Ref.~\cite{Bohme_PRL_2022}, and is again captured qualitatively by DFT.

In essence, by implementing the direct perturbation approach into DFT, Zh.~Moldabekov \emph{et al.}~\cite{Moldabekov_JCTC_2023}
have demonstrated an alternative way to obtain the static XC-kernel $K_\textnormal{XC}(\mathbf{q})$ within DFT itself without the need for numerical derivatives of the XC-functional. In particular, this method is straightforward even for hybrid functionals on higher rungs on Jacob's ladder of functional approximations~\cite{Perdew_AIP_2001}, for which numerical derivatives are known to be unstable.
A particularly interesting application of $K_\textnormal{xc}(\mathbf{q})$ is given by LR-TDDFT simulations to compute the dynamic structure factor $S(\mathbf{q},\omega)$, which is discussed briefly in Sec.~\ref{sec:TDDFT} below.

\subsubsection{Imaginary-time correlation functions (ITCF)\label{sec:ITCF}}

A well-known bottleneck of the \textit{ab initio} PIMC method is its restriction to the imaginary-time domain, see Sec.~\ref{sec:PIMC} above. While dynamic PIMC simulations are, in principle, possible~\cite{kleinert2009path}, the complex exponent of the time-evolution operator would lead to an additional \emph{phase problem}, limiting its application to ultra-short time scales.

A possible way to circumvent this limitation is to consider imaginary-time correlation functions, which can be easily computed within PIMC. A particularly important example is given by the imaginary-time density--density correlation function (ITCF)~\cite{Boninsegni1996,Filinov2012,Dornheim_JCP_ITCF_2021,Dornheim_MRE_2023,Dornheim_PTR_2023}
\begin{eqnarray}\label{eq:ITCF}
    F_{ee}(\mathbf{q},\tau) = \braket{\hat{n}_e(\mathbf{q},0)\hat{n}_e(-\mathbf{q},\tau)}\ ,
\end{eqnarray}
corresponding to the usual intermediate scattering function $F(\mathbf{q},t)$~\cite{siegfried_review}, but evaluated at the imaginary-time argument $t=-i\hbar\tau$.
For example, $F(\mathbf{q},\tau)$ is directly connected to the dynamic structure factor via a two-sided Laplace transform,
\begin{eqnarray}\label{eq:Laplace}
    F_{ee}(\mathbf{q},\tau) = \mathcal{L}\left[S_{ee}(\mathbf{q},\omega)\right] = \int\limits_{-\infty}^\infty \textnormal{d}\omega\ S_{ee}(\mathbf{q},\omega) e^{-\tau\hbar\omega}\ .
\end{eqnarray}
The numerical inversion of Eq.~(\ref{eq:Laplace}) to compute $S_{ee}(\mathbf{q},\omega)$ is often denoted as \emph{analytic continuation} in the literature~\cite{JARRELL1996133}, and constitutes a difficult, ill-conditioned inverse problem. Although several methods exist, e.g.~Refs.~\cite{Vitali_PRB_2010,Mishchenko_PRB_2000,Ferre_PRB_2016,Boninsegni_maximum_entropy,PhysRevB.94.245140,Otsuki_PRE_2017,Goulko_PRB_2017}, the reliability of these methods is difficult to judge in the absence of additional constraints. 
A benign exception is given by the UEG, where it has been possible to sufficiently constrain the space of possible trial spectra by imposing a number of exactly known conditions~\cite{Dornheim2018b,groth_prb_19}. The generalization of these ideas to warm dense hydrogen is a topic of ongoing efforts and will be covered in dedicated future works.

Very recently, T.~Dornheim \emph{et al.}~\cite{Dornheim_MRE_2023} have argued that, due to the formally unique mapping inherent to Eq.~(\ref{eq:Laplace}), often no analytic continuation is necessary to understand the physical mechanisms in a given system of interest. In fact, both $S_{ee}(\mathbf{q},\omega)$ and $F_{ee}(\mathbf{q},\tau)$ contain the same information, but in different representations. Surprisingly, considering $F_{ee}(\mathbf{q},\tau)$ might even be more practical for the interpretation of experimental 
XRTS measurements due to the well-known convolution theorem~\cite{Dornheim_T_2022},
\begin{eqnarray}\label{eq:convolution2}
    \mathcal{L}\left[S_{ee}(\mathbf{q},\omega)\right] = \frac{  \mathcal{L}\left[S_{ee}(\mathbf{q},\omega)\circledast R(\omega) \right] }{  \mathcal{L}\left[R(\omega)\right] }\ .
\end{eqnarray}
In other words, it is straightforward---and very robust with respect to experimental noise~\cite{Dornheim_T_follow_up,dornheim2023xray}---to obtain the ITCF from an XRTS measurement, given accurate knowledge for $R(\omega)$, whereas the corresponding deconvolution of Eq.~(\ref{eq:convolution2}) to get $S_{ee}(\mathbf{q},\omega)$ is notoriously unstable.
The task at hand is thus to understand the physical manifestation of effects such as temperature, density, or Coulomb coupling in the imaginary-time domain~\cite{Dornheim_MRE_2023}.

For example, the detailed balance relation 
\begin{eqnarray}\label{eq:detailed_balance}
    S_{ee}(\mathbf{q},-\omega) = e^{-\beta\hbar\omega} S_{ee}(\mathbf{q},\omega)\,,
\end{eqnarray}
provides a direct relation between the DSF at positive and negative frequencies in thermal equilibrium, which exclusively depends on the temperature~\cite{DOPPNER2009182,Sperling_PRL_2015}. Unfortunately, its direct application for temperature diagnostics is generally prevented by the convolution with the source function.
Yet, it is straightforward to translate Eq.~(\ref{eq:detailed_balance}) into the $\tau$-domain, resulting in the symmetry relation~\cite{Dornheim_T_2022,Dornheim_T_follow_up,Dornheim2023}
\begin{eqnarray}\label{eq:symmetry}
    F_{ee}(\mathbf{q},\tau) = F_{ee}(\mathbf{q},\beta-\tau)\ .
\end{eqnarray}
In other words, $F(\mathbf{q},\tau)$ is symmetric around $\tau=\beta/2=1/(2k_\textnormal{B}T)$, where it attains a minimum. This allows one to directly extract the temperature of an XRTS measurement without the need for simulations and approximations~\cite{Dornheim_T_2022}. It can thus be expected that Eq.~(\ref{eq:symmetry}) will give direct insights into upcoming experiments with hydrogen jets~\cite{Zastrau,Fletcher_Frontiers_2022}, and potential ambitious spherical implosion experiments at the NIF~\cite{Moses_NIF}.

\begin{figure}\centering
\includegraphics[width=0.5\textwidth]{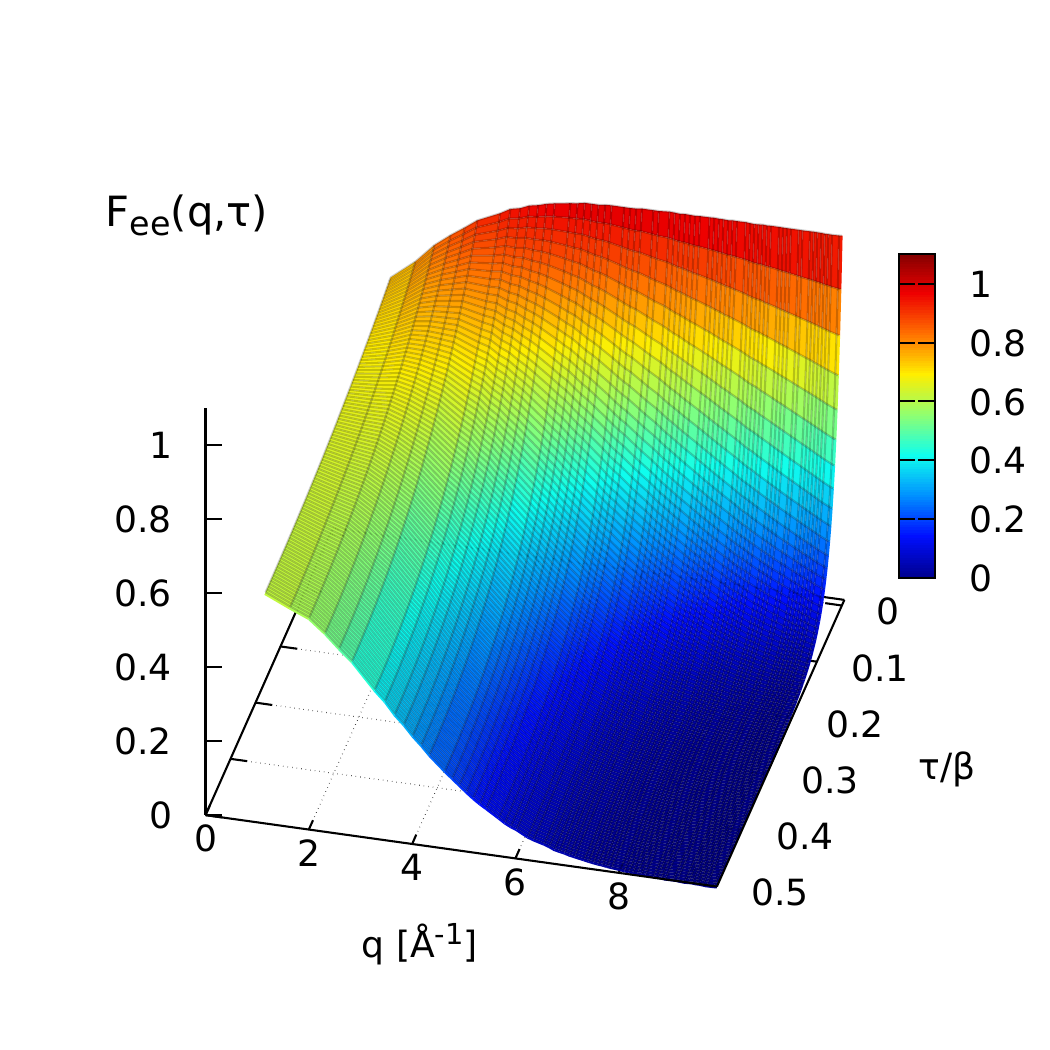}
\caption{\label{fig:ITCF_full_N32_rs2_theta1} \textit{Ab initio} PIMC results for the imaginary-time density--density correlation function $F_{ee}(\mathbf{q},\tau)$ of hydrogen with $N=32$ atoms at $r_s=2$ and $\Theta=1$ shown in the $q$-$\tau$-plane. Taken from Ref.~\cite{Dornheim_LFC_2024} with the permission of the authors.
}
\end{figure}

\begin{figure}\centering
\includegraphics[width=0.48\textwidth]{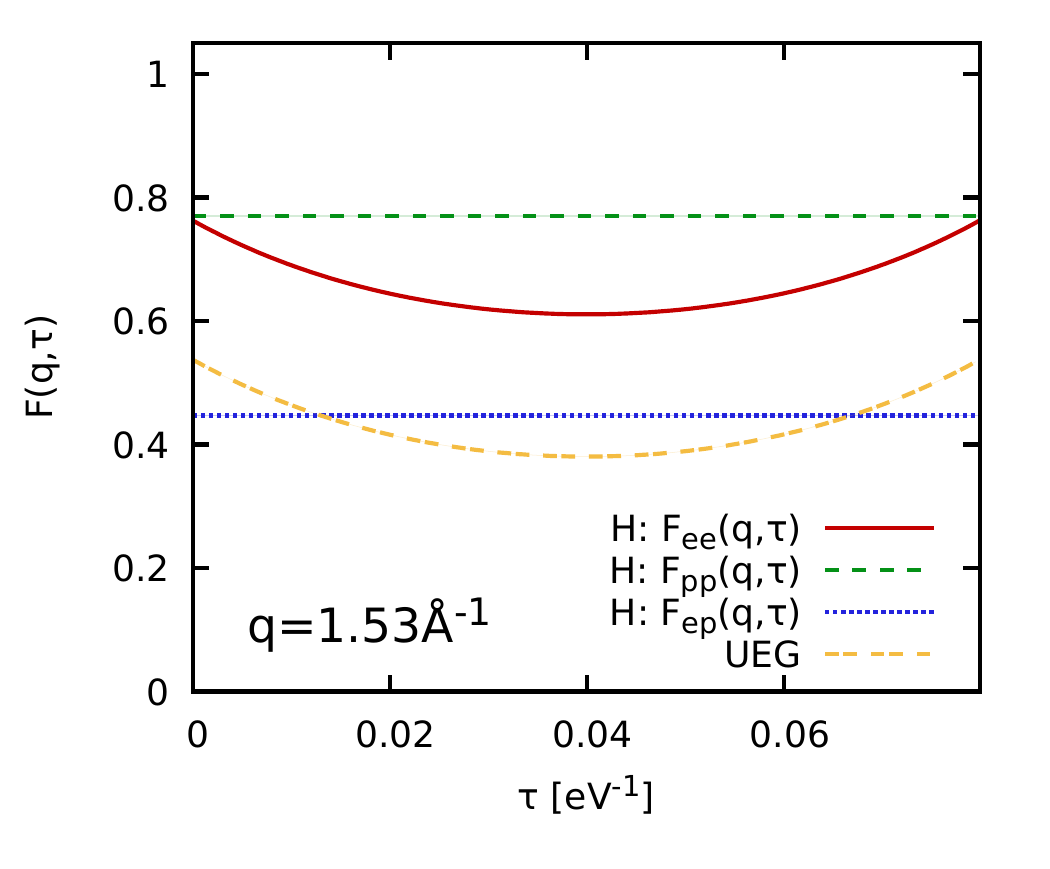}\\\vspace*{-1.3cm}\includegraphics[width=0.48\textwidth]{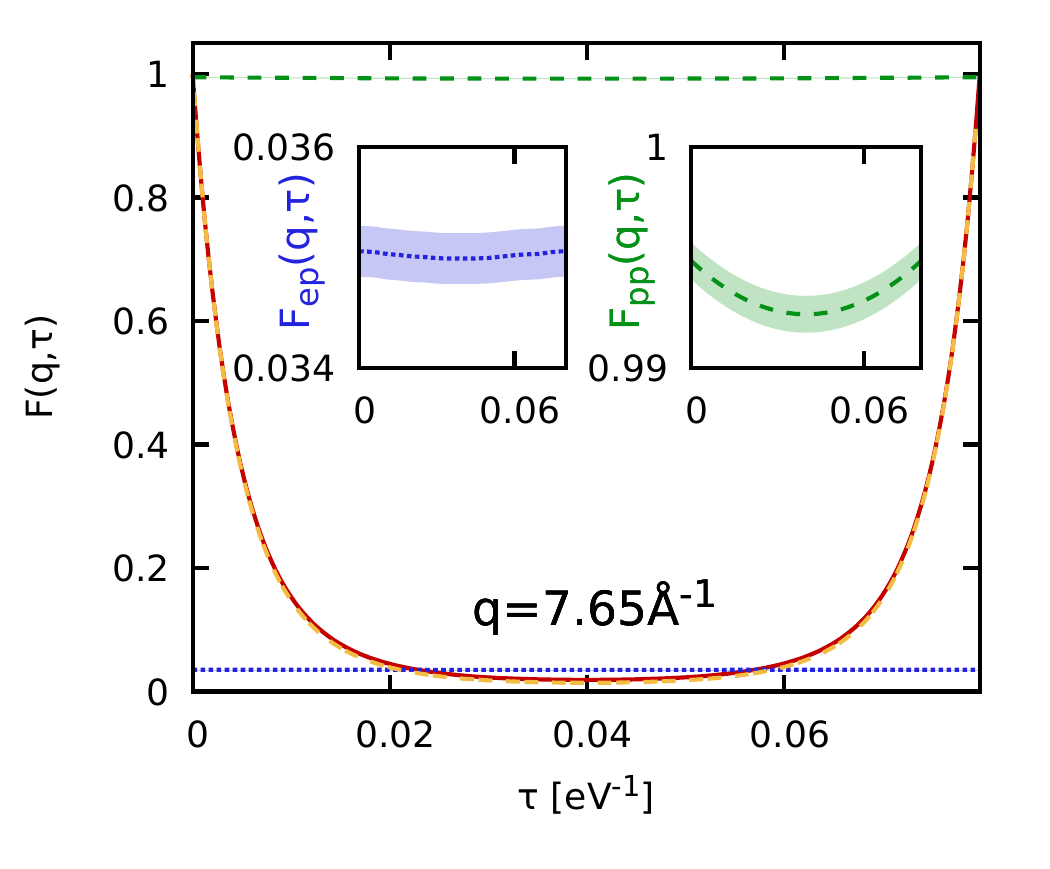}
\caption{\label{fig:H_N14_rs2_theta1_ITCF} \textit{Ab initio} PIMC results for various ITCFs of hydrogen at $r_s=2$ and $\Theta=1$ for $q=\SI{1.53}{\angstrom^{-1}}$ [top] and $q=\SI{7.65}{\angstrom^{-1}}$ [bottom]; solid red: $F_{ee}(\mathbf{q},\tau)$, dashed green: $F_{pp}(\mathbf{q},\tau)$, dotted blue: $F_{ep}(\mathbf{q},\tau)$, double-dashed yellow: UEG~\cite{Dornheim_MRE_2023}. The shaded intervals correspond to $1\sigma$ error bars. The insets in the right panel show magnified segments around $F_{ep}(\mathbf{q},\tau)$ and $F_{pp}(\mathbf{q},\tau)$. Taken from Ref.~\cite{Dornheim_LFC_2024} with the permission of the authors.
}
\end{figure}

In Fig.~\ref{fig:ITCF_full_N32_rs2_theta1}, we show \textit{ab initio} PIMC results for the ITCF of hydrogen for $N=32$ atoms at $r_s=2$ and $\Theta=1$ taken from Ref.~\cite{Dornheim_LFC_2024}.
Clearly, $F_{ee}(\mathbf{q},\tau)$ displays a rich structure, and its main features can be summarized as follows. For $\tau=0$, the ITCF is equal to the static structure factor, $F_{ee}(\mathbf{q},0)=S_{ee}(\mathbf{q})$; the latter is, by definition, the normalization of $S_{ee}(\mathbf{q},\omega)$. The decay of $F_{ee}(\mathbf{q},\tau)$ with $\tau$ becomes increasingly steep for large wavenumbers. It is a direct consequence of the quantum delocalization of the electrons, which is more pronounced on shorter length scales $\lambda=2\pi/q$; it would be entirely absent in a classical system and is substantially less pronounced for the heavier protons, cf.~Fig.~\ref{fig:H_N14_rs2_theta1_ITCF}.
Interestingly, the main trends of $F_{ee}(\mathbf{q},\tau)$ can be explained by a simplified single-particle imaginary-time Gaussian diffusion model~\cite{Dornheim_PTR_2023}.
To get additional insights into the behavior of $F_{ee}(\mathbf{q},\tau)$, we consider the frequency moments of $S_{ee}(\mathbf{q},\omega)$, which are defined as~\cite{Dornheim2023}
\begin{eqnarray}\label{eq:moments}
M_\alpha^{S}  = \int_{-\infty}^\infty \textnormal{d}\omega\ S_{ee}(\mathbf{q},\omega)\ \omega^\alpha\ .
\end{eqnarray}
It is easy to see that the $M_\alpha^{S}$ can alternatively be computed from the ITCF via~\cite{Dornheim2023,Dornheim_PRB_2023,Dornheim_MRE_2023}
\begin{eqnarray}\label{eq:moments_derivative}
M_\alpha^{S}= \frac{\left( -1 \right)^\alpha}{\hbar^\alpha}  \left. \frac{\partial^\alpha}{\partial\tau^\alpha} F_{ee}(\mathbf{q},\tau) \right|_{\tau=0} \ ,
\end{eqnarray}
i.e., as its derivatives with respect to $\tau$ around $\tau=0$.
This relation has recently been used to extract the first five moments of the DSF from \textit{ab initio} PIMC results for the warm dense UEG over a broad range of wave numbers~\cite{Dornheim_moments_2023}. In addition to being interesting in their own right, accurate knowledge of various $M_\alpha^{S}$ is important to constrain the aforementioned analytic continuation [cf.~Eq.~(\ref{eq:Laplace})]~\cite{Filinov2012,tkachenko_book}; the application of such methods to warm dense hydrogen based on highly accurate PIMC input data will be explored in dedicated future works.
A particularly important example is given by the universal f-sum rule~\cite{quantum_theory,Dornheim_LFC_2024}
\begin{eqnarray}\label{eq:fsum}
   \left. \frac{\partial}{\partial\tau}F_{ab}(\mathbf{q},\tau)\right|_{\tau=0} = -\delta_{a,b}\frac{\hbar^2 q^2}{2m_a}\ .
\end{eqnarray}
It implies that the slope of $F(\mathbf{q},\tau)$ around the origin increases quadratically in magnitude with the wave number $q$.
In practice, Eq.~(\ref{eq:fsum}) is also very useful to obtain the proper normalization of a measured XRTS signal, which is a-priory unknown~\cite{dornheim2023xray}, but strictly required to use the ITCF for other purposes, such as the study of electron--electron correlations based on the static structure factor $S_{ee}(\mathbf{q})=F_{ee}(\mathbf{q},0)$, or for the evaluation of the imaginary--time version of the fluctuation--dissipation theorem [Eq.~(\ref{eq:chi_static})].

In Fig.~\ref{fig:H_N14_rs2_theta1_ITCF}, we show various species-resolved ITCFs as a function of $\tau$ for two selected wave numbers. The double-dashed yellow lines correspond to the UEG~\cite{Dornheim_PRL_2020_ESA,Dornheim_PRB_ESA_2021} and have been included as a reference. 
For $q=\SI{1.53}{\angstrom^{-1}}$, $F_{ee}(\mathbf{q},\tau)$ of hydrogen (solid red) is substantially shifted upwards compared to the UEG at the same conditions, whereas this shift is nearly entirely absent for $q=7.65\,$\AA$^{-1}$.
To understand this observation, we decompose the full dynamic structure factor into (quasi-)elastic and inelastic contributions~\cite{Vorberger_PRE_2015},
\begin{eqnarray}\label{eq:Chihara}
    S_{ee}(\mathbf{q},\omega) = \underbrace{S_\textnormal{el}(\mathbf{q},\omega)}_{W_R(\mathbf{q})\delta(\omega)} + S_\textnormal{inel}(\mathbf{q},\omega)\ .
\end{eqnarray}
The former is conventionally attributed to bound electrons and an additional screening cloud of free electrons that follows the protons, whereas the latter contains bound-free (and its reverse, free-bound~\cite{boehme2023evidence}) and free-free transitions within the widely employed Chihara picture~\cite{Chihara_1987,siegfried_review,Gregori_PRE_2003}.
The spectral weight of the elastic feature is given by the Rayleigh weight $W_R(\mathbf{q})=S_{ep}^2(\mathbf{q})/S_{pp}(\mathbf{q})$, which constitutes a direct measure for the degree of electronic localization around the ions.
Clearly, the elastic contribution to the full DSF manifests in the ITCF as a constant shift, which explains the observed differences between the UEG and hydrogen in Fig.~\ref{fig:H_N14_rs2_theta1_ITCF}.

The dashed green and dotted blue curves correspond to $F_{pp}(\mathbf{q},\tau)$ and $F_{ep}(\mathbf{q},\tau)$, which appear to be constant on the depicted scale. The insets in the bottom panel show magnified plots for the larger wave number. Evidently, no $\tau$-dependence can be observed for $F_{ep}(\mathbf{q},\tau)$ within the given uncertainty interval. This is unsurprising as Eq.~(\ref{eq:fsum}) implies that $F_{ep}(\mathbf{q},\tau)$ is constant in linear order.
In contrast, we find a small yet significant $\tau$-decay for $F_{pp}(\mathbf{q},\tau)$; its reduced magnitude compared to the electrons directly follows from the heavier mass.

\subsubsection{Static density response of hydrogen\label{sec:full_hydrogen_response}}

\begin{figure}\centering
\includegraphics[width=0.44\textwidth]{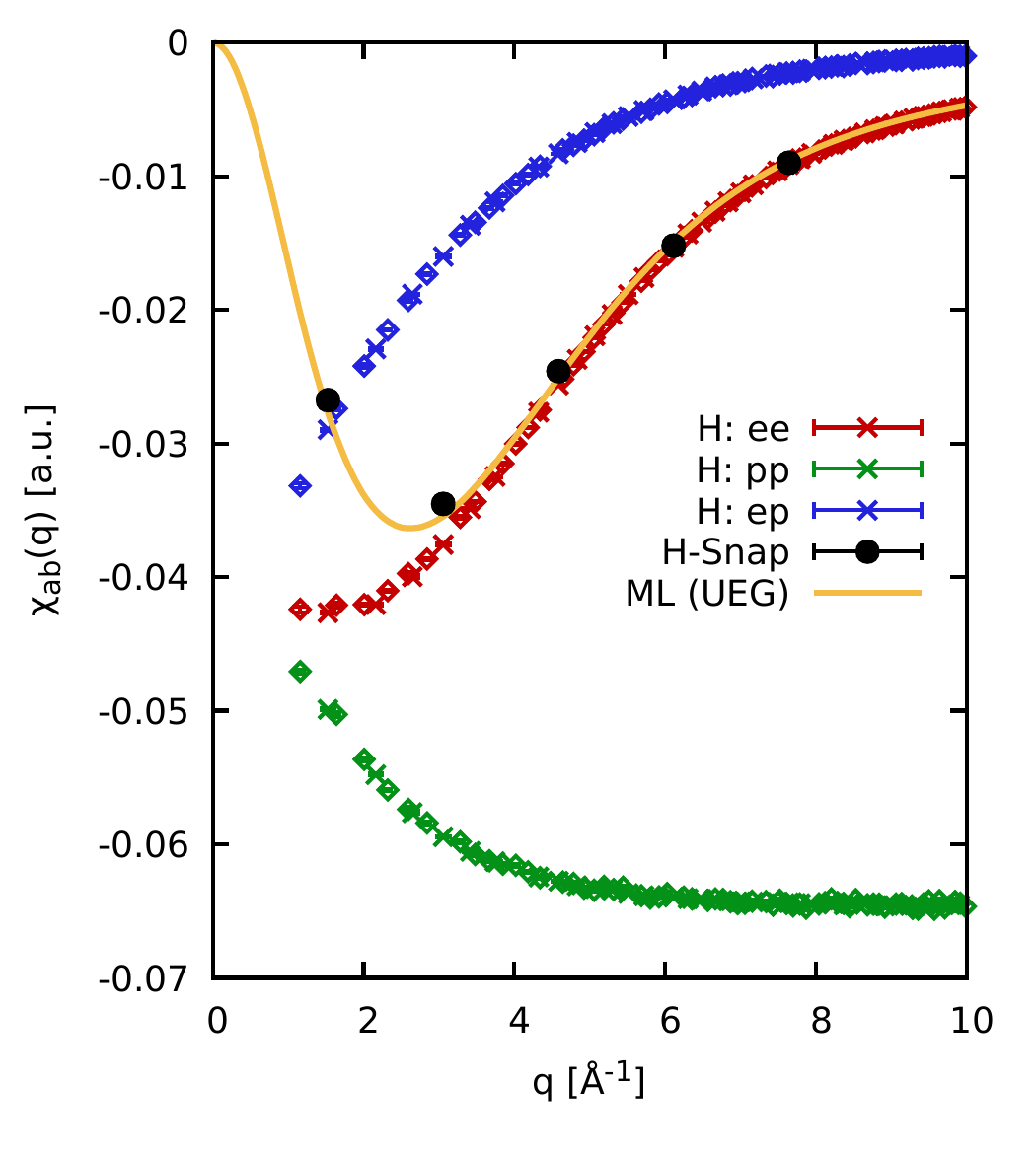}
\caption{\label{fig:Chi_ab} \textit{Ab initio} PIMC results for the species-resolved static density response of hydrogen at $r_s=2$ and $\Theta=1$. Red, green, and blue symbols: $\chi_{ee}(\mathbf{q})$, $\chi_{pp}(\mathbf{q})$, and $\chi_{ep}(\mathbf{q})$ for full hydrogen evaluated from the ITCF via Eq.~(\ref{eq:chi_static}); black circles: electronic density response of fixed proton snapshot, see Ref.~\cite{Bohme_PRL_2022}; solid yellow line: machine-learning representation of $\chi_{ee}(\mathbf{q})$ of the UEG model~\cite{dornheim_ML} at the same conditions. Adapted from Ref.~\cite{Dornheim_LFC_2024} with the permission of the authors.
}
\end{figure}

A key application of the ITCF is given by the imaginary-time version of the fluctuation--dissipation theorem~\cite{bowen2,Dornheim_MRE_2023},
\begin{eqnarray}\label{eq:chi_static}
\chi_{ab}(\mathbf{q}) = - \sqrt{n_a n_b} \int_0^\beta \textnormal{d}\tau\ F_{ab}(\mathbf{q},\tau)\ .
\end{eqnarray}
In practice, it is thus possible to obtain the full $q$-dependence of the static density response from a single PIMC simulation of the unperturbed system. This is a considerable advantage over the direct perturbation approach governed by the modified Hamiltonian Eq.~(\ref{eq:H_perturbed}).
For completeness, we note that similar relations exist for various nonlinear response functions~\cite{Dornheim2023,Dornheim_JCP_ITCF_2021,dornheim_cpp22}.

In Fig.~\ref{fig:Chi_ab}, we show the species-resolved static density response of hydrogen for $r_s=2$ and $\Theta=1$, i.e., for the same conditions explored in Sec.~\ref{sec:ITCF}. 
Let us first consider the blue symbols that show PIMC results for $\chi_{ep}(\mathbf{q})$, i.e., the response of the electrons to a perturbation of the protons (or vice versa). It monotonically decays with $q$, as the electronic localization around the ions becomes less important on small length scales.
The green symbols show the static proton response function, $\chi_{pp}(\mathbf{q})$. It exhibits the opposite trend compared to $F_{ep}(\mathbf{q},\tau)$, which can be understood as follows: in the limit of large $q$, $F_{pp}(\mathbf{q},\tau)\approx S_{pp}(\mathbf{q})$ approaches one; its small $\tau$-decay shown in Fig.~\ref{fig:H_N14_rs2_theta1_ITCF} is negligible. Equation~(\ref{eq:chi_static}) then implies the classical short wavelength limit, $\lim_{q\to\infty}\chi_{pp}^\textnormal{cl}(\mathbf{q})=-n\beta$, which explains the apparent convergence of the green symbols in Fig.~\ref{fig:Chi_ab}.
For smaller $q$, electronic screening of the effective proton--proton interaction leads to a reduction of the density response, which attains a finite value in the limit of $q\to0$.

\begin{figure*}\centering
\includegraphics[width=0.85\textwidth]{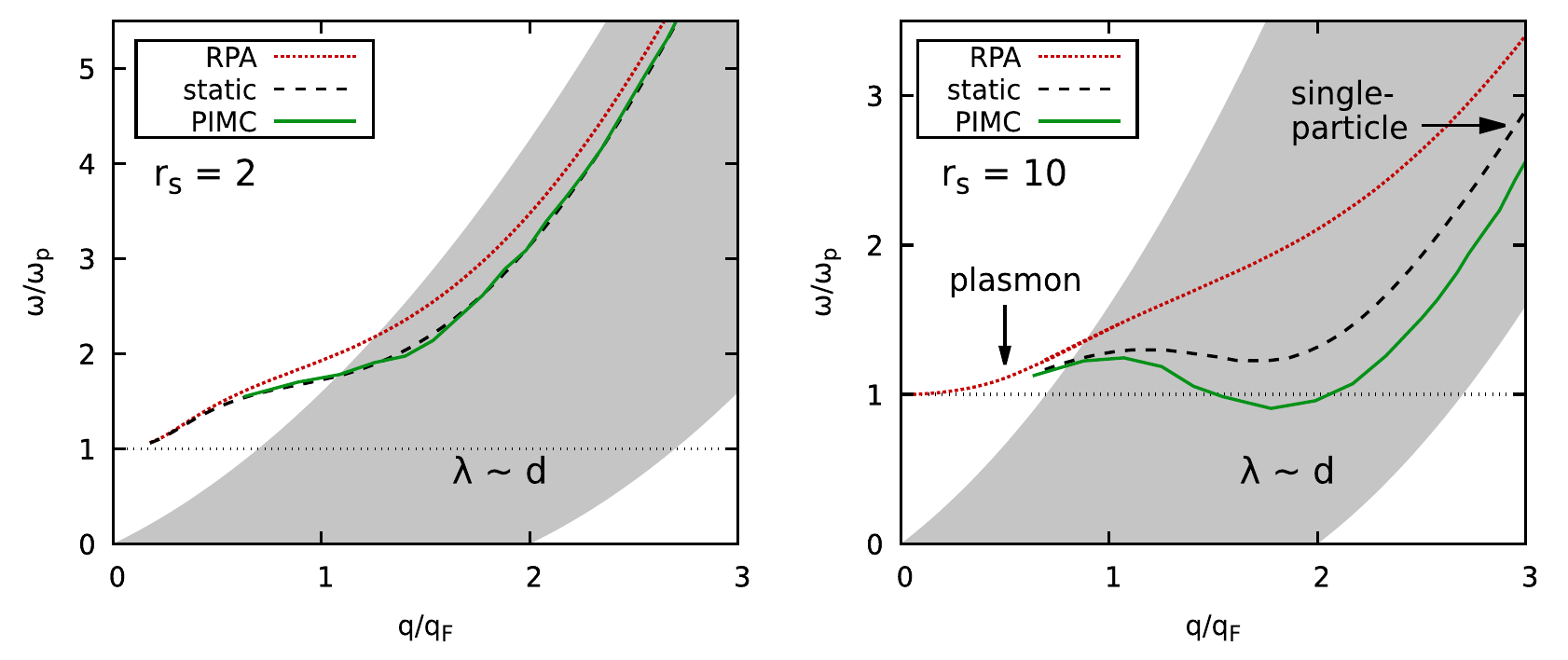}
\caption{\label{fig:roton_UEG}  Position of the maximum of $S(q,\omega)$ for jellium at $\Theta=1$ computed within the RPA (dotted red line), the \emph{static approximation} (cf.~Eq.~(\ref{eq:static_approximation}), Ref.~\cite{Dornheim2018b}, dashed black line), and from the full analytic continuation of PIMC results~\cite{Dornheim2018b,groth_prb_19} (solid green line) for $r_s=2$ (left) and $r_s=10$ (right). 
The shaded grey area indicates the pair continuum~\cite{mahan1990many}.
In all cases, the DSF exhibits the well-defined collective plasmon excitation for $q\to0$ and a parabolic single-particle dispersion for $q\gg q_\textnormal{F}$. The \emph{roton feature} occurs at intermediate $q$ where the wavelength $\lambda=2\pi/q$ is comparable to the average interparticle distance $d$. It is a direct consequence of the exchange--correlation induced alignment of pairs of electrons, cf.~Fig.~\ref{fig:pair_alignment}, and is, therefore, not captured by the RPA.
Taken from P.~Hamann \emph{et al.}~\cite{hamann_prr_23} with the permission of the authors. 
}
\end{figure*}

Let us next consider the red symbols that display the static electronic density response, $\chi_{ee}(\mathbf{q})$, of full two-component hydrogen.
In stark contrast to $\chi_{pp}(\mathbf{q})$, it monotonically decays for $q\gtrsim2\,$\AA$^{-1}$. This is exclusively due to the quantum delocalization of the electrons, as $F_{ee}(\mathbf{q},\tau)$ vanishes increasingly fast with $\tau$ for large wave numbers.
In practice, the response of the electrons vanishes when the wavelength of the external perturbation becomes comparable to their thermal wavelength $\lambda_\beta$.
In fact, the protons are subject to the same effect, but they manifest only for substantially larger $q$ due to their heavier mass.
In the limit of small $q$, $\chi_{ee}(\mathbf{q})$ plateaus; a hypothetical minimum at a finite value of $q$ is possible but cannot be resolved with the considered $N=32$ hydrogen atoms. 
To understand the rich physics that guides the electronic density response in this regime, it is convenient to compare it with other models to isolate a particular effect.
The solid yellow curve shows $\chi_{ee}(\mathbf{q})$ of the UEG~\cite{dornheim_ML} at the same conditions. It agrees well with the red symbols for $q\gtrsim4\,$\AA$^{-1}$, but increasingly deviates for long wavelengths.
Indeed, it completely vanishes for $q\to0$ due to the perfect screening of the UEG~\cite{kugler_bounds}.
The latter breaks down for full two-component hydrogen as the unperturbed protons, on average, follow the harmonically perturbed electrons; this leads to an effective screening of the electron--electron interaction that facilitates the larger response of hydrogen compared to the UEG on large-length scales.
An alternative, though fully equivalent, way to understand this is to recall the direct relation of $\chi_{ee}(\mathbf{q})$ to the inverse frequency moment of the DSF~\cite{Vitali_PRB_2010},
\begin{eqnarray}\label{eq:inverse_moment}
    \chi_{ee}(\mathbf{q}) = - 2n_e \int_{-\infty}^\infty \textnormal{d}\omega\ \frac{S_{ee}(\mathbf{q},\omega)}{\omega}\ .
\end{eqnarray}
The static density response is thus highly sensitive to the low-frequency behavior of $S_{ee}(\mathbf{q},\omega)$, which, for an electron--ion system, is dominated by the elastic feature, cf.~Eq.~(\ref{eq:Chihara}).
The larger density response of hydrogen is thus a direct consequence of the electronic localization around the ions described by the Rayleigh weight, $W_R(\mathbf{q})$.
Finally, the black dots in Fig.~\ref{fig:Chi_ab} correspond to the linear density response of the electrons in a fixed proton snapshot that has been discussed in Sec.~\ref{sec:static_density_response} above. They are in close qualitative agreement with the UEG at these conditions, but, unsurprisingly, miss the rich interplay between the electrons and protons of the full electronic density response, i.e., the red symbols.

In addition to being interesting in their own right, highly accurate PIMC results for hydrogen, and potentially other light elements, are important for several reasons, e.g.: i) as a basis to compute the exchange--correlation kernel $K_\textnormal{xc}(\mathbf{q})$---a key input for TDDFT~\cite{Moldabekov_PRR_2023,ullrich2012time}, ionization potential depression models~\cite{Zan_PRE_2021}, and the construction of advanced XC-functionals for thermal DFT simulations~\cite{pribram}; ii) to benchmark dynamic simulations and widely used approximate models such as the Chihara decomposition~\cite{Gregori_PRE_2003,boehme2023evidence}, cf. Sec.~\ref{sssub:chihara}; and iii) as an unambiguous prediction of upcoming XRTS measurements with fusion plasmas and hydrogen jets~\cite{Fletcher_Frontiers_2022,Zastrau}.

\subsubsection{Prediction of a roton-type feature for jellium at low density\label{sec:roton}}

In Sec.~\ref{sec:ITCF}, we have discussed the dynamic density response of hydrogen based on PIMC results in the imaginary-time domain.
%For jellium, on the other hand, 
For jellium, it is possible to carry out a reliable analytic continuation, cf. T.~Dornheim \emph{et al.}~\cite{Dornheim2018b,groth_prb_19,Dornheim_PRE_2020,hamann_prb_20}, 
to obtain highly accurate results for the DSF over a broad range of parameters.
This study has revealed two important points: 1) the \emph{static approximation} Eq.~(\ref{eq:static_approximation}) is remarkably accurate in the case of the UEG for high to moderate densities and 2) the position of the maximum in $S_{ee}(\mathbf{q},\omega)$---here simply denoted as $\omega(q)$---exhibits a non-monotonous dependence on the wave number $q$ in the low-density regime, with a pronounced minimum around $q\approx 2q_\textnormal{F}$.
Both points are illustrated in Fig.~\ref{fig:roton_UEG}, where we show $\omega(q)$ at $\Theta=1$, for $r_s=2$ (left) and $r_s=10$ (right). More specifically, the dotted red, dashed black, and solid green curves have been obtained based on the RPA [$K_\textnormal{xc}(\mathbf{q},\omega)\equiv0$], \emph{static approximation} [$K_\textnormal{xc}(\mathbf{q},\omega)\equiv K_\textnormal{xc}(\mathbf{q},0)$], and the full, reconstructed results for $K_\textnormal{xc}(\mathbf{q},\omega)$, respectively. For both conditions, all curves exactly reproduce the collective plasmon excitation in the limit of $q\to0$. Similarly, we find the same parabolic dependence of $\omega(q)$ on $q$ in the single-particle limit of $q\gg q_\textnormal{F}$.

Due to the role of $r_s$ as the quantum coupling parameter, electronic exchange--correlation effects are of moderate importance in the regime of metallic densities, and the systematic deviation of RPA to the other curves is comparably small; the \emph{static approximation} can, in fact, hardly be distinguished from the exact results in this regime.
This situation changes dramatically at $r_s=10$, which is located at the margins of the strongly coupled electron liquid regime~\cite{dornheim_electron_liquid}. Indeed, the exact PIMC-based results for $\omega(q)$ exhibit a pronounced minimum for intermediate wave numbers, which phenomenologically resembles the well-known \emph{roton feature} of quantum liquids such as ultracold helium~\cite{Trigger,Godfrin2012,cep,Ferre_PRB_2016,Dornheim_SciRep_2022}.
The \emph{static approximation} qualitatively captures this nontrivial feature, whereas it is completely absent from the mean-field based RPA results.

\begin{figure}\centering
\includegraphics[width=0.49\textwidth]{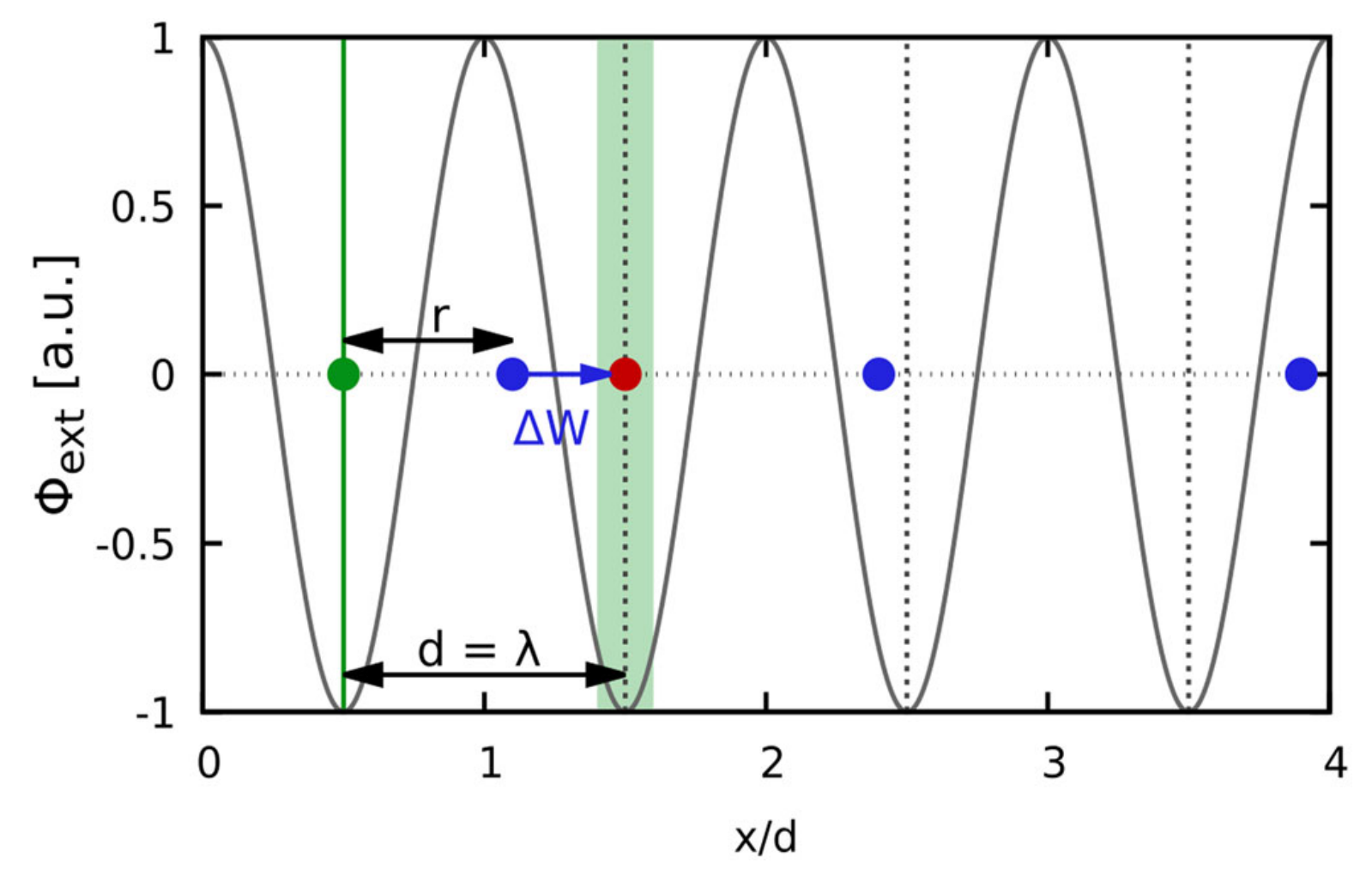}
\caption{\label{fig:pair_alignment} Schematic illustration of electronic pair alignment causing the \emph{roton feature} shown in Fig.~\ref{fig:roton_UEG}. The green bead shows an arbitrary reference particle, and the blue bead other electrons which are, on average, disordered. Applying an external harmonic perturbation [cf.~Eq.~(\ref{eq:H_perturbed}), grey curve] with a wavelength $\lambda$ that is commensurate to the average interparticle distance $d$ changes the free energy landscape
and reinforces the minimization of the interaction energy $W$. The corresponding change $\Delta W$ of the latter qualitatively explains the negative \emph{roton shift}. Taken from T.~Dornheim \emph{et al.}~\cite{dornheim_comphys_22}. 
}
\end{figure}

To understand the physical origin of the roton minimum in the UEG, T.~Dornheim \emph{et al.}~\cite{dornheim_comphys_22} have proposed to consider the fluctuation--dissipation theorem [cf.~Eq.~(\ref{eq:FDT})], which implies that the DSF is fully determined by the linear density response of the system to an infinitesimal external harmonic perturbation.
In Fig.~\ref{fig:pair_alignment}, the latter situation is schematically illustrated, with the green bead corresponding to an arbitrarily chosen reference particle. In addition, the blue beads correspond to other particles in the system, which are, on average, disordered at $r_s=10$ and $\Theta=1$ [note the absence of a pronounced peak in $S_{ee}(\mathbf{q})$ discussed e.g.~in Ref.~\cite{dornheim_comphys_22}].
The dark grey cosinusoidal curve shows an external harmonic perturbation along the $x$-direction, where the wavelength is of the same order as the average interparticle distance, i.e., $\lambda=2\pi/q\approx d$.
For any value of $q$, the electrons are energetically incentivized to align themselves to the minima of the perturbation. For small $q$ where $\lambda\gg d$, this trend is suppressed by the perfect screening in the UEG~\cite{kugler_bounds}. Conversely, the alignment is negligible for $q\gg q_\textnormal{F}$ as it happens on increasingly small length scales, $\lambda\ll d$.
For the situation that is depicted in Fig.~\ref{fig:pair_alignment}, the alignment of the leftmost blue bead to the cosine perturbation coincides with a minimization of the interaction energy $W$, i.e., a negative change $\Delta W$.
It has been shown in Refs.~\cite{dornheim_comphys_22,Dornheim_JCP_2022} that this alignment of pairs of electrons explains the \emph{roton minimum} shown in Fig.~\ref{fig:roton_UEG}.
Further, this shift in the interaction energy is an electronic XC-effect and, therefore, fundamentally not captured by the RPA that entails a mean-field description of the density response.
We note that these findings are consistent with similar observations for jellium in the ground-state~\cite{Takada_PRL_2002,Takada_PRB_2016,koskelo2023shortrange}.

While being interesting in their own right, the results discussed in this section have been obtained for the UEG model, cf. Sec.~\ref{subsec:jellium}. It is thus important to ask if the \emph{roton minimum} can be observed in a real system, with warm dense hydrogen being a promising candidate~\cite{hamann_prr_23}.

\subsubsection{Prediction of a roton-type feature in warm dense hydrogen\label{sec:roton_H}}

Roton excitations are well-known from superfluid helium, but a similar ``roton feature'' was recently also observed in simulations of the electron liquid~\cite{dornheim_comphys_22}, cf. Sec.~\ref{sec:roton}. Interestingly, it was predicted by P.~Hamann \textit{et al.} that this feature should be observable also in the plasmon dispersion of hydrogen -- in the metallic (ionized plasma) phase~\cite{hamann_prr_23}. In the following, we briefly outline this result.

As a first step to investigate how the dispersion behavior changes when considering a two-component partially ionized plasma compared to jellium, we treat electron-ion interactions in the relaxation time approximation (RTA). This corresponds to introducing a BGK collision operator as discussed in Sec.~\ref{sss:qke}:
%\[ \left(\frac{\partial f}{\partial t}\right)_{\rm coll} = -\frac{1}{\tau} ( f - f_{\rm eq} ) \]
which contains a characteristic relaxation time $\tau$ for the electron momentum distribution which is the inverse
of the electron-ion collision frequency,
$\nu = \tau^{-1}$.

Solving the quantum kinetic equation of the electrons, Eq.~(\ref{eq:boltzmann-eq}) with the collision integral (\ref{eq:i-rta}) in linear response, i.e. for a weak monochromatic field,
 one derives the following expression for the dielectric function \cite{ludwig_jpcs_10}, which is due to N.~D.~Mermin \cite{mermin_prb_70} and is referred to as the Mermin dielectric function,
\begin{equation}\label{eq:mermin_df}
\epsilon^{\rm M}(\vec{q},\omega) = 1 + \frac{(1+i\nu/\omega)(\epsilon(\vec{q},\omega+i\nu)-1)}{1+i\,\frac{\nu}{\omega}\,
\frac{\epsilon(\vec{q},\omega+i\nu)-1}{\epsilon(\vec{q},0)-1}
}\,,
%(\epsilon(\vec{q},\omega+i\nu)-1)/(\epsilon(\vec{q},0)-1)
\end{equation}
were, $\epsilon(\vec{q},\omega)$ is the dielectric function of jellium, which is recovered in the limit of vanishing electron-ion collisions, $\nu \rightarrow 0$. The original derivation uses the Lindhard function (random-phase approximation), further research has shown this approach to be compatible with dielectric functions that include electronic correlations, e.g. are described using local field corrections $G(\vec{q},\omega)$ \cite{Fortmann_PRE_2010}.
There exist various extensions of the Mermin dielectric function that introduce complex and frequency-dependent collision frequencies $\nu(\omega)$, i.e. collision rates that depend on the energy scale \cite{selchow2001,Reinholz_pre_2000}.
Expressions for the collision frequency can be derived using approximations of quantum kinetic theory -- either from reduced density operators (BBGKY-hierarchy) or nonequilibrium Green functions \cite{bonitz_qkt,kwong_prl_00}, cf. Sec.~\ref{sss:qke}. Several approximations were discussed in Ref.~\cite{Moldabekov2019}. The analysis of Ref.~\cite{hamann_prr_23}  used the dynamically screened Born approximation but neglected the plasmon peak in the integrand, for numerical reasons,
\begin{equation}\label{eq:collision}
    i\nu(\omega) = \frac{\omega_p}{6\pi^2 n_e\omega} \int\limits_0^\infty \mathrm{d}q\, q^6 V_s^2(q) S_{ii}(q) \left[\epsilon(q,\omega) - \epsilon(q,0) \right].
\end{equation}
Here the ion-ion static structure factor $S_{ii}(q)$ enters in the calculation of the potential at which electrons are scattered.

Another approach interprets the Mermin dielectric function as a generalized version of the Drude formula, which matches in the optical limit
\begin{equation}
    \lim\limits_{q\to 0} \sigma^M(q,\omega) = \frac{\omega_p^2}{-i\omega + \nu(\omega)}.
    \nonumber
\end{equation}
This is particularly useful in the context of DFT-MD simulations where it is possible to obtain results for the dynamic conductivity via the Kubo-Greenwood (KG) formula \cite{Gajdo_vasp_2006}, cf. Sec.~\ref{sss:dft-kubo-greenwood}, from which collision frequencies can be extracted and used in Eq.~(\ref{eq:mermin_df}), thereby extrapolating such results to finite wave vectors \cite{plagemann2012dynamic}.  
However, in the absence of true first-principles calculations providing electron-ion local field corrections, such an extrapolation is of unknown quality and has, in some cases, been shown to be not reliable~\cite{ramakrishna2019}. Further research into transport properties of hydrogen under high pressure and, in particular, comparisons between KG approaches and linear response TDDFT will shed light on this important issue, see Sec.~\ref{sec:transport}.

\begin{figure}\centering
\includegraphics[width=0.42\textwidth]{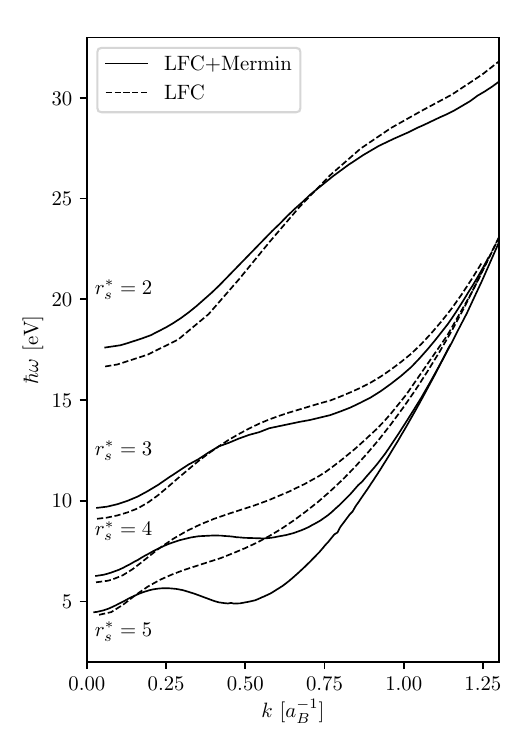}
\caption{Emergence of the negative dispersion of the peak of $S(q,\omega)$ for partially ionized hydrogen at $T=50\,000$ K. Results for jellium using the static LFC (dashed) are compared to results which additionally account for collisions with ions using the Mermin dielectric function, Eq.~\eqref{eq:mermin_df} [solid]. In contrast to Fig.~\ref{fig:roton_UEG}, here absolute units are used. At this temperature, a local minimum would start to emerge for jellium at significantly lower densities only, corresponding to $r_s^* > 6$.}
\label{fig:Mermin}
\end{figure}

The dispersion of the peak position of the dynamic structure factor is shown in Fig.~\ref{fig:Mermin}. There we compare results for jellium, using the static local field correction (dashed line, labeled LFC), and calculations including, in addition, collisions with the ionic background based on the Mermin dielectric function, Eq.~\eqref{eq:mermin_df} [solid line]. Collision frequencies were calculated using Eq.~(\ref{eq:collision}) while setting $S_{ii}(q)=1$~\footnote{Changes in the collision frequency due to the ion-ion structure factor cause only minor changes in the dynamic structure factor, in general, and no visible changes for the peak position, in particular, for the considered densities and temperatures.}.

\begin{figure}\centering
\includegraphics[width=0.45\textwidth]%{ROTON_SHIFT.png}
{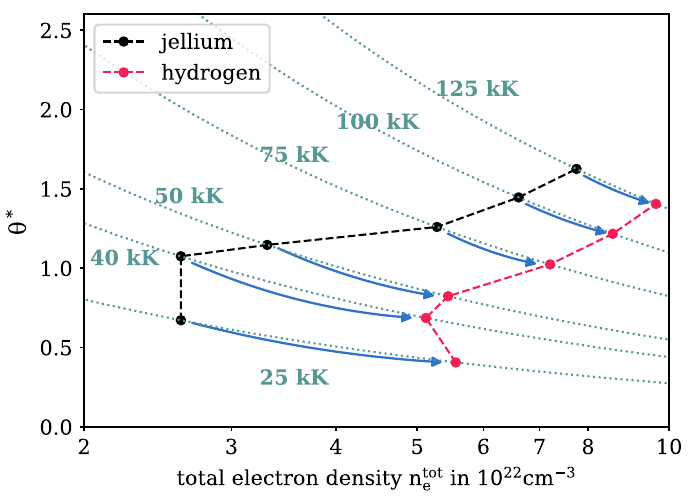}
\caption{\label{fig:roton_shift_h} 
Change of the line indicating roton behavior (to its left) from jellium to hydrogen. The arrows illustrate the effect of partial ionization at a given temperature. Updated version of figure from P.~Hamann \emph{et al.}~\cite{hamann_prr_23}, using degree of ionization data from Ref.~\cite{filinov_pre_23}, cf. Fig.~\ref{fig:ion-degree-mil-fil}.
}
\end{figure}

The expected location of the roton feature in jellium and dense hydrogen is shown in Fig.~\ref{fig:roton_shift_h}; the location is to the left of the black (for jellium) or red dashed (for hydrogen) curves, respectively. As the roton is expected to occur in the unbound part of the electrons that form the free electrons gas within the hydrogen sample, the partial ionization was taken into account, based on PIMC simulations~\cite{filinov_pre_23}. The red data points for several isotherms indicate the maximum total electron densities to be around $6\ldots 10\times 10^{22}$~cm$^{-3}$ ($0.1\ldots \SI{0.17}{\gram\per\cubic\centi\meter}$). The necessary densities can be achieved in experiments for instance using a hydrogen jet~\cite{Zastrau,Fletcher_Frontiers_2022}.

The resolution in angular and energy space that is necessary to observe the predicted roton feature in XRTS experiments~\cite{hamann_prr_23} will only be available at modern XFEL facilities that combine a brilliant, narrow-band X-ray laser with fast detectors and a high repetition optical laser and target delivery system~\cite{Tschentscher_2017,LCLS_2016}. In addition, advanced spectral techniques~\cite{mcbride_2018,gawne2024ultrahigh} might be necessary and, naturally, a theory-free determination of basic parameters such as temperature and density~\cite{Dornheim_T_2022,Dornheim_T_follow_up,dornheim2023xray,dornheim2024unraveling}.
%}

\subsubsection{Ion-acoustic mode in partially ionized hydrogen}\label{sssec:i-acoustic_H}
After analyzing correlation effects in the electron plasmon dispersion in dense hydrogen, we now turn to the low-frequency collective oscillations of the protons.\\

\begin{figure}
\includegraphics{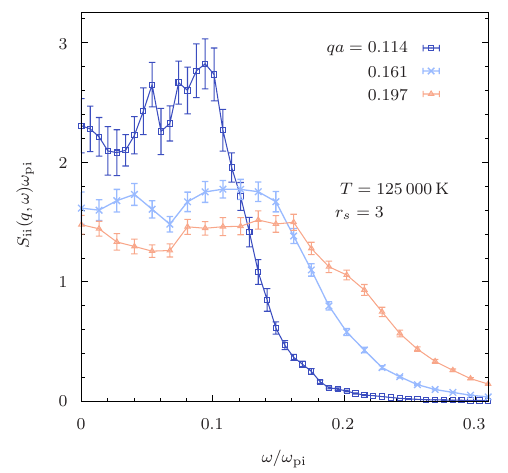}
\caption{\label{fig:kelbg_dsf} Ion-ion dynamic structure factor $S_\text{ii}(q,\omega)$ at $r_s=3$ and $T=\SI{125000}{\kelvin}$ for various wave vectors $q$, where $a$ is the Wigner Seitz radius. Frequencies are given in units of the ion plasma frequency, $\omega_\text{pi}$.}
\end{figure}

We have used the LAMMPS code~\cite{thompson2022} to perform molecular dynamics simulations of a hydrogen plasma with the improved Kelbg potential (Sec.~\ref{ssubsec:pair-pot-md}), similar to previous simulations in Refs.~\cite{filinov_pre04,bonitz_cpp05}. Compared to DFT-MD simulations (Sec.~\ref{subsec:dft}), they are computationally much cheaper and allow one to treat much larger system sizes with thousands of protons and electrons. In addition, they do not rely on the Born-Oppenheimer approximation and treat electrons dynamically (non-adiabatic). On the other hand, they are applicable only at sufficiently high temperatures and low densities. A comparison with PIMC simulations~\cite{filinov_pre04} indicates that the improved Kelbg potential (IKP) is applicable for temperatures well above \SI{60000}{\kelvin} at $r_s=4$ and $r_s=6$. 
New tests of the IKP were presented in Sec.~\ref{sss:rpimc-vs-fpimc} and provided more details on the accessible density-temperature range. They also confirm the MD simulations should be accurate for the parameters of Fig.~\ref{fig:kelbg_dsf}.

The MD simulations give direct access to various time correlation functions and their spectra, including the ion-ion dynamic structure factor. When calculating dynamic properties, one must keep in mind, however, that the (improved) Kelbg potential was designed to reproduce thermodynamic properties, and its use for the calculation of dynamic quantities remains to be tested. 

In Fig.~\ref{fig:kelbg_dsf}, we show preliminary results from a simulation with $40\,000$ protons and electrons for the ion-ion DSF of a hydrogen plasma at $r_s=3$ and $T=\SI{125000}{\kelvin}$. At the smallest wave number, the data indicate the formation of an ion-acoustic mode well below the ion plasma frequency, which appears to vanish at larger $q$. Longer simulations are still required to reduce the noise and to verify the peak formation. This would provide access to the ion-acoustic speed, see also Ref.~\cite{bonitz_cpp05}. Experimentally, acoustic modes in methane under WDM conditions have recently been resolved in experiments using inelastic x-ray scattering~\cite{white2023speed}.

\subsection{Transport and optical properties}\label{sec:transport}
We now turn to the transport (e.g. electrical and thermal conductivity) and optical (e.g. opacity) properties of dense hydrogen in thermodynamic equilibrium. Here, a variety of simulation methods is available. For a summary of the relevant properties and simulation approaches, see Sec.~\ref{sss:simulations-hierarchy} and  Tab.~\ref{tab:observables}.

\subsubsection{Previous comparisons of transport coefficients}\label{sss:previous-comparisons-transport}
At a transport coefficients comparison workshop that took place in 2016 \cite{grabowski2020}, participants were invited to compute electrical conductivity, thermal conductivity, viscosity, diffusion, and stopping power for three plasma compositions (H, C, and CH) with densities varying from \SI{0.1}{\gram\per\cubic\centi\meter} to \SI{100}{\gram\per\cubic\centi\meter} and temperatures from 0.2 eV to 2 keV. The models included KS-DFT, OF-DFT, TDDFT, average atom models, and various analytical models.
All models tested showed significant variation (factors of three or more) for a portion of the density-temperature space. Agreement among models was generally higher in the classical weakly coupled regime. At low temperatures and densities, uncertainties in the ionization state were the major source of disagreement. For most transport coefficients, typical best-case variations among codes were 20 percent in the weakly coupled regime, factors of two in warm dense matter, and worsening to factors of ten or more, at low temperatures. 
A follow up transport workshop took place in 2023 and significantly refined the analysis for a broad group of materials., cf. Ref.~\cite{stanek_pop_24}. One conclusion was that the difference in the DC electrical conductivity was at worst a factor of seven between all models and a factor of two between similar models. 

The workshops mentioned above did not attempt to rank the different simulations or to identify the most accurate ones. In contrast, in the present paper, we have attempted to evaluate the accuracy of different approaches -- for hydrogen and a limited range of parameters -- and presented comparisons of the associated thermodynamics results (which are also the basis for transport applications) in Secs.~\ref{sec:td}. A clear preference has been observed for first-principles methods, i.e. QMC and DFT. In the following sections we focus on DFT results and theoretical issues associated with the method since QMC does not give access to transport and optical properties [see Tabs.~\ref{tab:methods} and \ref{tab:observables}]. In Sec.~\ref{sss:dft-kubo-greenwood} we discuss the Kubo-Greenwood formula that is the basis of the KS-DFT approach to transport quantities. This is followed by state of the art results for the electrical and thermal conductivity in Sec.~\ref{sss:results-sigma} and opacity, in Sec.~\ref{sss:results-opacity}. We conclude with a discussion of the important issue of scattering effects in DFT simulations, in Sec.~\ref{sss.collisions-dft}. 

\subsubsection{KS-DFT approach. Kubo-Greenwood formula}%\label{sss:dft-conductivity}
\label{sss:dft-kubo-greenwood}

The electronic transport properties such as electrical and thermal conductivity can be derived from linear response theory ~\cite{kubo1957statistical,greenwood1958boltzmann}. The resulting frequency-dependent Onsager coefficients can be expressed as~\cite{Holst2011}
\begin{align}
\MoveEqLeft L_n(\omega) = \frac{2\pi(-e)^{2-n}}{3V\omega} \sum_{\textbf{k}\mu\nu} \left| \bra{\textbf{k}\nu}\hat{\textbf{v}}\ket{\textbf{k}\mu}\right|^2 (f_{\textbf{k}\nu}-f_{\textbf{k}\mu}) \nonumber\\ 
&\times \left(\frac{E_{\textbf{k}\mu}+E_{\textbf{k}\nu}}{2} - h\right)^n \delta(E_{\textbf{k}\mu}-E_{\textbf{k}\nu}-\hbar\omega)\,,
\label{eq:onsager}
\end{align}
where $\omega$ is the frequency and $V$ denotes the volume of the simulation box. $E_{\textbf{k}\mu}$ and $f_{\textbf{k}\mu}$ describe the energy eigenvalue and the occupation number of the respective Bloch state $\ket{\textbf{k}\mu}$. The states are calculated within the Kohn-Sham framework of DFT together with the transition matrix elements with the velocity operator $\bra{\textbf{k}\nu}\hat{\textbf{v}}\ket{\textbf{k}\mu}$. The enthalpy per electron $h$ can be directly obtained from the chemical potential and the entropy, which are self-consistently calculated within the DFT as the free-energy is minimized. 
The real parts of the electrical conductivity, $\sigma$, and and the thermal conductivity, $\lambda$,
are subsequently calculated via
\begin{align}
\sigma &= L_0(\omega)  \,,
\label{eq:sigma-kg}\\
\lambda &= \frac{1}{T}\left(L_2(\omega) - \frac{L_{1}^2(\omega)}{L_0(\omega)}\right)  \,.
\label{eq:lambda-kg}
\end{align}
When calculating the DC limit, $\omega \to 0$, of these quantities, great care has to be taken of the representation of the $\delta$ function in Eq.~(\ref{eq:onsager}), which is usually done by a gaussian whose width has to be chosen to the typical distance between bands in the eigenvalue spectrum. Additionally, Eqs.~(\ref{eq:sigma-kg}) and (\ref{eq:lambda-kg}) are generally tensor quantities, so components need to be combined, which is only trivial for cubic systems. Also, note that most DFT codes are only calculating transitions at the same $\textbf{k}$ point ($\textbf{k} \rightarrow \textbf{k}$) and might not be applicable in their standard version for non-local potentials. Many of these technical aspects are carefully described in recent literature~\cite{Gajdo_vasp_2006, french_PRE_2022, Demyanov2022, Melton2024}. 

The above Kubo-Greenwood formalism is the standard approach to obtain electronic transport properties from DFT and has led to many studies of the electrical and thermal conductivity of hydrogen. Most of the pressure-temperature range covered in these studies are typical for the interior of giant planets~\cite{Holst_PRB_2008, Lorenzen2010, Holst2011, vandeBund2021}, where Kohn-Sham DFT is very efficient. Other studies pushed the calculations to extremely high temperatures in the ICF domain~\cite{Recoules2009, Lambert2011, desjarlais_PRE_2017, french_PRE_2022}, which is only possible when the density is sufficiently high so that the number of Bloch states remains computationally tractable or by combining Kohn-Sham DFT with orbital-free DFT. On the low density end of the warm dense matter region, data are scarce, because the calculations are computationally very demanding due to the large simulation box sizes~\cite{french_PRE_2022}.

\subsubsection{DFT results for the electrical conductivity}\label{sss:results-sigma}
\begin{figure}\centering
\includegraphics[width=0.49\textwidth]{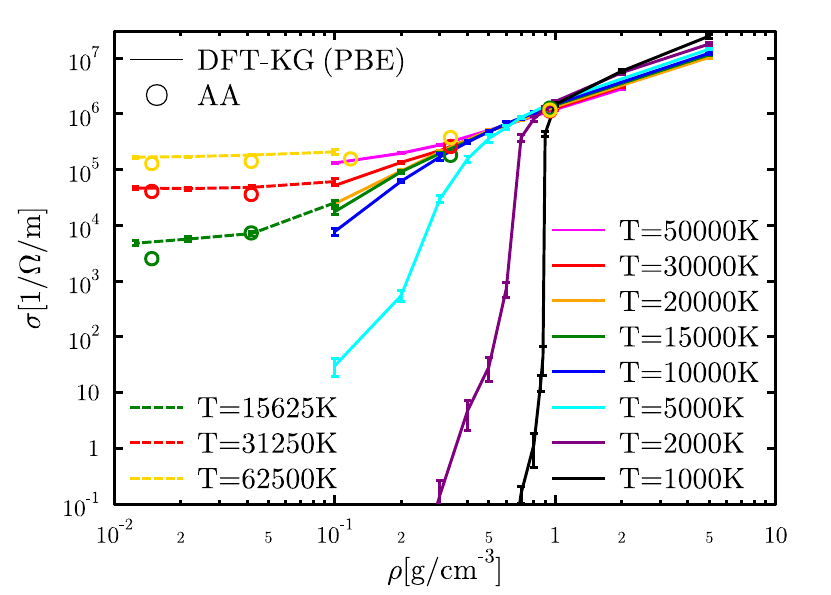}
\caption{\label{fig:sigma_BH}
Results for the electrical conductivity, Eq.~(\ref{eq:sigma-kg}), of dense hydrogen, as a function of density for different temperatures. Average atom data (AA, circles) and DFT-KG using the PBE functional (lines) are compared. Some DFT data (full lines) were taken from B.~Holst \emph{et al.}~\cite{Holst2011}, new results of this paper that correspond to the EOS results of Fig.~\ref{fig:rpimc-pimc-FVT2-pressure} are shown by the dashed lines. }
\end{figure}

The electrical conductivity of dense hydrogen plasma is shown in Fig.~\ref{fig:sigma_BH} as computed by B.~Holst \textit{et al.} \cite{Holst2011} using the PBE XC functional when evaluating the Kubo-Greenwood formula, cf. Sec.~\ref{sss:dft-kubo-greenwood}. The electrical conductivity increases with density, which is a result of pressure ionization, cf. Sec.~\ref{sec:wd-hydrogen}. For temperatures below \SI{1500}{\kelvin}, i.e.\ in the dense fluid phase, the electrical conductivity shows jumps which are typical of a first-order phase transition -- the LLPT, cf. Sec.~\ref{sssec:hydrogen-lowt}. Above the critical point of the LLPT, the increase is steep but continuous and becomes less pronounced with increasing temperature. Here, thermal ionization leads to a broadening of the nonmetal-to-metal transition. Note that the temperature trend of the electrical conductivity is inverted above the transition density which is a typical metal-like behavior. Here, increasing temperature broadens the Fermi function which allows for more scattering processes of the electrons at ions so that their mobility is reduced. 

For the three temperatures considered at the lower densities shown in Fig.~\ref{fig:sigma_BH}, we can observe satisfactory agreement between DFT-KG results and AA data. For these conditions, the conductivity never drops to zero as there always is some remnant, fluctuating partial ionization, in agreement with the PIMC results, cf. Fig.~\ref{fig:ion-degree-mil-fil}.

\begin{figure}\centering
\includegraphics[width=0.47\textwidth]{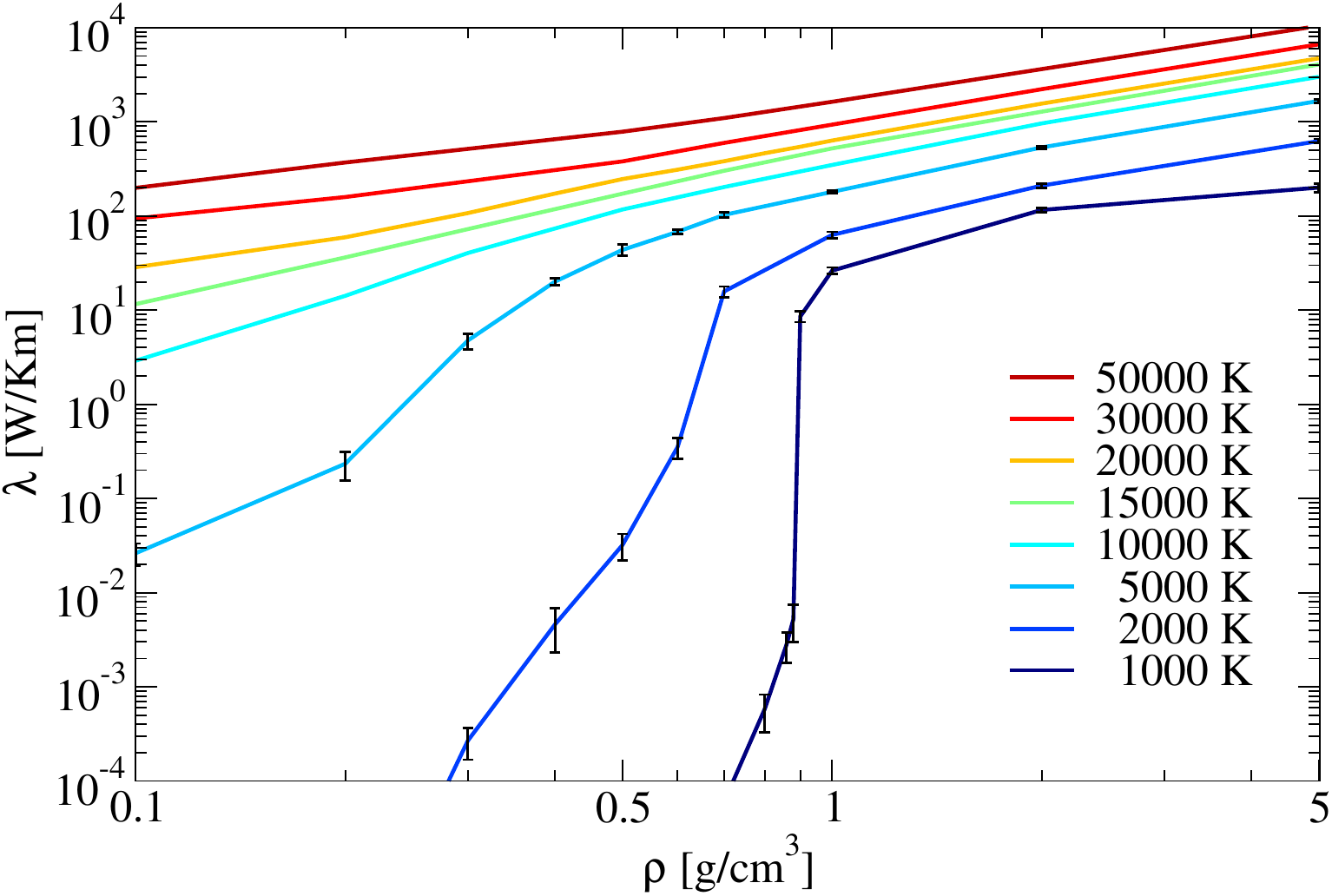}
\caption{\label{fig:lambda_BH}
Results for the electronic thermal conductivity, Eq.~(\ref{eq:lambda-kg}), of dense hydrogen plasma as a function of the density for different temperatures. Taken from B.~Holst \emph{et al.}~\cite{Holst2011} with the permission of the authors.
}
\end{figure}

The electronic thermal conductivity shows a very similar behavior, see Fig.~\ref{fig:lambda_BH}. In the dense fluid below \SI{1500}{\kelvin}, a sharp rise over several orders of magnitude appears at about \SI{0.9}{\gram\per\cubic\centi\meter}. This is a result of an abrupt pressure-driven ionization. Above the critical temperature, the nonmetal-to-metal transition is thermally broadened as in the case of the electrical conductivity. Contrary to the electrical conductivity, the isotherms of the thermal conductivity increase systematically with temperature for all densities.    

\subsubsection{DFT results for the opacity}\label{sss:results-opacity}

The radiative properties of dense hydrogen (deuterium) can be derived from the frequency-dependent electrical conductivity calculated by DFT-based molecular-dynamics (quantum molecular dynamics, QMD) simulations combined with the linear-response theory of Kubo and Greenwood, cf. Sec.~\ref{sss:dft-kubo-greenwood}. Starting from the real part of electric conductivity, $\sigma_1 (\omega)$, Eq.~(\ref{eq:sigma-kg}), we can use the  Kramers-Kronig relation to obtain its imaginary part, $\sigma_2 (\omega)$, by principle-value integration. The obtained complex conductivity is directly related to the frequency-dependent dielectric function, $\epsilon (\omega)=\epsilon_1 (\omega) + i\epsilon_2 (\omega)$:
\begin{eqnarray}
\epsilon_1 (\omega) & = & 1 - \frac{4\pi}{\omega}\sigma_2(\omega)\,, \\
\epsilon_2 (\omega) & = & \frac{4\pi}{\omega}\sigma_1(\omega)\,. 
\end{eqnarray}
which allows one to calculate the refractive index, $n(\omega) + ik(\omega)$:
\begin{eqnarray}
n(\omega) & = & \sqrt{ \frac{|\epsilon(\omega)|+\epsilon_1(\omega)}{2}}\,, \\
k(\omega) & = & \sqrt{ \frac{|\epsilon(\omega)| -\epsilon_1(\omega)}{2}}\,. 
\end{eqnarray}
Finally, the frequency-grouped Rosseland mean opacity of dense hydrogen can be computed for photon energies in between $\hbar \omega_1$ and $\hbar \omega_2$, by using the following integration formula:
\begin{equation}
K_R(\omega_1:\omega_2) = \frac{\int\limits_{\omega_1}^{\omega_2} n(\omega)^2 \frac{\partial B(\omega, T)}{\partial T} d\omega}  
{\int\limits_{\omega_1}^{\omega_2} n(\omega)^2 \frac{1}{\alpha_m(\omega)} \frac{\partial B(\omega, T)}{\partial T} d\omega},
\label{eq:opacity}
\end{equation}
with the mass absorption coefficient being defined as $\alpha_m(\omega)=4\pi \sigma_1(\omega)/[n(\omega) c \rho]$, where $c$ and $\rho$ are the speed of light and mass density, respectively. Here, the Planck function, 
\begin{align}
B(\omega, T)=\frac{\hbar \omega^3}{4\pi^3 c^2}\, \frac{1}{e^{\hbar \omega/k_B T}-1}\,,     
\end{align}
gives the radiation spectrum of a black-body system at temperature $T$.

\begin{figure}
    \centering
    \includegraphics[width=0.49\textwidth]
    %{Figure_Opacity_Comparison_QMD_vs_AOT_vs_DFTAA.pdf}
    {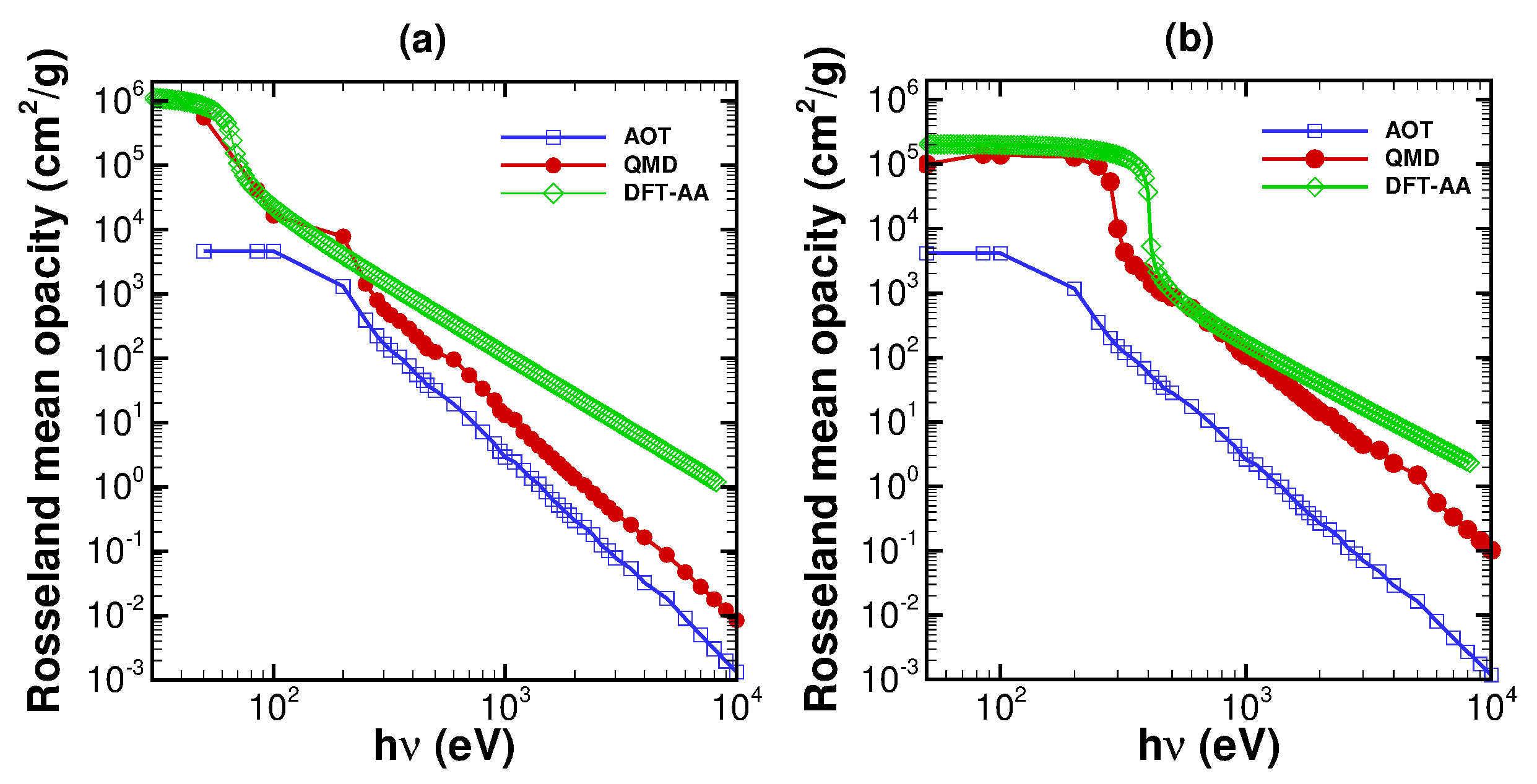}
    \caption{Comparison of the deuterium 48-group Rosseland mean opacity among QMD, AOT, and DFT-AA calculations as a function of the group photon energy. (a) $\rho=\SI{5.388}{\gram\per\cubic\centi\meter}$ and $k_BT=\SI{10.8}{\eV}$. (b) $\rho=\SI{199.561}{\gram\per\cubic\centi\meter}$ and $k_BT=\SI{43.1}{\eV}$. }
    \label{opacity_D2}
\end{figure}

This method of combining DFT with Kubo-Greenwood linear response theory has been applied to build first-principles opacity tables (FPOT) of warm-dense hydrogen (deuterium) \cite{PRE_Hu_2014} and polystyrene \cite{PRB_Hu_2017} in a wide range of densities and temperatures. One example is shown in Fig.~\ref{opacity_D2} for the opacity of deuterium where we compare FPOT (marked as ``QMD'' as it was derived from QMD+Kubo-Greenwood calculations where the PBE XC functional was used) and the astrophysics opacity table (AOT) created by Los Alamos National Lab \cite{huebner_lanl_77}. Figure~\ref{opacity_D2} displays the 48-group Rosseland mean opacity of deuterium in the warm dense matter range for two different conditions: (a) $\rho=\SI{5.388}{\gram\per\cubic\centi\meter}$ and $k_BT=\SI{10.8}{\eV}$, and (b) $\rho=\SI{199.561}{\gram\per\cubic\centi\meter}$ and $k_BT=\SI{43.1}{\eV}$. It indicates that the opacity 
is sensitive to strong coupling and degeneracy effects and can differ significantly from traditional atomic code calculations in the WDM regime. The difficulty of extending first-principles DFT-MD/QMD+KG calculations of the opacity to high temperature (beyond the Fermi temperature) lies in the number of bands needed for convergence. It is noted that the DFT-based average-atom (DFT-AA, Sec.~\ref{ss:aa-models}) model calculations (green diamonds in Fig.~\ref{opacity_D2}), which are more readily extended to high temperatures, give good agreement with QMD calculations, in particular in the low and mid frequency ranges. High-frequency opacity calculations with the QMD+KG method invoke extrapolations, which can result in some uncertainties.      

 In conclusion, we note that the opacity of hydrogen has recently been measured at the NIF for conditions that are relevant for the interior of red dwarf stars. The conceptual paper~\cite{Lutgert_POP_2022} shows also DFT-MD predictions for a (25\%/75\%) hydrogen-tritium mixture at $\rho=\SI{200}{\gram\per\cubic\centi\meter}$ and temperatures of $k_BT=\SI{100}{\eV}$ and $k_BT=\SI{150}{\eV}$, in comparison with an AA model [cf. Sec.~\ref{ss:aa-models}], very similar to those displayed in Fig.~\ref{opacity_D2}.

\subsubsection{Correlation and collision effects in DFT and TDDFT}\label{sss.collisions-dft}

While e-e \textit{correlations} are treated on the level of the chosen XC functional, it has long been debated whether or not e-e \textit{collisions} are properly included in calculations of electronic transport using KS-DFT for the evaluation of the Kubo-Greenwood formula (Sec.~\ref{sss:dft-kubo-greenwood}); see, e.g., Refs.~\cite{reinholz_PRE_2015,desjarlais_PRE_2017,shaffer_PRE_2020}. The Spitzer theory~\cite{spitzer_PR_1953} gives exact values for the transport coefficients in the non-degenerate ($\Theta\gg 1$) and weakly coupled ($\Gamma\ll 1$) limit, both for considering and neglecting e-e collisions, besides e-i collisions. This information can be used to determine the coefficients of a virial expansion of the electrical conductivity $\sigma(n,T)$, Eq.~(\ref{eq:sigma-kg}), and to benchmark DFT results, see Ref.~\cite{roepke_PRE_2021}. In this limit, the electrical ($\sigma$) and thermal conductivity ($\lambda$) as well as the thermopower ($\alpha$) can simply be expressed by prefactors $f$, $a$, and $L$ as follows: 
\begin{eqnarray}
   \sigma &=& f \frac{ (4\pi\epsilon_0)^2 (k_BT)^{3/2} }{ m_e^{1/2}e^2 } \frac{1}{\ln{\Lambda_c}} \,,\nonumber\\
   \alpha &=& a \frac{k_B}{e} \,,\\
   \lambda &=& L \left( \frac{k_B}{e} \right)^2 T \sigma \,.\nonumber
\end{eqnarray}\label{eq:Spitzer}
Here, $\ln{\Lambda_c}$ is the well-known Coulomb logarithm, cf. Eq.~(\ref{eq:coulomb-log}). For fully ionized hydrogen plasma with $Z=1$, the Spitzer values for the prefactors with and without considering e-e collisions are $f^{ei}=1.0159$, $f^{ei+ee}=0.5908$, $a^{ei}=1.5$, $a^{ei+ee}=0.7003$, $L^{ei}=4.0$, and $L^{ei+ee}=1.5966$~\cite{spitzer_PR_1953}. Overall, e-e collisions reduce the electronic transport coefficients considerably in the non-degenerate limit. 

\begin{figure}\centering
\includegraphics[width=0.48\textwidth]{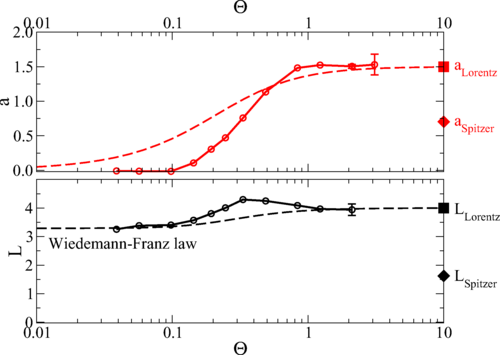}
\caption{\label{fig:a-L-French2022}
Thermopower coefficient $a$ (red) and Lorenz number $L$ (black) from DFT (full) and the relaxation time approximation (RTA, dashed) along the $\Gamma^8 \Theta^7 = 1$ line.
Taken from M.~French \emph{et al.}~\cite{french_PRE_2022} with the permission of the authors.
}
\end{figure}

Recently, M.~French \textit{et al.} \cite{french_PRE_2022} performed extensive calculations of the transport properties of hydrogen across the plasma plane along the line $\Gamma^8 \Theta^7 = 1$, in order to check if the correct Spitzer values are reproduced in the non-degenerate ($\Theta\gg 1$) and weakly coupled ($\Gamma\ll 1$) limit. Evaluating the Kubo-Greenwood formula in that limit, using KS-DFT, is numerically demanding since huge simulation cells and a large number of bands are required in order to converge the calculations. This has so far prevented obtaining conclusive results. M.~French \textit{et al.} \cite{french_PRE_2022} could show explicitly that the KS-DFT results approach the Lorentz plasma values without e-e collisions, but not the Spitzer values, so this approach does not capture e-e scattering processes, see Fig.~\ref{fig:a-L-French2022} for $a$ and $L$. 

In the degenerate limit ($\Theta\ll 1$), the influence of e-e collisions vanishes, due to strong screening of the Coulomb interactions. More importantly, Pauli blocking prevents e-e scattering processes except for a small energy range near the Fermi energy. Inspecting Fig.~\ref{fig:a-L-French2022}, the correct values for $L=\pi^2/3$ (Wiedemann-Franz law) and $a=0$ follow from KS-DFT. For a more detailed discussion of the influence of e-e collisions on the transport coefficients, see Refs.~\cite{redmer_phys-rep_97,roepke_PRE_2021}. Note that H.~Reinholz \textit{et al.}~\cite{reinholz_PRE_2015} have derived a correction factor $R_{ee}(T,\Theta,Z)$ for the influence of e-e collisions on the electrical conductivity which can be applied in a wide temperature and density range.

A potential alternative route towards optical properties from KS-DFT simulations is given by LR-TDDFT~\cite{ullrich2012time,Moldabekov_JCTC_2023} and RT-TDDFT~\cite{PhysRevB.107.115131,ramakrishna2022electrical_dft_u,dynamic2}, cf. Sec.~\ref{sss:tddft}. In principle, both methods allow one to consistently include e-e correlation effects, although both the dynamic XC-kernel and the dynamic XC-potential are usually treated in an adiabatic (i.e., static) approximation in practice.

\subsection{Outlook: Time-dependent simulations}\label{ss:t-dep-results}
After discussing simulations of thermodynamic, dynamic, transport, and optical properties of dense hydrogen in thermal equilibrium, we now briefly outline the prospects for time-dependent simulations. We consider two examples of quantities: in Sec.~\ref{sec:TDDFT} we discuss the application of linear-response TDDFT to the dynamic structure factor. Then, in Sec.~\ref{sss:stopping-g1-g2} we discuss the application of quantum kinetic equations to the electronic stopping power.

\subsubsection{Dynamic structure factor from linear-response TDDFT simulations\label{sec:TDDFT}}

\begin{figure}\centering
\includegraphics[width=0.41\textwidth]{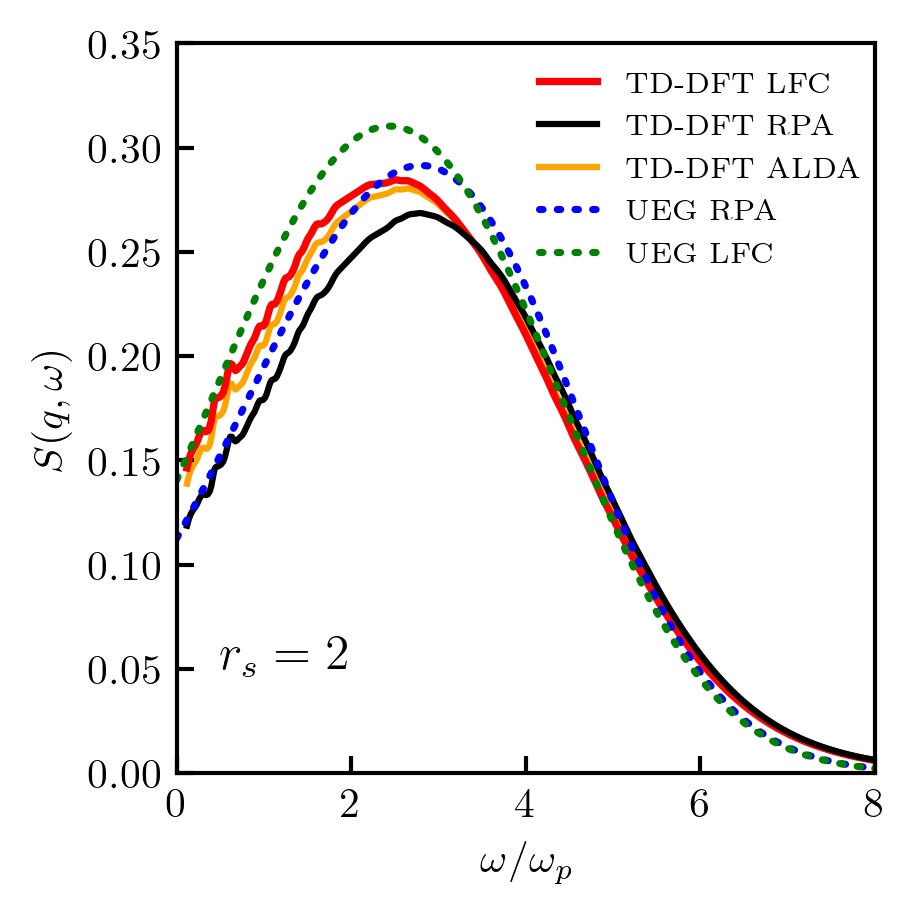}
\caption{\label{fig:DSF_rs2_theta1}
Linear response TDDFT results for the inelastic component of the dynamic structure factor [cf.~Eq.~(\ref{eq:Chihara})] of hydrogen at $r_s=2$ and $\Theta=1$ for $q=3.06\,$\AA$^{-1}$. Taken from M.~B\"ohme \emph{et al.}~\cite{Bohme_PRL_2022} with the permission of the authors.
}
\end{figure}

A promising route towards \textit{ab initio} results for dynamic properties of hydrogen is given by linear-response TDDFT. As explained above, here the central quantity is the dynamic electronic density response function $\chi_{ee}(\mathbf{q},\omega)$, which is expressed in terms of the dynamic KS response $\chi_0(\mathbf{q},\omega)$ and the, in principle, also dynamic XC-kernel $K_\textnormal{xc}(\mathbf{q},\omega)$.
In practice, both the XC-functional and the XC-kernel have to be approximated, with the latter being particularly difficult. Very recently, Z.~Moldabekov \emph{et al.}~\cite{Moldabekov_JCTC_2023,Moldabekov_PRR_2023} have suggested a new way that allows one to compute the static XC-kernel $K_\textnormal{xc}(\mathbf{q})=K_\textnormal{xc}(\mathbf{q},0)$ for arbitrary XC functionals on any rung of Jacob's ladder of functional approximations~\cite{Perdew_AIP_2001} without the need to explicitly evaluate functional derivatives.
This, in turn, makes it possible to compute the dynamic density response function within the \emph{static approximation}~\cite{Dornheim2018b,Dornheim_PRB_ESA_2021,Moldabekov_PRR_2023} 
\begin{eqnarray}\label{eq:static_approximation}
    \chi_\textnormal{static}(\mathbf{q},\omega) = \frac{\chi_0(\mathbf{q},\omega)}{1 - \left[v_q+K_\textnormal{xc}(\mathbf{q})\right]\chi_0(\mathbf{q},\omega)}\ ,
\end{eqnarray}
which can be generalized to a Dyson's type equation, cf.~Eq.~(\ref{eq:Dyson}), for inhomogeneous systems.
The fluctuation--dissipation theorem [cf.~Eq.~(\ref{eq:FDT})] then opens the way to estimate $S_{ee}(\mathbf{q},\omega)$ from Eq.~(\ref{eq:static_approximation}).
The \emph{static approximation} is expected to work well for moderate densities $r_s\lesssim4$ and moderate temperatures $\Theta\gtrsim1$ based on previous investigations of the UEG~\cite{Dornheim2018b,groth_prb_19,Dornheim_PRB_ESA_2021}.

In Fig.~\ref{fig:DSF_rs2_theta1}, we show corresponding linear-response TDDFT results for the inelastic component of the DSF [cf.~Eq.~(\ref{eq:Chihara})], for hydrogen at $r_s=2$ and $\Theta=1$, for $q=3.06\,$\AA$^{-1}$. The dotted green and blue curves have been obtained for the UEG, where the inclusion of the local field correction leads to a well-known red shift~\cite{Dornheim2018b}. 
For hydrogen, the RPA curve that has been obtained by setting $K_\textnormal{xc}(\mathbf{q})\equiv0$, in Eq.~(\ref{eq:static_approximation}), is qualitatively close to the RPA result for the UEG and exhibits a peak at the same frequency. Similarly, using the static XC-kernel from Ref.~\cite{Bohme_PRL_2022} (solid black), or the adiabatic LDA (ALDA) kernel (solid yellow) leads to the same red shift as for the UEG.

The further exploration of this framework to study the dynamic density response and dynamic structure factor of hydrogen (and potentially other elements) constitutes an important topic that will be pursued in dedicated future works.

\subsubsection{Capabilities of real-time-dependent DFT simulations}
Real-time TDDFT simulations will be of much use for improved descriptions, in particular of conductivities and stopping power~ \cite{GRABOWSKI2020100905,stanek_pop_24}. The direct time-dependent simulation of the reaction of the system to an external perturbation, be it via radiation or particles, constitutes an advantage over linear response methods such as LR-TDDFT. Thus, the dynamic structure of warm dense matter at finite wavenumbers, i.e. the XRTS signal, can be computed~\cite{dynamic2}. In combination with Ehrenfest dynamics, the stopping of fast beam particles in warm dense hydrogen can be investigated~\cite{Magyar_CPP_2016,Ding_PRL_2018,Hentschel_pop_2023,Kononov2023,kononov_2024}. A special advantage of a direct simulation is the capability to include non-linear effects and, in principle, also electron-electron collisions. In addition, the optical limit of the conductivity can be accessed more straightforwardly as via the DFT-Kubo-Greenwood approach.

\subsubsection{Time-dependent charged particle stopping from quantum kinetic theory simulations\label{sss:stopping-g1-g2}}
The stopping power, $S_x=\frac{d\langle E\rangle}{dx}$ -- the energy loss of energetic particles per unit length in a plasma -- is of crucial importance for many applications. Many simulation techniques have been used to compute $S_x$, including molecular dynamics, e.g. Ref.~\cite{zwicknagel_pre_00}, or time-dependent DFT, for recent results, see Refs.~\cite{Kononov2023,kononov_2024}.

%\jan{
On the other hand, the recent advances in the description of correlated quantum plasmas, as described in this manuscript, have made it possible to obtain the density fluctuations, $\delta n({\bf r})$, in a WDM system from first principles. This naturally includes non-linear fluctuations~\cite{Dornheim2020,Dornheim2021}. On this basis, it is possible to determine the stopping power from the consideration of the friction experienced by a beam particle due to the nonlinear (electronic) density fluctuations~\cite{PhysRevB.37.9268}
\begin{equation}
S_x=-Ze\iint d^3{\bf r}d^3{\bf r}'\; 
\delta({\bf r}-{\bf v}t) \frac{{\bf v}\cdot\nabla}{v}\frac{1}{|{\bf r}-{\bf r}'|}\delta n({\bf r}',t)\,,
\end{equation}
where $Ze$ is the charge of the beam ions moving at velocity $\bf v$. The density fluctuation can then be obtained in linear response theory via the inverse dielectric function, which leads to 
\begin{eqnarray}\label{eq:sp}
    S_x(v) &=&\frac{2Z^2e^2}{\pi v^2}\int_0^{\infty} \, \frac{{\rm d}k}{k} \, \int_{\omega_-}^{\omega_+} {\rm d} \omega  ~{\rm Im} \left[\frac{1}{\epsilon(k,\omega)}\right] \nonumber\\
    && \times n_B(\hbar\omega)\left( \hbar \omega - \frac{k^2}{2m} \right)\,,
\end{eqnarray}
where $n_B$ is the Bose function and $\omega_{\pm}= k^2/2m \pm kv$.
\begin{figure}    \centering
    \includegraphics[width=0.45\textwidth]{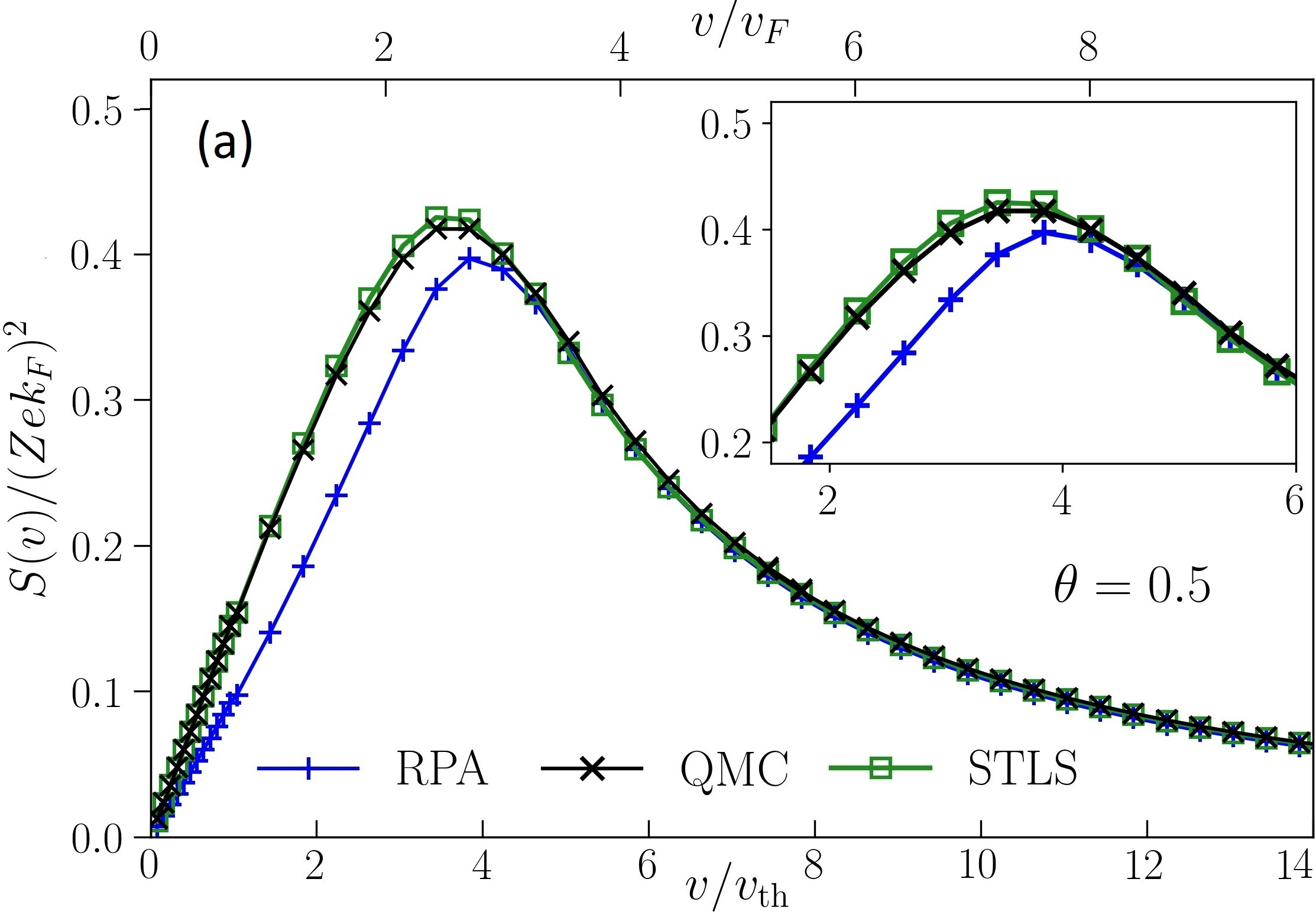}
    \caption{Stopping power as a function of projectile velocity (in units of the thermal velocity and the Fermi velocity, respectively) for $r_s=4$ and $\Theta=0.5$. Uncorrelated results  (RPA) are compared to QMC results and the STLS model. Taken from Ref.~\cite{moldabekov_pre_20} with the permission of the authors.}
    \label{fig:stopping-rs4}
\end{figure}
It follows from Eq.~(\ref{eq:chi-correlated-def}) that the inverse dielectric function allows to systematically account for correlation effects via the dynamic local field correction $G$ that is accurately available from QMC simulations \cite{moldabekov_pre_20}. The effect of electronic correlations is demonstrated in Fig.~\ref{fig:stopping-rs4} where we compare RPA results ($G\equiv 0$) to QMC results and the analytical STLS model \cite{SingwiTosi_Review}. While for large projectile velocities, $v\gg v_{th}$, correlation effects are not important, for velocities, $v\lesssim 3v_{th}$ they cause a significant increase of the electronic stopping power.

Independently from linear response theory, the stopping power can be obtained from quantum kinetic theory which directly benefits from the first-principles simulation results shown in this paper. The equation for the stopping power of the beam particles interacting with a warm dense matter system reads
\begin{equation}\label{eq:stopping-sx}
S_x(t) =\frac{1}{n_b}
\int\frac{d{\bf p}_b}{(2\pi\hbar)^3}
\frac{{\bf p}_b\cdot{\bf v}}{v}
\frac{\partial f_b({\bf p}_b,t)}{\partial t}\,.
\end{equation}
The time derivative of the beam Wigner function, $f_b$,  is given by a kinetic equation which reads, for the homogeneous case,~\cite{kremp2005quantum}
\begin{equation}
\frac{\partial f_b({\bf p}_b,t)}{\partial t}=\sum_{c}{\bf I}_{bc}({\bf p}_b,t)\,.
\end{equation}
This equation has to be solved in tandem with a similar one for the plasma particles. The collision integral can be of, e.g., Boltzmann type, Lenard-Balescu like, or have any other shape~\cite{Gericke1999,Gericke_1999b,Gericke_2002}, cf. Sec.~\ref{sss:g1-g2}. Nonlinear and strong coupling effects can be accounted for via higher order (beyond second order Born) contributions to the selfenergies (scattering rates) $\Sigma^{\gtrless}$~\cite{dornheim_2024ab,dornheim_2024}.
%}

\begin{figure}
    \centering
    \includegraphics[width=0.5\textwidth,trim=.5cm 1.5cm .5cm 1.5cm, clip]{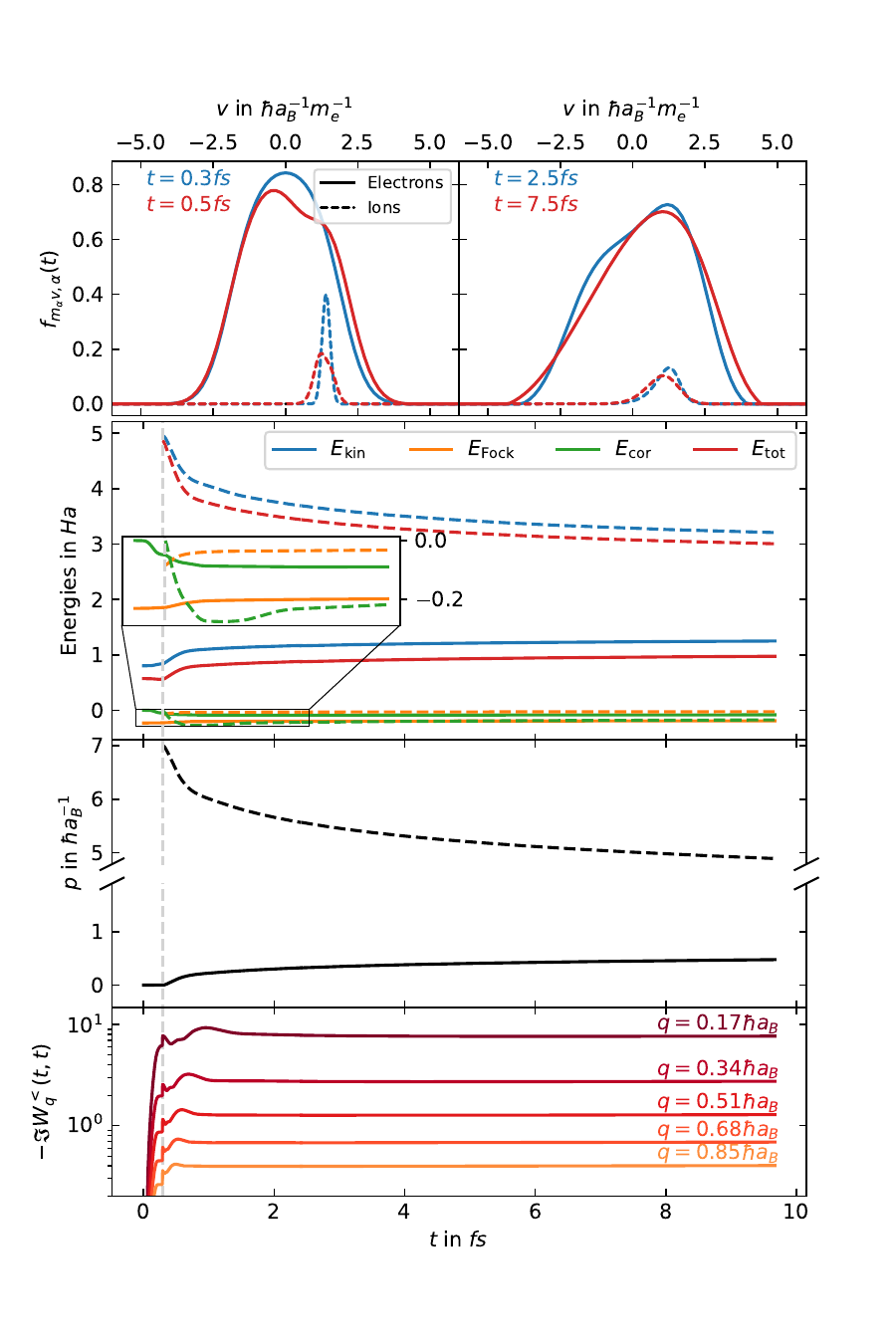}
    \caption{Time evolution of electronic (full lines) and ionic (dashed lines) distribution functions and observables during ion impact. The electronic parameters are $r_s=0.5$ and $\Theta=0.81.$ The ion impact starts at $t=0.3$fs, cf. vertical grey line.
    Top: total, kinetic, mean field (Fock), and correlation energy per particle. Middle figure: momentum per particle. Bottom: Electronic plasmon number for five wave numbers. For details, see text.}
    \label{fig:observables-g1g2}
\end{figure}

Here we present new results for the application of quantum kinetic equations. They have been obtained within the G1--G2 scheme that was introduced in Sec.~\ref{sss:g1-g2}.
An example of the solution of the G1--G2 equations for a dense quasi-1D plasma which extends previous simulations, Ref.~\cite{makait_cpp_23}, 
 is shown in Fig. \ref{fig:observables-g1g2}. We consider a fully ionized electron-ion plasma at high density and moderate temperature corresponding to $r_s=0.5$ and $\Theta=0.81$. For technical reasons, we consider a quasi-one-dimensional plasma (for details of the model, cf. Ref.~\cite{makait_cpp_23}) and reduce the ion mass to $5m_e$.
 We start with an electron plasma in thermal equilibrium which is impacted by a nearly monoenergetic ion beam (dashed blue curve in the upper left part of Fig.~\ref{fig:observables-g1g2}). The simulations demonstrate how the electron distribution changes: it shifts to the right (electrons gain momentum) and broadens (electrons are being heated). At the same time, ions lose energy, i.e. their distribution broadens. 

 From the time-dependent distributions, we also compute macroscopic observables, the time evolution of which is shown in Fig.~\ref{fig:observables-g1g2}. Note that the G1--G2 scheme has the same conservation properties as the exact many-particle system: it conserves particle number, total momentum, and total energy. The time evolution of the mean momentum (momentum per particle) of electrons and ions is depicted in the middle plot. When multiplied with the respective particle numbers, the momentum loss of ions is exactly compensated by the momentum gain of the electrons. In the top part of Fig.~\ref{fig:observables-g1g2}, we plot the different energy contributions of electrons (full lines) and ions (dashed lines) where the total energy consists of kinetic, exchange and correlation energy, $E^a_{\rm tot}=E^a_{\rm kin}+E^a_{\rm Fock}+E^a_{\rm cor}$, with $a=e, i$. Let us have a closer look at the dynamics. At $t=0$, we start with uncorrelated electrons (ideal Fermi gas). To obtain the correct initial state, which includes correlation effects due to the Coulomb interaction of the electrons, we apply the \textit{adiabatic switching} procedure, e.g. Ref.~\cite{hermanns_prb14, schluenzen_jpcm_19}, during which correlation energy forms which are negative, see the green curve in the inset (due to the weak coupling conditions the correlation energy is substantially smaller than the kinetic energy).

 At time $t=0.3\,$fs the ion beam impacts the correlated electron plasma giving rise to a rapid increase of kinetic, mean field, and (negative) correlation energy. Correlation and Fock energy saturate after about $1\,$fs, which can be identified with the correlation time in the system, see Refs.~\cite{bonitz-etal.96pla, bonitz96pla}. After this time, the energy exchange between ions and electrons continues but at a significantly slower rate. Also, the distribution functions continue to change their shape up to about $8\,$fs, cf. Fig.~\ref{fig:observables-g1g2}, which can be identified with the relaxation time. Subsequently, the dynamics are dominated by a kinetic energy exchange between the two components, i.e. by electron-ion temperature equilibration. Another interesting question is that on the mechanism of the energy exchange between electrons and ions -- are these two-particle collisions or are collective excitations involved? This question is answered in the bottom panel of Fig.~\ref{fig:observables-g1g2} where we plot the plasmon occupation number for five wave numbers. First, we observe a rapid build up of the plasmon occupations, during the adiabatic switching interval, $t \lesssim 0.3\,$fs. With the impact of the ions, the plasmon population increases and exhibits oscillations with the q-dependent plasmon frequency, $\omega(q)$, which are damped out when the correlation time is reached. The direct access to the time-dependent plasmon properties is provided by the G1--G2 scheme if it is solved with GW self-energies, cf. Eq.~(\ref{eq:g2-equation}), as in the present case. 

 This example demonstrates the capabilities of the G1--G2 scheme for dense quantum plasmas. It fully includes electronic correlations and their dynamics as well as electron-electron scattering effects which are missing in KS-DFT, cf. Sec.~\ref{sss.collisions-dft}.
 All relevant observables can be computed, including their time evolution. This includes the time-dependent stopping power and dielectric properties.
 Overcoming the memory bottleneck arising from the storage of the two-particle function and, with this, the extension of the G1--G2 scheme to 2D and 3D plasmas should become possible in the near future by applying the quantum fluctuations approach \cite{schroedter_cmp_22,schroedter_23,schroedter_pssb23}, see Sec.~\ref{sss:g1-g2}.

In conclusion, we note that the stopping power is not only of interest for dense plasmas but has recently been investigated also for other targets, such as correlated quantum materials for which NEGF-Ehrenfest dynamics simulations \cite{balzer_prb16, Schluenzen_CPP_2019, borkowski_pss_22}
as well as TDDFT-Ehrenfest simulations \cite{Kononov_2dmat_22} were reported. There not only the energy loss of the projectile is important but also the charge transfer from the target to the projectile, in particular in the case of highly charged ions \cite{balzer_cpp_21,niggas_prl_22}.

\section{Summary and Outlook\label{sec:summary}}
Hydrogen at high pressure continues to be the focus of research in many fields, including astrophysics, basic science, and technology,
in particular, for inertial confinement fusion. Our understanding of this simplest of all chemical elements remains far from complete. This may seem surprising since the basic equation governing its properties -- the Schrödinger equation -- has been known for nearly one century. However, strong compression gives rise to a complicated interplay of many particles --  a combination of quantum, spin, correlation, and thermal effects, which poses challenges both for experiments and simulations.

In this review article, we focused on important simulation approaches that are capable of substantially advancing our knowledge about dense hydrogen in the near future. Of particular interest was the question about the accuracy and predictive capability of commonly used simulation techniques. In many fields of physics, including atomic physics, semiconductor optics, or cold atoms in traps or optical lattices, this question is routinely answered by high-precision experiments. In the field of dense plasmas, this is not yet possible. Therefore, we outlined strategies to use first-principle simulations, i.e. ``computer experiments'', that can be used to benchmark models and simulations. In the following, we summarize the main results and discuss open questions.

\subsection{Summary}\label{ss:summary}
The first part of this review (Sec.~\ref{sec:phase-dia}) was devoted to the phase diagram of hydrogen at high pressure. We discussed the hypothetical plasma phase transition (PPT), Sec.~\ref{sssec:hydrogen-hight}, that has been predicted to occur in partially ionized hydrogen in the gas phase and concluded that there is neither experimental nor reliable theoretical evidence. We then turned to hydrogen at low temperature, in the liquid and solid phases, and reviewed the knowledge about the solid-liquid and insulator-metal transition (LLPT). 

A related topic is given by the additional metal-to-superconductor phase transition that was proposed by N.~Ashcroft~\cite{ashcroft_prl_68}, and further explored in subsequent works, e.g., Refs.~\cite{Babaev2004,Richardson_PRL_1997,McMahon_PRB_2011,Richardson_PRB_1997}. In this regard, accurate QMC simulations might help by giving insights into effective electronic interactions in the medium~\cite{Kukkonen_PRB_2021,Dornheim_JCP_2022}, which are important for the estimation of the critical temperature~\cite{Pellegrini_PRB_2023}. In the case of deuterium, it is further feasible to carry out either PIMD or PIMC simulations based on an effective ionic potential and, in this way, to directly estimate the superfluid fraction~\cite{cep}. This has recently allowed Myung \textit{et al.}~\cite{Myung_PRL_2022} to report a hypothetical supersolid phase~\cite{RevModPhys.84.759} at high pressure and low temperature, although further verification is required.

In Sec.~\ref{sec:theory} we reviewed the most important simulation methods. A particular emphasis was given to first principles approaches, such as quantum Monte Carlo (QMC) and density functional theory (DFT) simulations, but we also discussed average atom and chemical models, classical and quantum hydrodynamics, and semiclassical simulations with quantum potentials. 
The methods differ strongly in terms of computational effort, applicable parameter range, and resolution. Most importantly, there are fundamental differences between the methods in their theoretical level and approximations and, hence, their expected accuracy. These expectations are derived from the general (qualitative) validity range of the approximations (including self-energies, in the case of Green functions, or XC functionals, in the case of DFT). At the same time, the quantitative accuracy when an approximation is applied to dense hydrogen can only be established using strict benchmarks against known results. 

Therefore, in Sec.~\ref{sec:results} we presented extensive tests of various methods using fermionic PIMC results of Ref.~\cite{filinov_pre_23} as reference. The results are summarized as follows:
\begin{itemize}
    \item RPIMC with free particle nodes is the most accurate of the approximate methods, in the considered parameter range, with relative errors of the pressure not exceeding $2\%$, except for the lowest temperature, $T=15\,625$K. The accuracy of the results (and the quality of the nodes) at high density, $r_s\lesssim 3$, remains to be tested.
    \item KS-DFT-MD with PBE functionals exhibits pressure deviations from FPIMC of up to $7\%$. The agreement improves substantially with the KDT16 finite-temperature PBE functional, especially for $T \sim 60\,000$K, to about $2\%$, whereas for $T \sim 30\,000$K the deviations are larger. 
    \item Semiclassical MD simulations with the improved Kelbg potential achieve an accuracy of $1\dots 3\%$, for the pressure, for $T\gtrsim 60\,000$K and $r_s \gtrsim 7$, where the density range quickly widens with increasing temperature.
    \item DFT-AA models showed good agreement with FPIMC, for pair distributions and static structure factors, and with KS-DFT, for the conductivity. For quantitative benchmarks additional tests will be required.
    \end{itemize}

Further, we demonstrated that, aside from benchmarking other methods, first-principles simulations can also be combined with simpler models, such as chemical models, to compute 
quantities of interest that are difficult to access otherwise. An example is
the ionization potential depression (IPD) for which various models exist but only a few first-principles results have been reported \cite{hu_prl_17}. In  Sec.~\ref{sss:ipd-qmc} we showed that, by using the fractions of free particles and atoms from FPIMC simulations as input, allows one to directly obtain the IPD. Similar approaches can also be developed for other quantities and more complex systems.

An interesting property of correlated quantum many-body systems is the non-exponential decay of the momentum distribution, $n(\textbf{k})$, at large values of the argument, which we demonstrated for jellium in Sec.~\ref{sec:n(k)}. It is expected that similar behavior exists in dense hydrogen, both for electrons and ions. Quantification of this effect could have implication for fusion rates.

\subsection{Outlook}\label{ss:outlook}
\subsubsection{Upcoming experiments and suggested model developments}\label{sec:outlook-exp-models}
The importance of dense hydrogen is reflected in a large number of ambitious upcoming experiments at top facilities around the world. This includes
  experiments with liquid hydrogen jets at the European XFEL and LCLS, equation of state measurements at FAIR (GSI Darmstadt),  compression experiments at the Omega laser facility in Rochester, and compression experiments at Sandia's Z machine via the Fundamental Science Program using pulsed power; see, e.g. Ref.~\cite{Knudson_Science_2015}. Furthermore, at the National Ignition Facility extensive implosion experiments are planned based on proposals that are submitted via the NIF Discovery Science Program (e.g.~Ref.~\cite{Celliers_Science_2018}), see also the newly developed colliding planar shock platform at the NIF~\cite{MacDonald_POP_2023}.

These experiments pose tremendous challenges for theory and simulations and require a long-term strategy in the field.
Experience shows that there is no ``silver bullet'' \footnote{quoting P.C. Martin, in: Ref.~\cite{pngf1}, p. 2}, i.e., no single method that would allow one to compute all quantities precisely and to solve all problems. Instead, a smart combination of methods is required. For example, for ICF modeling, large length and time scales and spatial inhomogeneities have to be resolved which is currently only possible with a hydrodynamic description of the dense plasma. At the same time, hydrodynamic models have an unknown accuracy, in particular, in the warm and hot dense matter regime. Substantial improvements are possible with input for the equation of state and transport and optical properties from more accurate approaches, such as average atom models or semiclassical MD with quantum potentials. These methods, in general, also have an unknown accuracy and can be benchmarked and improved with input from more accurate simulations such as Kohn-Sham DFT. However, KS-DFT requires the choice of an exchange-correlation functional the accuracy of which is generally unknown. The use of FPIMC results as benchmarks -- even if they are available only for a limited set of parameters -- allows one to significantly improve KS-DFT results for dense hydrogen, cf. Sec.~\ref{ss:summary}. Such a combination of methods promises to be both accurate and efficient, not only for ICF modeling.
\begin{figure}[th]
    \centering    \includegraphics[width=0.48\textwidth]{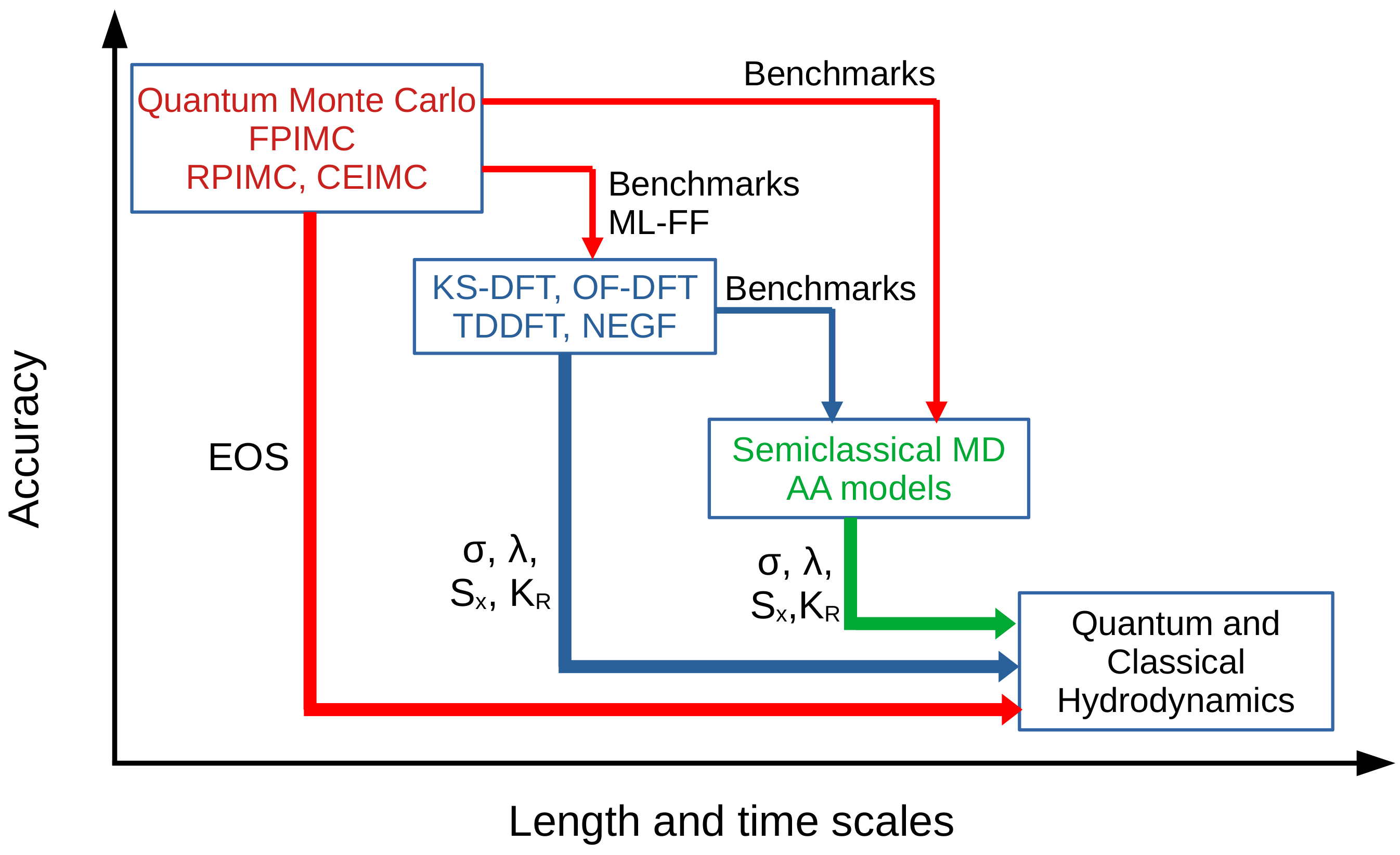}
    \caption{Schematic overview of important simulation methods for dense hydrogen, ordered by accuracy (and resolution) vs. maximum length and time scales that can be reached by the simulation. QMC methods can be used to benchmark other methods and to improve their accuracy. They also allow for the construction of accurate force fields using machine learning approaches (ML-FF). DFT can be used to benchmark more approximate methods, including AA models.
    Horizontal arrows: input for fluid simulations (e.g. for ICF modeling) which include EOS: equation of state; $\sigma$ ($\lambda$): electrical (thermal) conductivity; $S_x$: stopping power, Eq.~(\ref{eq:stopping-sx}); $K_R$: opacity, Eq.~(\ref{eq:opacity}).
    More details of the listed methods and quantities can be found in Sec.~\ref{sec:theory} and Tabs.~\ref{tab:methods}--\ref{tab:observables}. 
    }
    \label{fig:methods-outlook}
\end{figure}

 The approach we are putting forward is sketched in Fig.~\ref{fig:methods-outlook}. It involves the development of first-principles fermionic PIMC simulations which provide the highest accuracy results for a limited range of parameters, which can be used to test and improve (indicated by the vertical red arrows) more approximate methods, including RPIMC, Green functions, and KS-DFT. The latter, in turn, are capable of simulations for a significantly larger range of parameters and materials. Similarly, having accurate KS-DFT results available may substantially improve (vertical blue arrow) other less accurate simulations, including average atom models or classical and quantum hydrodynamics.

For such a hybrid approach to be successful, it is necessary to continuously develop and improve all key simulation methods, including FPIMC, RPIMC, CEIMC, Green functions, DFT, average atom models, and hydrodynamics which is motivated by their complementary strengths and limitations as was illustrated in Tab.~\ref{tab:observables}. There are many important developments going on in these fields, among which we highlight
\begin{itemize}
    \item Further improvement of fermionic PIMC (and potentially PIMD~\cite{Hirshberg_PNAS_2019,Hirshberg_JCP_2020,Feldman_JCP_2023,xiong2024gpu}) simulations for dense hydrogen and extension to stronger degeneracy. Application of the $\xi$-extrapolation method~\cite{Xiong_JCP_2022,Dornheim_JCP_xi_2023,Xiong_PRE_2023,Dornheim_JPCL_2024,dornheim_2024ab,dornheim2024unraveling,Dornheim_LFC_2024,xiong2024gpu} to hydrogen over broader range of parameters, and more observables, in particular static properties related to the EOS. 

    \item Improving RPIMC simulations of hydrogen (and beyond) by using variationally optimized nodal surfaces~\cite{militzer_pre_00}. We note that a somewhat related optimization of the thermal density matrix has recently been pursued by Xie \textit{et al.}~\cite{Xie_PRL_2023}, and was successfully benchmarked against FPIMC results for electrons in a 2D harmonic trap~\cite{JML-1-38}.

    \item Strongly degenerate fully ionized hydrogen plasmas can be simulated with high accuracy using configuration PIMC (CPIMC), as was demonstrated for jellium \cite{schoof_cpp15,schoof_prl15,groth_prb16}. Developing CPIMC for hydrogen appears to be promising.
    \item  Using highly accurate QMC simulations on the microscale to train machine-learning representations is likely to constitute a viable route towards larger systems.%

    \item Further development of finite temperature XC functionals for DFT~\cite{ksdt,groth_prl17,karasiev_importance_2016,kushal,moldabekov2023bound,kozlowski2023generalized,Karasiev_PRL_2018,Karasiev_PRB_2022,MIHAYLOV2024108931} on higher rungs of Jacob's ladder of functional approximations~\cite{Perdew_AIP_2001}, for example based on FPIMC results for the XC-kernel either of the UEG~\cite{hamann_prb_20,Tolias_JCP_Matsubara_2024,dornheim2024dynamic} or warm dense hydrogen~\cite{Dornheim_LFC_2024}, see e.g.~Ref.~\cite{pribram}.
    This might be complemented by the further exploration of high-$T$ DFT methods such as extended KS-DFT~\cite{Blanchet_cpc_2022,Zhang_POP_2016} or the spectral quadrature approach~\cite{Bethkenhagen_high_T_2021}.

    \item Alternative DFT schemes including e.g., \textit{ab initio} molecular simulations with numeric atom-centered orbitals (FHI-aims)~\cite{blum_CPC_2009}, and stochastic DFT which scales linearly with $N$ and is, therefore, suited for performing simulations with large particle numbers in order to answer ``large system questions''~\cite{Cytter_Rabani_prb2018,Baer_prl_2013,fabian_WIRE_2019}. Recently, a mixed stochastic-deterministic approach has been proposed~\cite{White_prl_2020} which combines the best aspects of both schemes.  
    \item Nonlinear effects can be analyzed efficiently either based on the LFC/the XC-kernel~\cite{Dornheim2021}, the ITCF~\cite{Dornheim_JCP_ITCF_2021}, or by direct simulation of the full density response~\cite{Dornheim2020}. Relevant observables include the stopping power~\cite{Echenique_PRB_1986,Nagy_PRA_1989} and effective potentials and forces~\cite{Dornheim_JCP_2022}.

    \item The combination of Green function methods (e.g.~Dyson and Bethe-Salpeter equation) with Kohn-Sham orbitals from a DFT simulation as input~\cite{vanSetten2013} should allow for high-quality spectral information (spectral function, DOS, cf.~Tab.~\ref{tab:observables}) also for partially ionized hydrogen.
\end{itemize}

\subsubsection{First-principles input for ICF modeling}\label{sec:outlook-icf-input}
Let us return to the discussion of ICF modeling of Sec.~\ref{sss:qhd} where we pointed out the importance of accurate data for the equation of state, transport, and optical properties. As we demonstrated in this paper, very accurate data are available from first-principle simulations that are suitable as input into hydrodynamic simulations. A general scheme illustrating this ``hand over'' of results is sketched in Fig.~\ref{fig:methods-outlook}.

    For improved ICF modeling, moreover,  it is important to understand 
    the validity of the hydrodynamic or diffusion approximation used in ICF simulations. 
    Codes such as those relying on kinetic theory (e.g. Vlasov-Fokker-Planck) can  provide useful information as to the accuracy of the hydrodynamics description during ICF burn \cite{huang2017, taitano2021}. This approach is valuable to better understand the micro-physics of diffusive and mixing processes at material interfaces \cite{stanton2018} which in turn contribute to Rayleigh-Taylor and Richtmyer-Meshkov instabilities. However, in the initial compression phase where quantum degeneracy effects of the electrons are important, classical kinetic theory should be replaced by more accurate kinetic theory, as was discussed in this paper, cf. Sec.~\ref{sss:qke}.

An example of uncertain input in the radiation matter equations that has gotten some attention \cite{dimonte2008, glosli2008} is the Coulomb logarithm, Eq.~(\ref{eq:coulomb-log}), which is used to determine the rate at which electrons and ions couple in a burning plasma. Several different model forms have been proposed to improve the Coulomb logarithm  \cite{gericke2002} and account for the different plasma length scales (e.g. Landau length, thermal de Broglie wavelength, Debye length, see Sec.~\ref{sec:wd-hydrogen}). 
At the same time, in the quantum degenerate regime, the concept of the Coulomb logarithm breaks down and has to be replace by scattering rates that follow from improved QKE, cf. Sec.~\ref{sss:qke}. Moreover, time-dependent simulations of the initial relaxation phase that include the nonlinear stopping power appear to be possible by using combinations of TDDFT, semiclassical MD, and quantum kinetic theory.

\subsubsection{Towards high precision diagnostics of warm and hot dense hydrogen}\label{sec:outlook-xrts}
The upcoming experimental developments that were mentioned in Sec.~\ref{sec:outlook-exp-models} crucially depend on accurate diagnostics and interpretation of the measurements.
Let us start with the XRTS technique, which is arguably one of the most valuable methods of diagnostics for WDM but, in the case of hydrogen, is severely hampered due to its comparably small scattering cross section. In this regard, we note exciting new capabilities to probe optically heated hydrogen jet targets~\cite{Zastrau,Fletcher_Frontiers_2022} with high-repetition rate XFEL X-ray sources. First, the high repetition rate (pump-probe experiments using the DiPOLE~\cite{Mason} and ReLaX~\cite{Laso_relax} laser can be performed with a repetition rate of $\sim10\,$Hz at the European XFEL) allows one to average over $\sim10^5$ individual shots to accumulate a sufficiently strong scattering signal. This is further facilitated by the high brightness of XFELs, see, e.g.~Ref.~\cite{Fletcher2015}.
Second, it is now routinely possible to produce X-ray sources with a bandwidth of $\Delta E\sim\SI{0.5}{\eV}$ by monochromating seeded XFELs. 
In combination with ultrahigh resolution Si (533) Diced Crystal Analysers at XFEL facilities~\cite{mcbride_2018,Wollenweber_RSI_2021}, this allows for extremely precise/high resolution measurements of the inelastic scattering. 
In fact, E.E.~McBride \emph{et al.}~\cite{mcbride_2018} have presented a specialized set-up with a resolution of $\Delta E\sim \SI{0.05}{\eV}$, which has allowed A.~Descamps \emph{et al.}~\cite{Descamps2020} to resolve the dynamic ion feature of single-crystal diamond. More recently, T.~Gawne \emph{et al.}~\cite{gawne2024ultrahigh} demonstrated a set-up for measuring the XRTS spectrum with a resolution of $\Delta E\sim\SI{0.1}{\eV}$ over a spectral window of tens of eV.
In combination, these developments will allow for unprecedented XRTS measurements of hydrogen at solid state density over a broad range of temperatures, $k_BT\sim (1-10^2)\,$eV. 

In particular, a narrow and well-characterized source-and-instrument function is key to deconvolve the XRTS intensity in the Laplace domain~\cite{Dornheim_T_follow_up}, cf.~Eq.~(\ref{eq:convolution2}) in Sec.~\ref{sec:ITCF}. This will allow for model-free temperature diagnostics~\cite{Dornheim_T_2022}, as well as for a host of other applications, such as extracting the normalization~\cite{dornheim2023xray,dornheim2024unraveling} and, in this way, the electron static structure factor $S_{ee}(\mathbf{q})$. Moreover, probing the hydrogen jet at multiple scattering angles allows for simultaneous measurements, in the collective and non-collective regimes. This is crucial to check the consistency of theoretical models and simulations, and to detect possible non-equilibrium and inhomogeneity effects~\cite{Vorberger_PLA_2024}.
Finally, the high resolution might allow one to detect the theoretically predicted \emph{roton-type} feature in warm dense hydrogen~\cite{hamann_prr_23} in a dedicated experiment (see Sec.~\ref{sec:roton_H}).
For completeness, we note that hydrogen jets are not limited to XRTS experiments, but can be combined with a multitude of other diagnostics. For example, L.~Yang~\emph{et al.}~\cite{Yang2023} have recently proposed to use optical shadowgraphy measurements to benchmark hydrodynamics and particle-in-cell (PIC) simulations~\cite{Obst2017}.

At the same time, XRTS measurements on capsule implosion experiments (e.g.~at the NIF or the OMEGA laser facility) remain challenging and, in fact, unlikely.
An alternative has recently been suggested by J.~L\"utgert \emph{et al.}~\cite{Lutgert_POP_2022}, who proposed a new platform to measure the opacity of hydrogen at strong compression and high temperatures ($T\sim \SI{200}{\eV}$), resembling the conditions encountered in red dwarfs.
In addition to their utility for diagnostics, dependable results for the opacity will help to further advance our understanding of the optical properties of dense hydrogen (cf.~Tab.~\ref{tab:observables}).
Moreover, it will allow for the benchmarking of theoretical methodologies such as the Kubo-Greenwood formula (Sec.~\ref{sss:dft-kubo-greenwood}) and corresponding improvements that take into account electronic collisions~\cite{reinholz_PRE_2015}, LR-TDDFT with different static and, potentially, dynamic XC-kernels (Sec.~\ref{sec:TDDFT}), RT-TDDFT simulations (Sec.~\ref{sss:tddft}), and potentially NEGF (e.g.~Sec.~\ref{sss:g1-g2}).

\subsubsection{Towards future benchmarks of \emph{ab initio} simulations and models}\label{sec:outlook-future-benchmarks}
The present work has clearly shown the importance of a critical comparison of different methods and demonstrated practical approaches.
We presented extensive comparisons of static properties, such as the pressure, pair correlation functions, and related observables using, as a reference, fermionic QMC results. 

It is highly desirable to extend the comparisons to dynamic observables, in order to assess the accuracy and expected utility of different methods for the description and interpretation of different types of experiments.
Naturally, this is less straightforward than for statical properties, since QMC methods generally do not give one direct access to dynamic properties, cf.~Tab.~\ref{tab:observables} in Sec.~\ref{sss:simulations-hierarchy}.
Similarly, DFT-based results, e.g.~for the dynamic structure factor $S_{ee}(\mathbf{q},\omega)$, are usually based on one of the following three approximations:  i) Linear-response TDDFT,  with an adiabatic approximation to the XC-kernel~\cite{Moldabekov_JCTC_2023} (Sec.~\ref{sec:TDDFT}); even though empirical non-adiabatic kernels do exist, their accuracy is generally unclear, in particular at WDM conditions. ii) Combination of a dynamic collision frequency, computed in the optical (i.e., $q\to0$) limit, with the Born-Mermin approach~\cite{Witte_PRL_2017}. iii) Real-time TDDFT with an adiabatic approximation for the dynamic XC-potential~\cite{dynamic2}, cf.~Sec.~\ref{sss:tddft}.

In lieu of unassailable frequency-resolved benchmark data, we are confident that the ITCF, $F_{ee}(\mathbf{q},\tau)$, will play an important role in such a project. 
From a theoretical perspective, it contains the same information as the dynamic structure factor $S_{ee}(\mathbf{q},\omega)$, due to the uniqueness of the two-sided Laplace transform, Eq.~(\ref{eq:Laplace}). In practice, it is straightforward to compute the ITCF either from FPIMC simulations~\cite{Dornheim_JCP_ITCF_2021}, or from any other approach that gives one access to $S_{ee}(\mathbf{q},\omega)$. Moreover, being a wavenumber-resolved property, it is known to exhibit a negligible dependence on the system size~\cite{Dornheim_PRE_2020}, which was recently substantiated for the case of warm dense hydrogen~\cite{Dornheim_LFC_2024}.
We thus propose to benchmark DFT simulations, average atom models, and other theoretical approaches, such as the usual Chihara decomposition [Sec.~\ref{sssub:chihara}], against highly accurate FPIMC results for $F_{ee}(\mathbf{q},\tau)$, both for collective and non-collective scattering regimes,  over a broad range of densities for $\Theta\gtrsim1$.

In addition, one might consider other properties such as the static density response, static structure factor, and the positive frequency moments of $S_{ee}(\mathbf{q},\omega)$ that can be extracted from the derivatives of $F_{ee}(\mathbf{q},\tau)$ with respect to $\tau$ around $\tau=0$~\cite{Dornheim_PRB_2023}. Finally, we mention the enticing possibility to develop a sufficiently constrained analytic continuation (possibly by extending the stochastic sampling of the dynamic local field correction introduced in Ref.~\cite{Dornheim2018b}) to invert Eq.~(\ref{eq:Laplace}), and, in this way, obtain FPIMC results for $S_{ee}(\mathbf{q},\omega)$ and a number of other dynamic observables~\cite{hamann_prb_20,hamann_cpp_20}.

These efforts might be complemented by considering known analytical limits (cf.~Sec.~\ref{sss.collisions-dft}), and to derive new relations that extend previous results for the more simple UEG model, e.g., Refs.~\cite{quantum_theory,Dornheim2018b}.
Finally, we note that, while unassailable benchmarks for real WDM systems in non-equilibrium are currently lacking, one can benchmark time-dependent methods such as NEGF or RT-TDDFT against each other for well controlled test cases, such as lattice models, e.g.~\cite{schluenzen_cpp16,schluenzen_prb17,schluenzen_jpcm_19}. This might be a viable way to derive and test improved non-adiabatic XC correlation potentials and to obtain insights into dynamic effects beyond the linear response regime.

To conclude this article, we are confident that the field of dense hydrogen is entering a new era -- the era of high-precision simulations. We have discussed the arsenal of simulation methods that are available to compute relevant observables, based on first principles, and achieve reliable results that have predictive power. We outlined possible strategies to combine the strengths of different approaches to develop high precision simulations for experimentally relevant situations.

\section*{Data availability}

The data that support the findings of this study are available from the corresponding author upon reasonable request.

\section*{Acknowledgments} 

This work was partially supported by the Center for Advanced Systems Understanding (CASUS) which is financed by Germany’s Federal Ministry of Education and Research (BMBF) and by the Saxon state government out of the State budget approved by the Saxon State Parliament. MB~acknowledges funding by the Deutsche Forschungsgemeinschaft via projects BO1366-13/2 and BO1366-16.
TD acknowledges funding from the European Research Council (ERC) under the European Union’s Horizon 2022 research and innovation programme
(Grant agreement No. 101076233, ``PREXTREME''). Views and opinions expressed are however those of the authors only and do not necessarily reflect those of the European Union or the European Research Council Executive Agency. Neither the European Union nor the granting authority can be held responsible for them.
TD~acknowledges funding from the European Union's Just Transition Fund (JTF) within the project \textit{R\"ontgenlaser-Optimierung der Laserfusion} (ROLF), contract number 5086999001, co-financed by the Saxon state government out of the State budget approved by the Saxon State Parliament.
DC was supported by DOE DE-SC0020177. CP was supported by the European Union - NextGenerationEU under the Italian Ministry of University and Research (MUR) projects PRIN2022-2022NRBLPT CUP E53D23001790006 and PRIN2022-P2022MC742PNRR, CUP E53D23018440001. 
PS acknowledges funding from OXPEG and AWE UK.
RR acknowledges funding by the Deutsche Forschungsgemeinschaft via the Research Unit FOR~2440.
SBH was supported by Sandia National Laboratories, a multimission laboratory managed and operated by National Technology and Engineering Solutions of Sandia, LLC, a wholly owned subsidiary of Honeywell International Inc., for the U.S. Department of Energy’s National Nuclear Security Administration under contract DE-NA0003525. This paper describes objective technical results and analysis. Any subjective views or opinions that might be expressed in the paper do not necessarily represent the views of the U.S. Department of Energy or the United States Government.
Some of the authors would like to thank the Institut Henri Poincar\'e (UAR 839 CNRS-Sorbonne Universit\'e) and the LabEx CARMIN (ANR-10-LABX-59-01) for their support. Portions of this work were performed under the auspices of the U.S. Department of Energy by Lawrence Livermore National
Laboratory under contract DE-AC52-07NA27344 Lawrence Livermore National Security, LLC. SXH and VVK acknowledge the support by the Department of Energy [National Nuclear Security Administration] University of Rochester “National Inertial Confinement Program” under Award Number DE-NA0004144 and U.S. National Science Foundation PHY Grant No. 2205521.

The PIMC and DFT calculations were partly carried out at the Norddeutscher Verbund f\"ur Hoch- und H\"ochstleistungsrechnen (HLRN) under grants shp00026, mvp00018, and mvp00024, on a Bull Cluster at the Center for Information Services and High Performance Computing (ZIH) at Technische Universit\"at Dresden, and on the HoreKa supercomputer funded by the
Ministry of Science, Research and the Arts Baden-Württemberg and by
the Federal Ministry of Education and Research.

\newpage

\section{Appendix: List of acronyms}
\begin{table}[h!]
    \centering
    \begin{tabular}{|l|l|l|}
    \hline
    Acronym & Explanation & Section \\
    \hline
    AA & average atom model & \ref{ss:aa-models}\\
    AIMD & Ab initio MD (= BOMD)& \ref{subsec:dft}\\
    AIRSS & Ab initio random structure search& \ref{sssec:hydrogen-lowt}\\
    ALDA & Adiabatic LDA & \ref{subsec:dft}\\
    BOA & Born-Oppenheimer Approximation & \ref{subsec:dft}\\
    BOMD & Born-Oppenheimer MD & \ref{subsec:dft}\\
    CEIMC & Coupled electron-ion MC & \ref{subse:ceimc}\\
    CPIMC & Configuration PIMC & \ref{sec:cpimc} \\
    DAC & Diamond anvil cell  & \ref{sec:introduction}\\
    DFT & Density functional theory & \ref{subsec:dft}\\
    DFT-AA & DFT-based average atom model & \ref{ss:aa-models}\\
    DFT-MD & DFT with MD for ions (=BOMD) & \ref{subsec:dft}\\
    DMQMC & Density matrix QMC & \ref{sec:cpimc}\\
        DSF & Dynamic structure factor & \ref{sec:dynamics}\\
        EOS & Equation of state & \ref{sec:td}\\
        FCIQMC & Full-CI QMC & \ref{subse:ceimc}\\
        FSP & Fermion sign problem & \ref{sec:FSP}\\
        FP-PIMC & Fermionic propagator PIMC & \cite{filinov_pre_23}\\
        FVT & Fluid variational Theory & \ref{sssec:saha}\\
        GDSMFB & Finite-T LDA by S.~Groth \textit{et al.}& \cite{groth_prl17}\\
        GGA & Generalized gradient approximation & \ref{subsec:dft}\\
        ICF & Inertial confinement fusion & \ref{sec:introduction}\\
        IPD & Ionization potential depression & \ref{sss:ipd-qmc}\\
        IMT & Insulator to metal transition & \ref{sec:wd-hydrogen} \\
        ITCF & Imaginary time correlation function & \ref{sec:ITCF}\\
        KG & Kubo-Greenwood & \ref{sec:transport}\\
        KS-DFT & Kohn-Sham DFT & \ref{subsec:dft} \\
        KSDT & Finite-T LDA by V.~Karasiev \textit{et al.}& \cite{ksdt}\\
        LDA & Local density approximation & \ref{subsec:dft}\\
        LFC & Local field correction & \ref{sec:dynamics}\\
        LLPT & Liquid-liquid phase transition & \ref{sssec:hydrogen-lowt}\\
        LR-TDDFT & Linear response TDDFT & \ref{sss:tddft}\\
        MC & Monte Carlo & \ref{sec:PIMC}\\
        MD & Molecular dynamics & \ref{subsec:dft}\\
        ML-QMC & Machine learning QMC & \ref{ssubsec:ml-md}\\
        NIF & National Ignition Facility & \ref{sec:introduction}\\
        OCP & One-component plasma model & \ref{sec:wd-hydrogen}\\
        PBE & Perdew-Burke-Ernzerhof & \ref{subsec:dft} \\
        OF-DFT & Orbital-free DFT & \ref{subsec:dft}\\
        PB-PIMC & Permutation blocking PIMC & \ref{ssub:fsp-reduction}\\
        PDF & Pair distribution function & \ref{sssec:pdf}\\
        PIMC & Path integral MC & \ref{sec:PIMC} \\
        PIMD & Path integral MD & \ref{subsec:md}\\
        PPT & Plasma phase transition & \ref{sssec:hydrogen-hight}\\
        RPA & Random phase approximation & \ref{sec:dynamics}\\
        QMC & Quantum Monte Carlo &
        \ref{sec:PIMC}\\
        RPIMC & Restricted PIMC & \ref{sec:alternatives}\\
        RT-TDDFT & Real-time TDDFT & \ref{sss:tddft}\\
        TB-MD & tight-binding molecular dynamics & \ref{sssec:saha}\\
        TDDFT & Time-dependent DFT & \ref{sss:tddft}\\
        UEG & Uniform electron gas model (jellium) & \ref{subsec:jellium}\\
        VMC & Variational MC & \ref{subse:ceimc}\\
        XC & Exchange-correlation & \ref{ss:basics}\\
        XFEL & X-ray free electron laser & \ref{sssec:phase-diagram-opn}\\
        XRTS & X-ray Thomson scattering & \ref{sec:introduction}\\
        \hline
    \end{tabular}
    \caption{Acronyms used in this article and place of first/detailed reference}
    \label{tab:acronyms}
\end{table}

%%%%%%%%%%%%%%%%%%%%%%%%%%%%%%%%%%%%%%%%%%%%%%%%%%%%%%%%%%%%%%%%%%%%%%%%%%%%%%%%
% literature
%%%%%%%%%%%%%%%%%%%%%%%%%%%%%%%%%%%%%%%%%%%%%%%%%%%%%%%%%%%%%%%%%%%%%%%%%%%%%%%%
\newpage
%\bibliography{bibliography, mb-ref, CDM/cdm-refs}
\bibliography{bibexport}

\begin{thebibliography}{811}
\bibitem [{BO_()}]{BO_note}%
  \BibitemOpen
  \href@noop {} {}\bibinfo {note} {{We note that, while DFT is formally exact given the correct exchange--correlation functional, DFT-MD simulations are based on the Born-Oppenheimer approximation and, strictly speaking, not a first-principles approach, according to our definition.}}\BibitemShut {Stop}%
\end{thebibliography}
\end{document}